\documentclass[aps,prl,reprint,superscriptaddress,nofootinbib,floatfix]{revtex4-2}
\usepackage{amsmath,amssymb,bm,graphicx}
\usepackage{CJKutf8}
\usepackage[colorlinks=true,linkcolor=blue,citecolor=blue,urlcolor=blue]{hyperref}
\usepackage{orcidlink}
\hypersetup{pdftitle={Entanglement between distant spins in a polarized magnet forms a d-wave clover},
  pdfauthor={Alan Zhang, Ye Zhou}}
\newcommand{\eps}{\varepsilon}
\newcommand{\cL}{\mathcal L}
\newcommand{\cI}{\mathcal I}
\newcommand{\br}{\mathbf r}
\newcommand{\be}{\mathbf e}
\newcommand{\nn}{\mathrm{nn}}
\DeclareMathOperator{\Neg}{Neg}
\DeclareMathOperator{\sgn}{sgn}
\newcommand{\runin}[1]{\textit{#1}---}
\newcommand{\SM}{SM}
\begin{document}

\title{Entanglement between distant spins in a polarized magnet forms a \textit{d}-wave clover}
\author{Alan Zhang\,\orcidlink{0009-0000-1187-0755}}
\thanks{These authors contributed equally to this work}
\email{Contact author: zhana142@McMaster.ca}
\affiliation{Department of Physics and Astronomy, McMaster University, Hamilton, Ontario L8S 4M1, Canada}
\affiliation{Perimeter Institute for Theoretical Physics, Waterloo, Ontario N2L 2Y5, Canada}
\author{Ye Zhou (\begin{CJK*}{UTF8}{gbsn}周烨\end{CJK*})\,\orcidlink{0009-0004-7050-7736}}
\thanks{These authors contributed equally to this work}
\email{Contact author: ye.zhou.horizon@gmail.com}
\affiliation{Independent Researcher, Kunshan, Jiangsu, China}
\date{\today}

\begin{abstract}
Spin-orbit coupling makes a polarized magnet a source of virtual spin-flip pairs. These pairs
entangle distant spins: the symmetry of the coupling fixes the angular pattern, and the flips they
leave behind end it at a distance of order the inverse spin-orbit angle. For a spin-$1/2$
square-lattice ferromagnet we prove, at leading order, a $d$-wave inverse-square law, $1.376$ times
the spin-wave amplitude, a clover-shaped entangled region, exact without the Dzyaloshinskii-Moriya
term, and that each spin's entanglement is entirely pairwise. Four qubits can test the $d$-wave signs
at present error rates.
\end{abstract}

\maketitle

\runin{Introduction}Entanglement between two spins of a many-body ground state is usually confined
to near neighbors. In the quantum Ising chain it vanishes beyond next-nearest neighbors even at
criticality~\cite{Osterloh2002}, and in the two-dimensional model beyond nearest
neighbors~\cite{Wang2025}. The reason is geometric: a product of mixed local states lies in the
interior of the set of separable states~\cite{Parez2026}, and in gapped ground states the distance
beyond which two spins are separable is bounded in terms of the clustering length and the purity of
the one-spin state~\cite{Ferydouni2026}. Entanglement at long distance therefore requires nearly
pure local states and a long correlation length. Both occur in a magnet that a field holds close to
full polarization while its gap stays small.

That regime has been explored mainly in one dimension and within spin-wave theory. In spin chains
the range of pairwise entanglement diverges at factorizing and saturation
fields~\cite{Amico2006,Baroni2007}, whereas in two-dimensional antiferromagnets near their
factorizing field it stays very short~\cite{Roscilde2005}. In linear spin-wave theory, the
concurrence of anisotropic ferromagnets results from the anomalous pair correlation competing with
the magnon population and vanishes beyond a distance proportional to the exchange
length~\cite{Zou2020}; for spin spirals the range grows as the ferromagnet is
approached~\cite{Rozsa2025}.

\begin{figure*}[t]
 \centering
 \IfFileExists{figures/fig1_geometry.pdf}%
 {\includegraphics[width=\textwidth]{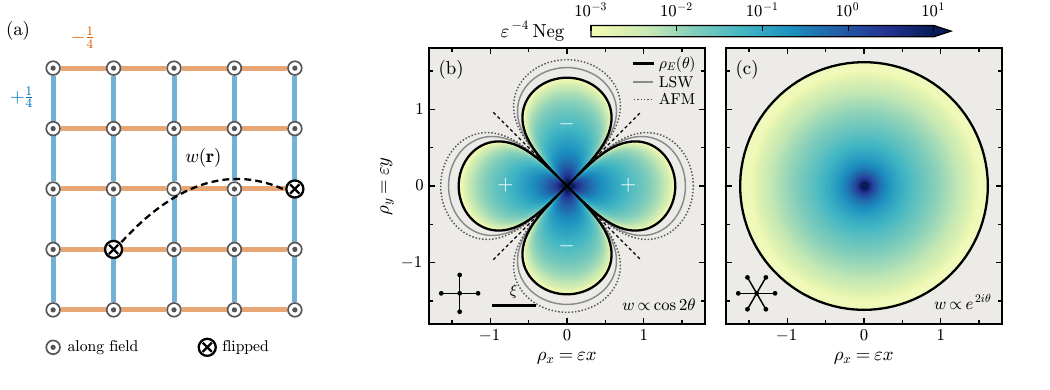}}%
 {\fbox{\parbox[c][2.6in][c]{0.9\textwidth}{\centering figure pending}}}
 \caption{Mechanism and geometry of two-spin entanglement near polarization. (a)~Spins along the
 field (dots) on the square lattice; the anisotropy creates pairs of neighboring flips (crosses) with
 amplitude $Jdf$, $f=-1/4$ on $x$ bonds (orange) and $+1/4$ on $y$ bonds (blue), linked by the pair
 amplitude $w(\br)$. (b)~Leading negativity $N_h$ of Eq.~\eqref{eq:crossover} at $h=2$ versus
 $\bm\rho=\eps\br$ (gray: $N_h=0$). Thick line: boundary $\rho_E(\theta)$; thin gray line: without
 the magnon interaction (spin-wave theory); dotted line: antiferromagnet above saturation; dashed
 lines: nodes of $w$, whose sign in each lobe is the $\eta$ of Eq.~\eqref{eq:witness}; bar: one
 correlation length. (c)~The same on the triangular lattice (two-magnon theory), where the sixfold
 joint symmetry selects a node-free $d+id$ amplitude and the boundary is a circle.}
 \label{fig:geometry}
\end{figure*}

We answer, for an interacting two-dimensional magnet, the questions an experiment would ask: in
which directions two spins are entangled, how strongly, how far, and how one can see it.
Spin-orbit coupling makes the exchange depend on the bond direction. In chiral magnets it produces
the Dzyaloshinskii-Moriya (DM) interaction that stabilizes helices and
skyrmions~\cite{Nagaosa2013,Fert2017}, together with a weaker symmetric
anisotropy~\cite{Shekhtman1992}; in Kitaev materials such symmetric bond-dependent exchange
dominates~\cite{Jackeli2009,Winter2017}. With the DM term alone the field-polarized phase is an exact
product state, so quantum treatments of chiral magnets find it nearly
unentangled~\cite{Haller2022,Haller2024}. The symmetric anisotropy, however, creates virtual pairs of
spin flips that entangle the spins they connect [Fig.~\ref{fig:geometry}(a)]. The symmetry of this
pair source fixes the angular pattern of the entanglement, the magnon correlation length its decay,
and the depletion caused by the pairs its boundary.

For the spin-$1/2$ square lattice these statements are theorems for the interacting model at leading
order in the spin-orbit angle $\eps$. The negativity follows a $d$-wave inverse-square law and fills
a four-leaf clover that ends a few correlation lengths away [Fig.~\ref{fig:geometry}(b)]; without the
DM term two spins off its nodal lines are entangled if and only if they lie inside it. Although the
flips are dilute, their interaction within a pair raises the amplitude $1.376$-fold over spin-wave
theory and pulls in the boundary. The pairs carry all the entanglement of each spin with the rest of
the magnet. Cylinder DMRG agrees with the two-magnon theory on the same cylinders. The two-magnon
theory gives a circle on the triangular lattice [Fig.~\ref{fig:geometry}(c)] and covers XXZ exchange
and antiferromagnets above saturation. Three measurement settings detect the entanglement in a
quantum simulator, and four qubits show its $d$-wave signs at present error rates.

\runin{Model and pair source}We consider spins $\mathbf S_i=\bm\sigma_i/2$ on the square lattice with
\begin{equation}
 \frac{H}{J}=\sum_{i}\sum_{\mu=x,y}\Bigl[\frac14-
 \mathbf S_i^{T}e^{\eps[\be_\mu]_\times}\mathbf S_{i+\be_\mu}\Bigr]
 -h\eps^2\sum_i S_i^z ,
 \label{eq:model}
\end{equation}
where $J>0$, $h$ is the field in units of $J\eps^2$, and $[\mathbf a]_\times\mathbf v=\mathbf
a\times\mathbf v$. Each bond carries isotropic ferromagnetic exchange in a frame rotated by $\eps$
about the bond direction: Heisenberg exchange $J\cos\eps$, a DM coupling $J\sin\eps$ with the DM
vector along the bond, and the symmetric anisotropy $-J(1-\cos\eps)S_i^\mu S_{i+\be_\mu}^\mu$, the
combination produced by superexchange with weak spin-orbit coupling~\cite{Shekhtman1992} and
observed in Ba$_2$CuGe$_2$O$_7$~\cite{Zheludev1998}. On the lattice the rotations about $x$ and $y$
do not commute, the
corresponding SO(3) gauge field has curvature~\cite{Hill2021}, and the anisotropy cannot be removed.
The model is invariant under translations and under a quarter turn
of the lattice combined with a quarter turn of the spins about the field axis (joint $C_4$ symmetry).
Lengths are in lattice units, and $k_B=1$.

The DM term changes the number of flipped spins by one. On the polarized state $|\Omega\rangle$ the
flips it creates from the two $x$ (or $y$) bonds of a site cancel, and its vertex between one and two
flips vanishes at zero total momentum. It therefore enters
neither the pair source nor the two-magnon problem below, and the proofs show that its virtual
processes, although its coupling $J\eps$ exceeds the gap, leave the leading order unchanged (\SM,
Secs.~S3--S7).

All pair entanglement stems from the symmetric anisotropy, a ferromagnetic compass
interaction~\cite{Nussinov2015} that creates two spin flips on neighboring sites with amplitudes
$Jdf$, where $d=1-\cos\eps$, $f(\pm\be_x)=-1/4$, and $f(\pm\be_y)=1/4$. The source is of
$d_{x^2-y^2}$ type, and the pair amplitude it induces, $w(\br)=\langle S_0^+S_{\br}^+\rangle$, is a
bond-nematic correlation of the kind that orders spontaneously in frustrated ferromagnets below
saturation~\cite{Shannon2006}. A pair with zero total momentum is governed by the two-magnon problem
of the Heisenberg ferromagnet~\cite{Wortis1963}. The separation $\br\neq0$ of the two flips hops on
the square lattice with the origin and its four bonds removed, because two flips cannot share a site,
and two neighboring flips gain one unit of exchange energy; together these make the pair Hamiltonian
the Laplacian $\cL$ of this punctured lattice.

The one-magnon gap is $Jg\eps^2$ at leading order, with $g=h-1/2$, so pairs have the correlation
length $1/(m\eps)$ with $m=\sqrt{2g}$. The proofs assume $h>3/2$, where exchange and Zeeman energy
dominate the DM term in every state; the bound is technical.

\runin{Inverse-square law}Let $F=-\cL^{-1}f$, let $\Neg(\rho)=(\|\rho^{T_2}\|_1-1)/2$ be the
negativity~\cite{Vidal2002} of the state $\rho_{0,\br}$ of two spins at separation $\br$, $r=|\br|$,
and let $\theta$ be the angle of $\br$ from the $x$ axis. For translation-invariant, jointly
$C_4$-symmetric ground states and $1\le r\ll1/\eps$ (Theorem~1 of the End Matter; proofs in the
Supplemental Material~(\SM)~\cite{SM}\nocite{Hoeffding1963,Corless1996,Horodecki1996,Delmotte1999,Rohart2013,Toth2007,Guhne2009}),
$\Neg(\rho_{0,\br})=\eps^2|F(\br)|/2+o(\eps^2/r^2)$, with
\begin{align}
 \frac{\Neg(\rho_{0,\be_x})}{\eps^2}&\longrightarrow\frac{4-\pi}{16(\pi-2)}=0.0470,\nonumber\\
 F(\br)&=\frac{\cos2\theta}{8(\pi-2)r^2}+O(r^{-4}).
 \label{eq:kernel}
\end{align}

Linear spin-wave theory replaces $\cL$ by the Laplacian $\cL_0$ of the full lattice. Because $\cL$ and
$\cL_0$ differ only at the four neighbors of the origin, where a $d$-wave amplitude has one
independent value, the two kernels are proportional, $F=-\cL_0^{-1}f/(1-k_\square)$ with
$k_\square=4/\pi-1$: the magnon attraction enhances the pair amplitude and the negativity by
$\pi/[2(\pi-2)]=1.376$ at every separation (\SM, Sec.~S4). If the Ising part of the exchange is $\Delta$
times its transverse part, at fixed gap, the same two-magnon theory gives the factor
$1/(1-\Delta k_\square)$: $1$ for XY exchange and $\pi/4$ for a bipartite antiferromagnet just above
saturation, which a sublattice rotation maps onto $\Delta=-1$ (\SM, Sec.~S15).

\runin{Where the entanglement ends}The inverse-square law cannot continue indefinitely, because the
pairs also deplete the polarization. Let $n_i=1/2-S_i^z$, $p=\langle n_i\rangle$, and
$q_{\br}=\langle n_0n_{\br}\rangle$. The partial transpose of $\rho_{0,\br}$ contains a $2\times2$
block with diagonal entries $p-q_{\br}$ and off-diagonal entries $w$ and $w^*$. Whatever the other
matrix elements, therefore,
\begin{equation}
 \Neg(\rho_{0,\br})\ge\bigl(|w(\br)|-p+q_{\br}\bigr)_+ ,
 \label{eq:block}
\end{equation}
with $(x)_+=\max(x,0)$. Two spins are entangled when their pair coherence exceeds the flip
population, the competition identified in spin-wave theory~\cite{Zou2020}, here with the
interactions included. Each spin is flipped whenever it belongs to a virtual pair, hence
$p\simeq\sum_{\br}|w(\br)|^2$, and we prove (Theorem~2)
\begin{equation}
 \frac{p}{\eps^4}\longrightarrow n_*=\frac14\sum_{\br\ne0}F(\br)^2
 =\frac{\pi}{256(\pi-2)}=0.01075 .
 \label{eq:population}
\end{equation}
Because $p\simeq\sum_{\br}|w(\br)|^2$, the squared concurrences of a spin with all others reach their
upper bound, the one-tangle $4p(1-p)$~\cite{Coffman2000,Osborne2006}, as $\eps\to0$ (End Matter).
The entanglement of each spin with the rest of the magnet is thus entirely pairwise, and a fraction
$4(4-\pi)^2/[\pi(\pi-2)]=0.82$ of it is shared with its four nearest neighbors.

\begin{figure}[t]
 \centering
 \IfFileExists{figures/fig2_crossover.pdf}%
 {\includegraphics[width=\columnwidth]{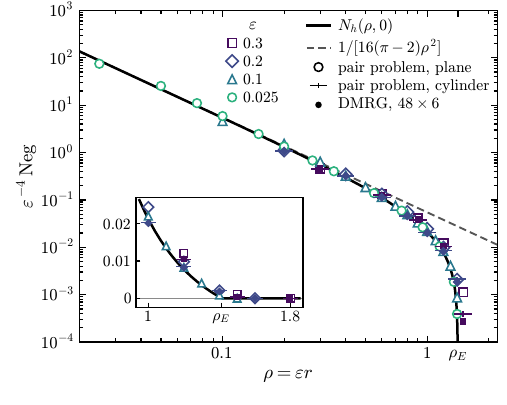}}%
 {\fbox{\parbox[c][2.7in][c]{0.9\columnwidth}{\centering figure pending}}}
 \caption{Crossover of the axial negativity at $h=2$. Solid line: leading negativity
 $N_h(\rho,0)$ of Eq.~\eqref{eq:crossover}, with $\rho=\eps r$. Dashed line: massless
 inverse-square law $1/[16(\pi-2)\rho^2]$. Open symbols: two-magnon pair problem on the plane,
 $(|w_2|-\sum_{\br}|w_2|^2)_+/\eps^4$, at the indicated $\eps$. Filled symbols: DMRG on a
 cylinder with $L_x=48$ and $L_y=6$. Flat crosses: pair problem on a cylinder of the same
 circumference. Inset: the region near $\rho_E$ on linear scales.}
 \label{fig:crossover}
\end{figure}

At $r\sim1/\eps$ the coherence has dropped to order $\eps^4$ and the magnon mass matters. The
response $w_2$ of the two-magnon problem with the gap is known in closed form (Theorem~3), and for
$\eps\br\to\rho(\cos\theta,\sin\theta)$
\begin{equation}
 \frac{w_2(\br)}{\eps^4}\longrightarrow
 A_h(\rho,\theta)=\frac{m^2\cos2\theta}{32(\pi-2)}K_2(m\rho),
 \label{eq:scaling}
\end{equation}
with $K_2$ the modified Bessel function; $A_h$ is the $d$-wave derivative
$(\partial_x^2-\partial_y^2)$ of the massive propagator $K_0(m\rho)$, matches
Eq.~\eqref{eq:kernel} for $\rho\to0$, and decays as $e^{-m\rho}$. We prove that the ground state of
the full model follows this form and saturates the bound~\eqref{eq:block} at this order (Theorem~6,
Fig.~\ref{fig:crossover}):
\begin{equation}
 \sup_{\eps r\ge c}\,\biggl|\frac{\Neg(\rho_{0,\br})}{\eps^{4}}-N_h(\eps\br)\biggr|
 \longrightarrow0,\qquad N_h=\bigl(|A_h|-n_*\bigr)_+ ,
 \label{eq:crossover}
\end{equation}
for every $c>0$. The leading negativity is positive inside the line $r_E(\theta)=\rho_E(\theta)/\eps$
and vanishes outside it, where
\begin{equation}
 K_2\bigl(m\rho_E(\theta)\bigr)=\frac{\pi}{8m^2|\cos2\theta|}.
 \label{eq:boundary}
\end{equation}
The entangled region is a four-leaf clover [Fig.~\ref{fig:geometry}(b)] that pinches at the
diagonals; its area, $4.53/\eps^2$ at $h=2$, is the number of spins each spin is entangled with. On
the axes $\rho_E=1.4128$ for $h=2$ and $1.2597$ for $h=3$, that is, $2.45$ and $2.82$ correlation
lengths. The interaction multiplies the coherence by $1.376$ and the depletion by its square, which
pulls the boundary in from $\rho_E=1.545$ in spin-wave theory to $1.413$ at $h=2$. The magnon repulsion of an antiferromagnet above
saturation pushes it out to $1.648$. As the gap closes, $\rho_E\to4/\sqrt\pi$, where depletion
alone ends the entanglement; with the DM term, textured states intervene first~\cite{Hill2021}.

Pairs with negativity above a threshold $\kappa\eps^4$ fill the region $|A_h|\ge n_*+\kappa$, which
tends to the clover as $\kappa\to0$ (Theorem~7). Outside the clover the negativity is at most
of order $\eps^{16/3}$, and the part of the state even under $\sigma^z_0\sigma^z_{\br}$ is
separable, so no measurement of such correlators can reveal
entanglement there (Theorem~S13). Whether the state itself is separable is decided at order $\eps^8$
by its odd part, which the DM term creates. Without the DM term the odd part vanishes and the clover
is exact: off the diagonals, two spins at a fixed scaled separation are entangled if and only if they
lie inside it (Theorem~9). With the DM term this is proved on the axes (Theorem~8).

Thermal magnons add to the depletion~\cite{Hofmann2012}; for $1\ll r\ll1/\eps$ a pair is predicted
to stay entangled up to $T_E\simeq Jg\eps^2/(2\ln r)$, the magnon gap divided by $2\ln r$ (\SM,
Sec.~S8), and Theorem~5 proves sufficient conditions of this form.

\begin{figure}[t]
 \centering
 \IfFileExists{figures/fig3_range.pdf}%
 {\includegraphics[width=\columnwidth]{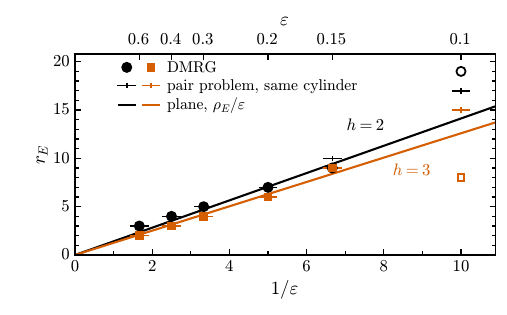}}%
 {\fbox{\parbox[c][2.3in][c]{0.9\columnwidth}{\centering figure pending}}}
 \caption{Largest axial separation $r_E$ with a negative partial transpose versus $1/\eps$ for $h=2$
 and $h=3$. Circles and squares: DMRG on cylinders of circumference $L_y=6$. Flat crosses: pair
 problem on the same cylinder without ends. Lines: boundary $\rho_E/\eps$ of the leading negativity on
 the plane. Open symbols: the flip population at the pair sites differs from its mid-cylinder value
 by more than $5\%$, an effect of the open ends.}
 \label{fig:range}
\end{figure}

Beyond perturbation theory, we computed ground states of Eq.~\eqref{eq:model} by
DMRG~\cite{White1992,Hauschild2018,Hauschild2024} on open cylinders of length $24$ or $48$ and
circumference $L_y\le12$, with bond dimensions up to $128$, and obtained the negativity from the
measured two-spin state (\SM, Sec.~S14). On a cylinder of fixed
circumference the pair problem has its own kernel, so the parameter-free comparison is with the
two-magnon problem on the same cylinder. For $L_y=6$ the largest axial separation with a negative
partial transpose, $r_E$, coincides with that prediction at nine of the ten points with $h=2$ or $3$
and $0.15\le\eps\le0.6$; the exception is affected by the open ends (Fig.~\ref{fig:range}). At these
angles $r_E$ and $\rho_E/\eps$ differ by less
than one lattice spacing, too little to tell the plane from the cylinder. Converged
profiles at $\eps=0.2$ and $0.3$ deviate from the pair problem on the same cylinder by $0.4\%$ to
$1.9\%$ (Fig.~\ref{fig:crossover}). Off the axes the circumference decides: at $\br=(3,2)$,
$\eps=0.25$, and $h=2$, inside the clover of the plane, a cylinder of circumference $6$ has a positive
partial transpose, as its own pair problem predicts, whereas wider cylinders are entangled and
approach the plane value (\SM, Sec.~S14).

\runin{Symmetry selects the geometry}A pair of spin flips carries two units of spin angular momentum
about the field axis. If the model and the state are invariant under a rotation of the lattice by
$\phi=2\pi/n$ combined with the same rotation of the spins, then $w(R_\phi\br)=e^{2i\phi}w(\br)$.
Since also $w(-\br)=w(\br)$, the angular harmonics $e^{i\ell\theta}$ of $w$ obey
\begin{equation}
 \ell\equiv2\pmod n,\qquad \ell\ \text{even}.
 \label{eq:selection}
\end{equation}
For $n\ge3$ the isotropic harmonic $\ell=0$, which would give a logarithmic instead of an
inverse-square response in two dimensions, is excluded by symmetry. For $n=4$ both $\ell=2$ and
$\ell=-2$ are allowed; the real source of Eq.~\eqref{eq:model} combines them into $\cos2\theta$,
with nodes on the diagonals. For $n=3$ or $6$ the lowest harmonic is $\ell=2$ alone: the amplitude
winds as $e^{2i\theta}$ and its modulus has no nodes. The rule uses only the joint symmetry and
therefore applies to any magnet near polarization that has it, such as honeycomb Kitaev materials with
the field along the threefold axis, whose pair source $J_{\pm\pm}(\gamma_{ij}S_i^+S_j^++\mathrm{h.c.})$
has $\gamma_{ij}=1$, $e^{\pm2\pi i/3}$ on the three bonds~\cite{Maksimov2020}.

For the triangular and honeycomb analogs of Eq.~\eqref{eq:model}, with each bond rotated about its
own direction, the two-magnon problem is again solvable and gives
$w\simeq\eps^2C_\Lambda e^{2i\theta}/r^2$, the $d+id$ form factor known from magnon pairing near
saturation on the triangular lattice~\cite{Momoi2012}, with $C_\triangle=0.02036$ there; the predicted
boundary is a circle, with $\rho_E=1.6154$ at $h=2$ [Fig.~\ref{fig:geometry}(c); \SM, Sec.~S13].
These two-magnon results lie outside the theorems, which concern the square lattice.

\runin{Proposed experiment}Every separable two-qubit state obeys
\begin{equation}
 \begin{split}
 W_{ij}={}&\eta\,\langle\sigma_i^x\sigma_j^x-\sigma_i^y\sigma_j^y\rangle\\
 &-\bigl(2-\langle\sigma_i^z\rangle-\langle\sigma_j^z\rangle\bigr)\le0
 \end{split}
 \label{eq:witness}
\end{equation}
for $\eta=\pm1$ (\SM, Sec.~S2). In the ground state $W=4(\eta\,\mathrm{Re}\,w-p)$, so at leading order
$W$ with $\eta=\sgn A_h$ is positive exactly inside the clover, and outside it no witness built from
these correlators can detect entanglement. Three settings therefore suffice~\cite{Guhne2002}: all
spins measured in $x$, in $y$, or in $z$, with the axes locked to the phase of the pair source. The
coherence term $4\,\mathrm{Re}\,w$ alone changes sign between bonds along $x$ and $y$ and vanishes on
the diagonals, a direct signature of the $d$-wave source. A collective $d$-wave witness built from
the same snapshots tolerates $1.83$ times larger error rates. In a magnet the pairs squeeze transverse
spin fluctuations at wave vectors on the $k_x$ and $k_y$ axes, which quantum Fisher information
extracts from neutron spectra~\cite{Scheie2021,Simeth2026} (End Matter).

\begin{figure}[t]
 \centering
 \IfFileExists{figures/fig4_experiment.pdf}%
 {\includegraphics[width=\columnwidth]{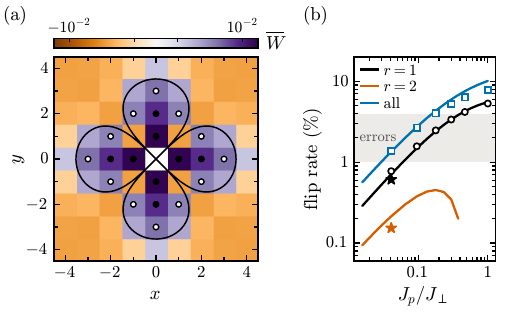}}%
 {\fbox{\parbox[c][2.1in][c]{0.9\columnwidth}{\centering figure pending}}}
 \caption{Proposed measurement on open arrays of Eq.~\eqref{eq:model} without the DM term.
 (a)~Witness $\overline W$ of Eq.~\eqref{eq:witness} on a $10\times10$ array at $\eps=0.4$ and $h=2$
 (pair theory), averaged over pairs with separation $(x,y)$ or its quarter turn away from the
 outermost ring (purple: $\overline W>0$; orange: $\overline W<0$). Filled (open) circles: $3\sigma$
 detection with at most $10^4$ ($10^6$) snapshots. Line: leading-order boundary $\rho_E/\eps$ (the
 pair at $r=4$ is still entangled). (b)~Tolerated rate of incoherent spin flips versus pair drive, at
 detuning $B=6J_p$, for the witness at $r=1$ and $2$ and the collective witness (all): lines, pair theory
 ($10\times10$); symbols, exact diagonalization ($4\times4$); star, panel (a). Band: reported
 preparation and readout errors~\cite{Karamlou2024,Andersen2025}.}
 \label{fig:experiment}
\end{figure}

The target is the model without the DM term, for which Theorem~9 holds. In qubit language it is an
XXZ model with hopping $J_\perp$ and Ising coupling $J_z$, plus pair creation $\mp J_p$ on $x$ and
$y$ bonds and a detuning $B=hJ\eps^2$ (End Matter); in a simulator these are independent. On a
$10\times10$ array at $\eps=0.4$ and $h=2$ the axial negativity falls from $6.1\times10^{-3}$ at $r=1$
to $1.9\times10^{-5}$ at $r=4$ and vanishes on the diagonals [Fig.~\ref{fig:experiment}(a)]; a
$3\sigma$ detection at $r=1$ needs about $600$ snapshots. Incoherent spin flips from preparation,
readout in $z$, or heating add to $p$ one for one, so at this drive their rate must stay below $0.6\%$
at $r=1$, less than the $1\%$--$4\%$ errors of present arrays~\cite{Karamlou2024,Andersen2025}.
Stronger drives trade range for margin [Fig.~\ref{fig:experiment}(b)]. At $J_p=J_\perp$, where $J_z$
vanishes and the model is the quantum compass model in a transverse field, exact diagonalization of
an open $4\times4$ array and the pair theory of a $10\times10$ array give nearest-neighbor
negativities of $6\%$ at $B=1.5J$, with the same sign change between $x$ and $y$ bonds and separable
diagonal pairs. A $2\times2$ plaquette thus tests the $d$-wave signs at present error rates.
Superconducting circuits with parametric pair drives~\cite{Poletto2012,Roth2017} and gate-based
processors~\cite{Haghshenas2026} can realize the model (End Matter).

\runin{Conclusion}Near polarization, spin-orbit coupling turns a magnet into a source of entangled
pairs, which carry all the entanglement of each spin. The symmetry of the source fixes the angular
pattern of two-spin entanglement, two-magnon scattering its amplitude, and depletion its boundary,
at a distance of order the inverse spin-orbit angle. On the square lattice these statements are
theorems at leading order, with an exact boundary when the DM term is absent. Measurements in three
bases on a programmable magnet can test this link between the symmetry of a pair source and the
geometry of entanglement, on four qubits now and on larger arrays as errors fall.

\begin{acknowledgments}
\runin{Acknowledgments}A.Z. acknowledges graduate scholarship and research support through
McMaster University. Research at Perimeter Institute is supported in part by the Government of
Canada through the Department of Innovation, Science and Economic Development and by the
Province of Ontario through the Ministry of Colleges, Universities, Research Excellence and
Security.
\end{acknowledgments}

\runin{Data availability}The code and data that support the findings of this Letter (solver
and DMRG scripts, measured two-spin density matrices, and figure data) are provided as
ancillary files of the Supplemental Material~\cite{SM}.

\runin{AI Use Disclosure}Claude Opus~5.5 and Claude Fable~5.2 (Anthropic), GPT-6-Astra,
GPT-6.2-Astra, and ChatGPT Pro (OpenAI), and Gemini~4 Neon and Carbon (Google) assisted with
derivations and checks, numerical code, literature searches, and writing. The authors directed this
work, verified all derivations, numbers, and conclusions, and take full responsibility for the
content.

\bibliography{refs}

\section*{End Matter}
\appendix
\setcounter{equation}{0}
\renewcommand{\theequation}{A\arabic{equation}}
\makeatletter\@ifundefined{theHequation}{}{\renewcommand{\theHequation}{A\arabic{equation}}}\makeatother

\runin{Appendix A: Precise statements}We set $J=1$ and write $b_i=S_i^+$,
$n_i=b_i^\dagger b_i$, $d=1-\cos\eps$, and $\ell_\eps=1+\ln(2+1/\eps)$. A \emph{symmetric
thermodynamic ground state} is a state $\omega$ of the infinite square lattice that is
invariant under translations and under the joint quarter turn and minimizes the energy per
site of Eq.~\eqref{eq:model} among translation-invariant states; the minimum is
$e_0(\eps,h)$. Such states exist and satisfy $\omega([H,A])=0$ for every local observable
$A$. We write
$p=\omega(n_0)$, $w(\br)=\omega(b_0b_{\br})$, $q_{\br}=\omega(n_0n_{\br})$,
$v_{\br}=\omega(b_0^\dagger b_{\br})$, and $e_{\rm pol}=(d-h\eps^2)/2$. $\cI$ is a compact
subset of $(3/2,\infty)$. All limits are $\eps\downarrow0$, taken after the thermodynamic
limit, and are uniform in $h\in\cI$ and in the states considered. Constants depend only on
$\cI$ and are not optimized. Section numbers refer to the \SM~\cite{SM}.

\textit{Theorem 1 (inverse-square law; Theorem~S1, Secs.~S3, S4, S6, and S7).} For every
$\delta\in(0,1)$,
\begin{equation}
 \sup_{1\le|\br|\le\eps^{-\delta}}|\br|^2
 \left|\frac{w(\br)}{\eps^2}-\frac{F(\br)}{2}\right|\longrightarrow0 ,
 \label{eq:A-w}
\end{equation}
and, for translation-invariant, jointly $C_4$-symmetric ground states,
\begin{equation}
 \sup_{1\le|\br|\le\eps^{-\delta}}|\br|^2
 \left|\frac{\Neg(\rho_{0,\br})}{\eps^2}-\frac{|F(\br)|}{2}\right|\longrightarrow0 .
 \label{eq:uniform}
\end{equation}
Both statements remain true on
$1\le|\br|\le\eps^{-1}\ell_\eps^{-a}$ for every $a>3/2$. They follow from the bound
\begin{equation}
 \begin{split}
 |w(\br)-w_2(\br)|\le C\bigl[&\eps^4\ell_\eps^3+\eps^5(1+|\br|)\ell_\eps^4\\
 &+\eps^6(1+|\br|)^2\ell_\eps^6\bigr],
 \end{split}
 \label{eq:A-bound}
\end{equation}
valid for $1\le|\br|\le\eps^{-1}$, with $w_2$ the pair response of Theorem~3, and from
$|\Neg(\rho_{0,\br})-|w(\br)||\le7p$. Moreover $F(\be_x)=(4-\pi)/[8(\pi-2)]$ and $F$ has the
asymptotic form~\eqref{eq:kernel}.

\textit{Theorem 2 (energy and depletion; Theorems~S2 and S3, Sec.~S5).} With
$F_1=F(\be_x)$, $c_*=F_1/8=(4-\pi)/[64(\pi-2)]$, and $n_*=\pi/[256(\pi-2)]$,
\begin{align}
 e_0-e_{\rm pol}&=-c_*\eps^4+g_6(h)\eps^6+o(\eps^6),\nonumber\\
 \frac{p}{\eps^4}&\longrightarrow n_* ,
 \label{eq:A-energy}
\end{align}
where $g_6(h)=n_*(h-\tfrac12)-F_1/96+F_1^2/8-\Gamma_3/4$ and $\Gamma_3$ is a finite
three-magnon constant independent of $h$, so that $g_6'(h)=n_*$.

\textit{Corollary (monogamy).} For translation-invariant, jointly $C_4$-symmetric ground states let
$C$ be the concurrence, $\tau_2=\sum_{\br\ne0}C(\rho_{0,\br})^2$, and $\tau_1=4\det\rho_0=4p(1-p)$ the
one-tangle. Then $\tau_2/\tau_1\to1$ (the ratio of Ref.~\cite{Roscilde2004}), and
$C(\rho_{0,\br})/\eps^2\to|F(\br)|$ for every $\br$. The bound $\tau_2\le\tau_1$ holds for every state
of finitely many qubits~\cite{Osborne2006} and hence on the lattice; $C\ge2\Neg$~\cite{Verstraete2001}
and Theorem~1 give $\liminf\eps^{-4}\sum_{|\br|\le R}C^2\ge\sum_{|\br|\le R}F^2$ for every $R$; and
$\sum_{\br\ne0}F^2=4n_*$.

\textit{Theorem 3 (massive pair operator; Theorem~S5, Sec.~S9).} Let $h$ range over a compact
subset of $(1/2,\infty)$, let $\cL_0$ be the Laplacian of the full lattice, and for $\sigma\ge0$ let
$G_\sigma=-(\cL_0+\sigma)^{-1}f$ and $t_\sigma=4G_\sigma(\be_x)$. The two-magnon block of $H/J$ at
zero total momentum is
\begin{equation}
 A_2=a_\eps\cL+2\Delta_\eps+\frac d2P_\nn,\qquad a_\eps=1-\frac d2,\quad \Delta_\eps=h\eps^2-d,
 \label{eq:pairblock}
\end{equation}
with $P_\nn$ the projector on the four neighbors of the origin. For small $\eps$, $A_2$, acting on
even functions on
$\mathbb Z^2\setminus\{0\}$, satisfies $A_2\ge2\Delta_\eps>0$, and the unique square-summable
solution of $A_2w_2=-df$ is
\begin{equation}
 w_2(\br)=\frac{d\,G_\sigma(\br)}{a_\eps-t_\sigma\cos\eps},\qquad
 \sigma=\frac{2\Delta_\eps}{a_\eps}.
 \label{eq:A-w2}
\end{equation}
In particular $F=G_0/(1-t_0)$, where $t_0=k_\square=4/\pi-1$ is a value of the lattice Green
function of the square lattice~\cite{Cserti2000}. Furthermore
$\sup_{\br}|w_2(\br)-dF(\br)|\le C\eps^4\ln(2/\eps)$,
$\eps^{-4}\sum_{\br}|w_2(\br)|^2\to n_*$, and for every compact
$X\subset\mathbb R^2\setminus\{0\}$
\begin{equation}
 \sup_{\eps\br\in X}\bigl|\eps^{-4}w_2(\br)-A_h(\eps\br)\bigr|\longrightarrow0 ,
 \label{eq:A-scaling}
\end{equation}
where $A_h(\bm\rho)$ is the function of Eq.~\eqref{eq:scaling} in polar coordinates.

\textit{Theorem 4 (two-qubit criterion; Theorem~S6, Secs.~S2 and S9).} In a
translation-invariant, jointly $C_4$-symmetric state the two-spin density matrix has, in the basis
$\uparrow\uparrow$, $\uparrow\downarrow$, $\downarrow\uparrow$, $\downarrow\downarrow$, the diagonal
$(a,b,b,q)$ with $a=1-2p+q$ and $b=p-q$, the coherences $w$ between $\uparrow\uparrow$ and
$\downarrow\downarrow$ and $v$ (real) between the one-flip states, and odd coherences $\pm\zeta$,
$\zeta=\omega(b_0n_{\br})^*$ (\SM, Eq.~(S274)); it is not an X state. Let $\rho_\eps$ be any family of
two-qubit states of this form such that, for some $\rho>0$ and $\theta$ with $\cos2\theta\neq0$,
\begin{equation}
 \begin{gathered}
 \frac{p}{\eps^4}\to n_*,\qquad \frac{w}{\eps^4}\to A_h(\rho,\theta),\\
 q=o(\eps^4),\qquad v=o(\eps^4),\qquad \zeta=o(\eps^6).
 \end{gathered}
 \label{eq:A-hyp}
\end{equation}
Then $\eps^{-4}\Neg(\rho_\eps)\to(|A_h(\rho,\theta)|-n_*)_+$, and if
$|A_h(\rho,\theta)|\neq n_*$ the states $\rho_\eps$ are entangled for all small $\eps$ if and
only if $|A_h(\rho,\theta)|>n_*$. Separability in the exterior follows from an exact Schur
complement of the partial transpose and from the positivity of $\rho_\eps$, which gives
$aq\ge|w|^2$; the decisive eigenvalue is of order $\eps^8$, so an approximation of the state to
$o(\eps^4)$ would not suffice.

\textit{Theorem 5 (thermal windows; Theorem~S4, Sec.~S8).} Let $\omega_T$ be a
translation-invariant,
jointly $C_4$-symmetric limit of Gibbs states of periodic lattices at temperature $T$, with
$h\in\cI$. For $\gamma>0$ and $\delta<1$ there are $c,c',c''>0$ and $r_\gamma$ such that, for
small $\eps$, $\Neg(\rho_{0,\br})\ge c'\eps^2/|\br|^2$ whenever
$r_\gamma\le|\br|\le\eps^{-\delta}$, $|\cos2\theta|\ge\gamma$, and
$T\le c''\eps^2/\ln(1/\eps)$; in this temperature range Eqs.~\eqref{eq:A-w} and
\eqref{eq:uniform} hold as well. With the free-energy bound of Ref.~\cite{Klippel2026} the
temperature condition for the lower bound on the negativity can be replaced by
$T\le c\,\eps^2/\ln(2+|\br|)$.

\textit{Theorem 6 (negativity at the scale $1/\eps$; Theorem~S8, Sec.~S10).}
For every $c>0$,
\begin{equation}
 \sup_{|\br|\ge c/\eps}\frac{|w(\br)-w_2(\br)|+|q_{\br}|+|v_{\br}|}{\eps^4}\longrightarrow0 ,
 \label{eq:A-mass}
\end{equation}
and Eq.~\eqref{eq:crossover} holds; the deviations of $w/\eps^4$ from $A_h$ and of
$\Neg/\eps^4$ from $N_h$ are at most $C_c(\eps\ell_\eps)^{2/13}$. If moreover
$n_*-|A_h(\eps\br)|\ge\delta>0$, then $|\zeta|\le C_{c,\delta}\eps^{14/3}\ell_\eps$ and
$\Neg(\rho_{0,\br})\le C_{c,\delta}\eps^{16/3}\ell_\eps^2$ (Proposition~S9, Sec.~S11).

\textit{Theorem 7 (range; Theorem~S10, Sec.~S11).} Let $\mathbf a\in\mathbb Z^2$ be
primitive with $\cos2\theta_{\mathbf a}\neq0$, and let $R(\kappa)$ be the largest
$|k\mathbf a|$, $k\in\mathbb N$, with $\Neg(\rho_{0,k\mathbf a})\ge\kappa\eps^4$. For fixed
$\kappa>0$, $\eps R(\kappa)\to\rho_\kappa$ with $|A_h(\rho_\kappa,\theta_{\mathbf a})|=n_*+\kappa$.
If $\kappa_\eps\to0$ and $\kappa_\eps/(\eps^{4/3}\ell_\eps^2)\to\infty$, then
$\eps R(\kappa_\eps)\to\rho_E(\theta_{\mathbf a})$, the largest such separation in any
direction is $[\rho_E(0)+o(1)]/\eps$, and the set
$\{\eps\br:\Neg(\rho_{0,\br})\ge\kappa_\eps\eps^4\}$ converges in Hausdorff distance to
$\{0\}\cup\overline{\{\bm\rho\neq0:|A_h(\bm\rho)|>n_*\}}$.

\textit{Theorem 8 (separability on the axes; Theorem~S16, Sec.~S12).} Suppose the state is also
invariant under complex conjugation in the $S^z$ basis combined with the reflection
$x\mapsto-x$; such states exist, by averaging over the finite symmetry group. Then for every $\nu>0$ and $\beta$ with $\beta\max_{h\in\cI}m<2/13$, on the axes
$\Neg(\rho_{0,\br})=(|w(\br)|-p+q_{\br})_+$ for $c\le\eps|\br|\le\rho_E(0)-\nu$, and the
pairs with $\rho_E(0)+\nu\le\eps|\br|\le\beta\ln(1/\eps)$ have a positive partial
transpose, with smallest eigenvalue at least $c_\beta\eps^{8+2\beta\max m}/\ln(1/\eps)$.

Without the reflection symmetry, the part of $\rho_{0,\br}$ even under
$\sigma^z_0\sigma^z_{\br}$ has a partial transpose with smallest eigenvalue at least
$c\,\eps^8$ on every compact set of scaled separations outside the clover and off the
diagonals (Theorem~S13); a single further correlation of a transverse spin with a flip decides
whether the full state is separable there (Proposition~S14).

\textit{Theorem 9 (model without the DM term; Theorem~S21, Sec.~S16).} Let $H_{\rm sym}$ be
Eq.~\eqref{eq:model} without the term proportional to $\sin\eps$, and let $\cI$ be a compact subset
of $(1/2,\infty)$. Theorems~1--3 and 6 hold for its symmetric thermodynamic ground states, with
$\Gamma_3$ replaced by $0$ in Theorem~2 and the rate $(\eps\ell_\eps^{3/2})^{1/2}$ in Theorem~6.
$H_{\rm sym}$ commutes with the rotation of all spins by $\pi$ about the field; ground states
invariant under it exist and have $\zeta=0$. For them and every compact set $K$ of scaled separations
on which $A_h\ne0$ and $|A_h|\ne n_*$, for small $\eps$ and $\eps\br\in K$, $\rho_{0,\br}$ is entangled
if and only if $|A_h(\eps\br)|>n_*$; otherwise its partial transpose has smallest eigenvalue at least
$c_K\eps^8$.

What is not proved for Eq.~\eqref{eq:model} is exact separability outside the clover off the axes,
on the diagonals, and at $\eps|\br|>\beta\ln(1/\eps)$. Theorem~4 would give it from
$\zeta=o(\eps^6)$, but Theorem~6 gives only $\zeta=O(\eps^{14/3}\ell_\eps)$, and the decisive
eigenvalue, of order $\eps^8$, depends on correlations of three and four spin flips that are not yet
controlled (\SM, Sec.~S12). For $H_{\rm sym}$ the diagonals and the far tail remain open. For the
largest separation $R_{\rm NPT}$ along a ray off the diagonals at which the negativity is nonzero,
only $\liminf\eps R_{\rm NPT}\ge\rho_E$ is known.

\runin{Appendix B: Measurement}In qubit language the bonds of Eq.~\eqref{eq:model} without the DM
term are $-J_\perp(\sigma_i^+\sigma_j^-+\mathrm{h.c.})-J_z\sigma_i^z\sigma_j^z\mp
J_p(\sigma_i^+\sigma_j^++\mathrm{h.c.})$ on $x$ and $y$ bonds, with $J_\perp=J(1+\cos\eps)/4$,
$J_z=J\cos\eps/4$, and $J_p=J(1-\cos\eps)/4$, and the field becomes the detuning $B=hJ\eps^2$. On
an open array the DM term would only tilt the edge spins. The predictions of Fig.~\ref{fig:experiment}
use the pair theory: the pair field $\phi$ solves $(H_{22}-E_\Omega)\phi=-H_{20}$, with $H_{22}$ the
Hamiltonian among states with two flips, $H_{20}$ its coupling to $|\Omega\rangle$, and
$E_\Omega=\langle\Omega|H|\Omega\rangle$, and the two-spin state is that of independent pairs. It
reproduces exact diagonalization of open $4\times4$ arrays to $0.6\%$ off the diagonals for
$\eps\le0.4$ and the DMRG range on $24\times6$ cylinders at six points, and at the strong drives of
Fig.~\ref{fig:experiment}(b) it is checked against exact diagonalization of open $2\times2$ to
$4\times4$ arrays. Switching the anisotropy on over a time $20/J$ from the polarized product state,
$i\partial_t\phi=(H_{22}-E_\Omega)\phi+H_{20}$, reproduces the ground-state witness at $r=1$ and $2$
to about $1\%$, as exact evolution of a $4\times4$ array confirms. In Fig.~\ref{fig:experiment}(a)
the axial negativities at $r=2$ and $3$ are $1.5\times10^{-3}$ and $3.7\times10^{-4}$. Averaging all
equivalent pairs away from the edges, with a tenth of the snapshots in $z$, a $3\sigma$ detection needs
about $600$ snapshots at $r=1$ and $10^4$ at $r=2$, where the tolerated flip rate is $0.15\%$.

Every separable state obeys $\sum_{i\ne j}M_{ij}\langle\sigma_i^x\sigma_j^x
-\sigma_i^y\sigma_j^y\rangle\le\|M\|\sum_i(1-\langle\sigma_i^z\rangle^2)$ for real symmetric $M$ with
zero diagonal. The $d$-wave kernel $M_{\mathbf k}=\sgn(\cos k_x-\cos k_y)$ collects the coherence of
all pairs, $\pi\eps^2/[32(\pi-2)]=8n_*\eps^2$ per spin, and tolerates flip rates
$\pi/[2(4-\pi)]=1.83$ times larger than the nearest-neighbor witness. In momentum space every mode on
the $k_x$ ($k_y$) axis has its $x$ ($y$) quadrature squeezed by $1-0.344\eps^2$, the quantity that
quantum Fisher information extracts from neutron spectra~\cite{Scheie2021,Simeth2026}.

In superconducting circuits a parametric drive at the sum of two qubit frequencies creates pairs with
the phase of the drive, so the $d$-wave source needs opposite phases on $x$ and $y$ couplers. Pairing
has been combined with hopping in a lattice of microwave modes~\cite{Busnaina2024} and on every
coupler of a qubit chain~\cite{Liang2024}. With native XY exchange the interaction factor is $1$; the
factor $1.376$ requires $J_z$. Gate-based processors can also Trotterize the model, with trapped ions
on a periodic torus without edges. Details are given in Sec.~S15 of the \SM, and the numerical
methods, convergence data, and the scripts for all figures in Sec.~S14.

\end{document}


\title{Supplemental Material for ``Entanglement between distant spins in a polarized magnet forms a \textit{d}-wave clover''}
\author{Alan Zhang\,\orcidlink{0009-0000-1187-0755}}
\thanks{These authors contributed equally to this work}
\email{Contact author: zhana142@McMaster.ca}
\affiliation{Department of Physics and Astronomy, McMaster University, Hamilton, Ontario L8S 4M1, Canada}
\affiliation{Perimeter Institute for Theoretical Physics, Waterloo, Ontario N2L 2Y5, Canada}
\author{Ye Zhou (\begin{CJK*}{UTF8}{gbsn}周烨\end{CJK*})\,\orcidlink{0009-0004-7050-7736}}
\thanks{These authors contributed equally to this work}
\email{Contact author: ye.zhou.horizon@gmail.com}
\affiliation{Independent Researcher, Kunshan, Jiangsu, China}
\maketitle
\section*{Guide to the proofs}
\label{sm:guide}

The diagram shows how the proofs depend on each other; the table lists the estimates on which they
rest. L$n$ denotes Theorem~$n$ of the Letter, and Theorems~\ref{thm:range}--\ref{thm:rdmcriterion}
are stated in Sec.~\ref{sm:conventions}. The top row collects exact statements about operators,
two-qubit matrices, and arbitrary translation-invariant states; the rows below concern ground states
and, in Sec.~\ref{sm:thermal}, Gibbs states. Arrows mark the main line of each argument, and each
box lists all sections its proofs use. The monogamy corollary of the Letter follows from
Theorems~\ref{thm:range} and \ref{thm:population}. Section~\ref{sm:symmetry} proves
Proposition~\ref{prop:selection} from the definitions alone; Secs.~\ref{sm:numerics}
and~\ref{sm:experiment} are not used in any proof.

\medskip
\noindent\begin{tikzpicture}[x=1pt,y=1pt,font=\footnotesize,
  box/.style={draw=black!55,rounded corners=1.5pt,align=flush center,inner sep=2.5pt},
  inp/.style={box,fill=black!6,text width=87.4pt,minimum height=44pt},
  res/.style={box,text width=137.4pt,minimum height=34pt},
  hdr/.style={font=\footnotesize\itshape,anchor=east},
  dep/.style={-{Stealth[length=4.5pt,width=3.6pt]},line width=0.6pt,black!70}]
 \node[hdr] at (510,0) {Exact inputs};
 \node[inp] (S2) at (46.5,-28) {\textbf{\ref{sm:rdm}} Two-site state\\
   $\bigl|\Neg-|w|\bigr|\le7p$;\\ three-setting witness};
 \node[inp] (S3) at (150.75,-28) {\textbf{\ref{sm:normalform}} Coercivity and\\ normal form,
   $h>3/2$\\ uses \ref{sm:green}};
 \node[inp] (S4) at (255,-28) {\textbf{\ref{sm:green}} Punctured\\ Green function\\
   $F$, $k_0=4/\pi-1$};
 \node[inp] (S6) at (359.25,-28) {\textbf{\ref{sm:current}} Creation currents,\\ few-flip
   resolvents\\ uses \ref{sm:green}};
 \node[inp] (S9) at (463.5,-28) {\textbf{\ref{sm:massive}} Pair operator,\\ two-qubit criterion\\
   Thms \ref{thm:massive}, \ref{thm:rdmcriterion}: \textbf{L3, L4}\\ uses \ref{sm:green}};
 \node[hdr] at (510,-59) {Ground states, separations up to $\eps^{-1}\ell_\eps^{-a}$};
 \node[res] (S5) at (71.5,-84) {\textbf{\ref{sm:population}} Energy and population\\
   Thms \ref{thm:population}, \ref{thm:energy}: \textbf{L2}\\ uses \ref{sm:normalform},
   \ref{sm:green}, \ref{sm:current}};
 \node[res] (S7) at (255,-84) {\textbf{\ref{sm:range}} Pair amplitude\\
   Thm \ref{thm:range}: \textbf{L1}\\ uses \ref{sm:rdm}--\ref{sm:current}, \ref{sm:massive}};
 \node[res] (S8) at (438.5,-84) {\textbf{\ref{sm:thermal}} Finite temperature\\
   Thm \ref{thm:thermal}: \textbf{L5}\\ uses \ref{sm:rdm}--\ref{sm:green}, \ref{sm:range}};
 \node[hdr] at (510,-111) {Separations of order $1/\eps$ and beyond};
 \node[res] (S10) at (71.5,-136) {\textbf{\ref{sm:massscale}} Negativity for $|\br|\ge c/\eps$\\
   Thm \ref{thm:massscale}: \textbf{L6}\\ uses \ref{sm:rdm}--\ref{sm:range}, \ref{sm:massive}};
 \node[res] (S11) at (255,-136) {\textbf{\ref{sm:threshold}} Range and entangled region\\
   Thm \ref{thm:threshold}: \textbf{L7}\\ uses \ref{sm:normalform}--\ref{sm:current},
   \ref{sm:massive}, \ref{sm:massscale}};
 \node[res] (S12) at (438.5,-136) {\textbf{\ref{sm:exterior}} Exact separability\\
   Thms \ref{thm:ex-parity}, \ref{thm:ex-axial}: \textbf{L8}\\ uses \ref{sm:rdm}, \ref{sm:green},
   \ref{sm:population}, \ref{sm:range}, \ref{sm:massive}--\ref{sm:threshold}};
 \node[box,text width=499pt,minimum height=14pt] (S16) at (255,-170) {\textbf{\ref{sm:nodm}}
   Model without the DM term: Thm \ref{thm:nodm}, \textbf{L9}. Reuses
   Secs.~\ref{sm:rdm}--\ref{sm:range} and \ref{sm:massive}--\ref{sm:threshold} with the parts
   that treat the DM term removed.};
 \draw[dep] (S5) -- (S7);
 \draw[dep] (S7) -- (S8);
 \draw[dep] (S10) -- (S11);
 \draw[dep] (S11) -- (S12);
 \draw[dep] (S5.south) -- (S10.north);
 \draw[dep] (S7.south west) -- (S10.north east);
\end{tikzpicture}

\medskip
{\centering\small
\renewcommand{\arraystretch}{1.07}
\begin{tabular}{@{}lll@{}}
\hline\hline
Key estimate & Statement & Where\\
\hline
Coercivity & $H_0\succeq(g_*-d/2)Q+m_*\eps^2\Nop$, where $h>3/2$ enters
 & Eq.~\eqref{eq:coercivity}\\
A priori densities & $p,\widetilde p\le C\eps^4\ell_\eps^4$,\quad
 $\widetilde{\mathfrak q}\le C\eps^6\ell_\eps^4$ & Eq.~\eqref{eq:apriori}\\
Punctured lattice & $\cL G_0=-(1-k_0)f$, $k_0=4/\pi-1$; factor $1/(1-k_0)=1.376$
 & Eqs.~\eqref{eq:massless-defect}, \eqref{eq:t0}\\
Pair response & $w_2=dG_\sigma/(a_\eps-ct_\sigma)$, exact for the two-magnon block
 & Eq.~\eqref{eq:rankone}\\
Creation currents & creation of two to four flips bounded through $\mathfrak q$ and $p$
 & Eq.~\eqref{eq:creation-current}\\
Two-site comparison & $|\Neg(\rho_{0\br})-|w(\br)||\le7p$ & Eq.~\eqref{eq:rdm-seven}\\
Pair amplitude & $|w-w_2|\le C[\eps^4\ell_\eps^3+\eps^5(1+|\br|)\ell_\eps^4
 +\eps^6(1+|\br|)^2\ell_\eps^6]$, $1\le|\br|\le\eps^{-1}$ & Eq.~\eqref{eq:range-pointwise}\\
Energy sandwich & dressed densities $\widehat p\le C_M\eps^4\ell_\eps^{-2M}$,\quad
 $\widehat{\mathfrak q}\le C_M\eps^6\ell_\eps^{-2M}$ & Lemma~\ref{lem:ms-sandwich}\\
Residual budget & pair-identity residuals $\le C\eps^4(\ell_\eps^{-7}+\eps\ell_\eps^{52})$
 for $|\br|\ge c/\eps$ & Lemma~\ref{lem:ms-budget}\\
Partial transpose & $|\Neg-(|w|-p+q)_+|\le8\sqrt2\,|\zeta|+2v^2$ & Lemma~\ref{lem:ms-pt}\\
Exterior bound & $\Neg\le2v^2/a+8|\zeta|^2/(b-|w|)$ & Proposition~\ref{prop:th-exterior}\\
Schur complement & for $b>|w|$ and $a>\chi$: $\rho^{T_2}\succeq0\Leftrightarrow
 aq-v^2\ge\chi(a+q-2v)$ & Eq.~\eqref{eq:exterior}\\
Field pressure & $0\le-f_b\le e^{-b/t}(-f_0)/\log2$ & Lemma~\ref{lem:field-pressure}\\
Parity without DM & $\zeta=0$ in parity-invariant ground states of $H_{\rm sym}$
 & Lemma~\ref{lem:nodm-parity}\\
\hline\hline
\end{tabular}\par}
\clearpage

\section{Model, notation, state classes, and results}
\label{sm:conventions}

We consider spin $1/2$ on the square lattice, with one positively oriented
bond in each coordinate direction per site. With $k_B=1$,
\begin{equation}
 \frac{H_{\eps,h}}J=\sum_{i,\mu=x,y}
 \left[\frac14-\mathbf S_i^T e^{\eps[\mathbf e_\mu]_\times}
 \mathbf S_{i+\mathbf e_\mu}\right]
 -h\eps^2\sum_iS_i^z,\qquad
 \mathbf S_i=\bm\sigma_i/2,\quad [\mathbf a]_\times\mathbf v=
 \mathbf a\times\mathbf v .
 \label{eq:model}
\end{equation}
Unless units are explicitly restored, Hamiltonians and energy densities
are divided by $J$, and $t=T/J$ is the dimensionless temperature.
Distances are in lattice spacings. We write
\begin{gather}
 b_i=S_i^+,\quad b_i^\dagger=S_i^-,\quad n_i=b_i^\dagger b_i=
 \tfrac12-S_i^z,\quad \Nop=\sum_i n_i,\qquad
 Q=\sum_{\langle ij\rangle}P^-_{ij},\quad
 \Ncontact=\sum_{\langle ij\rangle}n_i n_j, \label{eq:basicnotation}\\
 c=\cos\eps,\quad d=1-c,\quad s=\sin\eps,\quad
 a_\eps=1-d/2,\quad\Delta_\eps=h\eps^2-d,\quad
 e_{\rm pol}=d/2-h\eps^2/2,\quad
 \ell_\eps=1+\log(2+\eps^{-1}).\label{eq:angles}
\end{gather}
$P^-_{ij}$ is the singlet projector. The DM operator is denoted by
$\VDM$, and its particle-sector blocks by $\mathcal V_{jk}$; neither
symbol denotes a field interval. The fixed field set is
$\cI=[h_-,h_+]\Subset(3/2,\infty)$. An arbitrary compact set can be
replaced by its convex hull. The exact number-conserving one-magnon gap
is $\Delta_\eps=(h-1/2)\eps^2+O(\eps^4)$; $h>3/2$ is the stronger
coercivity hypothesis used below, not a claimed transition field.

\subsection{States and limits}
Let $\tau_x$ be lattice translation. If $\Phi_Z$ are the local terms of
Eq.~\eqref{eq:model}, the energy density of a translation-invariant
state $\omega$ is the expectation of one positively oriented bond of
each direction and the field term at the origin. Equivalently it is
the thermodynamic limit of $|\Lambda|^{-1}\omega(H_\Lambda/J)$.
We call a state a thermodynamic ground state here if it minimizes this
energy density over all translation-invariant states. We select the
minimizers invariant under the joint quarter-turn of the lattice and
spins. Averaging a minimizer over that finite symmetry preserves its
energy. Minimizing in this symmetric class gives the same minimum as
minimizing over all translation-invariant states.

This definition supplies precisely the variational property used for
the energy and population results. It also supplies stationarity:
Lemma~\ref{lem:stationarity-density} proves
$\omega([H/J,A])=0$ for every local $A$ by spatially separated local
unitary variations. No uniqueness or differentiability assumption is
made. Finite-temperature states are translation- and jointly-fourfold-
invariant local limits of periodic Gibbs states at fixed $\eps,h,t$.
They are stationary as well.

Every physical limit below takes volume to infinity at fixed angle and
temperature before $\eps\downarrow0$. Finite-range dressing kernels are
held fixed during the volume limit. A compact triplet coefficient
kernel $\Theta$ is held fixed during each small-angle limit; its
approximation accuracy is improved only afterwards. Infinite coefficient
currents are truncated before the volume limit at fixed positive
regulator, and the truncation is removed before that regulator tends
to zero. Exact parity and reflection statements for selected families
of states are distinguished from covariance of the entire set of
minimizers in Sec.~\ref{sm:green}.

The physical one-site population and two-site quantities are
\begin{equation}
 p=\omega(n_0),\quad w(\br)=\omega(b_0b_\br),\quad
 q_\br=\omega(n_0n_\br),\quad v_\br=\omega(b_0^\dagger b_\br).
 \label{eq:correlators}
\end{equation}
The exchange energy density is $\mathfrak q=\sum_\mu\omega(P^-_{0,e_\mu})$,
distinct from $q_\br$. A tilde denotes the state after a pair dressing.
An extensive expectation divided by $N$ below denotes the corresponding
translation density after taking its indicated thermodynamic limit.
All constants are independent of volume, displacement and the selected
state unless explicit arguments say otherwise. Constants indexed by
$\cI,\chi$ may depend on the endpoints of the field interval and on the
finitely many derivative bounds of a fixed smooth cutoff $\chi$.

\subsection{Results}
Let $\cL$ be the graph Laplacian on $\mathbb Z^2\setminus\{0\}$ with
the origin edges removed, so its degree at the four nearest neighbors
is three. Define the even square-summable source solution
\begin{equation}
 f(\pm e_x)=-\tfrac14,\quad f(\pm e_y)=\tfrac14,\quad
 F=-\cL^{-1}f,\qquad
 F_1=\frac{4-\pi}{8(\pi-2)},\quad
 S_F=\frac{\pi}{64(\pi-2)},\quad c_*=F_1/8,\quad n_*=S_F/4.
 \label{eq:constants}
\end{equation}
Section~\ref{sm:green} proves $F(e_x)=F_1$ and $\|F\|_2^2=S_F$.
The following numbered results specify the scopes used throughout.

\begin{theorem}[Uniform pair-entanglement law]\label{thm:range}
For every fixed $0<\delta<1$, uniformly for $h\in\cI$ and all the
symmetric thermodynamic ground states defined above,
\begin{align}
 \sup_{1\le|\br|\le\eps^{-\delta}}|\br|^2
 \left|\frac{w(\br)}{\eps^2}-\frac{F(\br)}2\right|&\longrightarrow0,
 \label{eq:main-amplitude}\\
 \sup_{1\le|\br|\le\eps^{-\delta}}|\br|^2
 \left|\frac{\Neg(\rho_{0\br})}{\eps^2}-\frac{|F(\br)|}2\right|
 &\longrightarrow0.\label{eq:main-negativity}
\end{align}
Both conclusions hold on the larger window
$1\le|\br|\le\eps^{-1}\ell_\eps^{-a}$ for every fixed $a>3/2$.
These are weighted absolute statements, including nodes; relative
statements require a nonzero leading profile. The same weighted
comparison holds with the interacting pair-triplet effective profile
defined in Sec.~\ref{sm:current}.
\end{theorem}

\begin{theorem}[Physical population]\label{thm:population}
For the same states and limit order,
\begin{equation}
 \sup_{h\in\cI,\omega}\left|p/\eps^4-n_*\right|\longrightarrow0,
 \qquad n_* =\frac{\pi}{256(\pi-2)}.
 \label{eq:main-population}
\end{equation}
\end{theorem}

\begin{theorem}[Sixth-order ground-state energy]\label{thm:energy}
The actual thermodynamic ground energy satisfies, uniformly on $\cI$,
\begin{equation}
 e_0(\eps,h)-e_{\rm pol}(\eps,h)
 =-c_*\eps^4+\eps^6g_6(h)+o_\cI(\eps^6),\qquad
 g_6(h)=(h-\tfrac12)n_*-F_1/96+F_1^2/8-\Gamma_3/4.
 \label{eq:main-energy}
\end{equation}
Here $\mathcal T=\mathcal V_{32}F$,
$\Gamma_3=\langle\mathcal T,Q_3^{-1}\mathcal T\rangle_3<\infty$,
and the triplet inner product is $1/6$ times counting measure on
distinct relative configurations. This field-independent inverse
quadratic form is defined by the zero-mass limit. In particular,
$g_6'(h)=n_*$.
\end{theorem}

\begin{theorem}[Thermal sufficient NPT windows]\label{thm:thermal}
Fix $0<\delta<1$ and an angular margin $\gamma>0$. There are
$r_\gamma<\infty$ and positive constants depending only on
$\cI,\gamma,\delta$ and the fixed cutoffs such that the symmetric
thermodynamic Gibbs RDM is NPT, with
$\Neg(\rho_{0\br})\ge b\eps^2/|\br|^2$, for
$r_\gamma\le|\br|\le\eps^{-\delta}$ and
$|\cos2\theta_\br|\ge\gamma$, under either of the following
sufficient conditions, at sufficiently small angle:
\begin{align}
 0<T&\le c_0 J\eps^2/\log(1/\eps),\label{eq:thermal-window-elementary}\\
 0<T&\le c_1 J\eps^2/\log(2+|\br|).
 \label{eq:thermal-window-klippel}
\end{align}
Nearest-neighbor pairs satisfy the same conclusions after decreasing
the constants. Under condition~\eqref{eq:thermal-window-elementary}
the limits~\eqref{eq:main-amplitude} and \eqref{eq:main-negativity}
hold as well, uniformly in the temperature. The proof for
condition~\eqref{eq:thermal-window-elementary} uses only the entropy
bound $-f_0(t)\le t\log2$. The proof for
condition~\eqref{eq:thermal-window-klippel} additionally uses the
zero-field two-dimensional ferromagnet pressure consequence of
Ref.~\cite{Klippel2026}, stated exactly in Sec.~\ref{sm:thermal}.
\end{theorem}

\begin{theorem}[Massive pair operator]\label{thm:massive}
For $h$ in a compact subset of $(1/2,\infty)$ and small positive angle,
the specified operator
\begin{equation}
 A_2=a_\eps\cL+2\Delta_\eps I+(d/2)P_\nn
 \label{eq:massive-block-statement}
\end{equation}
is positive. Its source solution $w_2=-dA_2^{-1}f$ is unique in
$\ell^2$ and has the exact rank-one formula in
Sec.~\ref{sm:massive}. Its fixed-distance limit is $w_2/d\to F$,
also in $\ell^2$, and
\begin{equation}
 \eps^{-4}w_2(\br_\eps)\longrightarrow
 A_h(\rho,\theta)=\frac{m_h^2\cos2\theta}{32(\pi-2)}K_2(m_h\rho),
 \qquad m_h=\sqrt{2h-1},
 \label{eq:profile-statement}
\end{equation}
uniformly when $\eps\br_\eps$ lies in compact subsets of
$\mathbb R^2\setminus\{0\}$. This theorem is about $A_2$.
\end{theorem}

\begin{theorem}[Two-qubit asymptotic criterion]\label{thm:rdmcriterion}
Let physical RDMs have the symmetry-constrained form
\eqref{eq:rho}. At a nonnodal scaled separation suppose
\begin{equation}
 p/\eps^4\to n_*,\quad w/\eps^4\to A_h\ne0,\quad
 q_\br=o(\eps^4),\quad v_\br=o(\eps^4),\quad\zeta=o(\eps^6).
 \label{eq:criterion-hypotheses}
\end{equation}
Then $\Neg/\eps^4\to(|A_h|-n_*)_+$. Away from equality, the state
is entangled for all sufficiently small angle if and only if
$|A_h|>n_*$. Uniform hypotheses and strict margins give uniform
conclusions. The full proof, including the exterior Schur complement,
is in Sec.~\ref{sm:massive}.
\end{theorem}

\begin{proposition}[Rotational selection]\label{prop:selection}
For a translation-invariant state with a joint rotation
$U_\phi b_iU_\phi^\dagger=e^{-i\phi}b_{R_\phi i}$,
$\phi=2\pi/n$, one has
$w(R_\phi\br)=e^{2i\phi}w(\br)$ and $w(-\br)=w(\br)$.
In a scalar exchange-even far field the allowed angular harmonics
satisfy $l\equiv2\pmod n$ and $l$ even. For a lattice with a basis,
this scalar rule applies to the exchange-even acoustic projection;
the sublattice entries transform jointly. The proof and explicit
other-lattice pair kernels are in Sec.~\ref{sm:symmetry}.
\end{proposition}

Three further groups of results concern separations of order
$1/\eps$. Section~\ref{sm:massscale} proves the leading negativity on
the whole region $|\eps\br|\ge c$ (Theorem~\ref{thm:massscale}) and
the range at a fixed positive threshold
(Corollary~\ref{cor:ms-range}). Section~\ref{sm:threshold} extends the
range to thresholds that vanish more slowly than
$\eps^{4/3}\ell_\eps^2$ and proves the convergence of the entangled
region (Theorem~\ref{thm:threshold}). Section~\ref{sm:exterior}
collects what is known about exact separability outside that region,
on the axes, on the diagonals and in the far tail, and states the
open problems. The sharp activated thermal boundary obtained from the
dilute-density Lambert-$W$ equation remains a leading-order
prediction; Section~\ref{sm:massive} gives its formula.

\section{Two-site reduced state and a three-setting witness}
\label{sm:rdm}

We use the negativity convention
$\Neg(\rho)=(\|\rho^{T_2}\|_1-1)/2$. A translation-invariant state
has equal populations $p$ at the two selected sites. In the ordered
one-flip basis $\{|\uparrow\downarrow\rangle,
|\downarrow\uparrow\rangle\}$ the following is an exact principal
block of the partial transpose, without an X-state assumption:
\begin{equation}
 M_\br=\begin{pmatrix}p-q_\br&w(\br)^*\\w(\br)&p-q_\br\end{pmatrix}.
 \label{eq:principal-pt}
\end{equation}
Its smaller eigenvalue and the min-max principle give
\begin{equation}
 \Neg(\rho_{0\br})\ge(|w(\br)|-p+q_\br)_+
 \ge(|w(\br)|-p)_+.
 \label{eq:population-npt}
\end{equation}
The inequality remains valid when other coherences are present.

\begin{lemma}[Population-retaining comparison]\label{lem:rdm-seven}
If the one-site transverse means vanish and the two populations are
equal, then
\begin{equation}
 \left|\Neg(\rho_{0\br})-|w(\br)|\right|\le7p.
 \label{eq:rdm-seven}
\end{equation}
In particular it holds in every translation- and jointly-fourfold-
invariant state.
\end{lemma}
\begin{proof}
The joint rotation multiplies a one-site transverse mean by $i$ or
$-i$; translation invariance therefore makes that mean zero. In the
ordered basis $00,01,10,11$, taking a one-site partial trace shows
that each vacuum-to-one-flip entry is the negative of the
corresponding one-flip-to-two-flip entry. Positivity bounds the latter
by $\sqrt{(p-q_\br)q_\br}\le p$. The normal hopping entry has modulus
at most $p-q_\br\le p$.

Compare $\rho$ with the trace-one Hermitian matrix having entry one
at $00,00$, its actual $00,11$ and $11,00$ coherences, and zero
elsewhere. Positivity of this comparison matrix is not needed.
Its partial-transpose eigenvalues are $1,0,|w|,-|w|$, so its
trace-norm expression for negativity equals $|w|$. The entrywise
absolute sum of the difference after partial transposition is at
most $14p$: the diagonal entries cost at most $4p$, the normal
hopping pair at most $2p$, and the eight entries associated with
the four odd coherences at most $8p$. The trace norm is bounded by
this entrywise sum, since each matrix unit has trace norm one.
The reverse triangle inequality and the factor $1/2$ in the
definition of negativity prove Eq.~\eqref{eq:rdm-seven}.
\end{proof}

\subsection{A linear witness using three spin settings}
The measured variables in this paragraph are Pauli operators
$\sigma^\alpha=2S^\alpha$, whose outcomes are $\pm1$. For any
$\eta\in\{-1,1\}$ define
\begin{equation}
 \mathcal W_\eta(\br)=
 \eta\langle\sigma_0^x\sigma_\br^x-\sigma_0^y\sigma_\br^y\rangle
 -\bigl(2-\langle\sigma_0^z\rangle-\langle\sigma_\br^z\rangle\bigr).
 \label{eq:three-setting-witness}
\end{equation}
It uses only uniform $x$, $y$, and $z$ measurement settings. For a
translation-invariant state,
\begin{equation}
 \mathcal W_\eta=4[\eta\operatorname{Re}w-p],\qquad
 \Neg(\rho_{0\br})\ge(\mathcal W_\eta/4)_+.
 \label{eq:witness-negativity}
\end{equation}

\begin{lemma}[Separable bound]\label{lem:witness-separable}
For every separable two-qubit state, including unequal one-site
populations, $\mathcal W_\eta\le0$.
\end{lemma}
\begin{proof}
For a product state let $\mathbf u,\mathbf v$ be its Bloch vectors,
with lengths at most one. Cauchy-Schwarz in the transverse plane
and then the arithmetic-geometric mean inequality give
\begin{align}
 \eta(u_xv_x-u_yv_y)
 &\le\sqrt{1-u_z^2}\sqrt{1-v_z^2}
 \le2\sqrt{(1-u_z)(1-v_z)}\nonumber\\
 &\le(1-u_z)+(1-v_z).
\end{align}
The middle inequality uses $1+u_z\le2$ and $1+v_z\le2$, with
$u_z,v_z\in[-1,1]$. This is precisely the desired product-state
bound. A separable state is a convex mixture of product states,
and Eq.~\eqref{eq:three-setting-witness} is linear, so the bound
holds for every such mixture.
\end{proof}

In a nonnodal direction choose $\eta=\sgn F(\br)$. Translation and
joint-rotation averages of this witness remain valid separable
witnesses, because each term is a lattice relabeling and a product
rotation of a valid witness. The anomalous identity
$\langle\sigma^x\sigma^x-\sigma^y\sigma^y\rangle=4\operatorname{Re}w$
proves Eq.~\eqref{eq:witness-negativity} from
Eq.~\eqref{eq:population-npt}. The population term is essential;
an anomalous correlation alone is not a separability criterion.

\emph{Sampling.} Choose the setting $xx$, $yy$, or $zz$ with
probabilities $1/4$, $1/4$, and $1/2$, and record
$X=4\eta\sigma_0^x\sigma_\br^x-2$,
$X=-4\eta\sigma_0^y\sigma_\br^y-2$, or
$X=2(\sigma_0^z+\sigma_\br^z)-2$, respectively. Then
$\mathbb E X=\mathcal W_\eta$ and $-6\le X\le2$. For the mean
$\bar X$ of $N$ independent preparations, Hoeffding's
inequality~\cite{Hoeffding1963} gives
\begin{equation}
 \Pr\bigl(|\bar X-\mathcal W_\eta|\ge t\bigr)\le2\exp(-Nt^2/32).
 \label{eq:witness-hoeffding}
\end{equation}
With $t=M/2$ and $N\ge128\log(200)/M^2$ the right-hand side is at most
$0.01$. A state with $\mathcal W_\eta=M>0$ then gives $\bar X>M/2$
with probability at least $0.99$, whereas a separable state, for which
$\mathcal W_\eta\le0$, gives $\bar X\ge M/2$ with probability at most
$0.01$. This distribution-free bound is conservative; the snapshot numbers quoted in the Letter use
the variance of the estimator and a fixed number of snapshots per setting
(Sec.~\ref{sm:exp-errors}).

\section{A range-uniform normal form and excitation bounds}
\label{sm:normalform}

\subsection{Exact spin algebra and coercivity}
For distinct sites define
\begin{equation}
 T_{ij}=b_i^\dagger b_j+b_j^\dagger b_i,\quad
 Y_{ij}=b_i^\dagger b_j^\dagger+b_i b_j,\quad
 X_{ij}=b_i^\dagger b_j^\dagger-b_i b_j.
\end{equation}
Direct expansion of the bond rotation gives
\begin{align}
 H/J-Ne_{\rm pol}&=Q+s\VDM+d[B+W(f)]+h\eps^2\Nop,
 \label{eq:hardcore-decomposition}\\
 B&=-\Nop+\tfrac12\Ncontact-\tfrac12Q,\qquad
 Q=2\Nop-\Ncontact-\tfrac12\sum_{\langle ij\rangle}T_{ij},\nonumber\\
 W(u)&=\tfrac12\sum_{i,r\ne0}
 [u(r)b_i^\dagger b_{i+r}^\dagger+\overline{u(r)}b_i b_{i+r}],
 \nonumber\\
 \VDM&=\sum_{i,\mu}C_{i,i+e_\mu},\qquad
 C_{ij}=c_\mu(n_i b_j^\dagger-b_i^\dagger n_j)
 +\overline{c_\mu}(n_i b_j-b_i n_j),\quad
 c_x=i/2,\ c_y=-1/2.\label{eq:dm-cubic}
\end{align}
The part of the DM interaction linear in transverse spin is a
difference of one-site operators and telescopes on a torus. Thus
Eq.~\eqref{eq:dm-cubic} is exact, and $\VDM\Omega=0$ for the all-up
vacuum $\Omega$.

In the singlet/triplet basis, $C_{ij}$ connects the singlet with
$|\downarrow\downarrow\rangle$ with coefficient of modulus
$1/\sqrt2$. Completing a two-dimensional square gives, for
$\alpha>0$,
\begin{equation}
 \pm sC_{ij}\preceq\alpha P^-_{ij}
 +\frac{s^2}{2\alpha}n_i n_j.
\end{equation}
Since $\Ncontact\preceq2\Nop$, summing proves
$\pm s\VDM\preceq\alpha Q+s^2\Nop/\alpha$.
Set
\begin{equation}
 r_*=h_--\tfrac12>1,\quad
 \alpha_*=(r_*+1)/(2r_*),\quad g_*=1-\alpha_*,\quad
 m_*=r_*-\alpha_*^{-1}>0,\quad
 \kappa_*=g_*/2,\quad\mu_*=m_*/2.
 \label{eq:coercivity-constants}
\end{equation}
Using $d\le\eps^2/2$, $s^2\le\eps^2$ and $\Ncontact\succeq0$,
\begin{align}
 H_0:=Q+s\VDM+dB+h\eps^2\Nop
 &\succeq(g_*-d/2)Q+m_*\eps^2\Nop,\label{eq:coercivity}\\
 H_0&\preceq(1+\alpha_*)Q+(h_++\alpha_*^{-1})\eps^2\Nop.
 \nonumber
\end{align}
This identifies the role of $h>3/2$. The number-conserving
one-magnon dispersion, obtained directly from
Eq.~\eqref{eq:hardcore-decomposition}, is
\begin{equation}
 \omega_{\rm nc}(k)=\Delta_\eps+a_\eps(2-\cos k_x-\cos k_y).
\end{equation}
Its leading gap closes at $h=1/2$, independently of the sufficient
coercivity allocation in Eq.~\eqref{eq:coercivity}.

\subsection{Support-size estimates}
\begin{lemma}[Local vacuum and swap estimates]\label{lem:support-size}
Let $A_Z=A_Z^\dagger$ have support on at most $k$ actual sites,
and let $N_Z=\sum_{i\in Z}n_i$. If
$\langle0_Z,A_Z0_Z\rangle=0$, then
\begin{equation}
 |\omega(A_Z)|\le3\|A_Z\|\sqrt{\omega(N_Z)}.
 \label{eq:vacuum-diagonal-bound}
\end{equation}
If $A_Z$ is odd under the swap of the endpoints of an edge $e$,
then $|\omega(A_Z)|\le2\|A_Z\|\sqrt{\omega(P^-_e)}$.
If in addition $A_Z|0_Z\rangle=0$, then
\begin{equation}
 |\omega(A_Z)|\le2\|A_Z\|
 \sqrt{\omega(P^-_e)\omega(N_Z)}.
 \label{eq:swap-vacuum-bound}
\end{equation}
If $A_Z$ and its adjoint annihilate the local vacuum, without the
Hermiticity assumption, then
\begin{equation}
 |\omega(A_Z)|\le\|A_Z\|\omega(N_Z).
 \label{eq:vacuum-annihilation-bound}
\end{equation}
\end{lemma}
\begin{proof}
Let $P_0=|0_Z\rangle\langle0_Z|$ and $T=1-P_0\preceq N_Z$.
Split $A_Z$ into its two off-diagonal blocks relative to $P_0,T$
and its $T A_ZT$ block. The first two have total expectation at
most $2\|A_Z\|\sqrt{\omega(T)}$, and the last at most
$\|A_Z\|\omega(T)\le\|A_Z\|\sqrt{\omega(T)}$.
For a swap-odd Hermitian operator its two diagonal swap blocks
vanish. The off-diagonal-block Cauchy-Schwarz inequality gives
$2\|A_Z\|\sqrt{\omega(P^-_e)\omega(1-P^-_e)}$.
If the local vacuum is also annihilated, the same decomposition
takes place in its orthogonal complement; its second probability
is bounded by $\omega(T)$. Finally, when both local vacuum row
and column vanish, $A_Z=T A_ZT$ and positivity of the compressed
state gives $|\omega(A_Z)|\le\|A_Z\|\omega(T)$.
Purifying a mixed state proves the same bounds for arbitrary states.
\end{proof}

For translated local sums the geometric diameter is irrelevant:
a translated support of $k$ sites covers each site exactly $k$
times. Summing the preceding inequalities and using
Cauchy-Schwarz in the translation index gives, respectively,
$3\sqrt{k}J_A\sqrt{N\langle\Nop\rangle}$,
$C_kJ_A\sqrt{N\langle Q\rangle}$, and
$C_kJ_A\sqrt{\langle Q\rangle\langle\Nop\rangle}$,
where $J_A$ is the per-site absolute norm budget. In particular,
no box of side equal to the support diameter is filled in this
counting. Real and imaginary Hermitian parts give the corresponding
bounds for a complex local observable, with a harmless factor two.

\subsection{Pair dressing and the first residual}
Let $F_R=\chi(\cdot/R)F$ be the real even cutoff of
Sec.~\ref{sm:green}, and define
\begin{equation}
 X_R=\tfrac12\sum_{i,r\ne0}F_R(r)
 (b_i^\dagger b_{i+r}^\dagger-b_i b_{i+r}),\qquad U_R=e^{dX_R}.
 \label{eq:pair-dressing}
\end{equation}
The two-particle normalization is
$\|X_R\Omega\|^2=N\|F_R\|_2^2/2$ and
$\langle X_R\Omega,QX_R\Omega\rangle=N(F_R,\cL F_R)/2$.
Each particle hops with weight $1/2$; combining their two hops
in a relative coordinate gives the retained-edge Laplacian $\cL$.

For three distinct sites, the swap $\mathsf T_{ij}=1-2P^-_{ij}$
gives
\begin{equation}
 [P^-_{ij},X_{ik}]=\tfrac12(X_{ik}-X_{jk})\mathsf T_{ij},\qquad
 A_{ij;k}=[P^-_{ij},X_{ik}]-\tfrac12(Y_{ik}-Y_{jk}).
 \label{eq:first-local-commutator}
\end{equation}
One checks $A_{ij;k}\Omega=0$, $\|A_{ij;k}\|\le3$, and
$\mathsf T_{ij}A_{ij;k}\mathsf T_{ij}=-A_{ij;k}$:
the latter follows from
$[P^-_{ij},X_{ik}+X_{jk}]=0$. The same-edge commutator is zero.
Combining coefficients at the two endpoints therefore yields
\begin{equation}
 [Q,X_R]=W(\cL F_R)+A_R,\qquad
 |\omega(A_R)|\le64a_R
 \sqrt{\langle Q\rangle\langle\Nop\rangle}.
 \label{eq:first-residual-bound}
\end{equation}
Indeed each translated edge/third-site term satisfies
Eq.~\eqref{eq:swap-vacuum-bound}, with support three;
the coefficient sum per direction is at most $2a_R$, and
$\sqrt{q_x}+\sqrt{q_y}\le\sqrt{2(q_x+q_y)}$.
The resulting constant $12\sqrt6$ is below the displayed
conservative value. This proof also explains why combining the
coefficients first replaces $a_R$ by a constant multiple of $v_R$.

\begin{lemma}[Residual source bound]\label{lem:zero-mass-source}
For a finite real even $u$ with $\sum u=0$,
\begin{equation}
 |\omega(W(u))|\le8\|u\|_{1,1}\sqrt{N\langle Q\rangle}.
 \label{eq:zero-mass-source-bound}
\end{equation}
\end{lemma}
\begin{proof}
Fix $r_*=e_x$. In the punctured relative graph, a Manhattan path
with a bounded detour around zero joins $r_*$ to $r\ne0$ in at
most $4(1+|r|_1)$ steps. At one step $v\to v'$ the physical
endpoint swap takes $Y_{i,i+v}$ to $Y_{i,i+v'}$.
Since $\mathsf T-1=-2P^-$ and $\|Y\|=1$, the expectation
changes by at most $4\sqrt{\omega(P^-_{i+v,i+v'})}$.
Summing translations at a fixed step gives at most
$4\sqrt{N\langle Q\rangle}$. In $W(u)$ subtract the reference
pair using $\sum u=0$, sum the paths, and include the factor
$1/2$ in $W$. This gives the constant eight.
\end{proof}

\subsection{Cubic cancellation}
The operator $C_{ij}$ in Eq.~\eqref{eq:dm-cubic} is swap-odd.
For $k\notin\{i,j\}$ write $F_i=F_R(k-i)$, $F_j=F_R(k-j)$.
The two contributions to $D_R=[\VDM,X_R]$ are split exactly as
\begin{equation}
 \frac{F_i+F_j}{2}[C_{ij},X_{ik}+X_{jk}]
 +\frac{F_i-F_j}{2}[C_{ij},X_{ik}-X_{jk}].
 \label{eq:cubic-split}
\end{equation}
The plus commutator and the same-edge term are swap-odd and have
bounded local norm; Lemma~\ref{lem:support-size} controls their
sum by $C a_R\sqrt{N\langle Q\rangle}$. The remaining coefficients
have the exact cancellation
\begin{equation}
 \sum_{k\notin\{i,j\}}[F_R(k-i)-F_R(k-j)]=0.
 \label{eq:cubic-zero-sum}
\end{equation}
Evenness makes the two omitted endpoint terms equal.

Fix a site $k_0$ next to the edge, distinct from its endpoints.
Subtract the minus commutator with $k=k_0$ in the weighted sum;
Eq.~\eqref{eq:cubic-zero-sum} makes the subtraction exact. Move
$k_0$ to $k$ along a path avoiding $i,j$, of length at most
$8(1+|k-i|_1)$. Such a path follows coordinate steps, with at
most a rectangular detour around the two excluded sites. A swap
on each step transports the third site and changes an operator
of norm at most four by an expectation of at most
$16\sqrt{\omega(P^-_e)}$. Translation summation and the path
length then give, for example,
\begin{equation}
 |\omega(D_R)|\le4096(a_R+b_R)\sqrt{N\langle Q\rangle}.
 \label{eq:cubic-old-bound}
\end{equation}
The displayed constant exceeds the endpoint, two-orientation,
path-length and Hermitian-phase factors just listed. Only the
finiteness and range independence of this constant will be used.

\subsection{Finite Taylor expansion and its budget}
Use $\operatorname{ad}_{X_R}A=[A,X_R]$. A local operator on
$k$ sites satisfies the incident-pair bound
\begin{equation}
 \|\operatorname{ad}_{X_R}^jA_Z\|_{\mathrm{dec}}
 \le\|A_Z\|(2a_R)^j k(k+1)\cdots(k+j-1),
 \label{eq:pair-support-budget}
\end{equation}
where the left side is the sum of the norms in the local
commutator decomposition. Each nonzero commutator meets the
current support, costs at most twice the product of norms, and
adds at most one new site; induction proves the formula.
Taylor's integral remainder uses unitary conjugation, so its
operator norm does not acquire a volume factor beyond the sum
of the translated initial terms.

Expand $Q$ and $s\VDM$ through two commutators, and the terms
already carrying $d$ or $\eps^2$ through one. The exact result is
\begin{align}
 U_R^\dagger(H/J-Ne_{\rm pol})U_R
 ={}&H_0+dW(q_R)+dA_R+sdD_R+d^2Z_R
 +h\eps^2d[\Nop,X_R]\nonumber\\
 &+\tfrac12sd^2[ [\VDM,X_R],X_R]+\mathcal E_R,\label{eq:normalform-expansion}\\
 Z_R={}&[B+W(f),X_R]+\tfrac12[[Q,X_R],X_R].\nonumber
\end{align}
Equation~\eqref{eq:pair-support-budget} gives
$\|\mathcal E_R\|/N\le C_\cI\eps^6(1+a_R)^3$.
The retained double DM commutator has odd magnon parity, hence
zero local vacuum diagonal, and expectation at most
$C a_R^2\sqrt{N\langle\Nop\rangle}$.
The scalar and local remainder of $Z_R$ obey
\begin{equation}
 \langle Z_R\rangle_\Omega/N=\mathcal A(F_R),\qquad
 |\omega(Z_R-N\mathcal A(F_R))|
 \le C(1+a_R)^2\sqrt{N\langle\Nop\rangle}.
 \label{eq:quartic-coarse-bound}
\end{equation}
To prove the scalar, put $\Psi_R=X_R\Omega$ and
$\Phi=W(f)\Omega$. Then $B\Omega=Q\Omega=0$,
$\langle[W(f),X_R]\rangle_\Omega=2\operatorname{Re}\langle\Phi,\Psi_R\rangle
=N(f,F_R)$, and
$\langle[[Q,X_R],X_R]\rangle_\Omega/2
=\langle\Psi_R,Q\Psi_R\rangle=N(F_R,\cL F_R)/2$.
For the remainder subtract the vacuum expectation of each local
summand, whose support has at most four sites and per-site norm
sum $C(a_R+a_R^2)$. Apply Eq.~\eqref{eq:vacuum-diagonal-bound}
before summing. The field commutator is $2W(F_R)$ and obeys the
same estimate with coefficient $C a_R$.

Choose the exchange allocations $g_*Q/8$ separately for
$dA_R,dW(q_R),sdD_R$. The first uses at most
$m_*\eps^2\Nop/8$ once $\eps a_R$ is small. Assign
$m_*\eps^2\Nop/4$ to the combined quartic/field
vacuum-subtracted terms and $m_*\eps^2\Nop/8$ to the retained
double DM commutator. Young's inequality then gives the following
scalar costs divided by $N\eps^4$:
\begin{equation}
 C_\cI\left[
 R^{-4}+\|q_R\|_{1,1}^2
 +\eps^2(a_R+b_R)^2+\eps^2(1+a_R)^4
 +\eps^4a_R^4+\eps^2(1+a_R)^3\right].
 \label{eq:normalform-costs}
\end{equation}
Every estimate holds with both signs. Taking $d\le g_*/4$
leaves at least $\kappa_*Q+\mu_*\eps^2\Nop$ in the lower
bound. The upper coefficients can be
$\overline\kappa=2$ and
$\overline\mu=h_++\alpha_*^{-1}+m_*/2$.
Equations~\eqref{eq:green-cutoff-budgets}--\eqref{eq:green-cutoff-scalar}
therefore prove the all-state normal form
\begin{equation}
 \begin{split}
 N\mathcal S_\eps-N\eps^4\eta_R+\kappa_*Q+\mu_*\eps^2\Nop
 &\preceq U_R^\dagger(H/J-Ne_{\rm pol})U_R\\
 &\preceq N\mathcal S_\eps+N\eps^4\eta_R
 +\overline\kappa Q+\overline\mu\eps^2\Nop,
 \end{split}\qquad
 \begin{gathered}
 \mathcal S_\eps=-d^2F_1/2,\\
 \eta_R=C_{\cI,\chi}\{R^{-2}+\eps^2[1+\log(2+R)]^4\}.
 \end{gathered}
 \label{eq:normalform}
\end{equation}
Sufficient hypotheses are $R\ge4$, torus side lengths at least
$16(R+2)$, and $\eps[1+\log(2+R)]\le c_{\cI,\chi}$, together
with the preceding scalar allocation inequalities. They define
a positive angle threshold independent of the volume and state.

\subsection{Thermodynamic automorphism and a-priori bounds}
At fixed $\eps,R$, the local adjoint series converges in norm
when $2da_R<1$. Indeed Eq.~\eqref{eq:pair-support-budget},
divided by $j!$, gives the summable majorant
$\|A_Z\|\binom{k+j-1}{j}(2da_R)^j$. Each finite-order term
stabilizes as volume increases; the majorant also controls the
tail uniformly. At order $j$ the diameter grows at most linearly
in $jR$; multiplying the geometric majorant by this additional
factor is still summable. This also controls boundary energy
densities and defines the infinite-volume automorphism and its
inverse. Define the pair-dressed state by
$\widetilde\omega(A)=\omega(U_RA U_R^\dagger)$, so
$\omega(A)=\widetilde\omega(U_R^\dagger A U_R)$.
At fixed range the conjugated local energy sums have vanishing
boundary-to-volume error, and Eq.~\eqref{eq:normalform} passes
to their densities. The vacuum trial and the lower bound imply
for every energy-density minimizer
\begin{equation}
 \kappa_*\widetilde{\mathfrak q}
 +\mu_*\eps^2\widetilde p\le2\eps^4\eta_R.
 \label{eq:dressed-rigidity}
\end{equation}
Expanding the local observable $n_0$ gives
$p\le\widetilde p+Cda_R\sqrt{\widetilde p}+4d^2a_R^2$:
its first commutator has zero vacuum diagonal, and its second
Taylor remainder is bounded by $4d^2a_R^2$. Thus with
$R_0=\lceil\eps^{-1}\rceil$,
\begin{equation}
 p\le C_p\eps^4\ell_\eps^4,\qquad
 \widetilde p\le C_t\eps^4\ell_\eps^4,\qquad
 \widetilde{\mathfrak q}\le C_q\eps^6\ell_\eps^4,
 \qquad a_{R_0}\le C_F\ell_\eps.
 \label{eq:apriori}
\end{equation}
All constants depend only on $\cI,\chi$. The volume limit has
already been taken when the range is selected as a function of
the angle.

For later comparison, direct conjugation also gives
\begin{equation}
 w(r)=dF_R(r)+O\!\left(|r|\sqrt{\widetilde{\mathfrak q}}
 +da_R\sqrt{\widetilde p}+d^2a_R^2\right),\qquad |r|\le R.
 \label{eq:direct-local-conjugation}
\end{equation}
To see this, a punctured relative path comparing $r$ and its
quarter-turn has $O(|r|)$ steps. The swap estimate bounds the
change of the dressed pair expectation by
$C|r|\sqrt{\widetilde{\mathfrak q}}$; the two expectations
have opposite signs by joint symmetry. The first commutator
of $b_0b_r$ with $X_R$ has vacuum scalar $F_R(r)$.
Subtract that scalar locally, apply
Eq.~\eqref{eq:vacuum-diagonal-bound}, and bound the second
Taylor remainder by $Cd^2a_R^2$. The stronger range theorem
uses the stationary identities in Sec.~\ref{sm:range}.

\section{The punctured Green function and its symmetries}
\label{sm:green}

All Fourier averages use $(2\pi)^{-2}\dd^2k$ on $[-\pi,\pi]^2$.
The full-lattice Laplacian and source have symbols
\begin{equation}
 \lambda(\bk)=4-2\cos k_x-2\cos k_y,\qquad
 \widehat f(\bk)=\tfrac12(\cos k_y-\cos k_x),\qquad
 \mathfrak d(\bk)=\cos k_x-\cos k_y.
 \label{eq:green-symbols}
\end{equation}
Put $G_0=\mathcal F^{-1}[\mathfrak d/(2\lambda)]$ and
$k_0=4G_0(e_x)$. The multiplier is bounded since
$|\mathfrak d|\le\lambda/2$, hence $G_0\in\ell^2$.
Its reflection and exchange symmetries give $G_0(0)=0$ and
$P_\nn G_0=-k_0f$. At the four neighbors of the puncture,
deleting the origin edge subtracts the local value from the
full-lattice Laplacian. Therefore
\begin{equation}
 \cL G_0=-(1-k_0)f,
 \qquad F=\frac{G_0}{1-k_0}.
 \label{eq:massless-defect}
\end{equation}
A square-summable harmonic function for $\cL$ has zero Dirichlet
form, is constant on the connected punctured graph, and hence
vanishes. Thus this solution is unique.

\subsection{Exact defect and norm integrals}
Let $x=1-\cos k_x$ and $y=1-\cos k_y$. For $x>0$, the
tangent-half-angle substitution in the second momentum gives
\begin{equation}
 \left\langle\frac1{x+1-\cos k_y}\right\rangle_{k_y}
 =\frac1{\sqrt{x(x+2)}},\qquad
 \left\langle\frac1{(x+1-\cos k_y)^2}\right\rangle_{k_y}
 =\frac{x+1}{[x(x+2)]^{3/2}}.
 \label{eq:green-elementary-integrals}
\end{equation}
For the first identity the integral becomes
$\pi^{-1}\int_{-\infty}^{\infty}[x+(x+2)t^2]^{-1}\dd t$;
the second follows by differentiating at $x>0$. It follows that
\begin{align}
 k_0&=\left\langle\frac{\mathfrak d^2}{\lambda}\right\rangle_k
 =-1+2\left\langle\frac{x^{3/2}}{\sqrt{x+2}}\right\rangle_{k_x}
 =-1+\frac8\pi\int_0^1\frac{1-u^2}{\sqrt{2-u^2}}\dd u
 =\frac4\pi-1,\label{eq:green-k}\\
 I_0&=\left\langle\frac{\mathfrak d^2}{\lambda^2}\right\rangle_k
 =\frac14-\frac1\pi\int_0^1\frac{\dd u}{(2-u^2)^{3/2}}
 =\frac14-\frac1{2\pi}.\label{eq:green-I}
\end{align}
The substitutions are $k_x=2v$, $u=\cos v$. The antiderivatives
are $u\sqrt{2-u^2}/2$ and $u/[2\sqrt{2-u^2}]$, respectively.
The exceptional point $x=0$ has measure zero and all final
integrands are bounded. Consequently, with
$Z_0=1-k_0=2(\pi-2)/\pi$,
\begin{equation}
 F(e_x)=\frac{k_0}{4Z_0}=\frac{4-\pi}{8(\pi-2)},\quad
 F(e_y)=-F(e_x),\quad
 \|F\|_2^2=\frac{I_0}{4Z_0^2}=\frac{\pi}{64(\pi-2)}.
 \label{eq:green-constants}
\end{equation}
Since $\cL F=-f$, one also has
$(F,\cL F)=-(f,F)=F_1$ and $c_*=F_1/8$.

\subsection{Far field, differences, and cutoff budgets}
Near zero momentum,
\begin{equation}
 \widehat F(\bk)=-\frac{k_x^2-k_y^2}{4Z_0|\bk|^2}+a_2(\bk),
 \qquad |\partial^\alpha a_2(\bk)|\le C_\alpha|\bk|^{2-|\alpha|}.
 \label{eq:green-symbol-expansion}
\end{equation}
Choose a smooth periodic low-frequency cutoff. With the inverse
Fourier convention above, the continuum inverse of $|k|^{-2}$ is
$-(2\pi)^{-1}\log|x|$, modulo a constant distribution. Applying
the two indicated derivatives gives the leading term below.
For completeness, on the annulus $|k|\asymp2^{-j}$ the remainder
kernel is bounded, after $M$ integrations by parts, by
$C_M2^{-4j}(1+2^{-j}|\br|)^{-M}$. Splitting its sum at
$2^{-j}|\br|=1$ and taking $M>4$ gives $C|\br|^{-4}$.
The multiplier away from zero is smooth and periodic and has
arbitrarily fast Fourier decay. Thus
\begin{equation}
 F(\br)=\frac{\cos(2\theta_\br)}{8(\pi-2)|\br|^2}
 +O(|\br|^{-4}),\qquad
 |\nabla^jF(\br)|\le C_j(1+|\br|)^{-2-j},\quad j=0,1,2.
 \label{eq:green-tail}
\end{equation}
Each lattice difference inserts a factor $e^{ik_\mu}-1$ in
the annular multiplier and proves the stated difference bounds
by the same summation. Increasing the constants covers the
finitely many differences meeting the origin. In particular
$F$ vanishes on both diagonals by coordinate-exchange symmetry.

For a fixed smooth radial cutoff $\chi$, equal to one on the unit
disk and zero outside radius two, let $F_R=\chi(\cdot/R)F$.
The preceding bounds imply
\begin{align}
 a_R:=\sum_{r\ne0}|F_R(r)|&\le C_\chi\log(2+R),\nonumber\\
 b_R:=\sum_{\mu,k\notin\{0,e_\mu\}}(1+|k|_1)
 |F_R(k)-F_R(k-e_\mu)|&\le C_\chi\log(2+R),\label{eq:green-cutoff-budgets}\\
 v_R:=\sum_{\mu,k\notin\{0,e_\mu\}}
 |F_R(k)-F_R(k-e_\mu)|&=O_\chi(1),\qquad v_R\to v<\infty.
 \nonumber
\end{align}
To verify the residual estimate, use the discrete product rule
for $\cL(\chi_RF)$ outside the fixed source support. On the
cutoff annulus, $F=O(R^{-2})$, $\nabla F=O(R^{-3})$,
$\nabla\chi_R=O_\chi(R^{-1})$ and
$\cL\chi_R=O_\chi(R^{-2})$. Hence every residual term is
$O_\chi(R^{-4})$, on $O(R^2)$ sites. The weight $1+|r|_1$
then gives
\begin{equation}
 q_R=f+\cL F_R,\quad \sum_rq_R(r)=0,\quad
 \|q_R\|_{1,1}:=\sum_r(1+|r|_1)|q_R(r)|\le C_\chi/R.
 \label{eq:green-residual}
\end{equation}
Both the source and the finite graph Laplacian sum to zero.
Completing the Dirichlet square gives the further exact identity
and bound
\begin{equation}
 \mathcal A(F_R):=\tfrac12(F_R,\cL F_R)+(f,F_R)
 =-F_1/2+\tfrac12(F_R-F,\cL(F_R-F))
 =-F_1/2+O_\chi(R^{-4}).
 \label{eq:green-cutoff-scalar}
\end{equation}
The last estimate follows by summing the squared gradient tail
$O(r^{-6})$ and the cutoff-annulus contributions. These
constants depend only on finitely many derivatives of $\chi$.

\subsection{Comparison with linear spin-wave theory}
In the linear spin-wave pair equation the relative coordinate has
the full-lattice inverse, so the massless response is $dG_0$.
Equations~\eqref{eq:massless-defect} and \eqref{eq:green-k} give
the pointwise enhancement, with common zeros retained,
\begin{equation}
 \frac{F}{G_0}=Z_0^{-1}=\frac{\pi}{2(\pi-2)}.
 \label{eq:lsw-enhancement}
\end{equation}
It comes from the retained-edge boundary condition, which includes
the Heisenberg contact attraction as well as exclusion.

\subsection{Twist parity and axial reality}
Let $\mathsf I$ invert lattice coordinates without rotating spins.
It reverses the oriented DM terms and fixes the other terms, so
$\mathsf I H_{\eps,h}\mathsf I^{-1}=H_{-\eps,h}$.
Thus the set of minimizers is mapped bijectively between the two
signs of the twist. For a covariant family
$\omega_{-\eps}(A)=\omega_\eps(\mathsf I^{-1}A\mathsf I)$,
translation and commutation at distinct sites give
\begin{equation}
 w_{-\eps}(\br)=w_\eps(-\br)=w_\eps(\br).
 \label{eq:epsilon-parity}
\end{equation}
This includes finite periodic Gibbs states and their consistently
chosen limits. It does not identify independently selected,
possibly different minimizers at the two signs of $\eps$.

Let $\mathsf C$ be complex conjugation in the $S^z$ basis and
$P_x$ reflect $x\mapsto-x$. Complex conjugation changes the sign
of the $x$-bond DM term and leaves the $y$-bond term unchanged;
$P_x$ changes the former sign again. Therefore the antiunitary
$\mathsf A_x=\mathsf C P_x$ preserves $H_{\eps,h}$.
For a state invariant under this symmetry,
\begin{equation}
 w(x,y)=\overline{w(-x,y)}.
 \label{eq:antiunitary-reflection}
\end{equation}
Here the antiunitary transform of a state includes complex
conjugation of its expectation, so the transformed functional
is again complex linear and positive. On the $y$ axis the
reflection fixes the separation; on the $x$ axis it reverses it,
and $w(-\br)=w(\br)$. Thus $w$ is real on both axes for these
states. Joint quarter-turn symmetry additionally makes it purely
imaginary on a diagonal. Periodic Gibbs states and their limits
have this reflection symmetry; it may also be imposed by
averaging a minimizer with its antiunitary image. The range and
population theorems do not require this additional state symmetry.

\section{Sixth-order energy and physical population}
\label{sm:population}

We prove Theorems~\ref{thm:energy} and~\ref{thm:population}.
The argument uses the exchange coercivity of Sec.~\ref{sm:normalform},
the punctured-lattice estimates of Sec.~\ref{sm:green}, and the
infinite-volume creation-current inequality of Sec.~\ref{sm:current}.
Both conjugations used below are finite-range automorphisms at each
fixed angle.  The coefficient-current estimate is applied to
thermodynamic expectation densities, not to a fixed torus with its
mass regulator removed.

\subsection{Cutoff kernels, local norms, and vacuum extraction}

Fix $\cI=[h_-,h_+]\Subset(3/2,\infty)$ and use the positive constants
$g_*,m_*$ defined in Sec.~\ref{sm:normalform}.  Let $\chi$ be a fixed
smooth radial cutoff, equal to one on the unit disk and zero outside
the disk of radius two.  Throughout the energy proof set
\begin{align}
 R&=\lceil\eps^{-2}\rceil, &F_R(r)&=\chi(r/R)F(r),
 &X&=X_{F_R}, &U_R&=e^{dX},                                      \label{eq:pop-cutoff}\\
 a_R&=\|F_R\|_1, &S_R&=\|F_R\|_2^2,
 &v_R&=\sum_{\mu=x,y}\sum_{k\notin\{0,e_\mu\}}
             |F_R(k)-F_R(k-e_\mu)|.                              \label{eq:pop-kernel-budgets}
\end{align}
The estimates in Sec.~\ref{sm:green} imply
\begin{align}
 a_R&=O_\chi(\log R),& S_R&\longrightarrow S_F,
 &v_R&\longrightarrow v_F<\infty,                                \label{eq:pop-kernel-limits}\\
 q_R&:=f+\cL F_R,&\gamma_R&:=\|q_R\|_{1,1}=O_\chi(R^{-1}),
 &\mathcal A_R&:=\tfrac12(F_R,\cL F_R)+(f,F_R)
                =-F_1/2+O_\chi(R^{-4}).                         \label{eq:pop-cutoff-error}
\end{align}
Here and below relative-pair inner products use counting measure.
For example, the convergence of $v_R$ follows because the omitted
tail of $|\nabla F|$ is $O(R^{-1})$, while the cutoff-gradient term
has size $O(R^{-3})$ on $O(R^2)$ sites.  These estimates are independent
of $h$.

For a translation sum $A=\sum_\alpha\sum_i\tau_iA_\alpha$, denote
the sum $\sum_\alpha\|A_\alpha\|$ in a specified local decomposition
by $\mathfrak b(A)$.  The support size of a summand means the number
of sites on which it acts; it does not mean the area of its convex
hull.  In finite volume we require the torus to contain the displayed
finite supports and paths without aliasing, for example
$L\ge16(R+2)$ for the pair dressing.  We first establish the local
identities there.  Their norm-density bounds are uniform in such
volumes.

\begin{lemma}[Local commutators and extraction]
\label{lem:pop-local-extraction}
Let a translation-invariant state have population $p=\omega(n_0)$.
A local operator $A_Z$ on at most $k$ sites satisfies
\begin{align}
 |\omega(A_Z)|&\le3\|A_Z\|\sqrt{kp}
       &&\text{if }\langle0_Z,A_Z0_Z\rangle=0,                    \label{eq:pop-vacuum-diagonal}\\
 |\omega(A_Z)|&\le kp\|A_Z\|
       &&\text{if }A_Z0_Z=A_Z^\dagger0_Z=0.                      \label{eq:pop-vacuum-killing}
\end{align}
For a Hermitian $A_Z$, subtracting its vacuum scalar and all of its
vacuum-to-configuration matrix elements as pure creators and their
adjoints leaves an operator satisfying the second condition, of norm
at most $2^{k+1}\|A_Z\|$.

If the sum of norms of pair-generator terms incident at one site is
$a_R$, the $j$th commutator of $A_Z$ with $X$ has local norm budget
at most
\begin{equation}
 \|A_Z\|(2a_R)^j k(k+1)\cdots(k+j-1).                            \label{eq:pop-pair-comm-count}
\end{equation}
If $Y$ is a triple generator with incident norm sum $a_Y$, the
corresponding bound is
\begin{equation}
 \|A_Z\|(2a_Y)^j k(k+2)\cdots(k+2j-2).                           \label{eq:pop-triple-comm-count}
\end{equation}
\end{lemma}

\begin{proof}
Let $P_Z=1-|0_Z\rangle\langle0_Z|$.  Then
$P_Z\le\sum_{i\in Z}n_i$.  Decomposition into the vacuum and its
orthogonal complement, followed by Cauchy-Schwarz, gives
\eqref{eq:pop-vacuum-diagonal}.  Under the stronger hypothesis,
$A_Z=P_ZA_ZP_Z$, which gives \eqref{eq:pop-vacuum-killing}.
For the extraction statement write
\begin{equation}
 A_Z=c_Z I+\sum_{\varnothing\ne E\subset Z}
 \left[t_E\prod_{i\in E}b_i^\dagger+
       \overline{t_E}\prod_{i\in E}b_i\right]+A_Z^{\rm ann},
 \quad c_Z=\langle0_Z,A_Z0_Z\rangle,\quad
 t_E=\langle E|A_Z|0_Z\rangle.                                 \label{eq:pop-local-extract-formula}
\end{equation}
Both the vacuum row and column of $A_Z^{\rm ann}$ vanish.  Since
$|c_Z|,|t_E|\le\|A_Z\|$ and there are at most $2^k-1$ nonempty
configurations, its norm is at most
$[1+1+2(2^k-1)]\|A_Z\|=2^{k+1}\|A_Z\|$.
The bounds extend to mixed states by their GNS representation.

Only generator terms meeting the current support contribute to a
commutator.  Their norm sum is at most $ka_R$ for pairs, and the
commutator adds a factor two and at most one site.  Induction proves
\eqref{eq:pop-pair-comm-count}.  A triple adds at most two sites,
which proves \eqref{eq:pop-triple-comm-count}.  Summing translates
changes the right side of \eqref{eq:pop-vacuum-diagonal} to
$3\sqrt{k}\,\mathfrak b(A)\sqrt p$ per site, and that of
\eqref{eq:pop-vacuum-killing} to $k\mathfrak b(A)p$.
Thus these estimates remain valid when the geometric diameter grows
with $R$.
\end{proof}

All scalar and source extractions below are performed in
\eqref{eq:pop-local-extract-formula} before the local terms are
summed.  Regrouping the extracted pure creators is legitimate:
their coefficients are uniquely determined by the vacuum image,
because distinct finite occupied configurations are linearly
independent.  No global vacuum projector is used in an extensive
operator estimate.

\subsection{The first commutator and the fourth-order sources}

Write $X_{ik}=b_i^\dagger b_k^\dagger-b_i b_k$ and
$Y_{ik}=b_i^\dagger b_k^\dagger+b_i b_k$.  The two-site swap identity
$[P^-_{ij},X_{ik}+X_{jk}]=0$, with $k\notin\{i,j\}$, gives
\begin{equation}
 [Q,X]=\sum_{\langle ij\rangle}\sum_{k\notin\{i,j\}}
 \Delta_{ij}F_R(k)[P^-_{ij},X_{ik}],\qquad
 \Delta_{ij}F_R(k)=F_R(k-i)-F_R(k-j).                            \label{eq:pop-gradient-first-comm}
\end{equation}
Same-pair generators commute with $P^-_{ij}$.  Each displayed
commutator has norm at most two and support at most three, so
$\mathfrak b([Q,X])\le2v_R$.  The exact three-site identity is
\begin{equation}
 [P^-_{ij},X_{ik}]-\tfrac12(Y_{ik}-Y_{jk})
 =(n_i b_j^\dagger-n_j b_i^\dagger)b_k^\dagger+\mathrm{h.c.}
                                                                    \label{eq:pop-A-local}
\end{equation}
Consequently
\begin{equation}
 [Q,X]=W(\cL F_R)+A_R,\qquad
 \frac{|\omega(A_R)|}{N}\le12v_R\sqrt{\mathfrak q p},
 \quad\mathfrak q=\omega(Q)/N.                                 \label{eq:pop-A-bound}
\end{equation}
Indeed each residual in \eqref{eq:pop-A-local} is odd under swapping
$i,j$, kills its local vacuum, and has norm at most three.  Its two
swap blocks give a bound $6\sqrt{\omega(P^-_{ij})
\omega(n_i+n_j+n_k)}$.  Summing translates and applying
Cauchy-Schwarz gives $6\sqrt3\,v_R\sqrt{\mathfrak q p}$;
the constant 12 is sufficient.

Define
\begin{equation}
 D_R=[\VDM,X],\qquad J_R=[\Nop,X]=2W(F_R),\qquad
 Z_R=[B+W(f),X]+\tfrac12[[Q,X],X].                              \label{eq:pop-DJZ}
\end{equation}
The local norm sums of $[\Nop,X]$, $[\Ncontact/2,X]$,
$[-Q/2,X]$, and $[W(f),X]$ are bounded by
$a_R,4a_R,v_R,2a_R$, respectively.  A further pair commutator of
\eqref{eq:pop-gradient-first-comm} costs at most $6a_R$.
Thus $Z_R$ is supported on at most four sites per summand and
\begin{equation}
 \mathfrak b(Z_R)\le z_R:=v_R+(7+6v_R)a_R.                       \label{eq:pop-Z-budget}
\end{equation}

\begin{lemma}[Vacuum-changing sources]
\label{lem:pop-pair-quartet-sources}
The vacuum scalar of $Z_R$ is $N\mathcal A_R$.  Its remaining
vacuum image consists of the pair source $W(g_R)$ and the
four-creation source $C_{4,R}^++C_{4,R}^{+\dagger}$, where
\begin{align}
 g_R&=B_2F_R,& B_2&=-2I+\tfrac12P_{\nn}-\tfrac12\cL,
 &\|g_R\|_1&\le\tfrac{13}{2}a_R,                               \label{eq:pop-quartic-pair}\\
 C_{4,R}^+&=\tfrac12\sum_{\langle ij\rangle}
 \sum_{\substack{k,l\notin\{i,j\}\\ k\ne l}}
 \Delta_{ij}F_R(k)\Delta_{ij}F_R(l)
 b_i^\dagger b_j^\dagger b_k^\dagger b_l^\dagger.                \label{eq:pop-quartic-four}
\end{align}
The absolute coefficient sum of $C_{4,R}^+$ per site is at most
$v_R^2/2$.  After termwise local extraction, the residual has
expectation density at most $128z_Rp$.

The vacuum-changing part of $D_R$ is exactly
$\mathcal C_3(\mathcal T_R)$, where
$\mathcal T_R=\mathcal V_{32}F_R$.  Its local vacuum-annihilating
remainder has expectation density at most $384a_Rp$.
\end{lemma}

\begin{proof}
Let $\Psi_R=X\Omega$.  Then
$\|\Psi_R\|^2/N=S_R/2$, and direct vacuum contractions give
\begin{equation}
 \frac{\langle\Omega,[W(f),X]\Omega\rangle}{N}=(f,F_R),\qquad
 \frac{\langle\Omega,[[Q,X],X]\Omega\rangle}{2N}
       =\tfrac12(F_R,\cL F_R).                                 \label{eq:pop-fourth-scalar}
\end{equation}
The operator $B$ conserves magnon number and kills $\Omega$.
Its restriction to the zero-momentum pair sector is $B_2$, so
$[B,X]\Omega=\Psi_{B_2F_R}$.  Since $\|\cL\|_{1\to1}\le8$,
the bound in \eqref{eq:pop-quartic-pair} follows.

Pure pair creators commute with one another, as do pure pair
annihilators.  Hence $[W(u),X]$ conserves number and has only a
scalar vacuum image.  Using \eqref{eq:pop-A-local}, the annihilation
part of $A_R$ kills every two-magnon vector: it requires the two
annihilated sites and a third distinct occupied site.  The creation
part acting on $\Psi_R$ gives
\begin{equation}
 A_R\Psi_R=\sum_{\langle ij\rangle}
 \sum_{\substack{k,l\notin\{i,j\}\\k\ne l}}
 \Delta_{ij}F_R(k)\Delta_{ij}F_R(l)
 b_i^\dagger b_j^\dagger b_k^\dagger b_l^\dagger\Omega.          \label{eq:pop-A-pair-action}
\end{equation}
The two possibilities for the initially occupied endpoint of $ij$
give the second difference in this formula.  The ordered $(k,l)$
sum is therefore required.  Since $A_R\Omega=0$, the factor one
half in $Z_R$ gives \eqref{eq:pop-quartic-four}.  Its absolute
coefficient sum is at most
$\frac12\sum_\mu(\sum_k|\Delta_\mu F_R(k)|)^2\le v_R^2/2$.
Applying Lemma~\ref{lem:pop-local-extraction} to the original
four-site summands of $Z_R$ costs at most $32z_R$ in norm and
$4p$ in their expectations, proving the residual bound.

The per-site norm sum of $\VDM$ is at most two, so
$\mathfrak b(D_R)\le8a_R$, on supports of at most three sites.
The vacuum image is $\VDM\Psi_R$, since $\VDM\Omega=0$.
Its one-magnon part vanishes: the only translation-invariant
one-magnon vector is uniform, and $\VDM$ kills that vector.
For completeness, on a uniform single-magnon state the two
creation amplitudes of each oriented bond cancel, while the
annihilation part requires two occupied endpoints.  For a triple
$\{a,b,t\}$, the remaining coefficient is
\begin{equation}
 \mathcal T_R(\{a,b,t\})
 =\sum_{\substack{\{a,b\}\subset\{a,b,t\}\\|a-b|_1=1}}
 c_{ab}\,[F_R(a-t)-F_R(b-t)],                                  \label{eq:pop-triplet-vertex}
\end{equation}
with $c_{ba}=-c_{ab}$.  This follows by creating the second endpoint
from either of the two possible initial pairs.  It is the exact
coefficient of $\mathcal C_3(\mathcal T_R)$ in the normalization of
Sec.~\ref{sm:current}.  Termwise extraction on three sites multiplies
the norm budget by at most 16; \eqref{eq:pop-vacuum-killing} then
gives $3\cdot16\cdot8a_Rp=384a_Rp$.
\end{proof}

\subsection{First conjugation: every term through sixth order}

We use $\operatorname{ad}_XA=[A,X]$.  The exact unperturbed-number
part of the shifted Hamiltonian is
\begin{equation}
 H_0=Q+s\VDM+dB+h\eps^2\Nop
 \succeq(g_*-d/2)Q+m_*\eps^2\Nop.                              \label{eq:pop-H0-coercivity}
\end{equation}
The inequality is the coercivity estimate of
Sec.~\ref{sm:normalform}.  Taylor expansion of $Q$ through $d^3$,
of $s\VDM$ through $d^2$, of $d(B+W(f))$ through $d^2$, and of
$h\eps^2\Nop$ through $d^2$ gives
\begin{align}
 H_1:=U_R^\dagger(H-Ne_{\rm pol})U_R
 ={}&H_0+dW(q_R)+dA_R+sdD_R+d^2Z_R+h\eps^2dJ_R\notag\\
 &+\frac{sd^2}{2}\operatorname{ad}_X^2\VDM
  +\frac{d^3}{6}\operatorname{ad}_X^3Q
  +\frac{d^3}{2}\operatorname{ad}_X^2B\notag\\
 &+\frac{d^3}{2}\operatorname{ad}_X^2W(f)
  +\frac{h\eps^2d^2}{2}\operatorname{ad}_X^2\Nop
  +\mathcal E_R.                                               \label{eq:pop-first-full-expansion}
\end{align}
The exact trigonometric coefficients are retained.  In particular,
the order-five trigonometric correction within $sdD_R$ is included
and will be bounded explicitly below.

Here is the derivation of the norm remainder, including its numerical
constants.  From \eqref{eq:pop-gradient-first-comm} and
\eqref{eq:pop-pair-comm-count}, the budgets of the first through
fourth commutators of $Q$ are
\begin{equation}
 2v_R,\qquad12v_Ra_R,\qquad96v_Ra_R^2,\qquad960v_Ra_R^3.
                                                                    \label{eq:pop-Q-budgets}
\end{equation}
The initial per-site budgets and support sizes of
$\VDM$, $B+W(f)$, and $\Nop$ are respectively $(2,2)$,
$(7/2,2)$, and $(1,1)$.  Their third-commutator budgets are therefore
\begin{equation}
 2(2a_R)^3(2)(3)(4)=384a_R^3,\quad
 \tfrac72(2a_R)^3(2)(3)(4)=672a_R^3,\quad
 (2a_R)^3(1)(2)(3)=48a_R^3.                                    \label{eq:pop-other-third-budgets}
\end{equation}
The integral Taylor remainder uses unitary conjugation, whose norm
is one.  Multiplication by $d^4/24$, $|s|d^3/6$, $d^4/6$, and
$h_+\eps^2d^3/6$, respectively, proves
\begin{equation}
 \frac{\|\mathcal E_R\|}{N}\le\mathfrak e_R:=%
 40v_Rd^4a_R^3+64|s|d^3a_R^3+112d^4a_R^3
                  +8h_+\eps^2d^3a_R^3
 =O_{\cI,\chi}(\eps^7\ell_\eps^3)=o_{\cI}(\eps^6).
                                                                    \label{eq:pop-first-remainder}
\end{equation}
Thus the constants $40,64,112,8$ arise from finite local commutator
counts.  No estimate of the norm of the extensive generator is used.

We next identify every retained vacuum scalar.  Assign charge $+1$
to $b_i^\dagger$ and charge $-1$ to $b_i$.  The charges of
$Q,B,\Nop$ are zero, those of $X,W(f)$ are $\pm2$, and those of
$\VDM$ are $\pm1$.  A nonzero vacuum diagonal requires charge zero.
This gives the complete list in Table~\ref{tab:pop-first-orders}.
The charge sets may contain algebraically vanishing terms; listing
them only enlarges the possibilities and does not omit any scalar.

\begin{table}[t]
\caption{All retained terms of the first conjugation.  The scalar
column is the vacuum expectation per site, including the displayed
coefficient.  The four constituents of $H_0$ separately kill the
vacuum.}
\label{tab:pop-first-orders}
\small
\begin{tabular}{lcl}
\hline
Term & Leading angle order & Vacuum scalar per site\\
\hline
$Q,\ s\VDM,\ dB,\ h\eps^2\Nop$ & $0,1,2,2$ & $0$\\
$dW(q_R),\ dA_R$ & $2$ & $0$\\
$sdD_R$ & $3$ & $0$; odd magnon charge\\
$d^2Z_R$ & $4$ & $d^2\mathcal A_R$\\
$h\eps^2dJ_R$ & $4$ & $0$; charge $\pm2$\\
$(sd^2/2)\operatorname{ad}_X^2\VDM$ & $5$ & $0$; charges $\pm1,\pm3,\pm5$\\
$(d^3/6)\operatorname{ad}_X^3Q$ & $6$ & $0$; charges $\pm2,\pm6$\\
$(d^3/2)\operatorname{ad}_X^2B$ & $6$ & $(d^3/2)(F_R,B_2F_R)$\\
$(d^3/2)\operatorname{ad}_X^2W(f)$ & $6$ & $0$; charges $\pm2,\pm6$\\
$(h\eps^2d^2/2)\operatorname{ad}_X^2\Nop$ & $6$ & $h\eps^2d^2S_R$\\
\hline
\end{tabular}
\end{table}

The two additional nonzero contractions are
\begin{align}
 \frac{\langle\Omega,\operatorname{ad}_X^2B\,\Omega\rangle}{N}
 &=\frac{2\langle\Psi_R,B\Psi_R\rangle}{N}
   =(F_R,B_2F_R),\notag\\
 \frac{\langle\Omega,\operatorname{ad}_X^2\Nop\,\Omega\rangle}{N}
 &=\frac{2\langle\Psi_R,\Nop\Psi_R\rangle}{N}=2S_R.
                                                                    \label{eq:pop-sixth-first-scalars}
\end{align}
Consequently the common scalar in the lower estimate and the vacuum
trial is
\begin{equation}
 V_R=d^2\mathcal A_R+\frac{d^3}{2}(F_R,B_2F_R)
                                      +h\eps^2d^2S_R.           \label{eq:pop-VR}
\end{equation}
Since $(F,P_{\nn}F)=4F_1^2$ and $(F,\cL F)=F_1$, expansion of $d$
and \eqref{eq:pop-cutoff-error} give
\begin{align}
 V_R&=-c_*\eps^4+\eps^6g_{6,2}(h)+o_{\cI}(\eps^6),\notag\\
 g_{6,2}(h)&=\frac{F_1}{48}+\frac{(F,B_2F)}{16}
                                      +\frac{hS_F}{4}
          =(h-\tfrac12)n_*-\frac{F_1}{96}+\frac{F_1^2}{8}.
                                                                    \label{eq:pop-pair-sixth-scalar}
\end{align}
The notation $g_{6,2}$ denotes this contribution to the scalar, not
an approximation replacing the ground-state energy.

For later absorption it is useful to record explicit bounds on the
last five retained terms of \eqref{eq:pop-first-full-expansion}
after their local scalars have been subtracted.  Their expectation
density is at most $\mathfrak h_R\sqrt p$, where
\begin{equation}
 \mathfrak h_R=%
 144|s|d^2a_R^2+112v_Rd^3a_R^2+504d^3a_R^2
                         +48h_+\eps^2d^2a_R^2.                 \label{eq:pop-first-higher-bound}
\end{equation}
Indeed the respective local budgets before the Taylor prefactors
are $48a_R^2$ on four sites, $96v_Ra_R^2$ on five sites,
$84a_R^2$ on four sites for the combined $B,W(f)$ terms, and
$8a_R^2$ on three sites.  Use \eqref{eq:pop-vacuum-diagonal}; scalar
subtraction increases a local norm by at most two.  These counts
give precisely the safe constants in \eqref{eq:pop-first-higher-bound}.
For every fixed $\nu>0$,
\begin{equation}
 \mathfrak h_R\sqrt p\le\nu\eps^2p+
       \frac{\mathfrak h_R^2}{4\nu\eps^2},\qquad
 \frac{\mathfrak h_R^2}{\eps^2}
       =O_{\cI}(\eps^8\ell_\eps^4)=o_{\cI}(\eps^6).             \label{eq:pop-first-higher-young}
\end{equation}

\subsection{A fixed compact triplet correction}

Use the per-site triplet scalar product
$\langle u,v\rangle_3=\frac16\sum_{\Omega_3}\overline u v$.
Section~\ref{sm:current} proves
\begin{equation}
 \mathcal T=\mathcal V_{32}F,\qquad
 \Gamma_3=\langle\mathcal T,Q_3^{-1}\mathcal T\rangle_3<\infty.
                                                                    \label{eq:pop-Gamma}
\end{equation}
The inverse is a zero-mass quadratic form.  Its finiteness means
that $u\mapsto\langle\mathcal T,u\rangle_3$ is a bounded functional
on the completion of compact triplet kernels in the exchange-energy
norm.  By the Riesz representation theorem and density in this
completion, choose compact, permutation-covariant kernels $\Theta$
with
\begin{equation}
 \delta_\Theta^2=
 \|Q_3^{1/2}\Theta+Q_3^{-1/2}\mathcal T\|_3^2
 \quad\text{arbitrarily small}.                                \label{eq:pop-Theta-error}
\end{equation}
Finite symmetry averaging preserves compactness and cannot increase
this error.  In particular the kernels can be chosen in the symmetry
sector of $\mathcal T$.  Completion of the square gives
\begin{align}
 \mathcal J(\Theta)&:=\langle\Theta,Q_3\Theta\rangle_3
              +2\operatorname{Re}\langle\mathcal T,\Theta\rangle_3
              =-\Gamma_3+\delta_\Theta^2,\notag\\
 \langle\mathcal T+Q_3\Theta,
      Q_3^{-1}(\mathcal T+Q_3\Theta)\rangle_3&=\delta_\Theta^2.
                                                                    \label{eq:pop-completion-square}
\end{align}

The reverse inverse-form comparison of Sec.~\ref{sm:current} has
factor nine for triplets.  With $I_\eta$ denoting the free
four-dimensional inverse form of the zero-extended source, it yields
\begin{equation}
 I_\eta(\mathcal T_R+Q_3\Theta)
 \le18\delta_\Theta^2+2C_4\|\mathcal T_R-\mathcal T\|_1^2,
 \qquad C_4=\langle\delta_0,L_4^{-1}\delta_0\rangle<\infty.
                                                                    \label{eq:pop-small-triplet-source}
\end{equation}
The vertex \eqref{eq:pop-triplet-vertex} and the cutoff-gradient
estimate imply $\|\mathcal T_R-\mathcal T\|_1=O_\chi(R^{-1})$.
For fixed $\Theta$, the residual has bounded $\ell^1$ norm
uniformly in $R$.  The parameter $\Theta$ will remain fixed through
the thermodynamic and weak-angle limits.  No summability assertion
about a massless triplet generator is required.

Let $C_\Theta^+$ be the pure triplet creator with
$C_\Theta^+\Omega=\Psi_\Theta$ and
$N^{-1}\langle\Psi_\Theta,\Psi_\Phi\rangle=
\langle\Theta,\Phi\rangle_3$.  Define
\begin{equation}
 Y=Y_\Theta=C_\Theta^+-C_\Theta^{+\dagger},\qquad
 \tau=\eps^3/2,\qquad V_\Theta=e^{\tau Y},
 \qquad H_2=V_\Theta^\dagger H_1V_\Theta.                        \label{eq:pop-triplet-dressing}
\end{equation}
The physical vacuum trial is $U_RV_\Theta\Omega$, in this order.
Write $a_\Theta$ for the norm sum of triple-generator terms incident
at one site.  It is a finite constant depending only on the fixed
compact kernel.

\subsection{Second conjugation: all terms and its remainder}

Keep the linear commutator of every term of $H_1$ through order
$\eps^3$, and the quadratic commutator of $Q$.  The resulting exact
finite Taylor expansion is
\begin{align}
 H_2={}&H_1+\tau[Q,Y]+\tau s[\VDM,Y]
                 +\tau d[B,Y]+\tau d[W(f),Y]\notag\\
 &+\tau d[[Q,X],Y]+\tau h\eps^2[\Nop,Y]
                 +\tau sd[D_R,Y]
                 +\frac{\tau^2}{2}[[Q,Y],Y]
                 +\mathcal E_{R,\Theta}.                       \label{eq:pop-second-full-expansion}
\end{align}
In particular $\tau s[\VDM,Y]$ is retained with its exact
coefficient; its order-$\eps^6$ trigonometric correction is included.
The combination producing the three order-$\tau d$ terms is exactly
$d(B+W(f)+[Q,X])$, since $W(q_R)+A_R=W(f)+[Q,X]$.

For completeness an explicit second-remainder bound follows from
\eqref{eq:pop-triple-comm-count}.  Set $y=a_\Theta$, $a=a_R$,
$v=v_R$, $z=z_R$, and use $h_+$ for a uniform bound.  Then
\begin{align}
 \frac{\|\mathcal E_{R,\Theta}\|}{N}
 \le{}&2\mathfrak e_R+128|\tau|^3y^3+32|s|\tau^2y^2\notag\\
 &+\tau^2y^2\{d(56+60v)+6h_+\eps^2\}
                 +240|sd|\tau^2ay^2\notag\\
 &+8|\tau|d^2yz+4h_+\eps^2d|\tau|ya
                 +192|\tau s|d^2ya^2\notag\\
 &+160|\tau|vd^3ya^2+336|\tau|d^3ya^2
                 +24h_+\eps^2d^2|\tau|ya^2\notag\\
 ={}&O_{\cI,\Theta,\chi}(\eps^7(1+a_R)^3)
       =o_{\cI,\Theta}(\eps^6).                                \label{eq:pop-second-remainder}
\end{align}
Here the first term bounds
$V_\Theta^\dagger\mathcal E_RV_\Theta-\mathcal E_R$.
The next term is the third-order remainder of $Q$:
$2(2y)^3(2)(4)(6)/6=128y^3$.
The remainder of $s\VDM$ after its first commutator is
$2(2y)^2(2)(4)/2=32y^2$.
For $d(B+W(f)+[Q,X])+h\eps^2\Nop$, the second-commutator
counts are $112y^2$, $120vy^2$, and $12y^2$, respectively,
before the Taylor factor $1/2$.
For $sdD_R$, the count is $480ay^2$ before that factor.
The last two lines bound the first-commutator differences of the
retained fourth-, fifth-, and sixth-order terms of $H_1$.
Their budgets, in the same order, are
$8yz$, $4ya$, $192ya^2$, $160vya^2$, $336ya^2$, and $24ya^2$,
including the first Taylor prefactors from
\eqref{eq:pop-first-full-expansion}.  Thus every omitted term in
the second conjugation is accounted for by a finite local norm
estimate.

The triplet generator has charges $\pm3$.  Table~\ref{tab:pop-second-orders}
therefore exhausts the possible new scalars.  The only nonzero ones
are calculated directly as
\begin{align}
 \frac{\langle\Omega,[D_R,Y]\Omega\rangle}{N}
       &=2\operatorname{Re}\langle\mathcal T_R,\Theta\rangle_3,
       \notag\\
 \frac{\langle\Omega,[[Q,Y],Y]\Omega\rangle}{N}
       &=2\langle\Theta,Q_3\Theta\rangle_3.                     \label{eq:pop-second-scalars}
\end{align}
Compactness of $\Theta$ implies
$\langle\mathcal T_R,\Theta\rangle_3=
\langle\mathcal T,\Theta\rangle_3$ once $R$ contains all pair
separations in its support.  Hence the common retained scalar is
\begin{align}
 V_{\eps,\Theta}
   &=V_R+2\tau sd\operatorname{Re}\langle\mathcal T_R,\Theta\rangle_3
              +\tau^2\langle\Theta,Q_3\Theta\rangle_3,          \label{eq:pop-full-scalar}\\
   &=-c_*\eps^4+\eps^6
       [g_{6,2}(h)+\mathcal J(\Theta)/4]
       +o_{\cI,\Theta}(\eps^6).                               \label{eq:pop-full-scalar-asymptotic}
\end{align}
Both $\tau sd$ and $\tau^2$ have leading coefficient $\eps^6/4$.
This fixes the normalization of the triplet correction.

\begin{table}[t]
\caption{Every new retained term of the second conjugation.  Scalars
are stated before multiplication by the displayed prefactor.  The
coefficient of $\tau s[\VDM,Y]$ is exact, so its zero scalar also
excludes a contribution from its order-six trigonometric part.}
\label{tab:pop-second-orders}
\small
\begin{tabular}{lcl}
\hline
Term & Leading angle order & Vacuum scalar per site\\
\hline
$\tau[Q,Y]$ & $3$ & $0$; charge $\pm3$\\
$\tau s[\VDM,Y]$ & $4$ & $0$; charges $\pm2,\pm4$\\
$\tau d[B,Y]$ & $5$ & $0$; charge $\pm3$\\
$\tau d[W(f),Y]$ & $5$ & $0$; charges $\pm1,\pm5$\\
$\tau d[[Q,X],Y]$ & $5$ & $0$; charges $\pm1,\pm5$\\
$\tau h\eps^2[\Nop,Y]$ & $5$ & $0$; charge $\pm3$\\
$\tau sd[D_R,Y]$ & $6$ & $2\operatorname{Re}\langle\mathcal T_R,\Theta\rangle_3$\\
$(\tau^2/2)[[Q,Y],Y]$ & $6$ & $2\langle\Theta,Q_3\Theta\rangle_3$\\
\hline
\end{tabular}
\end{table}

\subsection{Bounds on every nonscalar term}

In this subsection $p$ and $\mathfrak q$ denote the number and
exchange densities of an arbitrary infinite-volume
translation-invariant state to which $H_2$ is applied.  We give the
estimates before specializing to an energy minimizer.

\paragraph{The quadratic residuals.}
For every fixed $\beta>0$, \eqref{eq:pop-A-bound} and the
zero-total-source path bound of Sec.~\ref{sm:normalform} give
\begin{align}
 d|\omega(A_R)|/N
 &\le\beta\mathfrak q+(36v_R^2d^2/\beta)p,\notag\\
 d|\omega(W(q_R))|/N
 &\le\beta\mathfrak q+16d^2\gamma_R^2/\beta.                   \label{eq:pop-quadratic-absorption}
\end{align}
The first number coefficient is $O(\eps^4)=o(\eps^2)$.
For $R=\lceil\eps^{-2}\rceil$ the second scalar is
$O(\eps^8)=o(\eps^6)$.

\paragraph{The leading cubic residual.}
By local extraction, $[Q,Y]$ has pure triple source
$\mathcal C_3(Q_3\Theta)$ and a local vacuum-annihilating
remainder.  Indeed $[Q,Y]\Omega=Q\Psi_\Theta$; its local norm
budget is at most $8a_\Theta$ on four sites.  Extraction and
\eqref{eq:pop-vacuum-killing} bound the residual density by
$1024a_\Theta p$.  Lemma~\ref{lem:pop-pair-quartet-sources}
therefore gives
\begin{equation}
 sdD_R+\tau[Q,Y]
 =\tau\mathcal C_3(\mathcal T_R+Q_3\Theta)
       +\tau R_{3,R,\Theta}+(sd-\tau)D_R,                       \label{eq:pop-combined-cubic}
\end{equation}
where
\begin{equation}
 |\omega(R_{3,R,\Theta})|/N
       \le(384a_R+1024a_\Theta)p.                              \label{eq:pop-cubic-ann}
\end{equation}
The coefficient $\tau$ makes this an $o_\Theta(\eps^2)p$ cost.
Moreover $|sd-\tau|\le\eps^5/8$ for $0<\eps\le1$:
use $|s-\eps|\le\eps^3/6$, $d\le\eps^2/2$, and
$|d-\eps^2/2|\le\eps^4/24$.
Every three-site summand of $D_R$ has zero vacuum diagonal, and
$\mathfrak b(D_R)\le8a_R$, so
\begin{equation}
 |(sd-\tau)\omega(D_R)|/N\le6\eps^5a_R\sqrt p.                 \label{eq:pop-trig-cubic-error}
\end{equation}
Young's inequality against any fixed part of $\eps^2p$ gives
scalar cost $O(\eps^8a_R^2)=o(\eps^6)$.

Apply \eqref{eq:creation-current} to the first term in
\eqref{eq:pop-combined-cubic}, with regulator $\eta=\eps^2$.
Its source has bounded $\ell^1$ norm for fixed $\Theta$.
Assign $g_*\mathfrak q/16$ to the first current term and another
$g_*\mathfrak q/16$ to the mixed current-population term.  The
former costs
$C_3^2\eps^6I_{\eps^2}(\mathcal T_R+Q_3\Theta)/g_*$.
The latter leaves an $O_\Theta(\eps^4)p$ number coefficient.
The regulator and remaining population terms fit within
$m_*\eps^2p/8$ for sufficiently small angle, with scalar
$O_\Theta(\eps^8\ell_\eps)$.  Inserting
\eqref{eq:pop-small-triplet-source} proves
\begin{equation}
 \frac{|\tau\omega(\mathcal C_3(\mathcal T_R+Q_3\Theta))|}{N}
 \le\frac{g_*}{8}\mathfrak q+\frac{m_*}{8}\eps^2p
       +B_\cI\delta_\Theta^2\eps^6+o_{\cI,\Theta}(\eps^6),
 \qquad B_\cI=100C_3^2/g_*.                                   \label{eq:pop-cubic-small-loss}
\end{equation}
The factor 18 from \eqref{eq:pop-small-triplet-source} is already
covered by this choice.  Crucially $B_\cI$ is independent of
$\Theta$; only the vanishing remainder and its angle threshold
may depend on the chosen compact kernel.

\paragraph{The fourth-order sources, including $\tau s[\VDM,Y]$.}
The pair source from $d^2Z_R+h\eps^2dJ_R$ is exactly
\begin{equation}
 W(d^2g_R+2h\eps^2dF_R).                                       \label{eq:pop-combined-fourth-pair}
\end{equation}
After extracting $\eps^4$ its coefficient $\ell^1$ norm is
$O_\cI(a_R)$.  The four-particle source is
$d^2(C_{4,R}^++C_{4,R}^{+\dagger})$; its normalized coefficient
$\ell^1$ norm is $O(v_R^2)=O(1)$.
The local remainder from $Z_R$ costs $128d^2z_Rp=o(\eps^2)p$.

The operator $[\VDM,Y]$ has local norm budget at most
$8a_\Theta$ and support at most four sites.  Its only vacuum
images are pair and quartet creators, because the charges are
$\pm2,\pm4$; in particular its scalar is exactly zero.
Termwise extraction gives compact source kernels with finite
$\ell^1$ norms depending only on $\Theta$, and a remainder
bounded in density by $1024a_\Theta p$.  The prefactor
$\tau s=O(\eps^4)$ makes this remainder $o_\Theta(\eps^2)p$.
The pair and quartet kernels can be added to the kernels in
\eqref{eq:pop-combined-fourth-pair} and
\eqref{eq:pop-quartic-four}.

The fourth-order consequence of \eqref{eq:creation-current}
therefore applies to their sum, using
$g_*\mathfrak q/8+m_*\eps^2p/8$ and scalar
$o_{\cI,\Theta}(\eps^6)$.  To display the powers explicitly,
the pair inverse form is $O_\Theta(\ell_\eps^3)$ and the
six-dimensional quartet inverse form is $O_\Theta(1)$.
Their first current terms cost $O_\Theta(\eps^8\ell_\eps^3)$
and $O_\Theta(\eps^8)$, respectively.  The mixed-current term
has number coefficient $O_\Theta(\eps^6\ell_\eps^2)$.
The regulator terms have scalar costs at most
$O_\Theta(\eps^8\ell_\eps^2)$ and $O_\Theta(\eps^{10})$.
All of these scalars are $o_\Theta(\eps^6)$.

\paragraph{Each remaining fifth- and sixth-order term.}
The five higher-order terms retained from the first conjugation
are bounded by \eqref{eq:pop-first-higher-bound} after their
scalars have been removed.  This includes
$\operatorname{ad}_X^2\VDM$, $\operatorname{ad}_X^3Q$,
$\operatorname{ad}_X^2B$, $\operatorname{ad}_X^2W(f)$, and
$\operatorname{ad}_X^2\Nop$ separately as listed in
Table~\ref{tab:pop-first-orders}.

The four new fifth-order commutators in
\eqref{eq:pop-second-full-expansion} have zero vacuum diagonal.
Their norm budgets, before their displayed prefactors, are
$12a_\Theta$, $2a_\Theta$, $12v_Ra_\Theta$, and $2a_\Theta$,
on at most four, four, five, and three sites, respectively.
For the two new sixth-order commutators, remove the scalars in
\eqref{eq:pop-second-scalars} from each local summand.
The norm budgets before scalar subtraction are
$48a_Ra_\Theta$ on five sites for $[D_R,Y]$, and
$64a_\Theta^2$ on six sites for $[[Q,Y],Y]$.
It follows from Lemma~\ref{lem:pop-local-extraction} that all six
new terms, with their scalars removed, have combined density
bounded by $\mathfrak h_{R,\Theta}\sqrt p$, where
\begin{align}
 \mathfrak h_{R,\Theta}={}&
 7a_\Theta|\tau|\{d(14+12v_R)+2h_+\eps^2\}
       +672|\tau sd|a_Ra_\Theta
       +512\tau^2a_\Theta^2.                                  \label{eq:pop-second-higher-bound}
\end{align}
The constants use $3\sqrt5<7$ and $3\sqrt6<8$, together with
the factor two for scalar subtraction.  Thus
\begin{equation}
 \frac{(\mathfrak h_R+\mathfrak h_{R,\Theta}
                         +6\eps^5a_R)^2}{\eps^2}
     =O_{\cI,\Theta}(\eps^8\ell_\eps^4)
     =o_{\cI,\Theta}(\eps^6).                                 \label{eq:pop-all-higher-young}
\end{equation}
One fixed allocation $m_*\eps^2p/8$ controls this entire group
by Young's inequality.  Equations
\eqref{eq:pop-first-remainder} and
\eqref{eq:pop-second-remainder} cover all remaining Taylor terms.
This enumeration supplies a bound for every term in both
conjugations through order six, including every term whose scalar
vanishes by its charge rather than by its norm.

\subsection{Energy sandwich and the approximation limit}

We now assemble the inequalities in infinite-volume expectation
density.  Allocate $g_*\mathfrak q/8$ separately to the two
quadratic residuals in \eqref{eq:pop-quadratic-absorption}, and
$g_*\mathfrak q/8$ to each of the cubic and combined fourth-order
current bounds.  These fixed allocations total $g_*\mathfrak q/2$.
The three fixed number allocations, for the cubic current, the
fourth-order currents, and \eqref{eq:pop-all-higher-young}, total
$3m_*\eps^2p/8$.  Require the sum of the other number
coefficients, all $o_{\cI,\Theta}(\eps^2)$ as shown above, to be
at most $m_*\eps^2/8$.  Also require $d/2\le g_*/8$.
The remaining positive coefficients in
\eqref{eq:pop-H0-coercivity} are then at least
\begin{equation}
 \kappa_6=g_*/4,\qquad \mu_6=m_*/4.                            \label{eq:pop-final-coercivity}
\end{equation}
These small-angle requirements hold for each fixed compact
$\Theta$, uniformly for $h\in\cI$.  We have proved
\begin{equation}
 \frac{\omega(H_2)}N\ge V_{\eps,\Theta}
       -B_\cI\delta_\Theta^2\eps^6-o_{\cI,\Theta}(\eps^6)
       +\kappa_6\mathfrak q+\mu_6\eps^2p.                    \label{eq:pop-full-lower}
\end{equation}
Here $\omega(H_2)/N$ denotes the transformed interaction's
expectation density, not an extensive operator on an infinite
Hilbert space.

We specify the limit justification.  At fixed $\eps$ and compact
$\Theta$, both generators are finite range.  The local commutator
bounds \eqref{eq:pop-pair-comm-count} and
\eqref{eq:pop-triple-comm-count} make their local series convergent
for sufficiently small $d a_R$ and $|\tau|a_\Theta$; the same
estimates give the thermodynamic limits of the conjugations.
Thus these automorphisms may be inserted into the translation-invariant
energy-density variational principle.  For the current estimate,
at fixed $\eta=\eps^2$ truncate its absolutely summable coefficient
current, take the thermodynamic limit of the finite-support
identities, and then remove the current truncation, using the
$\ell^1$ continuity established in Sec.~\ref{sm:current}.
This proves \eqref{eq:pop-full-lower} for the required thermodynamic
states.  It makes no assertion about a fixed torus as $\eta\to0$.

Let $E_\eps(h)=e_0(\eps,h)-e_{\rm pol}(\eps,h)$.
The vacuum trial $U_RV_\Theta\Omega$ has energy density
\begin{equation}
 E_\eps(h)\le V_{\eps,\Theta}+o_{\cI,\Theta}(\eps^6).         \label{eq:pop-full-upper}
\end{equation}
All retained nonscalar terms have zero vacuum expectation by
Tables~\ref{tab:pop-first-orders} and~\ref{tab:pop-second-orders};
the norm remainders justify the displayed error directly.
Conversely, apply \eqref{eq:pop-full-lower} to the twice-dressed
image of an energy-minimizing state and drop its nonnegative
densities.  Combining with \eqref{eq:pop-completion-square} and
\eqref{eq:pop-full-scalar-asymptotic} gives, uniformly on $\cI$,
\begin{equation}
 \left|\frac{E_\eps(h)+c_*\eps^4}{\eps^6}
       -\left[g_{6,2}(h)-\frac{\Gamma_3}{4}\right]\right|
 \le(B_\cI+\tfrac14)\delta_\Theta^2
                         +o_{\cI,\Theta}(1).                  \label{eq:pop-energy-accuracy}
\end{equation}
For a prescribed $\zeta>0$, first choose a compact $\Theta$ with
$\delta_\Theta^2\le\zeta/[2(B_\cI+1/4)]$.  Keep this kernel
fixed and choose the angle threshold so that the second term in
\eqref{eq:pop-energy-accuracy} is at most $\zeta/2$ and the
displayed allocations hold.  All choices are independent of the
selected field and ground state.  This proves Theorem~\ref{thm:energy}:
\begin{align}
 e_0(\eps,h)-e_{\rm pol}(\eps,h)
   &=-c_*\eps^4+\eps^6g_6(h)+o_\cI(\eps^6),                  \label{eq:energy-sixth}\\
 g_6(h)&=(h-\tfrac12)n_*-\frac{F_1}{96}
                        +\frac{F_1^2}{8}-\frac{\Gamma_3}{4},
 &g_6'(h)&=n_*.                                                \label{eq:pop-g6}
\end{align}
The thermodynamic limit precedes the weak-angle limit.  Improvement
of the compact-kernel accuracy is a final proof-accuracy limit
over fixed kernels.  It does not presume a quantitative relation
between the support of $\Theta$ and $\eps$.

\subsection{Population from concavity}

To control all fields of the desired compact interval, apply
\eqref{eq:energy-sixth} on a compact interval
$\cI^+\Subset(3/2,\infty)$ containing $\cI$ in its interior.
For an energy-minimizing state at $h$ with population $p$, use that
same state as a trial state at $h\pm t$.  The polarized energy has
field derivative $-\eps^2/2$, so subtraction of $e_{\rm pol}$ gives
the exact supporting-slope inequalities
\begin{equation}
 \frac{E_\eps(h+t)-E_\eps(h)}{t}
 \le\eps^2p\le
 \frac{E_\eps(h)-E_\eps(h-t)}{t}.                              \label{eq:pop-concavity-slopes}
\end{equation}
They hold for every minimizer, including at a nondifferentiable
field.  Choose one fixed $t>0$ such that $h\pm t\in\cI^+$ for
all $h\in\cI$.  Write the uniform remainder in
\eqref{eq:energy-sixth} as $\eps^6r_\eps(h)$ on $\cI^+$.
Because $g_6$ is affine, both difference quotients of its coefficient
equal $n_*$.  Equation~\eqref{eq:pop-concavity-slopes} therefore yields
\begin{equation}
 \sup_{h\in\cI,\,\omega}
 \left|\frac{\omega(n_0)}{\eps^4}-n_*\right|
 \le\frac{2}{t}\sup_{h\in\cI^+}|r_\eps(h)|\longrightarrow0.
                                                                    \label{eq:pop-concavity-conclusion}
\end{equation}
This proves Theorem~\ref{thm:population}, without assuming uniqueness
of the minimizer or differentiability of the exact energy.

\subsection{Independent population proof by local dressing}

There is also a direct proof from the positive number term in
\eqref{eq:pop-full-lower}.  Denote the physical state by $\omega$,
its pair-dressed image by
$\widetilde\omega(A)=\omega(U_RA U_R^\dagger)$, and its
twice-dressed image by
$\widehat\omega_\Theta(A)=
\widetilde\omega(V_\Theta A V_\Theta^\dagger)$.
Let their populations be $p$, $\widetilde p$, and
$\widehat p_\Theta$.  Equations \eqref{eq:pop-full-lower} and
\eqref{eq:pop-full-upper} give, for every fixed compact $\Theta$,
\begin{equation}
 \limsup_{\eps\downarrow0}\sup_{h\in\cI,\omega}
       \frac{\widehat p_\Theta}{\eps^4}
 \le\frac{B_\cI}{\mu_6}\delta_\Theta^2.                        \label{eq:pop-double-density}
\end{equation}
Only triple-generator terms meeting the origin contribute to
$[b_0,Y]$.  The exact integral conjugation formula gives
\begin{equation}
 \|V_\Theta^\dagger b_0V_\Theta-b_0\|
 \le2a_\Theta|\tau|,
 \qquad
 |\sqrt{\widetilde p}-\sqrt{\widehat p_\Theta}|
 \le2a_\Theta|\tau|.                                         \label{eq:pop-triplet-local-transfer}
\end{equation}
Divide by $\eps^2$, take the angle limit with $\Theta$ fixed, and
then make $\delta_\Theta$ arbitrarily small.  The quantity
$\widetilde p$ does not depend on $\Theta$, so this proves
\begin{equation}
 \widetilde p=o_\cI(\eps^4),\qquad R=\lceil\eps^{-2}\rceil.
                                                                    \label{eq:dressed-density-sharp}
\end{equation}

For completeness the remaining transfer to the physical population
is local and two sided.  The exact first commutator and its
second-order norm remainder are
\begin{align}
 C_0:=[b_0,X]&=(1-2n_0)\sum_rF_R(r)b_r^\dagger,\notag\\
 U_R^\dagger b_0U_R&=b_0+dC_0+\mathcal F_0,
 &\|\mathcal F_0\|&\le4d^2a_R^2.                             \label{eq:pop-pair-local-transfer}
\end{align}
The latter bound follows from
$\|\operatorname{ad}_X^2b_0\|\le8a_R^2$ and the integral
Taylor remainder.  Since $(1-2n_0)^2=1$,
\begin{equation}
 q_C:=\widetilde\omega(C_0^\dagger C_0)
 =\sum_{r,t}F_R(r)F_R(t)\widetilde\omega(b_r b_t^\dagger).
                                                                    \label{eq:pop-C0-Gram}
\end{equation}
The normal-coherence matrix
$\widetilde\omega(b_t^\dagger b_r)$ is positive, and each entry
has modulus at most $\widetilde p$.  Using
$b_r b_r^\dagger=1-n_r$ and commuting distinct sites gives
\begin{equation}
 (1-2\widetilde p)S_R\le q_C\le S_R+\widetilde p a_R^2,
 \qquad |q_C-S_R|\le\widetilde p(a_R^2+2S_R).                  \label{eq:pop-C0-bound}
\end{equation}
The reverse triangle inequality in the GNS norm, followed by
$|\sqrt x-\sqrt y|\le\sqrt{|x-y|}$, now gives
\begin{equation}
 |\sqrt p-d\sqrt{S_R}|
 \le\sqrt{\widetilde p}+4d^2a_R^2
             +d\sqrt{\widetilde p(a_R^2+2S_R)}.                 \label{eq:pop-final-local-bound}
\end{equation}
Divide by $\eps^2$, use \eqref{eq:dressed-density-sharp},
$a_R=O(\ell_\eps)$, $S_R\to S_F$, and $d/\eps^2\to1/2$.
It follows that $\sqrt p/\eps^2\to\sqrt{S_F}/2$ uniformly,
and hence $p/\eps^4\to S_F/4=n_*$.  This argument is independent
of field differentiation.  Neither population proof uses
stationarity or assumes that a physical ground state lies in a
fixed magnon-number sector.

\section{Creation currents and finite-particle resolvents}
\label{sm:current}

The estimates in this section apply to arbitrary infinite-volume
translation-invariant spin states.  The finite-particle resolvents are used
to construct coefficient currents; they do not replace the many-body state.
We write
\begin{equation}
 p=\omega(n_0),\qquad
 \mathfrak q_\mu=\omega(P^-_{0,e_\mu}),\qquad
 \mathfrak q=\mathfrak q_x+\mathfrak q_y.
 \label{eq:current-state-densities}
\end{equation}
All constants in the current inequality are independent of the state, field,
and volume.  Field-dependent constants in its applications are uniform for
\(h\in\cI\).

\subsection{Configuration graphs and creation densities}

For \(n\in\{2,3,4\}\), set \(m_n=n-1\), \(D_n=2m_n\), and define
\begin{equation}
 \Omega_n=\left\{(r_1,\ldots,r_{m_n})\in(\mathbb Z^2)^{m_n}:
 r_j\ne0,\quad r_j\ne r_k\text{ for }j\ne k\right\}.
 \label{eq:current-configuration-space}
\end{equation}
Thus the first particle is at the origin and the other particles have
labeled relative coordinates.  Let \(L_{\Omega_n}\) be the graph Laplacian
whose edges move one relative coordinate by one nearest-neighbor step,
provided that the endpoint remains in \(\Omega_n\).  Each undirected edge
has weight one.  The corresponding free graph is \(\mathbb Z^{D_n}\), with
Laplacian \(L_{D_n}\).  Their Dirichlet forms, denoted by
\(\mathcal E_{\Omega_n}\) and \(\mathcal E_{D_n}\), are sums of squared
differences over undirected edges.  Both use counting measure.

For \(R_n\in\ell^1(\Omega_n)\), the creator-only and Hermitian translation
sums have the conventions
\begin{align}
 \mathcal C_n^+(R_n)
 &=\frac1{n!}\sum_i\sum_{r\in\Omega_n}R_n(r)
       b_i^\dagger\prod_{j=1}^{m_n}b_{i+r_j}^\dagger,
 &\mathcal C_n(R_n)&=\mathcal C_n^+(R_n)
                         +\mathcal C_n^+(R_n)^\dagger.
 \label{eq:current-creator-convention}
\end{align}
No extensive operator is needed in infinite volume.  Its expectation
density is the absolutely convergent quantity
\begin{equation}
 \mathfrak c_n(R_n;\omega)
 =\frac{2}{n!}\operatorname{Re}
   \sum_{r\in\Omega_n}R_n(r)
       \omega\!\left(b_0^\dagger\prod_{j=1}^{m_n}b_{r_j}^\dagger\right).
 \label{eq:current-density-definition}
\end{equation}
\begin{lemma}[Continuity of creation densities]
\label{lem:current-density-continuity}
For every translation-invariant state and \(R_n,S_n\in\ell^1(\Omega_n)\),
\begin{equation}
 |\mathfrak c_n(R_n;\omega)-\mathfrak c_n(S_n;\omega)|
 \le\frac{2}{n!}\|R_n-S_n\|_1.
 \label{eq:current-density-continuity}
\end{equation}
The creator-only density has continuity constant \(1/n!\).
For a finite kernel on a translation-invariant torus containing its
relative support without identifications, the density equals
\(\omega(\mathcal C_n(R_n))/N\).
\end{lemma}
\begin{proof}
Each product of distinct spin creators has norm one, so its expectation
has absolute value at most one. The difference of the two series in
\eqref{eq:current-density-definition} is therefore bounded by
\(2\sum_r|R_n(r)-S_n(r)|/n!\); this also proves absolute convergence.
Omitting the Hermitian conjugate removes the factor two. For a finite
kernel, translation invariance makes each of the \(N\) anchor
expectations equal, proving the finite-volume identity.
\end{proof}
The kernel need not be permutation invariant for these estimates.
If it is invariant under permutations of the particles,
including a change of the anchor, the factor \(1/n!\) counts each unordered
configuration once per translation volume.

\subsection{Retraction, routing, and congestion}

We orient every free or allowed coordinate edge positively, once in each
of the \(D_n\) coordinate directions.  A current changes sign if its edge
orientation is reversed.  Its \(\ell^p\) norm is taken over these positive
edges.  A fixed ordering is used for every finite choice below; it makes
all constructions deterministic without affecting the bounds.

\begin{lemma}[Retraction and bounded routing]
\label{lem:current-routing}
There is a retraction \(\pi_n:\mathbb Z^{D_n}\to\Omega_n\), equal to the
identity on \(\Omega_n\), and a path in \(\Omega_n\) for each positively
oriented free edge, joining the retractions of its endpoints, with the
following properties:
\begin{enumerate}
 \item every configuration has at most \(5^{m_n}\) preimages;
 \item every path has length at most \(384\);
 \item each allowed undirected edge is traversed by at most
       \(D_n67^{D_n}\) paths, and no path traverses it more than once.
\end{enumerate}
Consequently, for \((E_nv)(z)=v(\pi_n z)\),
\begin{equation}
 \|E_nv\|_2^2\le5^{m_n}\|v\|_2^2,
 \qquad
 \mathcal E_{D_n}(E_nv)
 \le\mathsf C_{E,n}\mathcal E_{\Omega_n}(v),
 \qquad
 \mathsf C_{E,n}=384D_n67^{D_n}.
 \label{eq:current-extension}
\end{equation}
\end{lemma}

\begin{proof}
Write \(m=m_n\le3\).  The ordered offset list is
\(\mathcal O_5=(0,e_x,-e_x,e_y,-e_y)\).  If \(z=(z_1,\ldots,z_m)\)
already belongs to \(\Omega_n\), set \(\pi_nz=z\).  Otherwise choose the
coordinates successively: at step \(j\), choose the first
\(z_j+\delta\), \(\delta\in\mathcal O_5\), outside the origin and the
\(j-1\) coordinates already chosen.  There are at most \(j\le3\)
forbidden candidates among five.  This defines a legal configuration and
moves each planar coordinate by at most one lattice step.  For a specified
image coordinate, its original coordinate is one of its five inverse
shifts.  Thus a specified configuration has at most \(5^m\) preimages,
including any identity preimage.  This proves the first norm bound.

We first construct a planar path avoiding at most three forbidden vertices.
Let \(u,v\) be allowed endpoints and \(D=|u-v|_1\).  If \(D=0\), use the
empty path.  Otherwise consider the four disjoint square boundaries,
centered at \(u\), of radii \(D+1,D+2,D+3,D+4\) in the planar
\(\ell^\infty\) norm.  At least one contains no forbidden vertex.  Choose
the first such boundary, of radius \(R\).  Both endpoints lie strictly
inside it.  From an allowed endpoint a forbidden vertex can lie on at most
one of its four axis rays, since the rays meet only at that endpoint.
Therefore one ray from each endpoint reaches the chosen boundary without
meeting an obstacle.  Choose the first admissible rays and join their exit
points by the shorter boundary arc.  The ray from \(u\) has length \(R\);
the ray from \(v\) has length at most \(R+D\).  The shorter arc has length
at most half the square perimeter, \(4R\).  The total length is therefore
at most
\begin{equation}
 6R+D\le7D+24,
 \label{eq:current-planar-path-length}
\end{equation}
and the entire path lies within radius \(D+4\) of \(u\).  A fixed rule
resolves ties in ray and arc choices.

Now let \(z\to z+e_\ell\) be a positive free edge, and write its image
endpoints as \(a=(a_1,\ldots,a_m)\) and \(b=(b_1,\ldots,b_m)\).
Because each retraction changes a coordinate by at most one step,
\(|a_j-b_j|_1\le3\) for every \(j\).  Choose parking positions
\begin{equation}
 p_j\in a_j+\{-2,-1,0,1,2\}^2
 \setminus\bigl(\{0\}\cup\{a_1,\ldots,a_m\}
                 \cup\{b_1,\ldots,b_m\}\cup\{p_1,\ldots,p_{j-1}\}\bigr).
 \label{eq:current-parking-choice}
\end{equation}
The displayed square has 25 sites.  The number of forbidden sites is at
most \(1+2m+j-1\le3m\le9\), so the first available choice always exists.
Move particles \(1,\ldots,m\) successively from \(a_j\) to \(p_j\),
then successively from \(p_j\) to \(b_j\).  During each move the origin
and the other \(m-1\) particles are the only obstacles.  Both endpoints
are free: parking sites avoid all original and target sites and each other;
target sites are distinct and avoid every parking site.  Thus the planar
construction applies at every step with at most three obstacles.

A parking move has Manhattan distance at most four and length at most
52.  An un-parking move has distance at most \(4+3=7\) and length at most
73.  The configuration path has length at most
\begin{equation}
 m(52+73)\le375<384.
 \label{eq:current-configuration-path-length}
\end{equation}
For later counting we also record its location.  A parking path is within
eight of its center \(a_j\), which is within one of \(z_j\), so it is
within nine of \(z_j\).  An un-parking path is within eleven of \(p_j\),
which is within three of \(z_j\), so it is within fourteen of \(z_j\).
The unmoved coordinates satisfy the same bound.  Every path vertex is
therefore within coordinatewise distance fourteen, and hence within the
looser radius 32, of \(z\).  Deterministic loop erasure does not increase
length or change confinement.  Use the empty path if the two images coincide.

To count congestion, fix an allowed undirected edge and fix one of its
endpoints once and for all.  If a selected path uses that edge, an endpoint
on the path is within 32 of the originating free point \(z\).  The fixed
endpoint is at distance at most one from it.  Hence
\(z\) lies in the radius-33 cube about this \emph{fixed} endpoint, with
at most \(67^{D_n}\) lattice points.  There are \(D_n\) choices of the
positive free-edge direction.  This gives \(D_n67^{D_n}\) paths, without
an additional factor for choosing an endpoint.  Loop erasure ensures that
a path uses each edge at most once.

For each free edge, telescope \(E_nv\) along its allowed path and apply
Cauchy-Schwarz, with the path length bounded by 384.  Sum over free edges
and use the congestion count.  This proves the second inequality in
\eqref{eq:current-extension}.  It holds first for finitely supported
functions and then by closure.  Bounded displacement of \(\pi_n\) ensures
that extension preserves finite support.
\end{proof}

\begin{lemma}[Pushforward of currents]
\label{lem:current-push}
Push a free-edge current \(J\) along the paths of
Lemma~\ref{lem:current-routing}, with the sign of the traversal, and denote
the resulting allowed current by \(P_nJ\).  For
\(J\in\ell^1\cap\ell^2\),
\begin{align}
 \|P_nJ\|_1&\le384\|J\|_1,
 &\|P_nJ\|_2^2&\le\mathsf C_{E,n}\|J\|_2^2,\label{eq:current-push-norms}\\
 \operatorname{div}(P_nJ)&=(\pi_n)_*\operatorname{div}J,
 &((\pi_n)_*g)(r)&=\sum_{z:\pi_nz=r}g(z),\label{eq:current-push-divergence}\\
 \| (\pi_n)_*g\|_1&\le\|g\|_1,
 &\| (\pi_n)_*g\|_2&\le5^{m_n/2}\|g\|_2.
 \label{eq:current-push-vertices}
\end{align}
Here divergence is outgoing minus incoming current.
\end{lemma}

\begin{proof}
The \(\ell^1\) edge bound follows by summing path lengths.  At an allowed
edge, Cauchy-Schwarz bounds the square of the pushed current by the number
of paths using the edge times the sum of the squared contributing free
currents.  Summing and then using the length bound gives the \(\ell^2\)
constant \(384D_n67^{D_n}\).  The divergence of the current on one selected
path is its initial endpoint mass minus its final endpoint mass; all
interior contributions cancel.  Summing proves
\eqref{eq:current-push-divergence} for finite currents.  The divergence map
has \(\ell^1\) norm at most two, so this identity extends to absolutely
summable currents.  Finally the vertex \(\ell^1\) bound is the triangle
inequality, and its \(\ell^2\) bound is Cauchy-Schwarz on fibers containing
at most \(5^{m_n}\) points.
\end{proof}

\subsection{Free resolvent and current estimates}

Write \(\bar R_n\) for zero extension to the free lattice and, for
\(0<\eta\le1\), define
\begin{equation}
 G_\eta=(L_{D_n}+\eta)^{-1}\bar R_n,\qquad
 I_\eta(R_n)=(\bar R_n,(L_{D_n}+\eta)^{-1}\bar R_n),\qquad
 M_n(R_n)=\|R_n\|_1.
 \label{eq:current-free-definitions}
\end{equation}
Inner products in \eqref{eq:current-free-definitions} use counting measure.
The use of \(G_\eta\) here is restricted to the free resolvent function;
the triplet dressing kernel below is denoted by \(\Theta\).

\begin{lemma}[Free current bounds]
\label{lem:current-free-resolvent}
Let \(J_\eta(z,z+e_j)=G_\eta(z)-G_\eta(z+e_j)\).  Then
\begin{align}
 \operatorname{div}J_\eta&=\bar R_n-\eta G_\eta,
 &\|J_\eta\|_2^2&\le I_\eta(R_n),\label{eq:current-free-divergence}\\
 \|G_\eta\|_1&\le\eta^{-1}M_n(R_n),
 &\|J_\eta\|_1&\le\frac{D_n\pi}{2\sqrt\eta}M_n(R_n),
 \label{eq:current-free-lone}\\
 \|G_\eta\|_2&\le H_{D_n}(\eta)M_n(R_n),\label{eq:current-free-ltwo}
\end{align}
where
\begin{equation}
 H_2(\eta)=\frac{\sqrt\pi}{4\sqrt\eta},\qquad
 H_4(\eta)=\frac\pi{16}\sqrt{\log(1+16/\eta)},\qquad
 H_6(\eta)=\sqrt{\frac{3\pi^3}{1024}}.
 \label{eq:current-H-constants}
\end{equation}
\end{lemma}

\begin{proof}
The divergence identity is the resolvent equation.  Its quadratic form gives
\(\|J_\eta\|_2^2+\eta\|G_\eta\|_2^2=I_\eta(R_n)\).
For the \(\ell^1\) bounds use
\(G_\eta=\int_0^\infty e^{-\eta t}e^{-tL_{D_n}}\bar R_n\,\dd t\).
The free heat kernel is positive, has total mass one, and factors over
coordinates; this proves the bound on \(\|G_\eta\|_1\).

For completeness, the one-dimensional heat kernel \(p_t(k)\) is symmetric
and nonincreasing in \(|k|\).  To see this without using an asymptotic
formula, approximate it by powers of the lazy nearest-neighbor kernel
with masses \(t/M,1-2t/M,t/M\), taking \(M\ge3t\), and then let
\(M\to\infty\).  Such a kernel is symmetric unimodal.  A nonnegative
symmetric unimodal summable sequence is a nonnegative combination of the
indicators of centered integer intervals: its coefficients are the
successive differences of its nonincreasing half-sequence.  The convolution
of two centered interval indicators is symmetric unimodal, since it counts
their overlap under a translation.  This proves preservation under
convolution and hence the asserted property of the limit.  It follows by
telescoping the two half-lines that
\begin{equation}
 \sum_k|p_t(k+1)-p_t(k)|=2p_t(0)
 \le\frac{\sqrt\pi}{2\sqrt t}.
 \label{eq:current-heat-gradient}
\end{equation}
The inequality follows from
\begin{equation}
 p_t(0)=\frac1{2\pi}\int_{-\pi}^{\pi}e^{-t(2-2\cos k)}\,\dd k
 \le\frac1{2\pi}\int_{\mathbb R}e^{-4tk^2/\pi^2}\,\dd k
 =\frac{\sqrt\pi}{4\sqrt t}.
 \label{eq:current-heat-diagonal}
\end{equation}
The first-difference norm in each coordinate of the product heat kernel is
the left side of \eqref{eq:current-heat-gradient}; the other factors have
\(\ell^1\) norm one.  Integrating in time and summing the \(D_n\)
coordinates gives
\(D_n(\sqrt\pi/2)\int_0^\infty e^{-\eta t}t^{-1/2}\dd t
=D_n\pi/(2\sqrt\eta)\), proving the current \(\ell^1\) bound.

The Fourier symbol in dimension \(D\) is
\(\lambda_D(k)=\sum_{j=1}^D(2-2\cos k_j)\), with normalized measure
\(\dd^Dk/(2\pi)^D\).  Since \(|\widehat{\bar R}_n|\le M_n(R_n)\),
\begin{equation}
 \|G_\eta\|_2^2\le M_n(R_n)^2
 \int_{[-\pi,\pi]^D}\frac{\dd^Dk}{(2\pi)^D}
             \frac1{[\lambda_D(k)+\eta]^2}.
 \label{eq:current-free-square-integral}
\end{equation}
Use \(\lambda_D(k)\ge a_0|k|^2\), \(a_0=4/\pi^2\), and enlarge the
cube to the ball of radius \(\pi\sqrt D\).  For \(D=2\), enlarging to
all of \(\mathbb R^2\) gives
\(\int_0^\infty r\,\dd r/[2\pi(a_0r^2+\eta)^2]
=\pi/(16\eta)\).  For \(D=4\), the radial bound is
\begin{equation}
 \frac1{8\pi^2}\int_0^{2\pi}\frac{r^3\,\dd r}{(a_0r^2+\eta)^2}
 =\frac{\pi^2}{256}
 \left[\log(1+16/\eta)+\frac{\eta}{16+\eta}-1\right]
 \le\frac{\pi^2}{256}\log(1+16/\eta).
 \label{eq:current-four-dimensional-integral}
\end{equation}
For \(D=6\), dropping \(\eta\) gives
\(\int_0^{\pi\sqrt6}r^5\,\dd r/[64\pi^3(a_0r^2)^2]
=3\pi^3/1024\).  These are the squares of
\eqref{eq:current-H-constants}.
\end{proof}

The same radial comparison gives two additional estimates used below:
\begin{align}
 \mathcal J_2(\eta)&:=(\delta_0,(L_2+\eta)^{-1}\delta_0)
 \le\frac\pi{16}\log(1+8/\eta),\label{eq:current-two-dimensional-log}\\
 \mathcal G_D&:=(\delta_0,L_D^{-1}\delta_0)<\infty,
 \qquad D=4,6.\label{eq:current-transience}
\end{align}
In the second line the inverse is a zero-mass quadratic form.  Near zero
its Fourier integrand is bounded by \(a_0^{-1}|k|^{-2}\), which is
integrable in these dimensions.  In particular
\begin{equation}
 I_\eta(R_n)\le\mathcal G_{D_n}\|R_n\|_1^2
 \quad(n=3,4),
 \qquad
 I_\eta(R_2)\le\mathcal J_2(\eta)\|R_2\|_1^2.
 \label{eq:current-source-transience}
\end{equation}
For example the radial comparison gives \(\mathcal G_4\le\pi^2/16\).
The estimates hold for complex sources by the bound on their Fourier
transform, without a sign assumption on their coefficients.

\subsection{Lifting coefficient currents to spin expectations}

\begin{lemma}[Creator Gram and swap bounds]
\label{lem:current-creator-gram}
Let \(A,B\) be finite sets of the same positive cardinality and put
\(P_A^\dagger=\prod_{x\in A}b_x^\dagger\).  For distinct sets,
\begin{equation}
 \omega(P_AP_A^\dagger)\le1,\qquad
 |\omega(P_AP_B^\dagger)|\le p\quad(A\ne B),\qquad
 |\omega(P_A^\dagger P_B)|\le p.
 \label{eq:current-creator-gram}
\end{equation}
Consequently, if \(Z=\sum_A t_A P_A^\dagger\), with distinct sets in the
sum, and \(S_Z=\sum_A|t_A|^2\), \(A_Z=\sum_A|t_A|\), then in the GNS
representation of \(\omega\),
\begin{equation}
 \|Z\Omega_\omega\|\le\sqrt{S_Z+pA_Z^2},\qquad
 \|Z^\dagger\Omega_\omega\|\le\sqrt p\,A_Z.
 \label{eq:current-creator-vector-bounds}
\end{equation}
If the support of \(Z\) avoids a nearest-neighbor edge \(xy\), and
\(A_{xy}=b_x^\dagger-b_y^\dagger\), then
\begin{equation}
 |\omega(A_{xy}Z)|
 \le\sqrt2\sqrt{\omega(P^-_{xy})}
       \bigl(\|Z\Omega_\omega\|+\|Z^\dagger\Omega_\omega\|\bigr).
 \label{eq:current-swap-bound}
\end{equation}
The conclusions hold for finite sums and their absolutely summable limits.
\end{lemma}

\begin{proof}
The diagonal creator expectation is
\(\omega(\prod_{x\in A}(1-n_x))\le1\).  For the off-diagonal one,
write \(C=A\cap B\), \(E=A\setminus B\), \(F=B\setminus A\).
The sets \(E,F\) are nonempty and disjoint.  Normal ordering leaves the
common projector \(\Pi_C=\prod_{x\in C}(1-n_x)\), commuting with the
remaining operators, and
\begin{equation}
 \omega(P_AP_B^\dagger)=\omega(\Pi_C P_F^\dagger P_E).
\end{equation}
Apply Cauchy-Schwarz in the positive, not necessarily normalized,
functional \(X\mapsto\omega(\Pi_C X\Pi_C)\).  Each resulting squared
norm is at most \(\omega(\prod_{x\in E}n_x)\) or
\(\omega(\prod_{x\in F}n_x)\), and hence at most \(p\).  The normal
Gram bound follows directly from
\(\omega(P_A^\dagger P_A)\le p\) and Cauchy-Schwarz.  Expanding the
two squared norms of \(Z\) proves
\eqref{eq:current-creator-vector-bounds}.

For the last assertion let \(P=P^-_{xy}\) and \(P^\perp=1-P\).
The swap of \(x,y\) changes the sign of \(A_{xy}\), so
\(A_{xy}=PA_{xy}P^\perp+P^\perp A_{xy}P\).
The operator \(Z\) and its adjoint commute with these edge operators.
The first block is bounded by
\(\sqrt{\omega(P)}\|A_{xy}\|\|Z\Omega_\omega\|\).
Move \(Z\) through the edge factors in the second block and take its
adjoint; it is bounded by
\(\sqrt{\omega(P)}\|A_{xy}\|\|Z^\dagger\Omega_\omega\|\).
The local two-spin action gives \(\|A_{xy}\|=\sqrt2\): it maps the
vacuum to the unnormalized antisymmetric one-spin vector and that vector
to a multiple of the two-spin occupied vector, with largest singular
value \(\sqrt2\).  This proves \eqref{eq:current-swap-bound}.
Absolute convergence and the norm estimates extend the finite-sum result.
\end{proof}

\begin{lemma}[Exact aggregation multiplicity]
\label{lem:current-aggregation}
Fix a positively oriented physical edge \((x,x+e_\mu)\), and an unordered
set \(A\) of \(m_n\) other sites disjoint from its endpoints.  Exactly
\begin{equation}
 \mathsf A_n=m_n^2(m_n-1)!=m_n m_n!
 \label{eq:current-aggregation-multiplicity}
\end{equation}
labeled relative coordinate edges, together with their translation
anchors, have this moved edge and this set of remaining particles.
Thus \(\mathsf A_2=1\), \(\mathsf A_3=4\), and \(\mathsf A_4=18\).
\end{lemma}

\begin{proof}
The anchor is one of the \(m_n\) sites in \(A\); it cannot be the moved
particle because the allowed coordinate graph moves only nonanchor
labels.  Choose its site in \(m_n\) ways, choose the moving label among
the \(m_n\) relative labels, and biject the remaining \(m_n-1\) labels
with the other sites in \(A\).  These choices uniquely determine the
relative configuration and its positive coordinate edge.  Conversely each
such edge determines these choices.  Both endpoints are legal because
\(A\) is disjoint from the moved edge and has distinct sites.  This proves
the exact count, not merely an upper bound.
\end{proof}

For a finite allowed current \(J\), summation by parts in the creator
polynomial gives, on each relative edge, a factor
\begin{equation}
 (b_x^\dagger-b_{x+e_\mu}^\dagger)\prod_{y\in A}b_y^\dagger,
 \qquad A\cap\{x,x+e_\mu\}=\varnothing.
 \label{eq:current-physical-edge-factor}
\end{equation}
After summing translation anchors, group by the physical edge.  Denote the
resulting coefficients of the remaining sets by \(t_A^\mu\), and the
current components moving in physical direction \(\mu\) by \(J_\mu\).
Lemma~\ref{lem:current-aggregation} and Cauchy-Schwarz on each fiber give
\begin{equation}
 \sum_A|t_A^\mu|\le\|J_\mu\|_1,\qquad
 \sum_A|t_A^\mu|^2\le\mathsf A_n\|J_\mu\|_2^2.
 \label{eq:current-grouped-budgets}
\end{equation}
Translation invariance permits fixing the physical edge at
\((0,e_\mu)\).  Apply \eqref{eq:current-swap-bound}, include the Hermitian
conjugate and the factor \(1/n!\), and use
\(\sqrt{x+y}\le\sqrt x+\sqrt y\).  Cauchy-Schwarz over \(\mu=x,y\)
then yields the explicit estimate
\begin{align}
 |\mathfrak c_n(\operatorname{div}J;\omega)|
 &\le\frac{2\sqrt2}{n!}\sqrt{\mathfrak q}
       \left[\sqrt{\mathsf A_n}\|J\|_2+2\sqrt p\|J\|_1\right]
 \nonumber\\
 &\le8n!\sqrt{\mathfrak q}
          \left[\|J\|_2+\sqrt p\|J\|_1\right].
 \label{eq:current-lifted-flow}
\end{align}
The second line is a convenient common bound for \(n=2,3,4\).
For a source \(S_n\), isolate the creator at the anchor.  At most
\(m_n!\) labeled tuples give a specified unordered set of the remaining
sites.  A separate application of the Gram bound gives
\begin{equation}
 |\mathfrak c_n(S_n;\omega)|
 \le\frac{2}{n!}\sqrt p
       \left[\sqrt{m_n!}\|S_n\|_2+\sqrt p\|S_n\|_1\right].
 \label{eq:current-lifted-source}
\end{equation}
These derivations also control the absolute value of the creator-only
density, with no larger constants.

\begin{lemma}[Creation-current inequality]
\label{lem:creation-current}
Let \(\omega\) be an infinite-volume translation-invariant state and
\(R_n\in\ell^1(\Omega_n)\), \(n=2,3,4\).  For \(0<\eta\le1\),
\begin{equation}
 \boxed{\begin{aligned}
 |\mathfrak c_n(R_n;\omega)|\le C_n\bigl\{&
 \sqrt{\mathfrak q}\,[\sqrt{I_\eta(R_n)}
                   +\eta^{-1/2}M_n(R_n)\sqrt p]\\
 &+\eta H_{D_n}(\eta)M_n(R_n)\sqrt p+M_n(R_n)p\bigr\},
 \end{aligned}}
 \label{eq:creation-current}
\end{equation}
with the field- and state-independent choice
\begin{equation}
 C_n=8n!\max\left\{\sqrt{\mathsf C_{E,n}},
                    192\pi D_n,5^{m_n/2},1\right\}.
 \label{eq:current-universal-constant}
\end{equation}
The same bound is valid for the creator-only density. No smallness
assumption on \(p\) is required. In a finite-volume derivation the
valid limit order is: truncate the current and source at fixed
\(\eta>0\), take the thermodynamic limit, remove the truncations,
and only then decrease \(\eta\).
\end{lemma}

\begin{proof}
Use the free current from Lemma~\ref{lem:current-free-resolvent} and push
it to \(\Omega_n\).  Because \(\pi_n\) is the identity on the support
of \(\bar R_n\), the divergence identity becomes
\begin{equation}
 R_n=\operatorname{div}(P_nJ_\eta)+\eta(\pi_n)_*G_\eta.
 \label{eq:current-divergence-decomposition}
\end{equation}
Apply \eqref{eq:current-lifted-flow} to the first term and
\eqref{eq:current-lifted-source} to the second.  Equations
\eqref{eq:current-push-norms}--\eqref{eq:current-free-ltwo} bound the
four resulting contributions by constants no larger than
\eqref{eq:current-universal-constant} times the four terms displayed in
\eqref{eq:creation-current}.  In particular, the \(\ell^1\) current
coefficient is \(384D_n\pi/2=192\pi D_n\), and the vertex
\(\ell^2\) coefficient is \(5^{m_n/2}\).  The normalization factors and
the aggregation multiplicity were retained in
\eqref{eq:current-lifted-flow}--\eqref{eq:current-lifted-source}; the
larger common constant absorbs them for the stated three values of \(n\).

We spell out the passage to infinite currents.  At fixed \(\eta>0\),
\(J_\eta\in\ell^1\cap\ell^2\) and \(G_\eta\in\ell^1\cap\ell^2\).
Truncate \(J_\eta\) to finitely many free edges, push the truncated
current, and separately truncate \((\pi_n)_*G_\eta\).  The pushed
currents converge in \(\ell^1\) and \(\ell^2\) by
\eqref{eq:current-push-norms}.  Since
\(\|\operatorname{div}J\|_1\le2\|J\|_1\), the coefficient sources in
\eqref{eq:current-divergence-decomposition} converge in \(\ell^1\).
Equation~\eqref{eq:current-density-continuity} passes their physical
expectation densities to the limit.  This proves the result for an
arbitrary infinite-volume translation-invariant state using only its
local positive-state inequalities.

The finite-volume interpretation has a definite order.  For fixed
\(\eta>0\) and fixed coefficient-current truncation, all spin operators
above have finite relative support.  Apply the finite sums on tori larger
than that support and take their thermodynamic local limit.  Then remove
the current and source truncations, using
\eqref{eq:current-density-continuity}.  Only afterward may one set
\(\eta=\eps^2\) and let \(\eps\downarrow0\).  In particular, the
Fourier estimates in Lemma~\ref{lem:current-free-resolvent} are estimates
on the infinite lattice; they are not bounds uniform in
\(\eta\downarrow0\) on a fixed torus, whose zero Fourier mode would
contribute to the resolvent.  This order is maintained in every application
below.
\end{proof}

\subsection{Physical inverse forms and the contact source}

The physical \(n\)-particle coefficient space has scalar product
\begin{equation}
 \langle u,v\rangle_n=\frac1{n!}\sum_{r\in\Omega_n}\overline{u(r)}v(r)
 \label{eq:current-physical-inner-product}
\end{equation}
and consists of functions invariant under relabeling the \(n\) particles
and changing the anchor.  More explicitly, permute the ordered sites
\((0,r_1,\ldots,r_{m_n})\), subtract the new first site from the others,
and retain the resulting relative coordinates.  Each such transformation
is a counting-measure-preserving bijection of \(\Omega_n\).

The exchange \(Q_n\) is the exact restriction of \(Q\) to this
translation-zero coefficient space.  A bond with one occupied and one
unoccupied site has diagonal matrix element \(1/2\) and hopping element
\(-1/2\).  Bonds with equal occupations contribute zero.  Thus its
configuration operator is one half the graph Laplacian of allowed
single-particle hops: an excluded hop removes its diagonal degree as well
as its off-diagonal element.  In relative coordinates the mobile particles
give \(\mathcal E_{\Omega_n}\); a hop of the anchor gives a simultaneous
shift of all relative coordinates.  If \(\mathcal E_{\rm anc}\) denotes
the sum of the squared differences for these allowed simultaneous shifts,
then
\begin{equation}
 \langle v,Q_nv\rangle_n
 =\frac1{2n!}\bigl[\mathcal E_{\Omega_n}(v)
                             +\mathcal E_{\rm anc}(v)\bigr]
 \ge\frac1{2n!}\mathcal E_{\Omega_n}(v).
 \label{eq:current-physical-graph-form}
\end{equation}
The two edge families are counted with their particle labels, even if
they coincide as relative edges.  For \(n=2\), both families give the
same punctured graph, and \(Q_2=\cL\), as required.

For a nonnegative operator, an inverse form at zero is defined by
\begin{equation}
 \langle R_n,Q_n^{-1}R_n\rangle_n
 =\sup_{v\ \mathrm{compact,\ symmetric}}
 \left[2\operatorname{Re}\langle R_n,v\rangle_n
                              -\langle v,Q_nv\rangle_n\right],
 \label{eq:current-inverse-form-definition}
\end{equation}
possibly infinite.  Equivalently it is the increasing limit of the positive
mass inverse forms.  The same convention defines \(I_0\) on the free
lattice.

\begin{lemma}[Two inverse-form comparisons]
\label{lem:current-inverse-comparisons}
For any source on \(\Omega_n\), in counting measure,
\begin{equation}
 (R_n,(L_{\Omega_n}+\eta)^{-1}R_n)
 \le\mathsf C_{E,n}I_\eta(R_n),\qquad \eta>0.
 \label{eq:current-forward-inverse}
\end{equation}
For a permutation- and reanchoring-invariant source, the physical
zero-mass comparison is
\begin{equation}
 I_0(R_n)\le\frac{n n!}{2}
                  \langle R_n,Q_n^{-1}R_n\rangle_n.
 \label{eq:current-reverse-inverse}
\end{equation}
In particular the factor on the right is nine for \(n=3\).
\end{lemma}

\begin{proof}
The extension fixes the source support, so
\((R_n,v)=(\bar R_n,E_nv)\).  Since
\(\mathsf C_{E,n}\ge5^{m_n}\),
\begin{equation}
 \mathcal E_{D_n}(E_nv)+\eta\|E_nv\|_2^2
 \le\mathsf C_{E,n}
          [\mathcal E_{\Omega_n}(v)+\eta\|v\|_2^2].
\end{equation}
Insert this in the variational formula for the left side of
\eqref{eq:current-forward-inverse}.  Enlarging the extended test functions
to all free-lattice test functions gives the supremum with quadratic
penalty \((L_{D_n}+\eta)/\mathsf C_{E,n}\), which equals
\(\mathsf C_{E,n}I_\eta(R_n)\).  This proves the stated direction.

For the reverse direction, restrict a compact free test function \(w\)
to \(\Omega_n\).  Its mobile-coordinate energy is at most
\(\mathcal E_{D_n}(w)\).  Interpolate a simultaneous anchor shift by
\(m_n\) free coordinate steps.  The intermediate points are allowed to
lie on collision sets because \(w\) is a free-lattice function.
Cauchy-Schwarz costs \(m_n\).  Summing over the original relative
coordinates and the two planar directions, each intermediate translated
coordinate step contributes a subset of a free-edge sum.  Hence
\(\mathcal E_{\rm anc}(w|_{\Omega_n})\le
m_n\mathcal E_{D_n}(w)\), and
\begin{equation}
 \langle w|_{\Omega_n},Q_nw|_{\Omega_n}\rangle_n
 \le\frac{n}{2n!}\mathcal E_{D_n}(w).
 \label{eq:current-anchor-interpolation}
\end{equation}
If the restriction is not particle symmetric, average it over the finite
relabeling/reanchoring group.  This preserves its pairing with \(R_n\)
and does not increase the exchange form, because that form is invariant
and convex.  Denote the averaged function by \(v\).  The source pairing is
\((\bar R_n,w)=n!\langle R_n,v\rangle_n\).  Therefore
\begin{align}
 2\operatorname{Re}(\bar R_n,w)-\mathcal E_{D_n}(w)
 &\le2n!\operatorname{Re}\langle R_n,v\rangle_n
                    -\frac{2n!}{n}\langle v,Q_nv\rangle_n.
\end{align}
Taking the supremum and rescaling \(v\) gives the factor
\((n!)^2/(2n!/n)=n n!/2\), proving
\eqref{eq:current-reverse-inverse}.
\end{proof}

For later use we give the contact estimate in its coefficient form.  For
an even pair kernel \(u\), extend \(u(0)=0\) and define
\begin{equation}
 v(u)=\sum_{\mu=x,y}\sum_{k\notin\{0,e_\mu\}}
                        |u(k)-u(k-e_\mu)|.
 \label{eq:current-pair-gradient-budget}
\end{equation}
For an oriented nearest-neighbor edge let \(c_{ab}=c_x=i/2\) in the
positive horizontal direction, \(c_{ab}=c_y=-1/2\) in the positive
vertical direction, and \(c_{ba}=-c_{ab}\).  For an unordered triple
\(C\), let \(t_{ab}\) be its unique site outside the pair \(\{a,b\}\).
The exact creation vertex is
\begin{equation}
 (\mathcal V_{32}u)(C)
 =\sum_{\substack{\{a,b\}\subset C\\|a-b|_1=1}}
       c_{ab}\,[u(a-t_{ab})-u(b-t_{ab})],
 \label{eq:current-triplet-vertex}
\end{equation}
where either orientation can be chosen because reversing the edge changes
both factors' signs.  Formula~\eqref{eq:current-triplet-vertex} follows
directly by applying
\(c_{ab}(n_ab_b^\dagger-b_a^\dagger n_b)\) to a pair vector.  The two
pre-existing pairs give the two displayed coefficients.  This also fixes
the source normalization relative to \eqref{eq:current-creator-convention}.

\begin{lemma}[Contact tubes and the triplet source]
\label{lem:current-contact}
For an even kernel with \(v(u)<\infty\), put
\(\mathcal T_u=\mathcal V_{32}u\).  Then
\begin{equation}
 \|\mathcal T_u\|_1\le3v(u),\qquad
 I_\eta(\mathcal T_u)
 \le18\mathcal J_2(\eta)(u,\cL u),
 \qquad \eta>0,
 \label{eq:current-contact}
\end{equation}
where the pair form on the right uses counting measure and the quadratic
estimate is asserted whenever that form is finite, or by its form closure.
Consequently
\(\mathcal T=\mathcal V_{32}F\) satisfies
\begin{equation}
 \boxed{\Gamma_3:=\langle\mathcal T,Q_3^{-1}\mathcal T\rangle_3
 \le\frac{\mathsf C_{E,3}}3\,\mathcal G_4
                                      \|\mathcal T\|_1^2<\infty.}
 \label{eq:current-gamma-finite}
\end{equation}
\end{lemma}

\begin{proof}
In relative coordinates \((0,x,y)\), the source is a sum of twelve tubes:
\(x=e\), \(y=e\), and \(y-x=e\), with
\(e\in\{\pm e_x,\pm e_y\}\).  On a tube, its two-dimensional
coefficient is one half, in absolute value, of a retained nearest-neighbor
difference of \(u\).  Extend that coefficient by zero at the excluded
collision points.  For each of the three tube families the four directions
count each retained undirected pair edge twice.  Hence, writing the
individual tube coefficients as \(a_j\),
\begin{equation}
 \sum_{j=1}^{12}\|a_j\|_1\le3v(u),\qquad
 \sum_{j=1}^{12}\|a_j\|_2^2\le\frac32(u,\cL u).
 \label{eq:current-tube-coefficient-count}
\end{equation}
This proves the first assertion.

For a tube \(T(x,y)=\mathbf1_{x=e}a(y)\), Fourier transformation gives
\begin{equation}
 (T,(L_4+\eta)^{-1}T)
 =\int\frac{\dd^2p\,\dd^2q}{(2\pi)^4}
       \frac{|\widehat a(q)|^2}{\lambda_2(p)+\lambda_2(q)+\eta}
 \le\mathcal J_2(\eta)\|a\|_2^2.
\end{equation}
The tube \(y=e\) is identical.  For \(y-x=e\), its Fourier transform is
\(e^{-iq\cdot e}\widehat a(p+q)\).  Change variables on the Fourier
torus to \((k,q)=(p+q,q)\), a measure-preserving unimodular change.
The denominator becomes \(\lambda_2(k-q)+\lambda_2(q)+\eta\).
Dropping its first nonnegative term gives exactly the same bound.  Finally
the square norm associated with the positive inverse form satisfies
\(\|\sum_{j=1}^{12}T_j\|_{(L_4+\eta)^{-1}}^2
\le12\sum_j\|T_j\|_{(L_4+\eta)^{-1}}^2\).
Together with \eqref{eq:current-tube-coefficient-count}, this proves the
factor 18 in \eqref{eq:current-contact}.

Section~\ref{sm:green} gives
\(|\nabla F(r)|\le C(1+|r|)^{-3}\), so \(v(F)<\infty\) and
\(\mathcal T\in\ell^1(\Omega_3)\).  By
\eqref{eq:current-physical-graph-form} and the inverse variational formula,
\begin{equation}
 \langle\mathcal T,(Q_3+\beta)^{-1}\mathcal T\rangle_3
 \le\frac13(\mathcal T,(L_{\Omega_3}+2\beta)^{-1}\mathcal T)
 \le\frac{\mathsf C_{E,3}}3 I_{2\beta}(\mathcal T).
\end{equation}
The factor \(1/3\) is
\((1/6)^2/(1/12)\); it comes from the physical scalar product and the
exchange weight.  Let \(\beta\downarrow0\) and use four-dimensional
transience, \eqref{eq:current-source-transience}, to obtain
\eqref{eq:current-gamma-finite}.
\end{proof}

For completeness the same comparison controls the interacting massive
triplet channel.  Put
\(A_3=a_\eps Q_3+3\Delta_\eps I+(d/2)\mathcal D_{\mathrm{nn},3}\), where
\(\mathcal D_{\mathrm{nn},3}\ge0\), and \(\zeta=6\Delta_\eps/a_\eps\).
Equation~\eqref{eq:current-physical-graph-form} gives
\begin{equation}
 \langle v,A_3v\rangle_3
 \ge\frac{a_\eps}{12}
          [\mathcal E_{\Omega_3}(v)+\zeta\|v\|_2^2].
\end{equation}
Consequently
\begin{equation}
 0\le\mathcal V_{23}A_3^{-1}\mathcal V_{32}
 \le\frac{12\mathsf C_{E,3}}{a_\eps}\mathcal J_2(\zeta)\cL
 \le10\pi\mathsf C_{E,3}\log(2/\eps)\cL
 \label{eq:current-triplet-form-bound}
\end{equation}
as physical pair forms.  The middle factor follows from
\(1/(3a_\eps)\) in the inverse variational formula, the factor 18 in
\eqref{eq:current-contact}, and
\((u,\cL u)=2\langle u,\cL u\rangle_2\).
For the last inequality use \(a_\eps\ge1/2\),
\(\zeta\ge6\eps^2\) on the stated field interval for \(0<\eps\le1\),
and \(\log(1+8/\zeta)\le3\log(2/\eps)\).
The resolvent here contains the exact exchange exclusion graph.

\begin{corollary}[Pair-triplet effective profile]
\label{cor:current-pair-triplet-profile}
Let \(A_2\) and \(\phi_2=-dA_2^{-1}f\) be as in
\eqref{eq:range-pair-operator}, and set
\begin{equation}
 \Sigma_\eps=\mathcal V_{23}A_3^{-1}\mathcal V_{32},\qquad
 B_\eps=A_2-s^2\Sigma_\eps,\qquad
 \phi_{23}=-dB_\eps^{-1}f.
 \label{eq:current-pair-triplet-profile}
\end{equation}
Write \(C_\Sigma=10\pi\mathsf C_{E,3}\).  If \(0<\eps\le1\) and
\begin{equation}
 C_\Sigma\eps^2\log(2/\eps)\le\frac14,
 \label{eq:current-effective-smallness}
\end{equation}
then \(B_\eps\ge A_2/2>0\).  Moreover, uniformly for \(h\in\cI\)
and every \(r\ne0\),
\begin{equation}
 |\phi_{23}(r)-\phi_2(r)|
 \le C_{\cI}\eps^4\log(2/\eps)\sqrt{\log(2+|r|)}.
 \label{eq:current-pair-triplet-comparison}
\end{equation}
One may take \(C_{\cI}=4C_\Sigma\sqrt{F_1 C_{\rm pr}}\), where
\(C_{\rm pr}\) is any uniform constant in the massless probe estimate
\((g_r,\cL^{-1}g_r)\le C_{\rm pr}\log(2+|r|)\) proved in
Lemma~\ref{lem:range-test-bounds}.
\end{corollary}

\begin{proof}
All norms in this proof use the physical pair scalar product
\(\langle u,v\rangle_2=\tfrac12\sum\overline u v\).
By \eqref{eq:current-triplet-form-bound} and \(A_2\ge a_\eps\cL\),
the positive operator
\begin{equation}
 E_\eps=s^2A_2^{-1/2}\Sigma_\eps A_2^{-1/2}
\end{equation}
has norm at most
\(\theta_\eps=2C_\Sigma\eps^2\log(2/\eps)\le1/2\).
Thus
\(B_\eps=A_2^{1/2}(1-E_\eps)A_2^{1/2}\ge A_2/2\).
In particular its inverse exists at each positive angle, since
\(A_2\ge2\Delta_\eps I\).  Resolvent factorization gives
\begin{align}
 \|A_2^{1/2}(\phi_{23}-\phi_2)\|_2
 &\le d\frac{\theta_\eps}{1-\theta_\eps}
                          \|A_2^{-1/2}f\|_2
 \nonumber\\
 &\le2C_\Sigma\sqrt{F_1}\,
                          \eps^4\log(2/\eps).
 \label{eq:current-effective-energy-bound}
\end{align}
For the last step, Sec.~\ref{sm:green} gives
\(\langle f,\cL^{-1}f\rangle_2=F_1/2\), and hence
\(\|A_2^{-1/2}f\|_2^2\le F_1/(2a_\eps)\le F_1\).
We also used \(d\le\eps^2/2\) and
\(\theta_\eps/(1-\theta_\eps)\le2\theta_\eps\).

The joint lattice/spin quarter-turn intertwines the two- and three-particle
DM blocks.  Under a pure lattice quarter-turn the block
\(\mathcal V_{32}\) acquires a phase and \(\mathcal V_{23}\) its
conjugate; \(A_3\) commutes with the rotation.  Therefore \(\Sigma_\eps\)
preserves the even pair sector of rotation eigenvalue \(-1\).  Both
profiles and their difference lie in that sector.  No restriction to its
real cosine component is imposed.

For the quadrupole probe \(g_r\) of \eqref{eq:range-test}, evaluation in
this sector is \(\psi(r)=2\langle g_r,\psi\rangle_2\).  The probe
estimate cited in the statement and positivity give
\begin{equation}
 \|A_2^{-1/2}g_r\|_2^2
 \le\frac{1}{2a_\eps}(g_r,\cL^{-1}g_r)
 \le C_{\rm pr}\log(2+|r|).
\end{equation}
Cauchy-Schwarz with \eqref{eq:current-effective-energy-bound} proves
\eqref{eq:current-pair-triplet-comparison} with the displayed constant.
\end{proof}

For every fixed \(\delta<1\), multiplying
\eqref{eq:current-pair-triplet-comparison} by \(|r|^2/\eps^2\) makes its
supremum on \(1\le|r|\le\eps^{-\delta}\) tend to zero, with upper bound
\(C_{\cI}\eps^{2-2\delta}\log^{3/2}(2/\eps)\).
Thus the effective pair-triplet profile has the same weighted matching
limit as \(\phi_2\), and the actual-state comparison follows by the
triangle inequality and Sec.~\ref{sm:range}.  This comparison does not
give an \(o(\eps^4)\) error at a fixed nonzero value of \(\eps r\).

\subsection{Compact triplet approximation}

\begin{lemma}[Approximation in the triplet exchange norm]
\label{lem:current-compact-triplet}
There are compact, particle-symmetric triplet kernels \(\Theta_M\), in
the symmetry sector of \(\mathcal T\), for which
\begin{align}
 \delta_M^2
 &:=\|Q_3^{1/2}\Theta_M+Q_3^{-1/2}\mathcal T\|_3^2
       \longrightarrow0,\label{eq:current-triplet-approximation}\\
 \langle\mathcal T+Q_3\Theta_M,
              Q_3^{-1}(\mathcal T+Q_3\Theta_M)\rangle_3
 &=\delta_M^2,\label{eq:current-triplet-dual-error}\\
 \langle\Theta_M,Q_3\Theta_M\rangle_3
       +2\operatorname{Re}\langle\mathcal T,\Theta_M\rangle_3
 &=-\Gamma_3+\delta_M^2.
 \label{eq:current-triplet-completion}
\end{align}
With \(F_R=\chi(\cdot/R)F\), \(\mathcal T_R=\mathcal V_{32}F_R\),
and \(R_{M,R}=\mathcal T_R+Q_3\Theta_M\), one has
\begin{equation}
 \boxed{I_\eta(R_{M,R})
 \le18\delta_M^2+2\mathcal G_4\|\mathcal T_R-\mathcal T\|_1^2,
 \qquad \eta>0,}
 \label{eq:current-triplet-residual}
\end{equation}
where \(\|\mathcal T_R-\mathcal T\|_1=O(R^{-1})\).  For fixed \(M\),
\(\sup_{R\ge4}\|R_{M,R}\|_1<\infty\).
\end{lemma}

\begin{proof}
Complete the compact physical triplet functions in the exchange norm
\(\langle v,Q_3v\rangle_3^{1/2}\).  This is a norm on compact functions:
the configuration graph is connected by the routing construction, and a
zero Dirichlet form is constant; a compact constant is zero.  Finiteness
of \(\Gamma_3\) means that \(v\mapsto-\langle\mathcal T,v\rangle_3\)
is bounded in this norm, with squared dual norm \(\Gamma_3\).
The Riesz vector \(W\) in the energy completion satisfies
\begin{equation}
 \langle Q_3^{1/2}W,Q_3^{1/2}v\rangle_3
 =-\langle\mathcal T,v\rangle_3,
 \qquad \|Q_3^{1/2}W\|_3^2=\Gamma_3.
\end{equation}
The notation \(W=-Q_3^{-1}\mathcal T\) is interpreted in this energy
completion; no \(\ell^2\) vector inverse is assumed.  By definition of
the completion, choose compact \(\Theta_M\to W\) in energy norm.
Average over particle relabelings and the required finite spatial symmetry
sector.  These projections preserve compactness and do not increase the
energy error, since \(Q_3\) commutes with them and \(\mathcal T\) lies
in that sector.  This proves \eqref{eq:current-triplet-approximation}.

For a compact test \(v\), the residual functional is
\begin{equation}
 \langle\mathcal T+Q_3\Theta_M,v\rangle_3
 =\langle Q_3^{1/2}(\Theta_M-W),Q_3^{1/2}v\rangle_3.
\end{equation}
Its squared dual norm is \(\delta_M^2\), proving
\eqref{eq:current-triplet-dual-error}.  Expanding the squared energy norm
of \(\Theta_M-W\) proves \eqref{eq:current-triplet-completion}.

The gradient estimate from Sec.~\ref{sm:green} gives
\begin{equation}
 v(F_R-F)\le C R^{-1}.
 \label{eq:current-cutoff-gradient}
\end{equation}
Indeed the tail sum of \((1+|r|)^{-3}\) outside radius \(R\) is
\(O(R^{-1})\).  The additional cutoff term has first difference
\(O(R^{-1})\), multiplies \(F=O(R^{-2})\), and is summed over an
annulus containing \(O(R^2)\) sites.  Near the puncture the cutoff is
identically one.  The first inequality of \eqref{eq:current-contact}
therefore gives the asserted \(\ell^1\) convergence of the triplet
sources.  Compactness of \(\Theta_M\) makes \(Q_3\Theta_M\) compact,
so the residual \(\ell^1\) norm is bounded uniformly in \(R\) at fixed
\(M\).

Apply \eqref{eq:current-reverse-inverse} with \(n=3\) to
\(\mathcal T+Q_3\Theta_M\), use
\(I_\eta\le I_0\) and
\(I_0(A+B)\le2I_0(A)+2I_0(B)\), and bound the remaining free source by
\(\mathcal G_4\|\mathcal T_R-\mathcal T\|_1^2\).  The factor is
\(2\times9=18\), giving \eqref{eq:current-triplet-residual}.
\end{proof}

The kernels \(\Theta_M\) and the number \(\Gamma_3\) are independent
of \(h\).  The order relevant to the energy expansion is to fix a compact
\(\Theta_M\), take the thermodynamic limit, then take
\(\eps\downarrow0\) with the chosen pair cutoff \(R(\eps)\to\infty\),
and finally improve the compact approximation by letting \(M\to\infty\).
There is no assumption that a growing triplet support has a uniform
commutator budget.

\subsection{Absorption of cubic and quartic sources}

We give the error estimates in a form that can be inserted directly into
an expectation-density bound with positive exchange and number terms.
They hold for every infinite-volume translation-invariant state and do
not require a preliminary small-density estimate.  In this subsection
\(\eta=\eps^2\) and \(\ell_\eps=1+\log(2+\eps^{-1})\).

\begin{lemma}[Cubic residual absorption]
\label{lem:current-cubic-absorption}
Fix a compact triplet kernel \(\Theta_M\), let \(R(\eps)\to\infty\),
and put \(R_{3,\eps}=\mathcal T_{R(\eps)}+Q_3\Theta_M\).  Suppose
\(|\gamma_\eps|\le A\eps^3\).  For any fixed \(\kappa,\nu>0\),
there is a sufficiently small angle, depending on these parameters and
the fixed kernel, such that
\begin{align}
 |\gamma_\eps\mathfrak c_3(R_{3,\eps};\omega)|
 &\le\kappa\mathfrak q+\nu\eps^2p
       +\frac{A^2C_3^2}{2\kappa}\eps^6 I_{\eps^2}(R_{3,\eps})
       +O_{A,\nu,\Theta_M}(\eps^8\ell_\eps)
 \nonumber\\
 &\le\kappa\mathfrak q+\nu\eps^2p
       +\frac{9A^2C_3^2}{\kappa}\delta_M^2\eps^6
       +o_{A,\kappa,\nu,\Theta_M}(\eps^6).
 \label{eq:current-cubic-absorption}
\end{align}
The scalar error is uniform in the state and in \(h\in\cI\) whenever
the coefficient bound is uniform on that interval.
\end{lemma}

\begin{proof}
Let \(\mathsf M=\sup_{R\ge4}\|\mathcal T_R+Q_3\Theta_M\|_1<\infty\).
The four terms in \eqref{eq:creation-current}, after multiplication by
\(A\eps^3\), are bounded by constants times
\begin{equation}
 \eps^3\sqrt{\mathfrak q I},\qquad
 \eps^2\mathsf M\sqrt{\mathfrak q p},\qquad
 \eps^5\mathsf M\sqrt{p\ell_\eps},\qquad
 \eps^3\mathsf M p,
 \qquad I=I_{\eps^2}(R_{3,\eps}).
 \label{eq:current-cubic-four-terms}
\end{equation}
For the first two terms retain the coefficient \(AC_3\) and use
Young's inequality with exchange allocation \(\kappa\mathfrak q/2\)
for each.  Their remaining costs are respectively
\begin{equation}
 \frac{A^2C_3^2}{2\kappa}\eps^6 I,
 \qquad
 \frac{A^2C_3^2}{2\kappa}\eps^4\mathsf M^2p.
\end{equation}
The second is at most \(\nu\eps^2p/3\) once
\(A^2C_3^2\eps^2\mathsf M^2/(2\kappa)\le\nu/3\).
By \eqref{eq:current-H-constants},
\(H_4(\eps^2)\le C\sqrt{\ell_\eps}\) for \(0<\eps\le1\).
The third term of \eqref{eq:current-cubic-four-terms} is bounded by
\(\nu\eps^2p/3+C_{A,\nu}\mathsf M^2\eps^8\ell_\eps\).
The last is at most \(\nu\eps^2p/3\) for small enough
\(\eps\mathsf M\).  This proves the first line of
\eqref{eq:current-cubic-absorption}, uniformly in the state.
Insert \eqref{eq:current-triplet-residual}; its extra cutoff term is
\(O(\eps^6R(\eps)^{-2})=o(\eps^6)\), and
\(\eps^8\ell_\eps=o(\eps^6)\).  The factor
\(18A^2C_3^2/(2\kappa)=9A^2C_3^2/\kappa\) gives the second line.
\end{proof}

For the coefficient \(\tau=\eps^3/2\) and an exchange allocation
\(\kappa=g_*/8\), the surviving sixth-order cost in
\eqref{eq:current-cubic-absorption} is at most
\begin{equation}
 \frac{18C_3^2}{g_*}\,\delta_M^2\eps^6+o_{\cI,\Theta_M}(\eps^6).
 \label{eq:current-cubic-energy-cost}
\end{equation}
Its prefactor is independent of the compact approximation index.  Only
the admissible angle and the vanishing remainder may depend on
\(\Theta_M\).  This separation is what allows the final approximation
limit after the weak-angle limit.

\begin{lemma}[Quartic pair and four-spin absorption]
\label{lem:current-quartic-absorption}
Suppose either \(n=2\) and
\(\|R_{2,\eps}\|_1\le B\ell_\eps\), or \(n=4\) and
\(\|R_{4,\eps}\|_1\le B\), with a fixed \(B<\infty\).
If \(|\gamma_\eps|\le A\eps^4\), then, for any fixed
\(\kappa,\nu>0\), sufficiently small \(\eps\) gives
\begin{equation}
 |\gamma_\eps\mathfrak c_n(R_{n,\eps};\omega)|
 \le\kappa\mathfrak q+\nu\eps^2p+
 \begin{cases}
 C_{A,B,\kappa,\nu}\eps^8\ell_\eps^3,&n=2,\\
 C_{A,B,\kappa,\nu}\eps^8,&n=4.
 \end{cases}
 \label{eq:current-quartic-absorption}
\end{equation}
In both cases the scalar error is \(o(\eps^6)\), uniformly in the state
and in any field dependence covered by the stated coefficient budgets.
\end{lemma}

\begin{proof}
Put \(M=\|R_{n,\eps}\|_1\).  Multiplying the first current term by its
coefficient and applying Young's inequality gives the two costs
\[
 \frac{\kappa\mathfrak q}{2},\qquad
 \frac{A^2C_n^2\eps^8I_{\eps^2}(R_{n,\eps})}{2\kappa}.
\]
Equations~\eqref{eq:current-two-dimensional-log} and
\eqref{eq:current-source-transience} bound this inverse form by
\(C M^2\ell_\eps\) when \(n=2\), and by \(C M^2\) when \(n=4\).
The stated source budgets therefore give the respective scalar costs
\(O(\eps^8\ell_\eps^3)\) and \(O(\eps^8)\).

The mixed current term is \(AC_n\eps^3M\sqrt{\mathfrak q p}\).
The other exchange allocation \(\kappa\mathfrak q/2\) leaves
\(A^2C_n^2\eps^6M^2p/(2\kappa)\), which is absorbed in
\(\nu\eps^2p/3\) because \(\eps^4M^2\to0\).
The regulator term is \(AC_n\eps^6M H_{D_n}(\eps^2)\sqrt p\).
For \(n=2\), \(H_2(\eps^2)=\sqrt\pi/(4\eps)\), so Young's inequality
with allocation \(\nu\eps^2p/3\) leaves \(O(\eps^8M^2)\).
For \(n=4\), \(H_6\) is bounded and the analogous cost is
\(O(\eps^{10}M^2)\).  Finally the last term is
\(AC_n\eps^4Mp\), bounded by \(\nu\eps^2p/3\) since
\(\eps^2M\to0\).  The indicated scalar remainders dominate all these
costs, proving \eqref{eq:current-quartic-absorption}.
\end{proof}

In particular, the scalar at order \(\eps^6\) in the two-dressing energy
expansion receives no unavoidable loss from its vacuum-changing quartic
pair or four-spin sources.  The cubic residual has only the arbitrarily
small loss \eqref{eq:current-cubic-energy-cost}.  These are expectation
bounds for the physical infinite-volume state, with the same order of
limits as \eqref{eq:creation-current}; no finite-particle ansatz for that
state is used.

\section{Uniform pair amplitudes in the actual states}
\label{sm:range}

We prove Theorem~\ref{thm:range} by a linear stationarity identity.
The bounds are first stated in terms of the physical population and
the dressed densities, so they also apply in Sec.~\ref{sm:thermal}.
Pair-kernel norms use counting measure on the punctured lattice.
An expectation of a translation sum divided by $N$ denotes its
infinite-volume density.
Constants may depend on $\cI$, the one fixed spatial cutoff, and
the one fixed Fourier partition used below, but not on the state,
volume, or observation displacement.

\subsection{Stationarity from local energy variations}

\begin{lemma}[Local stationarity of energy-density minimizers]
\label{lem:stationarity-density}
For a finite-range translation-invariant interaction with bounded
local terms, every translation-invariant energy-density minimizer
satisfies, for each local $A$,
\begin{equation}
 \omega([H,A])=0,\qquad
 [H,A]=\sum_{Z:Z\cap\operatorname{supp}A\ne\varnothing}[h_Z,A].
 \label{eq:local-stationarity}
\end{equation}
The statement remains true if minimization additionally imposes a
finite symmetry group of the interaction. Every local weak limit of
finite-volume Gibbs states is stationary independently.
\end{lemma}
\begin{proof}
Take $A=A^\dagger$, supported on $S$, and $U_t=e^{itA}$.
Choose a square cell side $M$ so large that $S+Mz$ are disjoint and
no interaction term meets two of them. If interaction supports have
diameter at most $\rho$, it suffices that
$M>\operatorname{diam}_\infty S+2\rho+2$.
The translated unitaries commute and define an automorphism on
local observables. Put
\[
 \Delta_A(t)=\sum_{Z:Z\cap S\ne\varnothing}
       \omega(U_t^\dagger h_ZU_t-h_Z).
\]
Each changed interaction term belongs to exactly one cell.
Translation invariance makes the energy-density change of the
periodically transformed state exactly $\Delta_A(t)/M^2$.
Averaging over the $M^2$ cell translations, and then over any imposed
finite symmetry group, gives an admissible trial state with the same
energy density. Minimality gives $\Delta_A(t)\ge0$ for both signs
of $t$, with $\Delta_A(0)=0$. This finite differentiable sum has
$0=\Delta_A'(0)=i\omega([H,A])$.
Apply this argument to the Hermitian real and imaginary parts of
a general $A$.

For a finite-volume Gibbs density matrix, cyclicity and
$[\rho_\Lambda,H_\Lambda]=0$ give
$\operatorname{Tr}\rho_\Lambda[H_\Lambda,A]=0$.
For sufficiently distant boundaries the commutator is the fixed
local operator in \eqref{eq:local-stationarity}; the identity passes
to every local weak limit. No identification with another
ground-state definition is required.
\end{proof}

\subsection{The exact pair equation}

Set
\begin{equation}
 A_2=a_\eps\cL+\mu_\eps I+\frac d2P_{\nn},\qquad
 \mu_\eps=2\Delta_\eps,\qquad \phi_2=-dA_2^{-1}f.
 \label{eq:range-pair-operator}
\end{equation}
The function $\phi_2$ is the pair response $w_2$ of
Theorem~\ref{thm:massive} and Sec.~\ref{sm:massive}.
The exact Hamiltonian decomposition is equivalently
\begin{equation}
 H-Ne_{\rm pol}=a_\eps Q+\Delta_\eps\Nop
       +\frac d2\Ncontact+dW(f)+s\VDM.
 \label{eq:range-hamiltonian}
\end{equation}
For an even finite kernel define
$B(u)=\frac12\sum_{i,r\ne0}u(r)b_i b_{i+r}$ and
\begin{equation}
 a_u=\|u\|_1,\quad S_u=\|u\|_2^2,\quad
 v_u=\sum_{\mu=x,y}\sum_{k\notin\{0,e_\mu\}}
                   |u(k)-u(k-e_\mu)|.
 \label{eq:range-kernel-budgets}
\end{equation}
Nearest-neighbour bonds below are counted once. Define
\begin{align}
 R_Q(u)&=\sum_{\langle ij\rangle}\sum_{k\notin\{i,j\}}
 [u(k-i)-u(k-j)](n_i b_j-n_j b_i)b_k,\label{eq:range-RQ}\\
 R_D(u)&=\sum_{\langle ij\rangle}\sum_{k\notin\{i,j\}}
 [u(k-i)n_jb_i b_k+u(k-j)n_i b_j b_k],\label{eq:range-RD}\\
 R_P(u)&=[B(u),W(f)]-\frac N2(u,f)_{\rm b}I,\qquad
 R_V(u)=[B(u),\VDM],\label{eq:range-RPV}
\end{align}
where $(u,v)_{\rm b}=\sum_{r\ne0}u(r)v(r)$ is bilinear.
For the real tests used below it is the usual Hilbert-space pairing.

\begin{lemma}[Stationary pair identity]
\label{lem:range-pair-identity}
Every translation-invariant stationary state satisfies
\begin{align}
 \tfrac12(u,A_2w+df)_{\rm b}
 ={}&-a_\eps\frac{\omega(R_Q(u))}{N}
 -\frac d2\frac{\omega(R_D(u))}{N}\notag\\
 &-d\frac{\omega(R_P(u))}{N}
 -s\frac{\omega(R_V(u))}{N}.
 \label{eq:range-stationarity}
\end{align}
With $p=\omega(n_0)$,
\begin{equation}
 |\omega(R_D(u))|/N\le4pa_u,\qquad
 |\omega(R_P(u))|/N\le128pa_u.
 \label{eq:range-even-residuals}
\end{equation}
\end{lemma}
\begin{proof}
For three distinct sites,
\begin{align}
 [b_i b_k,Q_{ij}]&=\tfrac12(b_i-b_j)b_k
                    +(n_i b_j-n_j b_i)b_k,\notag\\
 [b_i b_j,Q_{ij}]&=0.
 \label{eq:range-local-exchange}
\end{align}
Combining endpoint coefficients gives
$[B(u),Q]=B(\cL u)+R_Q(u)$.
Similarly $[B(u),\Ncontact]=B(P_{\nn}u)+R_D(u)$ and
$[B(u),\Nop]=2B(u)$.
The pair-source vacuum contraction is $N(u,f)_{\rm b}/2$.
Lemma~\ref{lem:stationarity-density}, applied to each local pair,
and $\omega(B(v))/N=(v,w)_{\rm b}/2$ prove the identity.

For distinct sites $|\omega(n_i b_jb_k)|\le p$: restrict the
positive state to the range of $n_i$ and use that $b_jb_k$ is a
contraction. The two endpoints and two directions give the first
bound. For the second, disjoint pairs commute, while
\[
 [b_i b_j,b_i^\dagger b_j^\dagger]=1-n_i-n_j,\qquad
 [b_i b_j,b_i^\dagger b_k^\dagger]=(1-2n_i)b_jb_k^\dagger.
\]
Equal pairs have absolute coefficient sum per site
$\frac12\sum_r|u(r)f(r)|$. Their residual expectation is at most
\[
 p\sum_r|u(r)f(r)|\le pa_u.
\]
For pairs sharing exactly one site, that common site is a unique
anchor. The coefficient sum per anchor is at most
$\sum_{j,k}|u(j)f(k)|=a_u\|f\|_1=a_u$.
The displayed three-site operator has norm at most one and,
together with its adjoint, kills its local vacuum. Thus
$|\omega(O_Z)|\le |Z|p\|O_Z\|$ gives $3pa_u$ for these terms.
Their sum is at most $4pa_u$, hence certainly the conservative
constant $128pa_u$ retained in \eqref{eq:range-even-residuals}.
\end{proof}

\subsection{Massive test kernels with uniform displacement bounds}

Let $\mathsf J(x,y)=(-y,x)$ and set
\begin{equation}
 g_r=\tfrac14(\delta_r+\delta_{-r}
             -\delta_{\mathsf Jr}-\delta_{-\mathsf Jr}),\qquad
 u_r=A_2^{-1}g_r,\qquad L_r=\log(2+|r|).
 \label{eq:range-test}
\end{equation}
Thus $u_r$ is real, even, and $u_r(\mathsf Jx)=-u_r(x)$.

\begin{lemma}[Massive test bounds]
\label{lem:range-test-bounds}
At sufficiently small angle, uniformly in $h\in\cI$ and $r\ne0$,
\begin{align}
 \|u_r\|_2+v_{u_r}&\le C_{\cI,\chi}(1+|r|),\notag\\
 a_{u_r}&\le C_{\cI,\chi}(1+|r|)^2\ell_\eps,\qquad
 (u_r,\cL u_r)\le C_{\cI,\chi}L_r .
 \label{eq:range-test-bounds}
\end{align}
Moreover,
\begin{equation}
 \sup_{r\ne0}|\phi_2(r)-dF(r)|
       \le C_{\cI,\chi}\eps^4\ell_\eps .
 \label{eq:range-massive-matching}
\end{equation}
\end{lemma}
\begin{proof}
Write $\lambda(k)=4-2\cos k_x-2\cos k_y$,
$\sigma=\mu_\eps/a_\eps$, and
$R_\sigma=(\cL_0+\sigma)^{-1}$. Set
$G_\sigma=-R_\sigma f$,
$t_\sigma=4G_\sigma(e_x)$, and
$Z_0=1-t_0=2(\pi-2)/\pi$.
The Fourier formulas in Sec.~\ref{sm:green} give
\begin{equation}
 t_\sigma=\int\frac{\mathfrak d(k)^2}{\lambda(k)+\sigma}
                  \frac{\dd^2k}{(2\pi)^2},\qquad
 |\widehat f(k)|\le\lambda(k)/4 .
 \label{eq:range-defect-data}
\end{equation}
In the rotation-odd sector the origin value vanishes and
$A_2=a_\eps\cL_0+\mu_\eps-cP_{\nn}$.
For $u_0=a_\eps^{-1}R_\sigma g_r$, the rank-one defect equation is
\begin{equation}
 u_r=u_0+\frac{4c}{a_\eps}
       \frac{u_0(e_x)}{1-(c/a_\eps)t_\sigma}G_\sigma,\qquad
 u_0(e_x)=a_\eps^{-1}G_\sigma(r).
 \label{eq:range-test-defect}
\end{equation}
Indeed $P_{\nn}u=-4u(e_x)f$, and evaluation of the full-lattice
resolvent equation at $e_x$ gives \eqref{eq:range-test-defect}.
The last identity in that equation follows by reciprocity:
the even rotation-odd projection of $\delta_{e_x}$ is $-f$,
and $(g_r,G_\sigma)=G_\sigma(r)$.
Here $|G_\sigma(r)|\le1/4$, $|t_\sigma-t_0|\le\sigma/4$, and
\[
 2(h_--1/2)\eps^2\le\sigma\le4h_+\eps^2.
\]
A sufficient explicit denominator condition is
\begin{equation}
 \eps^2\left(\tfrac14+\tfrac12t_0+h_+\right)\le Z_0/2,
 \qquad a_\eps-ct_\sigma\ge Z_0/2 .
 \label{eq:range-defect-threshold}
\end{equation}
Thus the defect coefficient is uniformly bounded.

Here is the summability argument for the full-lattice kernel.
Choose a smooth radial $\zeta$, equal to one for $|k|\le\pi/4$
and zero for $|k|\ge\pi/2$, and use the periodic partition
$1-\zeta(k)$ together with
$\zeta(2^j k)-\zeta(2^{j+1}k)$, $j\ge0$.
Let $H_{\kappa,\sigma}$ be the annular inverse kernel at
$\kappa=2^{-j}$. For $j_0=0,1,2,3$ nearest-neighbour differences,
\begin{equation}
 \|\nabla^{j_0}H_{\kappa,\sigma}\|_1
 \le C_{j_0,\zeta}\frac{\kappa^{j_0}}{\kappa^2+\sigma}.
 \label{eq:range-dyadic-l1}
\end{equation}
Derivatives through order six of the corresponding multiplier are
bounded by $C\kappa^{j_0-|\alpha|}/(\kappa^2+\sigma)$.
Fourier integration by parts bounds its kernel by
$C\kappa^{2+j_0}(\kappa^2+\sigma)^{-1}
(1+\kappa|x|)^{-6}$, whose lattice sum gives
\eqref{eq:range-dyadic-l1}. The constants use only finitely many
derivatives of the fixed partition, the elementary symbol, and a
convergent lattice sum. The smooth part away from zero is included
at scale one.

The source $g_r$ is the difference of two central second
differences. Telescope each long difference along a path of at
most $|r|_1$ nearest-neighbour steps. With $R_r=1+|r|$,
\begin{align}
 \|H_{\kappa,\sigma}*g_r\|_1
 &\le C\frac{\min(1,R_r^2\kappa^2)}{\kappa^2+\sigma},\notag\\
 \|\nabla(H_{\kappa,\sigma}*g_r)\|_1
 &\le C\frac{\kappa\min(1,R_r^2\kappa^2)}{\kappa^2+\sigma}.
 \label{eq:range-dyadic-source}
\end{align}
For the first sum, $\kappa\ge R_r^{-1}$ contributes $CR_r^2$.
Each scale $\sqrt\sigma\le\kappa<R_r^{-1}$ contributes at most
$CR_r^2$, and there are at most
$C[1+\log(2+\sigma^{-1})]$ such scales. Below both split scales,
$R_r^2\sum\kappa^2/\sigma\le CR_r^2$.
If the scales occur in the opposite order the estimate decreases.
For the gradient, the high scales give
$\sum_{\kappa\ge R_r^{-1}}\kappa^{-1}\le CR_r$; the low scales
give $R_r^2\sum_{\kappa<R_r^{-1}}\kappa^3/(\kappa^2+\sigma)
\le R_r^2\sum\kappa\le CR_r$.

For the square norm use
$|\widehat g_r(k)|\le\min(1,|r|^2|k|^2/4)$ and
$\lambda(k)\ge4|k|^2/\pi^2$. Dropping the mass, the radial
integrals below and above $R_r^{-1}$ are bounded by
$CR_r^4\int_0^{R_r^{-1}}k\,\dd k$ and
$C\int_{R_r^{-1}}^{\sqrt2\pi}k^{-3}\,\dd k$; both are
$O(R_r^2)$. The same estimates with $r=e_x$ apply to $G_\sigma$.
Together with \eqref{eq:range-test-defect}, they prove the first
three bounds in \eqref{eq:range-test-bounds}.

The massless probe energy is also controlled directly.
$E_0=(g_r,\cL_0^{-1}g_r)\le CL_r$ follows from the same Fourier
split, now giving a bounded low-momentum integral and a logarithm
above $R_r^{-1}$. If $v=\cL_0^{-1}g_r$ and $\beta=4(f,v)$,
its nearest-neighbour values give $P_{\nn}v=\beta f$.
Thus $\cL v=g_r-\beta f$ and
$\cL^{-1}g_r=v+\beta\cL^{-1}f$.
The identity $\cL^{-1}f=\cL_0^{-1}f/Z_0$ from
Sec.~\ref{sm:green} gives
$(g_r,\cL^{-1}f)=\overline\beta/(4Z_0)$. Consequently
\[
 (g_r,\cL^{-1}g_r)=E_0+\frac{|\beta|^2}{4Z_0}.
\]
Since $(f,\cL_0^{-1}f)=t_0/4$, Cauchy-Schwarz gives
$|\beta|^2\le4t_0E_0$. Thus the punctured probe energy is at most
$E_0/Z_0$. Positivity of $A_2-a_\eps\cL$ yields
\[
 (u_r,\cL u_r)\le a_\eps^{-1}(g_r,A_2^{-1}g_r)
 \le a_\eps^{-2}(g_r,\cL^{-1}g_r),
\]
proving the remaining bound.

Finally the same defect equation gives
$\phi_2=dG_\sigma/(a_\eps-ct_\sigma)$ and $F=G_0/Z_0$.
The estimates
\[
 \|G_\sigma-G_0\|_\infty
 \le\frac{\sigma}{4}
       \int\frac{\dd^2k}{(2\pi)^2(\lambda(k)+\sigma)},
 \qquad |t_\sigma-t_0|\le\sigma/4
\]
and \eqref{eq:range-defect-threshold} prove
\eqref{eq:range-massive-matching}, because
$\sigma=O_{\cI}(\eps^2)$ and the integral is $O(\ell_\eps)$.
\end{proof}

\subsection{The twelve contact tubes}

\begin{lemma}[Contact-source norms and inverse form]
\label{lem:range-contact}
For an even $u$ with $v_u<\infty$, let
$\mathcal T_u=\mathcal V_{32}u$ on
$\Omega_3=\{(x,y):0,x,y\text{ distinct}\}$ and extend it by zero
to $\mathbb Z^4$. With counting-measure norms,
\begin{align}
 \|\mathcal T_u\|_1&\le3v_u,\notag\\
 I_\eta(\mathcal T_u)&\le18I_2(\eta)(u,\cL u),\qquad
 I_2(\eta)\le\frac{\pi}{16}\log(1+8/\eta).
 \label{eq:range-contact-bound}
\end{align}
The inverse form is the free relative-lattice form in
\eqref{eq:creation-current}.
\end{lemma}
\begin{proof}
Put $c(e_x)=i/2$, $c(e_y)=-1/2$, and $c(-e)=-c(e)$.
An occupied bond $\{a,b\}$ with third site $t$ contributes
$c(b-a)[u(a-t)-u(b-t)]$; orientation reversal leaves this
product unchanged. In coordinates $(0,x,y)$ this is
\begin{align}
 \mathcal T_u(x,y)=\sum_{e=\pm e_x,\pm e_y}c(e)\bigl\{&
 \mathbf1_{x=e}[u(y)-u(y-e)]
 +\mathbf1_{y=e}[u(x)-u(x-e)]\notag\\
 &+\mathbf1_{y-x=e}[u(x)-u(x+e)]\bigr\}.
 \label{eq:range-twelve-tubes}
\end{align}
Each term is zero when a pair argument is zero. This imposes
exactly $x\ne0$, $y\ne0$, $x\ne y$.
There are three orientations and four directed bonds.

Define $D_eu(z)=u(z)-u(z-e)$ for $z\notin\{0,e\}$, and zero
otherwise. The coordinate tubes have coefficient $c(e)D_eu$;
the diagonal tube has a translate of $-c(e)D_eu$.
If $a_j$ are the twelve two-dimensional coefficients,
\begin{align}
 \sum_{j=1}^{12}\|a_j\|_1
 &=\tfrac32\sum_e\|D_eu\|_1=3v_u,\notag\\
 \sum_{j=1}^{12}\|a_j\|_2^2
 &=\tfrac34\sum_e\|D_eu\|_2^2
   =\tfrac32(u,\cL u).
 \label{eq:range-tube-count}
\end{align}
The directed sum counts each retained undirected edge twice.
These formulas use counting measure; the physical pair scalar
product would carry an additional factor $1/2$.

For $\mathbf1_{x=e}a(y)$, Fourier transformation gives
\[
 \int\frac{|\widehat a(q)|^2}{\lambda(p)+\lambda(q)+\eta}
       \frac{\dd^2p\,\dd^2q}{(2\pi)^4}
 \le I_2(\eta)\|a\|_2^2.
\]
The other coordinate orientation is identical.
For $\mathbf1_{y-x=e}b(x)$ the transform is
$e^{-iq\cdot e}\widehat b(p+q)$.
Set $k=p+q$ on the Brillouin torus; the denominator becomes
$\lambda(k-q)+\lambda(q)+\eta$.
Dropping its first nonnegative term gives the same bound,
uniformly in $k$. Cauchy-Schwarz in the positive inverse form
bounds the inverse form of the sum by twelve times the sum of
the individual forms. Equation \eqref{eq:range-tube-count}
gives the exact factor $12(3/2)=18$. Finally,
\[
 I_2(\eta)\le\frac1{2\pi}\int_0^{\sqrt2\pi}
       \frac{r\,\dd r}{4r^2/\pi^2+\eta}
       =\frac{\pi}{16}\log(1+8/\eta).
\]
\end{proof}
Consequently,
\begin{equation}
 I_{\eps^2}(\mathcal T_{u_r})\le C_{\cI}\ell_\eps L_r
       \le C_{\cI}\ell_\eps^2,\qquad 1\le|r|\le\eps^{-1}.
 \label{eq:range-contact-test}
\end{equation}

\subsection{Local extraction and observable conjugation}

Use $F_0=\chi(\cdot/R_0)F$, $R_0=\lceil\eps^{-1}\rceil$,
$X=X_{F_0}$, $U=e^{dX}$, and
$\widetilde\omega(A)=\omega(UAU^\dagger)$.
Write $a_F=\|F_0\|_1$ and $v_F=v_{F_0}$.
A local norm budget is the sum of local term norms per site.
For support size $k$, commutation with $X$ multiplies this
budget by at most $2ka_F$ and adds at most one site.
In particular,
\begin{equation}
\begin{array}{c|c|c|c|c}
 & A &[A,X]&[[A,X],X]&[[[A,X],X],X]\\ \hline
 R_Q(u)&2v_u&12v_ua_F&96v_ua_F^2&960v_ua_F^3\\
 R_V(u)&16a_u&96a_ua_F&768a_ua_F^2&7680a_ua_F^3\\
 \text{maximum support size}&3&4&5&6
\end{array}
\label{eq:range-bch-budgets}
\end{equation}
For the first row use $\|(n_i b_j-n_jb_i)b_k\|\le2$.
For the second, the four monomials in one DM bond have total
norm at most two, there are two such bonds per site, and the
pair coefficients incident at its endpoints sum to at most
$2a_u$. The commutator factor two gives $16a_u$.
The remaining columns multiply successively by $6a_F$,
$8a_F$, and $10a_F$.

Separating a local vacuum and its complement proves
\begin{align}
 O_Z|0_Z\rangle=O_Z^\dagger|0_Z\rangle=0
 &\ \Longrightarrow\
 |\widetilde\omega(O_Z)|\le |Z|\widetilde p\|O_Z\|,\notag\\
 \langle0_Z|O_Z|0_Z\rangle=0
 &\ \Longrightarrow\
 |\widetilde\omega(O_Z)|\le3\sqrt{|Z|\widetilde p}\|O_Z\|.
 \label{eq:range-local-vacuum}
\end{align}
For a swap-odd operator on an edge $ij\subset Z$ that
annihilates the local vacuum on both sides, the off-diagonal
swap blocks instead give
\begin{equation}
 |\widetilde\omega(O_Z)|\le2\|O_Z\|
 \sqrt{\widetilde\omega(P^-_{ij})\,
       \widetilde\omega\!\left(\sum_{k\in Z}n_k\right)}.
 \label{eq:range-swap-vacuum}
\end{equation}
These inequalities hold for complex expectations by their
two Hermitian phases, or directly by block Cauchy-Schwarz.

For an operator with zero local vacuum diagonal on at most four
sites, source extraction means subtracting the pure creator and
annihilator monomials with the same vacuum column and row.
There are at most fifteen
nonempty subsets on each side, each coefficient is at most the
original norm, and the remainder has norm at most $31$ times
that norm and annihilates its vacuum on both sides.
Conversion of an $n$-creator monomial to the kernel of
$\mathcal C_n^+$ costs at most $n!$.
Extraction is performed separately on each local term.

\begin{lemma}[Refined cubic residual]
\label{lem:range-refined-cubic}
For every even finite kernel,
\begin{equation}
 R_V(u)=\mathcal C_3^+(\mathcal T_{\bar u})^\dagger+\mathcal R_u,
 \qquad \mathcal T_{\bar u}=\mathcal V_{32}\overline u,
 \label{eq:range-cubic-decomposition}
\end{equation}
where the creator-only operator has coefficient $1/6$, and
\begin{equation}
 |\widetilde\omega(\mathcal R_u)|/N
 \le16a_u\sqrt{\widetilde{\mathfrak q}\widetilde p}
          +3v_u\widetilde p .
 \label{eq:range-cubic-remainder}
\end{equation}
\end{lemma}
\begin{proof}
Write a DM bond as
$C_{ij}=c_\mu(n_i b_j^\dagger-b_i^\dagger n_j)+\mathrm{h.c.}$
and set $U_\pm=[C_{ij},b_i^\dagger\pm b_j^\dagger]$.
Two-site algebra gives
\begin{align}
 U_+&=\overline c_\mu
       (b_i^\dagger b_j-b_i b_j^\dagger+n_i-n_j),\notag\\
 U_-&=2c_\mu b_i^\dagger b_j^\dagger+
       \overline c_\mu(T_{ij}-n_i-n_j+4n_in_j),
 \label{eq:range-U-identities}
\end{align}
with $T_{ij}=b_i^\dagger b_j+b_j^\dagger b_i$.
$U_+$ and the second term of $U_-$ kill the two-site vacuum
on both sides and have norm at most one.
The third-site commutator has coefficients
$(u_i+u_j)/2$ and $(u_i-u_j)/2$, with adjoints.
Its pure triple source is \eqref{eq:range-twelve-tubes}.
The plus term is swap-odd and vacuum-annihilating.
Equation \eqref{eq:range-swap-vacuum}, followed by
$\sqrt{\mathfrak q_x}+\sqrt{\mathfrak q_y}\le\sqrt{2\mathfrak q}$,
costs at most $4\sqrt6\,a_u\sqrt{\mathfrak q p}$.

For the same-edge term subtract
\[
 u(e_\mu)\overline c_\mu(b_i^\dagger-b_j^\dagger)
 +\overline{u(e_\mu)}c_\mu(b_i-b_j).
\]
Its translation sum is zero. The remaining swap-odd,
vacuum-annihilating term has norm at most $4|u(e_\mu)|$
and costs at most $4\sqrt2\,a_u\sqrt{\mathfrak q p}$.
The two constants sum to less than sixteen.
The remaining minus term has norm at most $|u_i-u_j|$
on three sites and costs $3v_up$.
The two phases $u,iu$ prove the complex annihilator version
with the same displayed upper bound.
\end{proof}

The first exchange-residual commutator obeys
\begin{align}
 [R_Q(u),X]\Omega_0&=0,\notag\\
 [R_Q(u),X]^\dagger\Omega_0
 &=\sum_{\langle ij\rangle}
 \sum_{\substack{k,l\notin\{i,j\}\\k\ne l}}
 \Delta_{ij}\overline u(k)\Delta_{ij}F_0(l)
 b_i^\dagger b_j^\dagger b_k^\dagger b_l^\dagger\Omega_0 ,
 \label{eq:range-quartet-source}
\end{align}
where $\Delta_{ij}u(k)=u(k-i)-u(k-j)$.
Every summand of $R_Q$ kills states with at most two occupied
sites. Its adjoint acting on $X\Omega_0$ selects an occupied
endpoint, and the two endpoint contributions give
$F_0(l-i)-F_0(l-j)$.
The ordered $(k,l)$ sum has no factor $1/2$.
Its absolute monomial coefficient sum per site is at most
$v_uv_F$, so the normalized quartet kernel has norm at most
$24v_uv_F$. After this single quartet row is extracted from
each local summand, the remaining norm budget is at most
$24v_u a_F$ and contributes at most $96v_u a_F\widetilde p$.

For $A_V=[R_V(u),X]$, support size is at most four and
parity is odd, so its vacuum images contain only one or three
flips. If $u(\mathsf Jx)=-u(x)$, then $B(u)$, $X$, and
$\VDM$ are invariant under the joint quarter-turn.
A translation-summed one-spin source is a multiple of
$\sum_i b_i^\dagger$ or $\sum_i b_i$, acquiring opposite
phases $i$ and $-i$. Its coefficient in $A_V$ is therefore
zero as an operator, independently of the selected state.
Local extraction leaves triple sources with combined kernel
norm at most $2(15)(3!)(96)a_ua_F=17280a_ua_F$.
The vacuum-annihilating remainder contributes at most
$4(31)(96)a_ua_F\widetilde p=11904a_ua_F\widetilde p$.

Expand both observables through two commutators:
\begin{equation}
 U^\dagger A U=A+d[A,X]+\frac{d^2}{2}[[A,X],X]+\mathcal E_A.
 \label{eq:range-observable-taylor}
\end{equation}
The second DM commutator is odd and has zero vacuum diagonal.
The second exchange commutator has possible magnon-number
changes $-6,-2,+2$, also excluding a vacuum diagonal.
Their expectations use \eqref{eq:range-local-vacuum}.
The third-order integral remainder uses unitary conjugation
and the last column of \eqref{eq:range-bch-budgets}, divided
by $3!$:
\begin{align}
 \|\mathcal E_{R_Q}\|/N&\le160d^3v_u a_F^3,\notag\\
 \|\mathcal E_{R_V}\|/N&\le1280d^3a_u a_F^3.
 \label{eq:range-observable-remainders}
\end{align}

\subsection{General bounds before imposing ground-state rates}

Let $\mathsf G_D=\int_{[-\pi,\pi]^D}
[\sum_{j=1}^D2(1-\cos k_j)]^{-1}\dd^Dk/(2\pi)^D$,
the finite diagonal massless free Green integral in
dimension $D=4,6$, and use $H_D,C_n$ from
\eqref{eq:creation-current}. Define
\begin{equation}
 \mathsf T_n(\widetilde p,\widetilde{\mathfrak q})
 =C_n\left[
 \sqrt{\mathsf G_{D_n}\widetilde{\mathfrak q}}
 +\eps^{-1}\sqrt{\widetilde{\mathfrak q}\widetilde p}
 +\eps^2H_{D_n}(\eps^2)\sqrt{\widetilde p}
 +\widetilde p\right],\quad n=3,4 .
 \label{eq:range-general-current-functional}
\end{equation}
Free transience gives $I_\eta(R)\le\mathsf G_{D_n}\|R\|_1^2$.
Thus a creator-only source of norm $M$ has expectation density
at most $M\mathsf T_n$. The same constants bound a complex
creator expectation by the two-phase decomposition of the
Hermitian current inequality.

For $\mathcal T=\mathcal T_{\bar u}$ put
$M_T=\|\mathcal T\|_1$, $I_T=I_{\eps^2}(\mathcal T)$, and
\begin{align}
 \mathcal V_0(u)={}&C_3\left[
 \sqrt{\widetilde{\mathfrak q}}\sqrt{I_T}
 +\eps^{-1}M_T\sqrt{\widetilde{\mathfrak q}\widetilde p}
 +\eps^2H_4(\eps^2)M_T\sqrt{\widetilde p}
 +M_T\widetilde p\right]\notag\\
 &+16a_u\sqrt{\widetilde{\mathfrak q}\widetilde p}
       +3v_u\widetilde p .
 \label{eq:range-V0}
\end{align}
Here $M_T\le3v_u$ and
$I_T\le18I_2(\eps^2)(u,\cL u)$.
The preceding identities give the following general bounds:
\begin{align}
 \mathcal E_Q(u)={}&8v_u\sqrt{\widetilde{\mathfrak q}\widetilde p}
 +d[24v_uv_F\mathsf T_4+96v_u a_F\widetilde p]\notag\\
 &+144\sqrt5\,d^2v_u a_F^2\sqrt{\widetilde p}
       +160d^3v_u a_F^3,\notag\\
 \mathcal E_V(u)={}&\mathcal V_0(u)
 +d[17280a_ua_F\mathsf T_3+11904a_ua_F\widetilde p]\notag\\
 &+1152\sqrt5\,d^2a_u a_F^2\sqrt{\widetilde p}
       +1280d^3a_u a_F^3,\notag\\
 |\omega(R_Q(u))|/N&\le\mathcal E_Q(u),\qquad
 |\omega(R_V(u))|/N\le\mathcal E_V(u).
 \label{eq:range-general-residuals}
\end{align}
For $\mathcal E_Q$, the swap-vacuum inequality and the
three-site norm bound give $4\sqrt3\,v_u
\sqrt{\widetilde{\mathfrak q}\widetilde p}$, bounded by the
displayed constant eight. The factors $144\sqrt5$ and $1152\sqrt5$
are the second-commutator budgets times $3\sqrt5/2$.
All densities inside $\mathcal E_Q,\mathcal E_V$ are dressed
densities. The population in \eqref{eq:range-even-residuals}
is the physical density. No ground-state or thermal rate
has entered these bounds.

For the real massive test, joint symmetry gives
$(g_r,w)=w(r)$, and $d(u_r,f)=-\phi_2(r)$.
The exact stationary identity consequently yields
\begin{equation}
 |w(r)-\phi_2(r)|
 \le2a_\eps\mathcal E_Q(u_r)+260dpa_{u_r}
                         +2|s|\mathcal E_V(u_r).
 \label{eq:range-master-bound}
\end{equation}
This holds for every stationary translation- and
jointly-fourfold-invariant state, with its actual dressed
densities.

\subsection{Ground-state range and cutoff limits}

\begin{proof}[Proof of Theorem~\ref{thm:range}]
Equation \eqref{eq:apriori}, with
$R_0=\lceil\eps^{-1}\rceil$, gives
\begin{equation}
 p,\widetilde p\le C_{\cI}\eps^4\ell_\eps^4,\qquad
 \widetilde{\mathfrak q}\le C_{\cI}\eps^6\ell_\eps^4,\qquad
 a_F\le C_\chi\ell_\eps,\quad v_F\le C_\chi .
 \label{eq:range-ground-inputs}
\end{equation}
These estimates use the fourth-order normal form, independently
of the sharp population coefficient. For sufficiently small
angle $\eps\ell_\eps^2\le1$, and
$\mathsf T_3+\mathsf T_4\le C_{\cI}\eps^3\ell_\eps^2$.
The contact bound \eqref{eq:range-contact-test} gives
\[
 |s|\mathcal V_0(u_r)\le C_{\cI}
 [\eps^4\ell_\eps^3+\eps^5v_{u_r}\ell_\eps^4
                         +\eps^6a_{u_r}\ell_\eps^4].
\]
The exchange source term in \eqref{eq:range-general-residuals}
is $O(\eps^5v_{u_r}\ell_\eps^2)$, its local remainder is
$O(\eps^6v_{u_r}\ell_\eps^5)$, and its higher commutators are
$O(\eps^6v_{u_r}\ell_\eps^4)$ and
$O(\eps^6v_{u_r}\ell_\eps^3)$.
Thus $\mathcal E_Q(u_r)\le
C_{\cI}\eps^5v_{u_r}\ell_\eps^4$.
For the DM transfer, the first-source term after multiplication
by $|s|d$ is $O(\eps^6a_{u_r}\ell_\eps^3)$; the other terms
are bounded by $O(\eps^7a_{u_r}\ell_\eps^5)$.
All constants are independent of $r$.
Equations \eqref{eq:range-master-bound} and
\eqref{eq:range-test-bounds} therefore prove
\begin{equation}
 |w(r)-\phi_2(r)|\le C_{\cI,\chi}
 [\eps^4\ell_\eps^3+
  \eps^5(1+|r|)\ell_\eps^4+
  \eps^6(1+|r|)^2\ell_\eps^6],
 \quad 1\le|r|\le\eps^{-1}.
 \label{eq:range-pointwise}
\end{equation}
The last logarithmic power is a common upper bound for the
explicit terms above.

We justify the noncompact test passage.
At fixed $\eps>0$ and $r$, put
$u^{(M)}=\chi(\cdot/M)u_r$ with a radial cutoff.
It preserves evenness and the rotation eigenvalue, and
\begin{align}
 \|u^{(M)}-u_r\|_1+\|u^{(M)}-u_r\|_2&\longrightarrow0,\notag\\
 v_{u^{(M)}}&\longrightarrow v_{u_r},\qquad
 (u^{(M)},\cL u^{(M)})\longrightarrow(u_r,\cL u_r).
 \label{eq:range-test-cutoff-limits}
\end{align}
The gradient statement follows from
$v_{u^{(M)}-u_r}\le4\|u^{(M)}-u_r\|_1$.
The energy statement follows from boundedness of $\cL$ on
$\ell^2$. Likewise $A_2$ is bounded on $\ell^1$, so
$A_2u^{(M)}\to g_r$ in $\ell^1$; since $|w|\le1$, the
left side of \eqref{eq:range-stationarity} converges.

Every residual and retained commutator has an absolutely
summable local coefficient decomposition with the displayed
norm budgets. Its expectation therefore converges.
The extracted quartet and triple kernels converge in $\ell^1$.
For fixed $\eta>0$, the resolvent norm is at most $\eta^{-1}$.
Thus, for any two coefficient sources,
\[
 |I_\eta(R)-I_\eta(S)|
 \le\eta^{-1}\|R-S\|_2(\|R\|_2+\|S\|_2)
 \le\eta^{-1}\|R-S\|_1(\|R\|_1+\|S\|_1).
\]
The first inequality follows by expanding the difference of the
two quadratic forms and using Cauchy-Schwarz. It proves their
continuity under the stated source convergence. The contact bound
and Dirichlet-energy convergence additionally retain the sharper
inverse-form estimate for the leading triple source.
Apply the finite-test identities, take the thermodynamic
limit at fixed $\eps,M$, and then let $M\to\infty$.
The coefficient-current cutoff in Sec.~\ref{sm:current}
is also removed at fixed positive mass after the
thermodynamic limit. Only afterwards is $\eps$ decreased.
There is no displacement-dependent probe smallness condition.

Combining \eqref{eq:range-pointwise},
\eqref{eq:range-massive-matching},
$d/\eps^2=1/2+O(\eps^2)$, and the decay of $F$, the weighted
amplitude error for $|r|\le\eps^{-\delta}$ is at most
\begin{equation}
 C_{\cI,\chi}
 [\eps^{2-2\delta}\ell_\eps^3+
  \eps^{3-3\delta}\ell_\eps^4+
  \eps^{4-4\delta}\ell_\eps^6]+C_{\cI,\chi}\eps^2 .
 \label{eq:range-weighted-error}
\end{equation}
It tends to zero for every fixed $\delta<1$.
The exact comparison in Sec.~\ref{sm:rdm},
\begin{equation}
 \bigl|\Neg(\rho_{0r})-|w(r)|\bigr|\le7p,
 \label{eq:range-rdm-comparison}
\end{equation}
adds at most $7C_{\cI}\eps^{2-2\delta}\ell_\eps^4$.
This proves the weighted negativity limit, including its
leading zeros at the quadrupole nodes.

For $|r|\le\eps^{-1}\ell_\eps^{-a}$, the three weighted
terms are respectively
$O(\ell_\eps^{3-2a})$,
$O(\ell_\eps^{4-3a})$, and
$O(\ell_\eps^{6-4a})$.
The matching term is $O(\ell_\eps^{1-2a})$.
Theorem~\ref{thm:population} makes the RDM contribution
$O(\ell_\eps^{-2a})$.
All tend to zero for $a>3/2$.
At fixed nonzero $\eps r$, the bound remains only
$O(\eps^4\ell_\eps^6)$; no $o(\eps^4)$ conclusion at that
scale is used.
\end{proof}

\section{Thermal sufficient NPT windows}
\label{sm:thermal}

\subsection{An exact field suppression of the reference pressure}
For a finite graph with nonnegative exchange strengths $\kappa_e$,
write $Q_\kappa=\sum_e\kappa_eP^-_e$ and
$Z_b(t)=\Tr\exp[-(Q_\kappa+b\Nop)/t]$. Let
$f_b(t)=-(t/N)\log Z_b(t)$. The following bound is independent
of a spin-wave approximation.

\begin{lemma}[Field-pressure inequality]\label{lem:field-pressure}
For $t>0$ and $b\ge0$, with $z=e^{-b/t}$,
\begin{equation}
 1\le Z_b(t)\le Z_0(t)^{\alpha(z)},\qquad
 0\le-f_b(t)\le\alpha(z)[-f_0(t)]
 \le\frac{e^{-b/t}}{\log2}[-f_0(t)],\quad
 \alpha(z)=\frac{\log(1+z)}{\log2}.
 \label{eq:field-pressure}
\end{equation}
\end{lemma}
\begin{proof}
The site transposition is $\mathsf T_e=1-2P^-_e$.
Independent Poisson clocks of rate $\kappa_e/(2t)$, run for unit time, generate
a random ordered product $U_\pi$ of those transpositions, with
\begin{equation}
 \mathbb E U_\pi=
 \exp\left[\frac1{2t}\sum_e\kappa_e(\mathsf T_e-1)\right]
 =e^{-Q_\kappa/t}.
\end{equation}
This follows either by the infinitesimal evolution equation for
the expectation or by its convergent ordered Poisson series.
The number operator commutes with every permutation. A spin
configuration contributes to $\Tr(U_\pi z^{\Nop})$ precisely
when it is constant on each permutation cycle. Therefore
\begin{equation}
 Z_b(t)=\mathbb E\prod_{c\in\mathrm{cycles}(\pi)}(1+z^{|c|}).
\end{equation}
For $0\le z\le1$ this product is at most
$(1+z)^{C(\pi)}=(2^{C(\pi)})^{\alpha(z)}$.
Jensen's inequality for $x^\alpha$, $0\le\alpha\le1$, gives
$Z_b\le(\mathbb E2^{C(\pi)})^\alpha=Z_0^\alpha$.
The all-up state has energy zero, so $Z_b\ge1$.
Finally $\log(1+z)\le z$. No monotonicity assertion about
exponentials of noncommuting operators enters this argument.
\end{proof}

There are two distinct zero-field inputs. The elementary one,
$Q_\kappa\succeq0$ and a two-dimensional on-site Hilbert space,
gives $Z_0(t)\le2^N$ and hence
\begin{equation}
 -f_b(t)\le t e^{-b/t}.
 \label{eq:trivial-pressure}
\end{equation}
For constant nearest-neighbor exchange $\kappa$ on the square
lattice, the additional input is Theorem 1.1 of
Ref.~\cite{Klippel2026}: for the Hamiltonian
$\sum_{\langle ij\rangle}(S^2-\mathbf S_i\cdot\mathbf S_j)$,
with each unoriented bond counted once and fixed spin $S$,
$\lim_{\beta\to\infty}\beta^2S f_2(\beta,S)=-\pi/24$.
At $S=1/2$, scaling the exchange by $\kappa$ gives
\begin{equation}
 f_{\kappa,0}(t)=-\frac{\pi}{12\kappa}t^2+o(t^2).
\end{equation}
Only the following consequence is used: with
$C_\kappa=\pi/(6\kappa)$ there is $t_\kappa>0$ such that
$-f_{\kappa,0}(t)\le C_\kappa t^2$ for $0<t\le t_\kappa$.
Applying Eq.~\eqref{eq:field-pressure} before taking the volume
limit yields, simultaneously for all $b\ge0$,
\begin{equation}
 0\le-f_{\kappa,b}(t)
 \le\frac{C_\kappa}{\log2}t^2e^{-b/t},\qquad 0<t\le t_\kappa.
 \label{eq:klippel-pressure}
\end{equation}
Thus an exponentially small factor multiplies the entire
zero-field bound, rather than an asymptotic leading term whose
additive remainder is uncontrolled at that precision.

\subsection{Gibbs excitation bounds retaining the physical population}
Use the normal form of Eq.~\eqref{eq:normalform}, with
reference Hamiltonian $H_{\rm ref}=\kappa_*Q+\mu_*\eps^2\Nop$.
For a finite-volume Gibbs state, the free-energy variational
principle gives $f_N(2t)\le2f_N(t)-u_N(t)$, and
$f_N(t)\le e_{0,N}$. Min-max comparison of eigenvalues in
Eq.~\eqref{eq:normalform} therefore gives
\begin{align}
 0\le u_N(t)-e_{0,N}
 &\le2\eta_R\eps^4-f_{{\rm ref},N}(2t),\label{eq:thermal-excess}\\
 \kappa_*\widetilde{\mathfrak q}
 +\mu_*\eps^2\widetilde p
 &\le4\eta_R\eps^4-f_{{\rm ref},N}(2t).
 \label{eq:thermal-rigidity}
\end{align}
For example, the first inequality follows by inserting the lower
free-energy bound and the vacuum upper bound on $e_{0,N}$ into
$u_N-e_{0,N}\le e_{0,N}-f_N(2t)$. The second also uses the
normal-form lower bound on the actual internal energy.
The comparison is spectral and does not require commutation of
the two Hamiltonians.

Take the thermodynamic limit at fixed $\eps,t,R$. Local energy
expectations pass to a Gibbs subsequence, and the reference
free energy has the same bulk limit for free and periodic
boundaries, since the missing-bond norm divided by volume
vanishes. The local population-conjugation bound in
Sec.~\ref{sm:normalform} then applies to the physical Gibbs
state. With $R_0=\lceil\eps^{-1}\rceil$, put
\begin{equation}
 \mathcal A_{\eps,t}=\eps^2\ell_\eps^4+
 \eps^{-4}[-f_{\kappa_*,\mu_*\eps^2}(2t)].
\end{equation}
After increasing constants depending only on $\cI,\chi$,
\begin{equation}
 p,\widetilde p\le C\eps^2\mathcal A_{\eps,t},\qquad
 \widetilde{\mathfrak q}\le C\eps^4\mathcal A_{\eps,t},\qquad
 a_{R_0}\le C\ell_\eps.
 \label{eq:thermal-general-densities}
\end{equation}
The quantities on the left are expectations in the actual Gibbs
state and its local dressing. Gibbs stationarity is proved in
Sec.~\ref{sm:range}, so the general residual estimates there
apply without a ground-state assumption.

\subsection{Self-contained temperature window}
Let $L_\eps=\log(1/\eps)$ and $0<\eps<e^{-1}$. Choose
$0<c_0\le\mu_*/12$. If $t\le c_0\eps^2/L_\eps$, then
Eq.~\eqref{eq:trivial-pressure} gives
\begin{equation}
 -f_{\kappa_*,\mu_*\eps^2}(2t)
 \le2t e^{-\mu_*\eps^2/(2t)}
 \le\frac{2c_0\eps^8}{L_\eps}.
 \label{eq:thermal-elementary-budget}
\end{equation}
The function $t\mapsto t e^{-b/t}$ is increasing for $t>0$,
which justifies evaluating at the endpoint. Thus
$\mathcal A_{\eps,t}\le C\eps^2\ell_\eps^4$ uniformly in this
temperature interval. By Eq.~\eqref{eq:thermal-general-densities}
the Gibbs densities then obey the bounds~\eqref{eq:range-ground-inputs}
that enter the ground-state proof, with larger constants, and the
stationary-state proof in Sec.~\ref{sm:range} applies unchanged. In
particular the weighted
amplitude and negativity convergence is uniform for every fixed
$\delta<1$ in this temperature window.

To obtain an explicit positive margin, choose $r_\gamma$ from
Eq.~\eqref{eq:green-tail} so that
$r^2|F(r)|\ge\gamma/[16(\pi-2)]$ for $r\ge r_\gamma$ and
$|\cos2\theta_r|\ge\gamma$. Set
\begin{equation}
 A_\gamma=\min\left\{F_1,\frac{\gamma}{16(\pi-2)}\right\}>0.
 \label{eq:thermal-angular-margin}
\end{equation}
The same bound with $A_\gamma$ includes nearest-neighbor pairs.
Equations~\eqref{eq:population-npt} and the uniform convergence
then give $\Neg\ge A_\gamma\eps^2/(8r^2)$, after reducing the
angle threshold. This proves
Eq.~\eqref{eq:thermal-window-elementary} without an external
free-energy theorem.

\subsection{Distance-dependent window using the two-dimensional pressure}
Now use Eq.~\eqref{eq:klippel-pressure}. Write
$t=\tau\eps^2$ and
\begin{equation}
 \vartheta(\tau)=\tau^2e^{-\mu_*/(2\tau)},\quad\vartheta(0)=0,
 \qquad \mathcal A=\eps^2\ell_\eps^4+\vartheta(\tau).
\end{equation}
For sufficiently small angle, Eq.~\eqref{eq:thermal-general-densities}
holds with $\mathcal A$ in place of $\mathcal A_{\eps,t}$,
since the constant $4C_{\kappa_*}/\log2$ is absorbed in $C$.
Let $L_r=\log(2+r)$, and choose $c_1\le\mu_*/32$.
For $\tau\le c_1/L_r$,
\begin{equation}
 \vartheta(\tau)\le\frac{c_1^2}{L_r^2}(2+r)^{-16},\qquad
 \sqrt{\vartheta(\tau)}\le\frac{c_1}{L_r}(2+r)^{-8}.
 \label{eq:thermal-power-envelope}
\end{equation}

We spell out the uniform residual estimate because retaining
the population alone would not suffice. In the general
stationary-state bounds of Sec.~\ref{sm:range}, insert
$p,\widetilde p\le C\eps^2\mathcal A$ and
$\widetilde{\mathfrak q}\le C\eps^4\mathcal A$.
Use $\sqrt{\mathcal A}\le\eps\ell_\eps^2+\sqrt\vartheta$,
$a_{u_r}\le Cr^2\ell_\eps$, $v_{u_r}\le Cr$, and
$(u_r,\cL u_r)\le C L_r$, enlarging constants for $r\ge1$.
The zero-temperature terms are exactly those already bounded
in Sec.~\ref{sm:range}. After multiplying by $r^2/\eps^2$,
the additional thermal terms are bounded by a constant times
the following list and terms of higher order in $\eps$:
\begin{gather}
 \eps r^2\sqrt{\vartheta\ell_\eps L_r},\quad
 \eps r^3\vartheta,\quad
 \eps^2r^3\ell_\eps^3\sqrt\vartheta,\quad
 \eps^2r^4\ell_\eps\vartheta,\nonumber\\
 \eps^3r^4\ell_\eps^2\sqrt\vartheta,\quad
 \eps^3r^4\ell_\eps^2\vartheta,\quad
 \eps^4r^4\ell_\eps^3\sqrt\vartheta.
 \label{eq:thermal-residual-monomials}
\end{gather}
The first term is the contact-tube inverse form of the DM
source. The terms with $r^3$ come from exchange residuals and
the mixed current-density terms; those with $r^4$ come from
the first observable commutator and the density/pair residuals.
The retained second commutators and third norm remainders are
covered by the higher powers indicated. These origins also
follow directly by substitution into the displayed general
residual bounds, before making any angle-dependent choice of
temperature.

For $r\le\eps^{-1}$, $L_r\le C\ell_\eps$.
Equation~\eqref{eq:thermal-power-envelope} bounds every term in
\eqref{eq:thermal-residual-monomials} by
$C c_1(1+c_1)\eps\ell_\eps^6$. Thus for every fixed
$\delta<1$, uniformly over the indicated temperatures and
$1\le r\le\eps^{-\delta}$,
\begin{equation}
 \frac{r^2}{\eps^2}|w(r)-dF(r)|
 \le C\left[\eps^{2-2\delta}\ell_\eps^3
 +\eps^{3-3\delta}\ell_\eps^4
 +\eps^{4-4\delta}\ell_\eps^6
 +c_1(1+c_1)\eps\ell_\eps^6\right]+C\eps^2
 \longrightarrow0.
 \label{eq:thermal-amplitude-control}
\end{equation}
Meanwhile the physical population, not a population in the
reference model, satisfies
\begin{equation}
 \frac{r^2p}{\eps^2}
 \le C\eps^{2-2\delta}\ell_\eps^4+
 \frac{C c_1^2}{(\log3)^2}.
 \label{eq:thermal-population-margin}
\end{equation}
Decrease $c_1$ until the last term is at most $A_\gamma/12$.
For sufficiently small angle $d/\eps^2\ge1/3$, and the
amplitude error in Eq.~\eqref{eq:thermal-amplitude-control}
is at most $A_\gamma/12$. The vanishing population remainder
can also be made at most $A_\gamma/12$.
Equation~\eqref{eq:population-npt} then gives, for example,
$\Neg\ge A_\gamma\eps^2/(12r^2)$ in the nonnodal region
specified in Theorem~\ref{thm:thermal}. This proves
Eq.~\eqref{eq:thermal-window-klippel}. The condition
$2t\le t_{\kappa_*}$ is uniform there because
$\tau\le c_1/\log3$ and $\eps\downarrow0$.

\subsection{Predicted thermal boundary}
\label{sm:thermal-boundary}
\begin{figure}[t]
 \centering
 \includegraphics[width=0.5\textwidth]{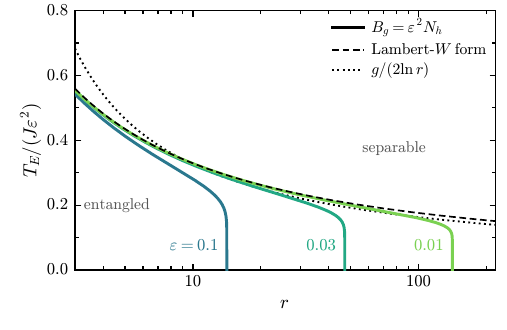}
 \caption{Predicted thermal boundary $T_E$ versus axial separation $r$ at $h=2$; pairs are entangled
 below the curves. Solid: $B_g(T_E/J\eps^2)=\eps^2N_h(\eps r,0)$ for the indicated $\eps$, ending at
 $\rho_E/\eps$. Dashed: Eq.~\eqref{eq:thermal-boundary-lambert}. Dotted: leading logarithm
 $Jg\eps^2/(2\ln r)$; in these units neither depends on $\eps$.}
 \label{fig:sm-thermal}
\end{figure}

The windows above are sufficient conditions. A prediction for the boundary itself follows from the
dilute magnon gas. At temperature $T=\tau J\eps^2$ the one-magnon band with gap $Jg\eps^2$ holds the
thermal population $\eps^2B_g(\tau)$, with $B_g(\tau)=-(\tau/2\pi)\ln(1-e^{-g/\tau})$~\cite{Hofmann2012},
which adds to the quantum depletion $n_*\eps^4$. The leading negativity then vanishes where
$|w(\br)|=n_*\eps^4+\eps^2B_g(\tau)$. For $1\ll r\ll1/\eps$ off the diagonals, where
$B_g\simeq(\tau/2\pi)e^{-g/\tau}$ and $|w|\simeq\eps^2C_\theta/r^2$ with
$C_\theta=|\cos2\theta|/[16(\pi-2)]$, this gives
\begin{equation}
 T_E(\br)\simeq\frac{Jg\eps^2}{W_0\bigl(gr^2/(2\pi C_\theta)\bigr)},
 \label{eq:thermal-boundary-lambert}
\end{equation}
with $W_0$ the principal branch of the Lambert function~\cite{Corless1996}; to leading logarithmic
order $T_E\simeq Jg\eps^2/(2\ln r)$, the magnon gap divided by $2\ln r$. Figure~\ref{fig:sm-thermal}
compares the full condition with these forms. This is a prediction of the two-magnon theory with a
free thermal magnon gas, not a theorem; Theorem~\ref{thm:thermal} proves sufficient conditions of the
same form.

\section{Massive response, reduced-state criterion, and activated boundary}
\label{sm:massive}

\subsection{Exact massive pair response}
\label{ss:massive-block}

\subsubsection{Conventions and the exact pair block}
Energies are measured in units of $J$, temperatures in units of $J/k_B$,
and distances in nearest-neighbor bond lengths. On the square lattice,
\begin{equation}
 \frac HJ=\sum_{i,\mu=x,y}\left[\frac14-
 \mathbf S_i^T e^{\eps[\mathbf e_\mu]_\times}
 \mathbf S_{i+\mathbf e_\mu}\right]
 -h\eps^2\sum_i S_i^z,\qquad \mathbf S_i=\bm\sigma_i/2.
 \label{eq:H}
\end{equation}
The thermodynamic limit precedes $\eps\downarrow0$. We use
$b_i=S_i^+$, $b_i^\dagger=S_i^-$, $n_i=b_i^\dagger b_i=1/2-S_i^z$,
$c=\cos\eps$, $d=1-c$, $s=\sin\eps$, and $a_\eps=(1+c)/2$.
The polarized energy per site is $e_{\rm pol}=(d-h\eps^2)/2$.
For $\Nop=\sum_i n_i$ and $\Ncontact=\sum_{\langle ij\rangle}n_i n_j$,
the exact hard-core identity on a periodic lattice is
\begin{align}
 H/J-Ne_{\rm pol}
 &=a_\eps Q+(h\eps^2-d)\Nop+\frac d2\Ncontact+dW(f)+s\VDM,\label{eq:hardcore}\\
 Q&=2\Nop-\Ncontact-\frac12\sum_{\langle ij\rangle}
 (b_i^\dagger b_j+b_j^\dagger b_i),\qquad
 W(f)=\frac12\sum_{i,\br\ne0}f(\br)
 (b_i^\dagger b_{i+\br}^\dagger+b_i b_{i+\br}),\nonumber\\
 \VDM&=\mathcal V_++\mathcal V_+^\dagger,\qquad
 \mathcal V_+=\sum_{i,\mu}c_\mu(n_i b_{i+\mathbf e_\mu}^\dagger
 -b_i^\dagger n_{i+\mathbf e_\mu}),\qquad
 c_x=i/2,\quad c_y=-1/2.\nonumber
\end{align}
Here $f(\pm\mathbf e_x)=-1/4$, $f(\pm\mathbf e_y)=1/4$, and
$f=0$ elsewhere. The linear DM terms telescope; the cubic operator $\VDM$
remains. The number-conserving one-magnon band is
\begin{equation}
 \omega_{\rm nc}(\bk)=h\eps^2-d+
 a_\eps(2-\cos k_x-\cos k_y).
 \label{eq:band}
\end{equation}
Thus its gap is $h\eps^2-d=g\eps^2+O(\eps^4)$, where
$g=h-1/2$; put $m_h^2=2g$. This calculation fixes the mass of the
specified pair block, without asserting a full interacting spectral-gap theorem.

Relative pair wavefunctions are even functions on
$\mathbb Z^2\setminus\{0\}$. Their inner product uses counting measure.
In particular, on an $N$-site torus,
$\Psi_u=\frac12\sum_{i,\br\ne0}u(\br)b_i^\dagger b_{i+\br}^\dagger\Omega$
has norm $\|\Psi_u\|^2=N\|u\|_2^2/2$.
Let $\cL$ be the retained-edge graph Laplacian on the punctured lattice,
and let $P_\nn$ multiply by the indicator of its four nearest neighbors.
The exact number-conserving two-magnon block at total momentum zero is
\begin{equation}
 A_2=a_\eps\cL+\mu_\eps I+\frac d2P_\nn,
 \qquad \mu_\eps=2h\eps^2-2d,
 \qquad w_2=-dA_2^{-1}f.
 \label{eq:A2}
\end{equation}
The absent hop into the origin and the nearest-neighbor attraction in
$Q$ together reduce the diagonal from four to three at the puncture's
neighbors. Exclusion alone would leave the diagonal equal to four.

\subsubsection{Rank-one defect solution}
Define
\begin{equation}
 \lambda(\bk)=4-2\cos k_x-2\cos k_y,\quad
 \mathfrak d(\bk)=\cos k_x-\cos k_y,\quad
 \av{u}=\int_{[-\pi,\pi]^2}\frac{\dd^2k}{(2\pi)^2}u(\bk),
 \label{eq:fourier}
\end{equation}
and, for $\sigma\ge0$,
\begin{equation}
 G_\sigma(\br)=\av{\frac{e^{i\bk\cdot\br}\mathfrak d(\bk)}{2[\lambda(\bk)+\sigma]}},
 \qquad t_\sigma=4G_\sigma(\mathbf e_x)
 =\av{\frac{\mathfrak d(\bk)^2}{\lambda(\bk)+\sigma}}.
 \label{eq:G}
\end{equation}
At $\sigma=0$ the multiplier is defined almost everywhere; it is bounded
because $|\mathfrak d|\le\lambda/2$. These functions are even in each coordinate
and odd under $x\leftrightarrow y$, so $G_\sigma(0)=0$ and their four
nearest-neighbor values alternate. Extending by zero at the origin,
$\cL=\cL_0-P_\nn$ in this sector and
$P_\nn G_\sigma=-t_\sigma f$. Consequently
\begin{equation}
 -(\cL+\sigma)^{-1}f=\frac{G_\sigma}{1-t_\sigma},\qquad
 \boxed{w_2(\br)=\frac{dG_\sigma(\br)}{a_\eps-ct_\sigma},
 \quad \sigma=\mu_\eps/a_\eps.}
 \label{eq:rankone}
\end{equation}
Indeed $A_2=a_\eps\cL_0+\mu_\eps-cP_\nn$ and
$A_2G_\sigma=-(a_\eps-ct_\sigma)f$. The four-site contact term
has only one independent component in this symmetry sector.

For completeness, integrating over one momentum and then setting
$u=\cos(k_x/2)$ gives
\begin{equation}
 t_0=-1+\frac8\pi\int_0^1\frac{1-u^2}{\sqrt{2-u^2}}\,\dd u
 =\frac4\pi-1,\qquad
 Z_0=1-t_0=\frac{2(\pi-2)}\pi>0.
 \label{eq:t0}
\end{equation}
The antiderivative in the integral is $u\sqrt{2-u^2}/2$.
In particular,
\begin{equation}
 F=-\cL^{-1}f=G_0/Z_0,\qquad
 F(\mathbf e_x)=\frac{4-\pi}{8(\pi-2)}.
 \label{eq:F}
\end{equation}
The inverse at zero mass denotes this $\ell^2$ source solution, not a
bounded inverse of $\cL$ on all of $\ell^2$. A square-summable solution
of $\cL u=0$ has zero Dirichlet form, is constant on the connected
punctured graph, and hence vanishes. This proves uniqueness.

\paragraph{Detailed assertions of Theorem~\ref{thm:massive}.}
Let $\cI_2\Subset(1/2,\infty)$.
For all sufficiently small positive $\eps$, uniformly in $h\in\cI_2$,
$A_2$ is strictly positive and Eq.~\eqref{eq:rankone} is its unique
square-summable source solution. At each fixed $\br\ne0$,
\begin{equation}
 w_2(\br)=dF(\br)+O_{\cI_2,\br}(\eps^4),\qquad
 \left\|\frac{w_2}{d}-F\right\|_2\longrightarrow0,
 \qquad \eps^{-4}\|w_2\|_2^2\longrightarrow n_*,
 \label{eq:Anorm}
\end{equation}
where $n_*=\pi/[256(\pi-2)]$.
For every compact $X\subset\mathbb R^2\setminus\{0\}$,
\begin{equation}
 \sup_{\substack{h\in\cI_2,\ \br\in\mathbb Z^2\setminus\{0\}\\
                         \eps\br\in X}}
 \left|\eps^{-4}w_2(\br)-A_h(\eps\br)\right|\longrightarrow0,
 \qquad
 A_h(\rho,\theta)=\frac{m_h^2\cos(2\theta)}{32(\pi-2)}K_2(m_h\rho).
 \label{eq:Alimit}
\end{equation}
Here $\rho=|\bx|$, $\theta=\arg\bx$, and $K_\nu$ is the modified
Bessel function on its positive-real branch.

\paragraph{Proof: positivity and the norm limit.}
For small $\eps$, $a_\eps>0$, $d\ge0$, and
$\mu_\eps\ge(2\min\cI_2-1)\eps^2>0$, so
$A_2\succeq\mu_\eps I$. Also $0\le t_\sigma\le t_0<1$ and
$a_\eps-ct_\sigma\to Z_0$ uniformly. Dominated convergence applied
to the bounded Fourier multipliers in Eq.~\eqref{eq:G} gives
$G_\sigma\to G_0$ in $\ell^2$, hence the norm limit.
Section~\ref{sm:green} evaluates $\|F\|_2^2=4n_*$.
For fixed $\br$, the derivative of $G_\sigma(\br)$ at $\sigma=0$
is integrable: evenness replaces the exponential by a cosine, and
the constant part cancels by antisymmetry of $\mathfrak d$. The remaining
numerator $(\cos(\bk\cdot\br)-1)\mathfrak d(\bk)$ is $O_\br(|\bk|^4)$.
The derivative of $t_\sigma$ is also integrable since $\mathfrak d^2=O(|\bk|^4)$.
Expanding the scalar prefactor then proves the fixed-distance bound.

\paragraph{Proof: the compact-uniform scaling limit.}
Choose a smooth radial cutoff $\chi$ equal to one near zero, supported
in a disk strictly inside the Brillouin square. Split Eq.~\eqref{eq:G}
with $1=\chi+(1-\chi)$. On the second part the multiplier and every
derivative are uniformly bounded as $\eps\downarrow0$ and $h\in\cI_2$.
Periodic integration by parts $N$ times in a coordinate with
$|r_j|\ge|\br|/\sqrt2$ therefore bounds that part by
$C_N|\br|^{-N}$. After multiplication by $\eps^{-2}$ it is
$O_{X,N}(\eps^{N-2})$.

In the first part set $\bk=\eps\bp$ and $\bx=\eps\br$. Then
\begin{equation}
 \eps^{-2}G_\sigma^{\rm low}(\br)
 =\int\frac{\dd^2p}{(2\pi)^2}e^{i\bp\cdot\bx}
       \chi(\eps\bp)q_{\eps,h}(\bp),\quad
 q_{\eps,h}(\bp)=\frac{\mathfrak d(\eps\bp)}{2[\lambda(\eps\bp)+\sigma]},
 \label{eq:scaledintegral}
\end{equation}
with
\begin{equation}
 \frac{\sigma}{\eps^2}=m_h^2+O_{\cI_2}(\eps^2),\qquad
 q_{\eps,h}(\bp)\longrightarrow
 q_{0,h}(\bp)=\frac{p_y^2-p_x^2}{4(|\bp|^2+m_h^2)}
 \label{eq:qsymbol}
\end{equation}
locally uniformly in $\bp,h$. Local convergence alone is insufficient
because $q_{0,h}$ does not decay radially. We control the oscillatory
tails explicitly. Taylor's formula, ellipticity
$\lambda(\bk)\ge c_0|\bk|^2$ on the cutoff support, and the quotient
rule imply, for each multi-index $\alpha$,
\begin{equation}
 |\partial_\bp^\alpha q_{\eps,h}(\bp)|
 \le C_\alpha|\bp|^{-|\alpha|},\qquad
 1\le|\bp|\le c_1/\eps.
 \label{eq:symbolestimate}
\end{equation}
The same estimate holds for $q_{0,h}$; all constants are uniform in
$h\in\cI_2$. To see the derivative powers, derivatives of the numerator
are $O(\eps^2|\bp|^{2-|\alpha|})$ and the denominator is bounded below
by $c\eps^2(|\bp|^2+1)$. Higher derivatives obey the same bound on
$\eps|\bp|\le c_1$, with their extra powers of $\eps$ absorbed by
$|\bp|^{-1}$.

Let $\psi$ be a second smooth radial cutoff, equal to one on the unit
disk and zero outside the disk of radius two. For $R\ge1$, integrate
the tail with factor $1-\psi(\bp/R)$ by parts $N>2$ times using
$\bx\cdot\nabla_\bp/(i|\bx|^2)$. A dyadic annulus of radius $s$
contributes at most $C_N|\bx|^{-N}s^{2-N}$. Derivatives of both cutoffs
obey the same bound on their annuli. Summing gives
\begin{equation}
 \left|\int\frac{\dd^2p}{(2\pi)^2}e^{i\bp\cdot\bx}
 [1-\psi(\bp/R)]\chi(\eps\bp)q_{\eps,h}(\bp)\right|
 \le C_{X,N}R^{2-N},\qquad \bx\in X.
 \label{eq:tail}
\end{equation}
This bound is uniform in sufficiently small $\eps$ and $h$.
It also holds for the limiting symbol with an outer smooth cutoff,
uniformly in its radius. At fixed $R$, dominated convergence in the
remaining compact integral is uniform in $(\bx,h)\in X\times\cI_2$.
Taking first $\eps\downarrow0$ and then $R\to\infty$ proves convergence
to the cutoff-defined oscillatory transform of $q_{0,h}$.

This transform equals the Abel limit obtained with
$e^{-\eta|\bp|^2}$ as $\eta\downarrow0$: derivatives of this factor
satisfy $|\partial^\alpha e^{-\eta|\bp|^2}|\le
C_\alpha|\bp|^{-|\alpha|}$ for $|\bp|\ge1$, uniformly in $\eta>0$,
so the same tail argument applies. The scalar transform follows from
\begin{equation}
 \int\frac{\dd^2p}{(2\pi)^2}
 \frac{e^{i\bp\cdot\bx}}{|\bp|^2+m_h^2}
 =\frac1{4\pi}\int_0^\infty\frac{\dd t}{t}
 e^{-m_h^2t-\rho^2/(4t)}=\frac{K_0(m_h\rho)}{2\pi},\qquad \rho>0.
 \label{eq:K0}
\end{equation}
The heat-kernel representation justifies differentiation away from
the origin; it includes no contact contribution there. Hence
\begin{align}
 \mathcal F^{-1}q_{0,h}(\bx)
 &=\frac1{8\pi}(\partial_x^2-\partial_y^2)K_0(m_h\rho)
 =\frac{m_h^2\cos2\theta}{8\pi}K_2(m_h\rho),\label{eq:Hessian}\\
 (\partial_x^2-\partial_y^2)f(\rho)
 &=\cos2\theta\,[f''(\rho)-f'(\rho)/\rho].\nonumber
\end{align}
The Bessel recurrence $K_2(z)=K_0(z)+2K_1(z)/z$ gives the last equality.
Multiplying by $d/\eps^2\to1/2$ and
$(a_\eps-ct_\sigma)^{-1}\to Z_0^{-1}$ proves Eq.~\eqref{eq:Alimit}.
\hfill$\square$

\subsubsection{Matching and the local correction}
The small- and large-argument expansions give
\begin{align}
 A_h(\rho,\theta)
 &=\frac{\cos2\theta}{16(\pi-2)\rho^2}
 -\frac{m_h^2\cos2\theta}{64(\pi-2)}
 +O_{\cI_2}(\rho^2|\log\rho|),\label{eq:smallrho}\\
 A_h(\rho,\theta)
 &=\frac{m_h^2\cos2\theta}{32(\pi-2)}
 \sqrt{\frac\pi{2m_h\rho}}e^{-m_h\rho}
 \left[1+\frac{15}{8m_h\rho}+O_{\cI_2}(\rho^{-2})\right].
 \label{eq:largerho}
\end{align}
The first expression matches
$dF(\br)\sim\eps^2\cos2\theta/[16(\pi-2)r^2]$.
A useful uniform lattice bound, which does not divide by an angular node, is
\begin{equation}
 \sup_{\br\ne0}|w_2(\br)-dF(\br)|
 \le C_{\cI_2}\eps^4\log(2/\eps).
 \label{eq:matching}
\end{equation}
Indeed $|G_\sigma-G_0|\le(\sigma/4)\av{1/(\lambda+\sigma)}
\le C\sigma\log(2/\sigma)$ and $|t_\sigma-t_0|\le\sigma/4$.
In particular the weighted absolute massless match holds for
$r\le\eps^{-\delta}$, every $\delta<1$, for this operator.
Equations~\eqref{eq:Alimit} and \eqref{eq:largerho} take the weak-angle
scaling limit first; the decay rate at fixed nonzero angle on the lattice
is direction dependent.

The local contact term is necessary even when only a fourth-order
fixed-distance correction is sought. Write $F_1=F(\mathbf e_x)$ and
$F_{m_h}=-(\cL+m_h^2\eps^2)^{-1}f$. Expanding
\begin{equation}
 A_2=\cL+\eps^2\left[-\frac14\cL+m_h^2 I+\frac14P_\nn\right]
 +O_{\cI_2}(\eps^4)
 \label{eq:Aexpand}
\end{equation}
and using $P_\nn F=-4F_1 f$ gives, at every fixed nonzero $\br$,
\begin{equation}
 \lim_{\eps\downarrow0}\frac{w_2(\br)-dF_{m_h}(\br)}{\eps^4}
 =\frac{1-4F_1}{8}F(\br).
 \label{eq:localcorrection}
\end{equation}
The fixed-distance differentiability established above justifies this
resolvent expansion despite the zero spectral threshold of $\cL$.
Theorem~\ref{thm:massive} concerns the block \eqref{eq:A2}; the full spin problem also
contains, among other connected terms, the pair-triplet self-energy
$-s^2\mathcal V_{23}(A_3-z)^{-1}\mathcal V_{32}$, where $\mathcal V_{32}$ is the two-to-three-magnon
block of $\VDM$ and $\mathcal V_{23}=\mathcal V_{32}^\dagger$. Its replacement by the microscopic RDM
requires the estimates specified in Theorem~\ref{thm:rdmcriterion} below.

\subsection{Microscopic reduced-state criterion}
\label{ss:massive-rdm}

\subsubsection{The symmetry-constrained state}
We use $\Neg(\rho)=(\|\rho^{T_2}\|_1-1)/2$, the sum of the
absolute values of negative partial-transpose eigenvalues.
Consider a translation-invariant spin-$1/2$ state invariant under the
joint quarter-turn of spin and lattice coordinates. The one-site
transverse mean acquires a nontrivial phase under this operation, so
$\langle b_i\rangle=0$. A joint half-turn followed by translation
exchanges the two selected sites; their RDM is invariant under
$\mathrm{Swap}(\sigma_z\otimes\sigma_z)$. With the additional zero
transverse means, its general form in the ordered basis
$\uparrow\uparrow,\uparrow\downarrow,\downarrow\uparrow,
\downarrow\downarrow$ is
\begin{equation}
 \rho_{0,\br}=
 \begin{pmatrix}
 a&\zeta&-\zeta&w^*\\
 \zeta^*&b&v&\zeta\\
 -\zeta^*&v&b&-\zeta\\
 w&\zeta^*&-\zeta^*&q
 \end{pmatrix},\qquad
 \begin{gathered}
 p=\langle n_0\rangle=\langle n_{\br}\rangle,\quad
 q=\langle n_0n_{\br}\rangle,\\
 a=1-2p+q,\quad b=p-q,\\
 w=\langle b_0b_{\br}\rangle,\quad
 v=\langle b_0^\dagger b_{\br}\rangle\in\mathbb R.
 \end{gathered}
 \label{eq:rho}
\end{equation}
To obtain the odd entries, half-turn invariance first gives
$\rho_{00,10}=-\rho_{00,01}$ and
$\rho_{10,11}=-\rho_{01,11}$. The equation
$\langle b_0\rangle=0$ then sets $\rho_{01,11}=\rho_{00,01}$.
Thus half-turn invariance alone is insufficient for
Eq.~\eqref{eq:rho}. Our convention is
$\zeta=\langle b_0n_{\br}\rangle^*$; neither $\zeta=0$ nor real $w$
is imposed.

For $\zeta=0$ the partial-transpose eigenvalues are
\begin{equation}
 b\pm|w|,\qquad
 \frac{a+q\pm\sqrt{(a-q)^2+4v^2}}2.
 \label{eq:Xeigen}
\end{equation}
For general $\zeta$, the first pair remains the spectrum of a principal
block, and therefore
\begin{equation}
 \Neg(\rho_{0,\br})\ge (|w|-p+q)_+.
 \label{eq:NPTsufficient}
\end{equation}
This is a sufficient condition; positivity of the other block must
also be resolved to prove separability.

\subsubsection{Exact exterior criterion}
Order the partial-transpose basis as $(00,11;01,10)$. It becomes
\begin{equation}
 \rho^{T_2}=
 \begin{pmatrix}
 a&v&\zeta^*&-\zeta\\
 v&q&\zeta^*&-\zeta\\
 \zeta&\zeta&b&w^*\\
 -\zeta^*&-\zeta^*&w&b
 \end{pmatrix},\qquad
 M=\begin{pmatrix}b&w^*\\w&b\end{pmatrix},\quad
 u=\begin{pmatrix}\zeta\\-\zeta^*\end{pmatrix}.
 \label{eq:PT}
\end{equation}
When $b>|w|$, $M$ is positive definite. Its Schur correction is the
same for every entry of the upper block:
\begin{equation}
 \chi=u^\dagger M^{-1}u
 =\frac{2[b|\zeta|^2+\Ree(w\zeta^2)]}{b^2-|w|^2}\ge0,
 \qquad
 \rho^{T_2}\succeq0\ \Longleftrightarrow\
 \begin{pmatrix}a-\chi&v-\chi\\v-\chi&q-\chi\end{pmatrix}\succeq0.
 \label{eq:Schur}
\end{equation}
For $a>\chi$ this is equivalent to the single determinant condition
\begin{equation}
 \boxed{aq-v^2\ge\chi(a+q-2v).}
 \label{eq:exterior}
\end{equation}
Indeed the determinant is
$(a-\chi)(q-\chi)-(v-\chi)^2=aq-v^2-\chi(a+q-2v)$;
a positive first diagonal entry and a nonnegative determinant also
force the second diagonal entry to be nonnegative. The strict conditions
$b>|w|$ and $a>\chi$ are part of this criterion. Singular cases are
tested directly from Eq.~\eqref{eq:PT}.

\paragraph{Detailed assertions of Theorem~\ref{thm:rdmcriterion}.}
Let $\rho_\eps$ be physical two-qubit states of the form
\eqref{eq:rho}, at displacements with
$\eps\br_\eps\to\rho(\cos\theta,\sin\theta)$,
$\rho>0$, $\cos2\theta\ne0$. Suppose
\begin{equation}
 \frac p{\eps^4}\to n_*,\qquad
 \frac w{\eps^4}\to A_h(\rho,\theta)\ne0,\qquad
 q=o(\eps^4),\quad v=o(\eps^4),\quad \zeta=o(\eps^6).
 \label{eq:Bhyp}
\end{equation}
Then
\begin{equation}
 \eps^{-4}\Neg(\rho_\eps)\longrightarrow
 (|A_h(\rho,\theta)|-n_*)_+.
 \label{eq:Bneg}
\end{equation}
Away from equality $|A_h|=n_*$, the state is entangled for all
sufficiently small $\eps$ if and only if $|A_h|>n_*$.
Uniform hypotheses imply uniform conclusions on compact nonnodal
sets with a fixed strict margin from equality.

\paragraph{Proof.}
Since $a=1+O(\eps^4)$ and
$b=\eps^4n_*+o(\eps^4)$, the smaller eigenvalue of $M$ is
$\eps^4(n_*-|A_h|)+o(\eps^4)$.
If $|A_h|>n_*$, this principal block has a negative eigenvalue and
the entire partial transpose cannot be positive.
To obtain the negativity limit, remove the $\zeta$ entries at an
operator-norm cost $O(|\zeta|)=o(\eps^4)$.
The smaller eigenvalue of the remaining $(00,11)$ block is
$q+O(v^2)=o(\eps^4)$, while its other eigenvalue tends to one.
Equation~\eqref{eq:Xeigen} and the eigenvalue perturbation bound
then give Eq.~\eqref{eq:Bneg}.

For $|A_h|<n_*$, the determinant of $M$ has the strictly positive
coefficient $n_*^2-|A_h|^2$ at order $\eps^8$.
The numerator of $\chi$ is $o(\eps^{16})$, so
$\chi=o(\eps^8)$ and $a>\chi$. Positivity of the \emph{original}
state gives the principal-minor inequality
\begin{equation}
 aq\ge|w|^2=\eps^8(|A_h|^2+o(1)).
 \label{eq:originalpositivity}
\end{equation}
Because $v^2=o(\eps^8)$ and
$\chi(a+q-2v)=o(\eps^8)$, Eq.~\eqref{eq:exterior} is strict for
small $\eps$. All four partial-transpose eigenvalues are therefore
nonnegative. Positive partial transpose is equivalent to separability
for two qubits~\cite{Horodecki}. The same argument is uniform when
the hypotheses and the nonzero lower bound on $|A_h|$ are uniform.
\hfill$\square$

The assumptions in Eq.~\eqref{eq:Bhyp} concern microscopic RDMs.
Theorem~\ref{thm:massive} supplies the asymptotics of $w_2$ and its pair norm;
it does not estimate the other entries of a microscopic RDM.
The missing full-model comparison at $r\asymp\eps^{-1}$, including
the odd coherence at precision $o(\eps^6)$, is necessary for a
thermodynamic crossover theorem. An $o(\eps^4)$ trace-distance
approximation by an X state does not control the smallest exterior
eigenvalue, whose decisive scale is $\eps^8$.

\subsubsection{Boundary, nodal cusp, and a sufficient far-tail condition}
For each nonnodal ray, $K_2(x)>0$ and
$K_2'(x)=-[K_1(x)+K_3(x)]/2<0$ for $x>0$.
It diverges as $2/x^2$ at zero and tends to zero at infinity.
Thus $|A_h|=n_*$ has exactly one positive solution,
\begin{equation}
 K_2(m_h\rho_E(\theta))=\frac\pi{8m_h^2|\cos2\theta|}.
 \label{eq:boundary}
\end{equation}
Its axial radii are $1.412821695745$ at $h=2$ and
$1.259664634696$ at $h=3$. If $c_\theta=|\cos2\theta|\downarrow0$,
Eq.~\eqref{eq:smallrho} gives
\begin{equation}
 \rho_E(\theta)=4\sqrt{\frac{c_\theta}{\pi}}
 \left[1-\frac{2m_h^2c_\theta}{\pi}
 +O(c_\theta^2|\log c_\theta|)\right]
 \sim4\sqrt{\frac{2|\theta-\pi/4|}{\pi}}.
 \label{eq:cusp}
\end{equation}
This is the cusp of the leading profile. It does not imply exact
diagonal separability: with the antiunitary reflection symmetry,
the microscopic $w$ is real on axes and purely imaginary, rather
than necessarily zero, on diagonals. Neither the nodes nor the equality
case is covered by the strict-margin conclusion of Theorem~\ref{thm:rdmcriterion}.

A supremal entanglement range also requires exclusion of more distant
entangled pairs. One sufficient uniform condition, for all
$r\ge C/\eps$ with some fixed sufficiently large $C$, is
\begin{equation}
 q\asymp p^2,\qquad q=o(p),\qquad
 |w|\le(1-\eta)p,\qquad v=o(p),\qquad
 \zeta=o(p^{3/2}),\quad \eta>0.
 \label{eq:tailcondition}
\end{equation}
Then $b^2-|w|^2\ge c_\eta p^2$ and
$\chi\le C_\eta|\zeta|^2/p=o(p^2)$, while
$aq-v^2\ge c p^2$ for small $\eps$. Equation~\eqref{eq:exterior}
proves PPT throughout that tail. Combined with uniform versions of
Eq.~\eqref{eq:Bhyp} in the intervening nonnodal scaling windows and
the interior matching, this gives the margin formulation: NPT for
$\rho<\rho_E-\eta'$ and PPT for $\rho>\rho_E+\eta'$, for every
fixed $\eta'>0$, and hence $\eps r_E\to\rho_E$ for the supremal
range. Condition~\eqref{eq:tailcondition} is an additional sufficient
hypothesis, not a consequence of the compact-window scaling limit.

\subsection{Thermal activation in the dilute-pair description}
\label{ss:activation}

\subsubsection{Dilute magnon density}
We use the dimensionless temperature $t=T/J$, with $k_B=1$. In the dilute soft-band
description the excitation energy is
$E_\bk=g\eps^2+|\bk|^2/2$, with $g=h-1/2>0$.
At $t=\tau\eps^2$, a radial integral gives
\begin{align}
 p_{\rm th}
 &=\int\frac{\dd^2k}{(2\pi)^2}
       \frac1{e^{(|\bk|^2/2+g\eps^2)/t}-1}
 =-\frac{t}{2\pi}\log(1-e^{-g\eps^2/t})
 =\eps^2 B_g(\tau),\label{eq:Bose}\\
 B_g(\tau)&=-\frac\tau{2\pi}\log(1-e^{-g/\tau})
 =\frac\tau{2\pi}e^{-g/\tau}
       [1+O(e^{-g/\tau})],\qquad \tau\downarrow0.
 \label{eq:Boltzmann}
\end{align}
For example, expanding the Bose occupation as
$\sum_{j\ge1}e^{-j(|\bk|^2/2+g\eps^2)/t}$ and integrating each
nonnegative term gives $t\sum_{j\ge1}e^{-jg\eps^2/t}/(2\pi j)$.
The coefficient $2\pi$ follows from the one-magnon stiffness $1/2$.
This is the leading finite-field ideal-ferromagnet density
associated with the spin-wave free energy~\cite{Hofmann}.

For the lattice band \eqref{eq:band}, the same scaling follows by
splitting off a neighborhood of zero and putting $\bk=\eps\bp$.
On compact positive $\tau,g$ intervals the complementary contribution
is exponentially small and dominated convergence applies to the
rescaled integral. In the activated regime, comparison at relative
precision also requires the correction to the gap divided by $t$ to
vanish. The number-conserving gap error is $O(\eps^4)$, so it satisfies
this requirement on the boundaries below, where
$\tau\asymp1/\log r$ or $1/\log(1/\eps)$.

The corresponding linear dilute-pair response at a nonzero scaled
separation is
\begin{align}
 A_{h,\tau}(\rho,\theta)
 =\frac{\cos2\theta}{32(\pi-2)}\bigg[
 m_h^2K_2(m_h\rho)
 +2\int_0^\infty\frac{u^3J_2(u\rho)}{u^2+m_h^2}
       \frac{\dd u}{e^{(u^2+m_h^2)/(2\tau)}-1}\bigg],
 \label{eq:thermalA}\\
 v_{\rm th}(\rho/\eps)
 =\eps^2 C_{g,\tau}(\rho),\qquad
 C_{g,\tau}(\rho)=\int_0^\infty\frac{u\,\dd u}{2\pi}
       \frac{J_0(u\rho)}{e^{(u^2/2+g)/\tau}-1}.
 \label{eq:thermalv}
\end{align}
The factor two in Eq.~\eqref{eq:thermalA} is the Bose enhancement
$\coth(E/2t)-1=2n_B(E)$. Its thermal integral is absolutely convergent.
For $\rho$ in a compact set and $g$ bounded away from zero, use
$|J_2(u\rho)|\le C_\rho u^2$ and the Gaussian Boltzmann bound to obtain
\begin{equation}
 |A_{h,\tau}-A_h|\le C\tau^3 e^{-g/\tau},\qquad \tau\downarrow0.
 \label{eq:thermalAsmall}
\end{equation}
This explicitly controls the thermal anomalous correction in the
activated scaling, while the population remains competitive.

\subsubsection{Interior inversion and convergence of the logarithm}
Fix a nonnodal direction and put
$C_\theta=|\cos2\theta|/[16(\pi-2)]$.
For $1\ll r\ll\eps^{-1}$ the leading zero-temperature anomalous
amplitude is $\eps^2C_\theta/r^2$; the quantum population
$n_*\eps^4$ is smaller by $O(\eps^2r^2)$.
The effective thermal boundary therefore solves
\begin{equation}
 B_g(\tau_E)=\frac{C_\theta}{r^2}.
 \label{eq:thermalinterior}
\end{equation}
$B_g$ is strictly increasing from zero to infinity, so this equation
has one positive solution. Replacing $B_g$ by its activated leading
term and writing $x=g/\tau_E$ yields
$xe^x=gr^2/(2\pi C_\theta)$. On the positive-real principal branch,
\begin{equation}
 \boxed{t_E^{\rm act}(r)=
 \frac{g\eps^2}{W_0\!\left(gr^2/(2\pi C_\theta)\right)}.}
 \label{eq:Lambert}
\end{equation}
This expression is exact for the Boltzmann equation, and asymptotic
for Eq.~\eqref{eq:thermalinterior}.
To quantify the latter statement, put $y=C_\theta/r^2$ and
$x=W_0(g/(2\pi y))$. At $\tau_W=g/x$ the fugacity is
$z=e^{-x}=2\pi yx/g$ and
$B_g(\tau_W)/y=1+O(z)$.
The logarithmic derivative of the Boltzmann density with respect to
$\tau$ is $1+g/\tau=1+x$. Monotonicity and the mean-value theorem
therefore give
\begin{equation}
 \frac{\tau_E}{\tau_W}=1+O\!\left(\frac{z}{1+x}\right)
 =1+O(r^{-2}),\qquad \tau_E\le\tau_W.
 \label{eq:Lambertconvergence}
\end{equation}
The comparison is uniform for fields in a compact positive-gap set
and angles bounded away from nodes. It addresses the effective-density
inversion; lattice matching and interacting-state errors are separate.

For $L=\log[gr^2/(2\pi C_\theta)]$, the principal-branch asymptotic
$W_0(e^L)=L-\log L+O(\log L/L)$~\cite{Corless1996} gives
\begin{equation}
 t_E^{\rm act}=
 \frac{g\eps^2}{2\log r+\log[g/(2\pi C_\theta)]
                 -\log(2\log r)+O(\log\log r/\log r)}
 \sim\frac{g\eps^2}{2\log r}.
 \label{eq:thermallog}
\end{equation}
The relative convergence of the last expression is only
$O(\log\log r/\log r)$. For the axial $h=2$ calculation using the
discrete massless kernel and the unexpanded Bose density,
$2\tau_E\log r/g$ is $1.011,1.079,1.084,1.080,1.074$ at
$r=10,10^2,10^3,10^4,10^5$, respectively.
Keeping $W_0$ avoids this slow leading-log approximation.

\subsubsection{Mass-scale activation and conditional RDM transfer}
At $r=\rho/\eps$ retain the quantum population. For a point strictly
inside the zero-temperature lobe define
$M_h(\rho,\theta)=|A_h(\rho,\theta)|-n_*>0$.
The full effective-density equation and its activated inversion are
\begin{equation}
 B_g(\tau_E)=\eps^2M_h,\qquad
 t_E^{\rm act}(\rho/\eps)=
 \frac{g\eps^2}{W_0\!\left(g/[2\pi\eps^2M_h]\right)}
 \sim\frac{g\eps^2}{2\log(1/\eps)}.
 \label{eq:massiveThermal}
\end{equation}
For fixed positive $M_h$, the relative Bose-versus-Boltzmann inversion
error is $O(\eps^2)$ by the same argument as
Eq.~\eqref{eq:Lambertconvergence}. There is no positive boundary
from this equation if $M_h\le0$.

The four population regimes are summarized in Table~\ref{tab:thermal}.
In the double scaling $B_g(\tau)/\eps^2\to\nu<\infty$,
Eq.~\eqref{eq:thermalAsmall} gives $A_{h,\tau}\to A_h$ and the
population-subtracted profile and boundary are
\begin{equation}
 \eps^{-4}\Neg_t\longrightarrow (|A_h|-n_*-\nu)_+,
 \qquad
 K_2(m_h\rho_E(\theta;\nu))
 =\frac{32(\pi-2)(n_*+\nu)}{m_h^2|\cos2\theta|},
 \label{eq:thermalboundary}
\end{equation}
provided the corresponding RDM estimates hold.

\begin{table}[htb]
\caption{Leading dilute-pair boundary in the thermodynamic-first limit.
The ranges refer to nonnodal directions and the RDM transfer specified
in the text.}\label{tab:thermal}
\begin{ruledtabular}
\begin{tabular}{ll}
Temperature scaling & Leading spatial boundary \\
\hline
Fixed $\tau>0$ & Fixed-distance criterion $|F(\br)|/2=B_g(\tau)$ \\
$\tau\to0$, $B_g(\tau)/\eps^2\to\infty$ &
$r_E\sim\sqrt{C_\theta/B_g(\tau)}\ll\eps^{-1}$ \\
$B_g(\tau)/\eps^2\to\nu\in(0,\infty)$ &
$r_E\sim\rho_E(\theta;\nu)/\eps$ \\
$B_g(\tau)/\eps^2\to0$ & Zero-temperature boundary \eqref{eq:boundary}
\end{tabular}
\end{ruledtabular}
\end{table}

There is a thermal distinction in the exterior proof.
From Eq.~\eqref{eq:thermalv},
$C_{g,\tau}(\rho)/B_g(\tau)\to1$ as $\tau\downarrow0$ for fixed
$\rho$: rescale $u=\sqrt\tau\,t$ and use $J_0(\sqrt\tau\,t\rho)\to1$.
Thus $v_{\rm th}/\eps^4\to\nu$, not zero, when $\nu>0$.
A sufficient thermal version of Theorem~\ref{thm:rdmcriterion} assumes
$p/\eps^4\to n_*+\nu$, $w/\eps^4\to A_h\ne0$,
$q=o(\eps^4)$, $v=O(\eps^4)$, $\zeta=o(\eps^6)$,
and, in the strict exterior, the determinant margin
\begin{equation}
 aq-v^2\ge c\eps^8,\qquad c>0.
 \label{eq:thermalmargin}
\end{equation}
The Schur proof then applies unchanged. The leading dilute Gaussian
expression $q=p^2+|w|^2+|v|^2+\cdots$ supplies this margin in that
approximation; a full-model proof must bound its connected correction.
At fixed $\tau>0$, the thermal population is $O(\eps^2)$ whereas
mass-scale anomalous coherence is $O(\eps^4)$, which removes the
mass-scale lobe in the dilute theory. Its second PT block has the
leading positive margin $p_{\rm th}^2$.

For interacting Gibbs states, Eqs.~\eqref{eq:Lambert} and
\eqref{eq:thermalboundary} require relative population control where
$p_{\rm th}\asymp\eps^4$, anomalous and odd-coherence remainders
smaller than the stated margins, and a thermal counterpart of the
far-tail bound \eqref{eq:tailcondition}. These estimates are not
provided by the ideal-gas integral or by an additive pressure error.
The sufficient thermal window of Theorem~\ref{thm:thermal} and the sharp activated
boundary therefore have different hypotheses.

\section{Leading negativity at the mass scale}
\label{sm:massscale}

Theorem~\ref{thm:range} controls the anomalous coherence for
$|\br|\le\eps^{-1}\ell_\eps^{-a}$. At a fixed nonzero value of
$\eps|\br|$ its error, Eq.~\eqref{eq:range-pointwise}, is only
$O(\eps^4\ell_\eps^6)$ and does not resolve the signal
$\eps^4A_h(\eps\br)$ of Theorem~\ref{thm:massive}. This section closes
that gap for the actual states on the whole unbounded region
$|\br|\ge c/\eps$. The proof uses three ingredients beyond
Secs.~\ref{sm:normalform}--\ref{sm:range}: a compact triplet correction
with explicit budgets on the exact three-particle exclusion graph
(Sec.~\ref{ss:ms-triplet}); a mass-adjusted point probe whose values
near the contact sources are $O(\eps^2)$ (Sec.~\ref{ss:ms-probe}); and a
transfer of the joint population and of the normal coherence through
the pair dressing (Sec.~\ref{ss:ms-transfer}). The exact stationarity
identity of Lemma~\ref{lem:range-pair-identity} then compares $w$ with
$w_2$ (Sec.~\ref{ss:ms-comparison}).

Throughout, the states are the symmetric thermodynamic ground states of
Sec.~\ref{sm:conventions}: translation-invariant energy-density
minimizers that are invariant under the joint quarter-turn, at fixed
$\eps$ and $h\in\cI=[h_-,h_+]\Subset(3/2,\infty)$. The thermodynamic
limit is taken at fixed positive $\eps$; finite probe and coefficient
current truncations are removed at that angle exactly as in
Secs.~\ref{sm:current} and~\ref{sm:range}; only then $\eps\downarrow0$.
No reflection symmetry, uniqueness of the minimizer or many-body gap is
assumed. The letter $c>0$ denotes a fixed lower bound on the scaled
separation $\eps|\br|$; in this section the cosine of the angle is
always written $\cos\eps$. Constants $C$ depend only on $\cI$ and on the
fixed cutoff $\chi$ of Sec.~\ref{sm:green}; a subscript records further
dependence, for example $C_c$. No constant depends on the state, the
displacement, the field in $\cI$, or $\eps$. For $\bx\in\mathbb
R^2\setminus\{0\}$ with polar coordinates $(\rho,\theta)$ we use
$A_h(\bx)=A_h(\rho,\theta)$ of Eq.~\eqref{eq:profile-statement},
$m_h=\sqrt{2h-1}$, and the leading profile
\begin{equation}
 \Phi_h(\bx)=\bigl(|A_h(\bx)|-n_*\bigr)_+ .
 \label{eq:ms-profile}
\end{equation}
The autocorrelation of the source solution $F$ of
Eq.~\eqref{eq:constants} is
$C_F(\br)=\sum_{\bk}F(\bk)F(\bk-\br)$, with $F(0)=0$; thus
$C_F(0)=\|F\|_2^2=4n_*$.

\subsection{Results}

\begin{theorem}[Leading negativity at the mass scale]
\label{thm:massscale}
Fix $c>0$. Uniformly for $h\in\cI$ and all symmetric thermodynamic
ground states, the following hold as $\eps\downarrow0$.
\begin{enumerate}
\item[(i)] The anomalous coherence follows the massive pair response,
\begin{equation}
 \sup_{|\br|\ge c/\eps}\frac{|w(\br)-w_2(\br)|}{\eps^4}\longrightarrow0,
 \qquad
 \sup_{|\eps\br|\ge c}\left|\frac{w(\br)}{\eps^4}-A_h(\eps\br)\right|
 \longrightarrow0 .
 \label{eq:ms-main-w}
\end{equation}
\item[(ii)] The joint population and the normal coherence satisfy
\begin{equation}
 \sup_{\br\ne0}\left|\frac{q_\br}{\eps^4}-\frac{F(\br)^2}4\right|
 \longrightarrow0,\qquad
 \sup_{\br\ne0}\left|\frac{v_\br}{\eps^4}-\frac{C_F(\br)}4\right|
 \longrightarrow0 .
 \label{eq:ms-main-qv}
\end{equation}
Consequently $\sup_{|\br|\ge r_\eps}(q_\br+|v_\br|+|\zeta_\br|)/\eps^4
\to0$ for every $r_\eps\to\infty$. The odd coherence $\zeta$ of
Eq.~\eqref{eq:rho} is used only through the positivity bound
$|\zeta_\br|\le\sqrt{(p-q_\br)q_\br}\le\sqrt{pq_\br}$.
\item[(iii)] The negativity has the leading profile~\eqref{eq:ms-profile}
on the entire unbounded region,
\begin{equation}
 \sup_{|\eps\br|\ge c}
 \left|\frac{\Neg(\rho_{0\br})}{\eps^4}-\Phi_h(\eps\br)\right|
 \longrightarrow0 .
 \label{eq:ms-main-neg}
\end{equation}
\item[(iv)] Put $\varpi_\eps=(\eps\ell_\eps)^{2/13}$. There are
$C_c$ and $\eps_c>0$ such that, for $0<\eps\le\eps_c$, each of the three
suprema in Eqs.~\eqref{eq:ms-main-w} and~\eqref{eq:ms-main-neg} is at
most $C_c\varpi_\eps$.
\end{enumerate}
\end{theorem}

The region in Eq.~\eqref{eq:ms-main-neg} contains the exterior
$|A_h|<n_*$, the nodal directions $\cos2\theta=0$, and the infinite
tail. There Eq.~\eqref{eq:ms-main-neg} states $\Neg(\rho_{0\br})=
o(\eps^4)$ uniformly. This bounds the negative partial-transpose weight
at order $\eps^4$; it is not a separability statement. Exact positivity
of the partial transpose is decided by the Schur condition
\eqref{eq:exterior}, whose decisive scale is $\eps^8$, and the present
estimates do not control it. At scaled points with $|A_h|>n_*$, the
principal block~\eqref{eq:principal-pt} gives a negative eigenvalue of
order $\eps^4$. The exponent $2/13$ in (iv) balances the two slowest
terms in a finite collection of bounds, Eq.~\eqref{eq:ms-rate-budget}; it
is not claimed to be the true error exponent.

\begin{corollary}[Fixed-threshold range and a positive-area NPT sector]
\label{cor:ms-range}
Let $h\in\cI$ and fix $\kappa>0$.
\begin{enumerate}
\item[(a)] Let $\mathbf a\in\mathbb Z^2\setminus\{0\}$ be primitive,
with polar angle $\theta$ and $\cos2\theta\ne0$, and put
\begin{equation}
 R_\kappa(\eps)=\sup\{|k\mathbf a|:k\in\mathbb N,\
 \Neg(\rho_{0,k\mathbf a})\ge\kappa\eps^4\}.
 \label{eq:ms-Rkappa}
\end{equation}
For sufficiently small $\eps$ this set is nonempty and bounded, and
$\eps R_\kappa(\eps)\to\rho_\kappa(h,\theta)$, the unique positive
solution of $|A_h(\rho_\kappa,\theta)|=n_*+\kappa$, that is,
\begin{equation}
 K_2(m_h\rho_\kappa)=\frac{32(\pi-2)(n_*+\kappa)}{m_h^2|\cos2\theta|}.
 \label{eq:ms-rhokappa}
\end{equation}
\item[(b)] The largest such distance over all lattice directions,
$R_\kappa^{\rm all}(\eps)=\sup\{|\br|:\br\ne0,\
\Neg(\rho_{0\br})\ge\kappa\eps^4\}$, satisfies
$\eps R^{\rm all}_\kappa(\eps)\to\rho_\kappa(h,0)$.
\item[(c)] Suppose $2\in\cI$. At $h=2$, for all sufficiently small
$\eps$, every lattice point with $0.9\le\eps|\br|\le1$ and
$|\theta_\br|\le0.15$ satisfies
$\Neg(\rho_{0\br})\ge0.0099\,\eps^4$.
\end{enumerate}
The limits in (a) and (b) are uniform over the states and over
$h\in\cI$.
\end{corollary}

The threshold $\kappa$ is fixed before $\eps\downarrow0$. Since
$\rho_\kappa\uparrow\rho_E$ as $\kappa\downarrow0$, with $\rho_E$ the
leading contour~\eqref{eq:boundary}, the supremal NPT distance along the
ray obeys $\liminf_{\eps\downarrow0}\eps R_{\rm NPT}(\eps)\ge
\rho_E(h,\theta)$: indeed $R_{\rm NPT}\ge R_\kappa$ for every
$\kappa>0$. No corresponding upper limit, and no exclusion of weak
remote NPT pairs, follows from Theorem~\ref{thm:massscale}. By the
symmetry $\theta\mapsto\theta+\pi/2$ of $|A_h|$, statement (c) also
holds on the images of the sector under lattice quarter-turns.
Table~\ref{tab:ms-radii} lists the solutions of
Eq.~\eqref{eq:ms-rhokappa} at $h=2$. The row $\kappa=0$ is the leading
contour $\rho_E$, which is not a proved farthest NPT distance.

\begin{table}[t]
\caption{Scaled radii $\rho_\kappa(2,\theta)$ solving
Eq.~\eqref{eq:ms-rhokappa} at $h=2$ ($m_h=\sqrt3$) along the rays
$\mathbf a=(1,0)$, $(2,1)$ and $(3,2)$, for which
$|\cos2\theta|=1$, $3/5$ and $5/13$. The value
$n_*=\pi/[256(\pi-2)]=0.010749759\ldots$. The first row is the
leading zero contour $\rho_E$.}
\label{tab:ms-radii}
\begin{ruledtabular}
\begin{tabular}{lccc}
$\kappa$ & $\mathbf a=(1,0)$ & $\mathbf a=(2,1)$ & $\mathbf a=(3,2)$\\
\hline
$0$ (contour $\rho_E$) & 1.412821696 & 1.211595138 & 1.047969139\\
$0.001$ & 1.376849603 & 1.177954500 & 1.016662270\\
$0.005$ & 1.261103795 & 1.070347647 & 0.917133677\\
$n_*$ & 1.143119235 & 0.961779667 & 0.817764814
\end{tabular}
\end{ruledtabular}
\end{table}

The proof of Theorem~\ref{thm:massscale} occupies
Secs.~\ref{ss:ms-triplet}--\ref{ss:ms-proof}; the rate (iv) is proved in
Sec.~\ref{ss:ms-rate}. The corollary is proved in
Sec.~\ref{ss:ms-proof}.

\subsection{A compact triplet correction with explicit budgets}
\label{ss:ms-triplet}

Section~\ref{sm:population} keeps a compact triplet kernel fixed through
the weak-angle limit and improves its accuracy only afterwards. That
order is not sufficient here. The probe of Sec.~\ref{ss:ms-probe} has a
logarithmically growing inverse form, and a fixed-kernel $o(1)$ cannot
be multiplied by it. We construct instead an explicit family of compact
kernels whose accuracy, norm and support are all quantified.

Let $\Omega_3$ be the labeled relative configuration space of
Eq.~\eqref{eq:current-configuration-space}. In this subsection the norms
$\|\cdot\|_p$ and the pairing $(\cdot,\cdot)$ on $\Omega_3$ use counting
measure, and $\langle u,v\rangle_3=(u,v)/6$ is the physical scalar
product. For $v:\Omega_3\to\mathbb C$ put
\begin{equation}
 (Q_3v)(z)=\tfrac12\sum_{z'\sim z}[v(z)-v(z')],
 \label{eq:ms-Q3}
\end{equation}
where $z'$ runs over the configurations obtained from $z$ by one
allowed nearest-neighbor hop of one of the three particles; a hop of
the anchor shifts both relative coordinates. On the symmetric subspace
this is the exchange operator $Q_3$ of Sec.~\ref{sm:current}, and for
every $v$ one has $(v,Q_3v)=[\mathcal E_{\Omega_3}(v)+\mathcal
E_{\rm anc}(v)]/2$, as in Eq.~\eqref{eq:current-physical-graph-form}.
Each configuration has at most twelve hops, so $0\le Q_3\le12$, and
$e^{-tQ_3}$ is a symmetric conservative Markov semigroup: it preserves
positivity and contracts every $\ell^p$ norm. The operator $Q_3$
commutes with relabelings, reanchorings and lattice symmetries. For
$S\in\ell^1(\Omega_3)$ let $\mu_S(B)=(S,\mathbf 1_B(Q_3)S)$ be its
spectral measure.

The exact source is $\mathcal T=\mathcal V_{32}F$. For $L\ge1$ its
cutoff is $\mathcal T_L=\mathcal V_{32}F_L$, $F_L=\chi(\cdot/L)F$.
Equations~\eqref{eq:current-contact} and~\eqref{eq:current-cutoff-gradient}
give
\begin{equation}
 \|\mathcal T\|_1+\sup_{L\ge1}\|\mathcal T_L\|_1\le C,\qquad
 \|\mathcal T-\mathcal T_L\|_1\le3v(F-F_L)\le CL^{-1}.
 \label{eq:ms-source-cutoff}
\end{equation}
The cutoff is radial, so $\mathcal T_L$ has the relabeling, reanchoring
and joint-rotation symmetry of $\mathcal T$.

\begin{lemma}[Nash bound on the exclusion graph]
\label{lem:ms-nash}
For every compact $v$ on $\Omega_3$,
\begin{equation}
 \|v\|_2^3\le C_{\rm N}\,(v,Q_3v)\,\|v\|_1 .
 \label{eq:ms-nash}
\end{equation}
Consequently, for $t,\lambda>0$ and $S\in\ell^1(\Omega_3)$,
\begin{align}
 \|e^{-tQ_3}\|_{1\to2}&\le C_{\rm N}t^{-1},\qquad
 \|e^{-tQ_3}\|_{1\to\infty}\le4C_{\rm N}^2t^{-2},
 \label{eq:ms-heat}\\
 \mu_S([0,\lambda])&\le4eC_{\rm N}^2\|S\|_1^2\lambda^2,\qquad
 \langle S,Q_3^{-1}S\rangle_3\le C\|S\|_1^2 .
 \label{eq:ms-spectral}
\end{align}
In particular $\mu_S$ has no atom at zero. The bounds hold for complex
sources and are uniform in their support.
\end{lemma}

\begin{proof}
Let $E=E_3$ be the extension of Lemma~\ref{lem:current-routing}. It
fixes $\Omega_3$ pointwise, so $\|Ev\|_2\ge\|v\|_2$. Each configuration
has at most $5^2$ preimages, so $\|Ev\|_1\le25\|v\|_1$. By
Eq.~\eqref{eq:current-extension},
$\mathcal E_{\mathbb Z^4}(Ev)\le\mathsf C_{E,3}\mathcal E_{\Omega_3}(v)
\le2\mathsf C_{E,3}(v,Q_3v)$; the anchor hops only add nonnegative
energy. On $\mathbb Z^4$ split the Fourier integral of a summable $g$
at $|\bk|=\varrho\le\pi$. With $\lambda_4(\bk)\ge4|\bk|^2/\pi^2$,
\begin{equation}
 \|g\|_2^2\le\frac{\varrho^4}{32\pi^2}\|g\|_1^2
 +\frac{\pi^2}{4\varrho^2}\mathcal E_{\mathbb Z^4}(g).
\end{equation}
For $\mathcal E_{\mathbb Z^4}(g)\le\pi^6\|g\|_1^2$ choose
$\varrho^6=\mathcal E_{\mathbb Z^4}(g)/\|g\|_1^2$; otherwise use
$\|g\|_2\le\|g\|_1$. Both cases give the four-dimensional Nash
inequality $\|g\|_2^3\le C\mathcal E_{\mathbb Z^4}(g)\|g\|_1$. Applied to
$g=Ev$ it proves Eq.~\eqref{eq:ms-nash}.

Let $f\in\ell^1$, $f_t=e^{-tQ_3}f$ and $y(t)=\|f_t\|_2^2$. Since $Q_3$ is
bounded, Eq.~\eqref{eq:ms-nash} extends to $\ell^1$ functions, and
$\|f_t\|_1\le\|f\|_1$. Hence
$y'=-2(f_t,Q_3f_t)\le-2y^{3/2}/(C_{\rm N}\|f\|_1)$, so
$\frac{\dd}{\dd t}y^{-1/2}\ge1/(C_{\rm N}\|f\|_1)$ and
$\|f_t\|_2\le C_{\rm N}\|f\|_1/t$. Self-adjointness and the semigroup
property give
$\|e^{-tQ_3}\|_{1\to\infty}\le\|e^{-tQ_3/2}\|_{2\to\infty}
\|e^{-tQ_3/2}\|_{1\to2}=\|e^{-tQ_3/2}\|_{1\to2}^2$, which proves
Eq.~\eqref{eq:ms-heat}. Since $\mathbf 1_{[0,\lambda]}(x)\le
e^{1-x/\lambda}$ for $x\ge0$,
\begin{equation}
 \mu_S([0,\lambda])\le e\,(S,e^{-Q_3/\lambda}S)
 \le e\|S\|_1\|e^{-Q_3/\lambda}S\|_\infty
 \le4eC_{\rm N}^2\lambda^2\|S\|_1^2 .
\end{equation}
Finally $\langle S,Q_3^{-1}S\rangle_3=\frac16\int\lambda^{-1}
\dd\mu_S$. The dyadic bands $(2^{-j-1},2^{-j}]$, $j\ge0$, contribute at
most $\sum_j2^{j+1}\cdot4eC_{\rm N}^2 2^{-2j}\|S\|_1^2$, and
$(1,12]$ contributes at most $\mu_S(\mathbb R)=\|S\|_2^2\le\|S\|_1^2$.
\end{proof}

For $L\ge2$ set
\begin{equation}
 \beta=L^{-2},\quad \mathsf N_L=\lceil13L^2\log L\rceil,\quad
 \mathsf P=I-\frac{Q_3}{12},\quad t_\beta=\frac{12}{12+\beta},\quad
 \Theta_L=-\frac1{12+\beta}\sum_{j=0}^{\mathsf N_L-1}
 t_\beta^j\,\mathsf P^j\mathcal T_L .
 \label{eq:ms-kernel}
\end{equation}
The matrix $\mathsf P$ is a lazy symmetric Markov kernel: its
off-diagonal entries are $1/24$ for each allowed hop, its diagonal is at
least $1/2$, its spectrum lies in $[0,1]$, and it contracts $\ell^1$.
Let $\mathsf S_L$ be the largest pair separation occurring in the
support of $\Theta_L$. Write $Y_L=\mathcal C_3^+(\Theta_L)-
\mathcal C_3^+(\Theta_L)^\dagger$ and let $y=a_{\Theta_L}$ be its
incident local norm budget, as in Sec.~\ref{sm:population}.

\begin{lemma}[Lazy-walk triplet kernels]
\label{lem:ms-poly}
For $L\ge2$ the kernel $\Theta_L$ is compact, invariant under
relabelings and reanchorings, and lies in the symmetry sector of
$\mathcal T$. With constants independent of $L$, $R$ and $\eta$,
\begin{align}
 \delta_L^2:=\|Q_3^{1/2}\Theta_L+Q_3^{-1/2}\mathcal T\|_3^2
 &\le CL^{-2},
 \label{eq:ms-kernel-accuracy}\\
 y\le\|\Theta_L\|_1\le\beta^{-1}\|\mathcal T_L\|_1&\le CL^2,
 \qquad \mathsf S_L\le2L+1+\mathsf N_L\le CL^2\log(2L),
 \label{eq:ms-kernel-norms}\\
 Q_3\Theta_L=-\mathcal T_L+t_\beta^{\mathsf N_L}\mathsf P^{\mathsf N_L}
 \mathcal T_L-\beta\Theta_L,&\qquad
 \sup_{R\ge4}\|\mathcal T_R+Q_3\Theta_L\|_1\le C,
 \label{eq:ms-kernel-residual}\\
 I_\eta(\mathcal T_R+Q_3\Theta_L)\le18\delta_L^2
 +2\mathcal G_4\|\mathcal T_R-\mathcal T\|_1^2
 &\le18\delta_L^2+CR^{-2},\qquad\eta>0 .
 \label{eq:ms-kernel-inverse}
\end{align}
\end{lemma}

\begin{proof}
The kernel $F_L$ vanishes beyond radius $2L$, and the vertex
\eqref{eq:current-triplet-vertex} adds one nearest-neighbor step, so
pair separations in the support of $\mathcal T_L$ are at most $2L+1$.
Each power of $\mathsf P$ moves one particle by at most one step, which
proves the support bound in Eq.~\eqref{eq:ms-kernel-norms}; $\mathsf
N_L\le13L^2\log L+1$. The operator $\mathsf P$ commutes with the
symmetries, which proves the symmetry statements. Since $\|\mathsf
P\|_{1\to1}\le1$, $\|\Theta_L\|_1\le(12+\beta)^{-1}\sum_{j\ge0}
t_\beta^j\|\mathcal T_L\|_1=\beta^{-1}\|\mathcal T_L\|_1$. In the
convention~\eqref{eq:current-creator-convention} each unordered triple
carries one monomial with coefficient $\Theta_L$; a fixed site belongs
to triples whose coefficients sum to $\|\Theta_L\|_1/2$, and the adjoint
doubles this. Hence $y=\|\Theta_L\|_1$; below, sums of $|\Theta_L|$ over
the triples meeting a fixed site are bounded by $y$.

From $Q_3+\beta=(12+\beta)(I-t_\beta\mathsf P)$ and the telescoping sum,
$(Q_3+\beta)\Theta_L=-(I-t_\beta^{\mathsf N_L}\mathsf P^{\mathsf N_L})
\mathcal T_L$, which is the identity in
Eq.~\eqref{eq:ms-kernel-residual}. Each term on its right side has
$\ell^1$ norm at most $\|\mathcal T_L\|_1$, and $\|\mathcal
T_R\|_1\le3v(F_R)\le C$. This bounds the residual uniformly in $L$ and
$R$; no commutator estimate in $y$ is used.

The Neumann series $(Q_3+\beta)^{-1}=(12+\beta)^{-1}\sum_{j\ge0}
t_\beta^j\mathsf P^j$ gives the exact difference
\begin{equation}
 -(Q_3+\beta)^{-1}\mathcal T_L-\Theta_L
 =-t_\beta^{\mathsf N_L}\mathsf P^{\mathsf N_L}(Q_3+\beta)^{-1}
 \mathcal T_L .
 \label{eq:ms-kernel-tail}
\end{equation}
Decompose $Q_3^{1/2}\Theta_L+Q_3^{-1/2}\mathcal T$ as
$Q_3^{1/2}[\Theta_L+(Q_3+\beta)^{-1}\mathcal T_L]+
[Q_3^{-1/2}-Q_3^{1/2}(Q_3+\beta)^{-1}]\mathcal T_L
+Q_3^{-1/2}(\mathcal T-\mathcal T_L)$. In the spectral representation
of $\mu=\mu_{\mathcal T_L}$, the square of the first term is
$\frac16\int t_\beta^{2\mathsf N_L}(1-\lambda/12)^{2\mathsf N_L}
\lambda(\lambda+\beta)^{-2}\dd\mu\le t_\beta^{2\mathsf N_L}
\langle\mathcal T_L,Q_3^{-1}\mathcal T_L\rangle_3$. The square of the
second is $\frac16\int\beta^2\lambda^{-1}(\lambda+\beta)^{-2}\dd\mu$.
Below $\beta$ its integrand is at most $\lambda^{-1}$, and the bands
$(2^{-j-1}\beta,2^{-j}\beta]$ contribute at most
$C\|\mathcal T_L\|_1^2\beta2^{-j}$ by Eq.~\eqref{eq:ms-spectral}. On
$[2^j\beta,2^{j+1}\beta]$ the integrand is at most $2^{-3j}\beta^{-1}$
and the mass at most $C\|\mathcal T_L\|_1^22^{2j+2}\beta^2$. Thus the
second square is at most $C\|\mathcal T_L\|_1^2\beta$. The third square
is at most $C\|\mathcal T-\mathcal T_L\|_1^2\le CL^{-2}$ by
Eqs.~\eqref{eq:ms-spectral} and~\eqref{eq:ms-source-cutoff}. Finally
$\log(1+x)\ge12x/13$ for $0\le x\le1/12$ gives
$t_\beta^{2\mathsf N_L}\le\exp(-2\mathsf N_L\beta/13)\le L^{-2}$. Adding
the three squares proves Eq.~\eqref{eq:ms-kernel-accuracy}.

For Eq.~\eqref{eq:ms-kernel-inverse}, $\langle\mathcal T+Q_3\Theta_L,
Q_3^{-1}(\mathcal T+Q_3\Theta_L)\rangle_3=\delta_L^2$ by the spectral
calculus. Apply Eq.~\eqref{eq:current-reverse-inverse} with $n=3$ to
this symmetric source, then use $I_\eta\le I_0$,
$I_0(A+B)\le2I_0(A)+2I_0(B)$ and
Eq.~\eqref{eq:current-source-transience} for $\mathcal T_R-\mathcal T$,
whose $\ell^1$ norm is $O(R^{-1})$ by
Eq.~\eqref{eq:ms-source-cutoff}.
\end{proof}

Two choices of the scale are used. For the limit statements (i)--(iii)
take $L=\ell_\eps^M$ with a fixed integer $M\ge1$, and $M=8$ below;
Lemma~\ref{lem:ms-poly} then gives
\begin{equation}
 \delta_L^2\le C\ell_\eps^{-2M},\qquad
 y\le\|\Theta_L\|_1\le C\ell_\eps^{2M},\qquad
 \mathsf S_L\le C_M\ell_\eps^{2M+1}.
 \label{eq:ms-log-choice}
\end{equation}
For the rate (iv) Sec.~\ref{ss:ms-rate} takes
$L=(\eps\ell_\eps)^{-4/13}$. At each fixed $\eps$ both choices are fixed
compact kernels; rounding $L$ upward changes no estimate.

\subsection{Energy sandwich with an angle-dependent kernel}
\label{ss:ms-sandwich}

Let $R=\lceil\eps^{-2}\rceil$, $X=X_{F_R}$ and $U_R=e^{dX}$ as in
Eq.~\eqref{eq:pop-cutoff}, and replace the fixed kernel of
Eq.~\eqref{eq:pop-triplet-dressing} by $\Theta_L$:
$V_L=e^{\tau Y_L}$ with $\tau=\eps^3/2$. For a state $\omega$ put
$\widetilde\omega(A)=\omega(U_RAU_R^\dagger)$ and
$\widehat\omega(A)=\widetilde\omega(V_LAV_L^\dagger)$, with populations
$\widetilde p,\widehat p$ and exchange density
$\widehat{\mathfrak q}=\widehat\omega(Q)/N$. The vacuum trial is
$U_RV_L\Omega$, in this order. Both generators commute with translations
and with the joint quarter-turn. At fixed $\eps$ both have finite range,
so the automorphisms, the finite-support identities and the
thermodynamic limit are justified exactly as in
Sec.~\ref{sm:population}. We write $a_R=\|F_R\|_1\le C\ell_\eps$,
$v_R\le C$ and $z_R\le C\ell_\eps$ for the pair budgets of
Eqs.~\eqref{eq:pop-kernel-budgets} and~\eqref{eq:pop-Z-budget}, and
\begin{equation}
 \mathsf M_L=\sup_{R\ge4}\|\mathcal T_R+Q_3\Theta_L\|_1\le C
 \label{eq:ms-ML}
\end{equation}
for the residual norm of Eq.~\eqref{eq:ms-kernel-residual}.

\begin{lemma}[Energy sandwich with an angle-dependent kernel]
\label{lem:ms-sandwich}
There is $c_0>0$, depending only on $\cI$ and $\chi$, such that the
following holds whenever $L\ge\ell_\eps$ and
$\eps(\ell_\eps+L^2)\le c_0$. For every symmetric thermodynamic ground
state and $h\in\cI$,
\begin{equation}
 \kappa_6\widehat{\mathfrak q}+\mu_6\eps^2\widehat p
 \le B_\cI\eps^6\delta_L^2+C\,\mathfrak B(\eps,L),
 \label{eq:ms-sandwich}
\end{equation}
where $\kappa_6=g_*/4$ and $\mu_6=m_*/4$ are the constants of
Eq.~\eqref{eq:pop-final-coercivity}, $B_\cI=100C_3^2/g_*$ is the
constant of Eq.~\eqref{eq:pop-cubic-small-loss}, and $\mathfrak B$ is
the sum of the scalar budgets of Table~\ref{tab:ms-sandwich}, evaluated
with $y\le CL^2$ and $\mathsf M_L\le C$. The constant $B_\cI$ does not
depend on the kernel. For $L=\ell_\eps^M$ the right side is at most
$C\eps^6\ell_\eps^{-2M}+C_M\eps^7\ell_\eps^{8M+4}$, and consequently,
for $0<\eps\le\eps_M$,
\begin{equation}
 \widehat p\le C_M\eps^4\ell_\eps^{-2M},\qquad
 \widehat{\mathfrak q}\le C_M\eps^6\ell_\eps^{-2M},\qquad
 \sqrt{\widetilde p}\le\sqrt{\widehat p}+2|\tau|y
 \le C_M(\eps^2\ell_\eps^{-M}+\eps^3\ell_\eps^{2M}).
 \label{eq:ms-densities-log}
\end{equation}
\end{lemma}

\begin{table}[t]
\caption{Every scalar budget of the energy sandwich, term by term, with
its dependence on the kernel budget $y=a_{\Theta_L}$, the residual norm
$\mathsf M_L$ and the inverse form $I_{\eps^2}$. Here $\ell=\ell_\eps$;
$M_2=C(\ell+y)$ and $M_4=C(1+y)$ are the $\ell^1$ norms of the
fourth-order pair and quartet sources after extraction of $\eps^4$.
Constants are omitted. The last column inserts $L=\ell^M$,
$y\le C\ell^{2M}$, $\mathsf M_L\le C$.}
\label{tab:ms-sandwich}
\begin{ruledtabular}
\begin{tabular}{llll}
Term & Origin & Budget & $L=\ell^M$\\
\hline
$\mathfrak e_R$, and $2\mathfrak e_R$ for its conjugation
 & Eqs.~\eqref{eq:pop-first-remainder}, \eqref{eq:pop-second-remainder}
 & $\eps^7\ell^3$ & $\eps^7\ell^3$\\
$128|\tau|^3y^3$ & Eq.~\eqref{eq:pop-second-remainder}
 & $\eps^9y^3$ & $\eps^9\ell^{6M}$\\
$32|s|\tau^2y^2$ & Eq.~\eqref{eq:pop-second-remainder}
 & $\eps^7y^2$ & $\eps^7\ell^{4M}$\\
$\tau^2y^2[d(56+60v_R)+6h_+\eps^2]$ & Eq.~\eqref{eq:pop-second-remainder}
 & $\eps^8y^2$ & $\eps^8\ell^{4M}$\\
$240|sd|\tau^2a_Ry^2$ & Eq.~\eqref{eq:pop-second-remainder}
 & $\eps^9\ell y^2$ & $\eps^9\ell^{4M+1}$\\
$8|\tau|d^2yz_R+4h_+\eps^2d|\tau|ya_R$ & Eq.~\eqref{eq:pop-second-remainder}
 & $\eps^7\ell y$ & $\eps^7\ell^{2M+1}$\\
$192|\tau s|d^2ya_R^2$ & Eq.~\eqref{eq:pop-second-remainder}
 & $\eps^8\ell^2y$ & $\eps^8\ell^{2M+2}$\\
$(160v_R+336)|\tau|d^3ya_R^2+24h_+\eps^2d^2|\tau|ya_R^2$
 & Eq.~\eqref{eq:pop-second-remainder}
 & $\eps^9\ell^2y$ & $\eps^9\ell^{2M+2}$\\
$(\mathfrak h_R+\mathfrak h_{R,\Theta_L}+6\eps^5a_R)^2/\eps^2$
 & Eqs.~\eqref{eq:pop-first-higher-bound}, \eqref{eq:pop-second-higher-bound}
 & $\eps^8(\ell^4+y^2)+\eps^{10}(\ell^2y^2+y^4)$
 & $\eps^8\ell^{4M}+\eps^{10}\ell^{8M}$\\
Cubic current, $\eps^6I_{\eps^2}(\mathcal T_R+Q_3\Theta_L)$
 & Eq.~\eqref{eq:current-cubic-absorption}
 & $B_\cI\eps^6\delta_L^2+\eps^6R^{-2}$ & $\eps^6\ell^{-2M}$\\
Cubic regulator & Eq.~\eqref{eq:current-cubic-absorption}
 & $\eps^8\ell\mathsf M_L^2$ & $\eps^8\ell$\\
Pair current, $\eps^8I_{\eps^2}(R_2)$
 & Eq.~\eqref{eq:current-two-dimensional-log}
 & $\eps^8\ell M_2^2$ & $\eps^8\ell^{4M+1}$\\
Quartet current, $\eps^8I_{\eps^2}(R_4)$
 & Eq.~\eqref{eq:current-source-transience}
 & $\eps^8M_4^2$ & $\eps^8\ell^{4M}$\\
Pair and quartet regulators & Eq.~\eqref{eq:current-quartic-absorption}
 & $\eps^8M_2^2+\eps^{10}M_4^2$ & $\eps^8\ell^{4M}$\\
$d\,W(q_R)$ & Eq.~\eqref{eq:pop-quadratic-absorption}
 & $\eps^4R^{-2}$ & $\eps^8$
\end{tabular}
\end{ruledtabular}
\end{table}

\begin{proof}
Every inequality of Sec.~\ref{sm:population} is stated with explicit
dependence on the triple-generator budget $a_\Theta$, on the residual
norm $\|\mathcal T_R+Q_3\Theta\|_1$ and on $\delta_\Theta$, before the
notation $o_\Theta(\eps^6)$ is introduced. We insert $\Theta=\Theta_L$
and $a_\Theta=y$, and keep these dependences.

\emph{Identities and the common scalar.} The expansions
\eqref{eq:pop-first-full-expansion} and
\eqref{eq:pop-second-full-expansion}, the charge
Tables~\ref{tab:pop-first-orders} and~\ref{tab:pop-second-orders}, and
the scalar $V_{\eps,\Theta_L}$ of Eq.~\eqref{eq:pop-full-scalar} hold
verbatim for the compact kernel $\Theta_L$. The same exact scalar
appears in the lower bound~\eqref{eq:pop-full-lower} and in the vacuum
trial~\eqref{eq:pop-full-upper}. It is kept exact and cancels upon
subtraction, so neither its expansion in $\eps$ nor its dependence on
$L$ is needed. The local series converge since $2da_R$ and
$|\tau|y\le c_0\eps^2$ are small.

\emph{Norm remainders.} The first eight rows of
Table~\ref{tab:ms-sandwich} are the terms of
Eqs.~\eqref{eq:pop-first-remainder} and~\eqref{eq:pop-second-remainder}
with $a_\Theta=y$; each is a finite local commutator count, and no
support-volume factor enters. They bound the norm of the omitted Taylor
remainders in both the lower bound and the vacuum trial.

\emph{Fifth- and sixth-order terms.} After removal of their scalars, the
retained terms of Tables~\ref{tab:pop-first-orders}
and~\ref{tab:pop-second-orders} and the trigonometric cubic term have
density at most $(\mathfrak h_R+\mathfrak h_{R,\Theta_L}+6\eps^5a_R)
\sqrt p$, by Eqs.~\eqref{eq:pop-first-higher-bound},
\eqref{eq:pop-second-higher-bound} and~\eqref{eq:pop-trig-cubic-error}.
Here $\mathfrak h_R=O(\eps^5\ell_\eps^2)$ and
$\mathfrak h_{R,\Theta_L}=7y|\tau|\{d(14+12v_R)+2h_+\eps^2\}
+672|\tau sd|a_Ry+512\tau^2y^2=O(\eps^5y+\eps^6\ell_\eps y+\eps^6y^2)$.
One allocation $m_*\eps^2p/8$ and Young's inequality leave the ninth
row.

\emph{Cubic current.} By Eq.~\eqref{eq:ms-ML} the source
$\mathcal T_R+Q_3\Theta_L$ has $\ell^1$ norm at most $C$, uniformly in
$L$. The first line of Eq.~\eqref{eq:current-cubic-absorption} with
$A=1/2$ and $\kappa=g_*/8$ costs
$(C_3^2/g_*)\eps^6I_{\eps^2}(\mathcal T_R+Q_3\Theta_L)
+C\eps^8\ell_\eps\mathsf M_L^2$, and its angle threshold depends on the
kernel only through $\mathsf M_L$. Equation~\eqref{eq:ms-kernel-inverse}
bounds the inverse form by $18\delta_L^2+CR^{-2}$, and
$18C_3^2/g_*\le B_\cI$. This is the only place where $\delta_L$ enters,
and its coefficient does not depend on $\Theta_L$.

\emph{Fourth-order currents.} The pair
source~\eqref{eq:pop-combined-fourth-pair} has $\ell^1$ norm
$O(\eps^4a_R)$ and the quartet source~\eqref{eq:pop-quartic-four} has
normalized norm $O(\eps^4)$. The operator $[\VDM,Y_L]$ has local budget
at most $8y$ on at most four sites; termwise extraction,
Lemma~\ref{lem:pop-local-extraction}, gives pair and quartet kernels of
$\ell^1$ norm at most $Cy$. Hence, after the factor $\eps^4$ is
extracted, the combined sources have norms $M_2\le C(\ell_\eps+y)$ and
$M_4\le C(1+y)$. Repeat the proof of
Lemma~\ref{lem:current-quartic-absorption} with these norms in place of
its fixed budgets. The first current terms cost $C\eps^8I_{\eps^2}(R_n)$,
with $I_{\eps^2}(R_2)\le CM_2^2\ell_\eps$ and $I_{\eps^2}(R_4)\le
CM_4^2$; the regulator terms cost $C\eps^8M_2^2$ and $C\eps^{10}M_4^2$.
These are rows twelve to fourteen.

\emph{Number coefficients.} The coefficients of $p$ that depend on the
kernel are $\tau(384a_R+1024y)$ from Eq.~\eqref{eq:pop-cubic-ann},
$1024|\tau s|y$ from the extraction remainder of $\tau s[\VDM,Y_L]$,
$C\eps^4\mathsf M_L^2$ and $C\eps^3\mathsf M_L$ from the cubic current,
and $C\eps^6M_n^2$ and $C\eps^4M_n$ from the fourth-order currents.
Divided by $\eps^2$ they are at most
$C[\eps(\ell_\eps+y)+\eps+\eps^4(\ell_\eps+y)^2]\le C(c_0+\eps)$.
The other number coefficients, $36v_R^2d^2/\beta$ from
Eq.~\eqref{eq:pop-quadratic-absorption} and $128d^2z_R$, are
$O(\eps^4\ell_\eps)$. For $c_0$ and $\eps$ small their sum is at most
$m_*\eps^2/8$, which leaves $\mu_6$ as in
Eq.~\eqref{eq:pop-final-coercivity}. The exchange allocations of
Sec.~\ref{sm:population} are unchanged; the quadratic residual $dW(q_R)$
costs the last row.

\emph{Conclusion.} Apply the lower bound to the twice-dressed image
$\widehat\omega$ of a minimizer, for which
$\widehat\omega(V_L^\dagger U_R^\dagger(H-Ne_{\rm pol})U_RV_L)/N
=\varpi_\eps(h)$, and compare with the vacuum trial. The common scalar
cancels and Eq.~\eqref{eq:ms-sandwich} follows. For $L=\ell_\eps^M$,
Eq.~\eqref{eq:ms-log-choice} gives the last column. Its largest entry
for small $\eps$ is $32|s|\tau^2y^2=O(\eps^7\ell_\eps^{4M})$; every
entry other than the $\delta_L$ term is at most
$C_M\eps^7\ell_\eps^{8M+4}$, a valid but not sharp common exponent.
Since $\eps\ell_\eps^{10M+4}\to0$, this is smaller than
$\eps^6\ell_\eps^{-2M}$ for small $\eps$, which gives the first two
bounds in Eq.~\eqref{eq:ms-densities-log}. The third follows from
Eq.~\eqref{eq:pop-triplet-local-transfer} with $a_\Theta=y$.
\end{proof}

The bound on $\widetilde p$ is consistent with, and quantifies,
Eq.~\eqref{eq:dressed-density-sharp}. It does not assert that the
single-dressed exchange density is $o(\eps^6)$: that density contains the
deterministic triplet contribution, and only the twice-dressed densities
in Eq.~\eqref{eq:ms-densities-log} enter below. For $M=8$,
\begin{equation}
 \sqrt{\widehat p}\le C\eps^2\ell_\eps^{-8},\qquad
 \sqrt{\widehat{\mathfrak q}}\le C\eps^3\ell_\eps^{-8},\qquad
 y\le C\ell_\eps^{16},\qquad \mathsf S_L\le C\ell_\eps^{17}.
 \label{eq:ms-M8}
\end{equation}

For polynomially large $L$ two entries of Table~\ref{tab:ms-sandwich}
are too large: $32|s|\tau^2y^2$ and the first-commutator differences
$8|\tau|d^2yz_R+4h_+\eps^2d|\tau|ya_R$ of the fourth-order terms are of
order $\eps^7L^4$ and $\eps^7\ell_\eps L^2$, while the target is
$\eps^6L^{-2}$. Three odd operators remove these
obstructions. Group the first conjugation~\eqref{eq:pop-first-full-expansion}
by leading angle order, using $W(q_R)+A_R=W(f)+[Q,X]$:
\begin{equation}
 \begin{gathered}
 U_R^\dagger(H-Ne_{\rm pol})U_R=Q+s\VDM+H_{[2]}+H_{[3]}+H_{[4]}
 +H_{[5]}+H_{[6]}+\mathcal E_R,\\
 H_{[2]}=d\{B+W(f)+[Q,X]\}+h\eps^2\Nop,\quad H_{[3]}=sdD_R,\quad
 H_{[4]}=d^2Z_R+h\eps^2dJ_R,\quad
 H_{[5]}=\tfrac12sd^2\operatorname{ad}_X^2\VDM ,
 \end{gathered}
 \label{eq:ms-grouping}
\end{equation}
where $H_{[6]}$ is the sum of the four sixth-order terms of
Eq.~\eqref{eq:pop-first-full-expansion}. The subscripts give the leading
angle order, not a particle sector, and the trigonometric coefficients
are exact. The per-site local norm sums of $Q,s\VDM,H_{[2]},\ldots,
H_{[6]}$ are $O(1)$, $O(\eps)$, $O(\eps^2)$, $O(\eps^3\ell_\eps)$,
$O(\eps^4\ell_\eps)$, $O(\eps^5\ell_\eps^2)$ and $O(\eps^6\ell_\eps^2)$;
for $H_{[2]}$ this uses $\mathfrak b([Q,X])\le2v_R$. Their magnon parities
alternate: $Q$, $H_{[2]}$, $H_{[4]}$, $H_{[6]}$ are even and $s\VDM$,
$H_{[3]}$, $H_{[5]}$ are odd.

\begin{lemma}[Sixth-order energy inequality for a general cutoff]
\label{lem:ms-energyL}
Let $c_0$ be as in Lemma~\ref{lem:ms-sandwich}, and suppose
$L\ge\ell_\eps$, $\eps(\ell_\eps+L^2)\le c_0$. In the second conjugation
retain, in addition to the terms of
Eq.~\eqref{eq:pop-second-full-expansion}, the three operators
\begin{equation}
 C_Q=\frac{\tau^3}6\operatorname{ad}_{Y_L}^3Q,\qquad
 C_V=\frac{s\tau^2}2\operatorname{ad}_{Y_L}^2\VDM,\qquad
 C_4=\tau[H_{[4]},Y_L].
 \label{eq:ms-odd-terms}
\end{equation}
Then, for every symmetric thermodynamic ground state and $h\in\cI$,
\begin{equation}
 \kappa_6\widehat{\mathfrak q}+\mu_6\eps^2\widehat p
 \le C\eps^6L^{-2}+C\bigl[\eps^7\ell_\eps^3+\eps^8\ell_\eps^3L^4
 +\eps^{10}L^8+\eps^{16}L^{12}\bigr].
 \label{eq:ms-energyL}
\end{equation}
If the four ratios $\eps\ell_\eps^3L^2$, $\eps^2\ell_\eps^3L^6$,
$\eps^4L^{10}$ and $\eps^{10}L^{14}$ of the bracket to $\eps^6L^{-2}$ are
bounded, then $\widehat p\le C\eps^4L^{-2}$,
$\widehat{\mathfrak q}\le C\eps^6L^{-2}$, and
$\sqrt{\widetilde p}\le C(\eps^2L^{-1}+\eps^3L^2)$.
\end{lemma}

\begin{proof}
The operators $Q$ and $H_{[4]}$ are even and $\VDM$, $Y_L$ are odd, term by
term in the local decomposition, so each local term of
Eq.~\eqref{eq:ms-odd-terms} changes the magnon number by an odd amount
and has zero vacuum diagonal. The retained operators therefore add no
scalar: the common scalar of Lemma~\ref{lem:ms-sandwich} is unchanged
and cancels as before. Their local supports have at most eight sites,
for $C_Q$, independently of $\mathsf S_L$. By
Eqs.~\eqref{eq:pop-triple-comm-count} and~\eqref{eq:pop-vacuum-diagonal},
\begin{equation}
 \frac{|\widehat\omega(C_Q+C_V+C_4)|}N
 \le C\bigl[\eps^9y^3+\eps^7y^2+\eps^7\ell_\eps y\bigr]\sqrt{\widehat p}.
 \label{eq:ms-odd-bound}
\end{equation}
This coefficient is added to that of the other fifth- and sixth-order
terms before the single Young allocation $m_*\eps^2p/8$; no further
fraction of the number term is spent. The additional scalar cost is
$C[\eps^{16}y^6+\eps^{12}y^4+\eps^{12}\ell_\eps^2y^2]$.

With these operators retained, $Q$ is expanded through three $Y_L$
commutators, $s\VDM$ through two, $H_{[4]}$ through one, and the other
groups as in Sec.~\ref{sm:population}. The omitted norm remainders of
Table~\ref{tab:ms-refined} replace rows 2--8 of
Table~\ref{tab:ms-sandwich}: the former third-order remainder
$128|\tau|^3y^3$ of $Q$, the remainder $32|s|\tau^2y^2$ of $s\VDM$ and
the first-commutator difference of $H_{[4]}$ are now retained
operators. Every Taylor remainder is conjugated by a unitary, and the
table accounts for every group in Eq.~\eqref{eq:ms-grouping}. The
fourth-order and cubic currents, the Young cost of the previously
retained terms, now written
$C[\eps^8(\ell_\eps^4+y^2)+\eps^{10}(\ell_\eps^4+\ell_\eps^2y^2+y^4)]$,
and the quadratic residuals are unchanged. The number coefficients are
those of Lemma~\ref{lem:ms-sandwich}.

Insert $y\le CL^2$, $L\ge\ell_\eps$, $\eps\ell_\eps\le1$ and
$\delta_L^2\le CL^{-2}$. Then $\eps^{12}y^4+\eps^{10}y^3\le C\eps^{10}L^8$;
$\eps^8y^2$, $\eps^9\ell_\eps y^2$, $\eps^{10}\ell_\eps y^2$,
$\eps^8\ell_\eps^2y$, $\eps^9\ell_\eps^2y$, $\eps^{12}\ell_\eps^2y^2$ and
the fourth-order current cost $\eps^8\ell_\eps^3(1+y)^2$ are at most
$C\eps^8\ell_\eps^3L^4$; $\eps^8\ell_\eps^4$, $\eps^{10}\ell_\eps^4$ and
the cubic costs
$\eps^8\ell_\eps+\eps^{10}$ are at most $C\eps^7\ell_\eps^3$;
$\eps^{10}(\ell_\eps^2y^2+y^4)\le C(\eps^8\ell_\eps^3L^4+\eps^{10}L^8)$;
and $\eps^{16}y^6\le C\eps^{16}L^{12}$. No term with a limiting exponent
is discarded. This proves Eq.~\eqref{eq:ms-energyL}. If the four ratios
are bounded, the bracket is at most $C\eps^6L^{-2}$, which gives the
density bounds, and Eq.~\eqref{eq:pop-triplet-local-transfer} gives the
bound on $\widetilde p$.
\end{proof}

\begin{table}[t]
\caption{Omitted norm remainders, per site, in the energy inequality of
Lemma~\ref{lem:ms-energyL}; $\ell=\ell_\eps$, constants omitted.}
\label{tab:ms-refined}
\begin{ruledtabular}
\begin{tabular}{ll}
Omitted remainder & Norm density\\
\hline
Pair remainder $\mathcal E_R$ and its conjugation & $\eps^7\ell^3$\\
Fourth $Y_L$ commutator of $Q$ & $\eps^{12}y^4$\\
Third $Y_L$ commutator of $s\VDM$ & $\eps^{10}y^3$\\
Second $Y_L$ commutator of $H_{[2]}$ & $\eps^8y^2$\\
Second $Y_L$ commutator of $H_{[3]}$ & $\eps^9\ell y^2$\\
Second $Y_L$ commutator of $H_{[4]}$ & $\eps^{10}\ell y^2$\\
First $Y_L$ difference of $H_{[5]}$ & $\eps^8\ell^2y$\\
First $Y_L$ difference of $H_{[6]}$ & $\eps^9\ell^2y$
\end{tabular}
\end{ruledtabular}
\end{table}

\subsection{Kernel decay and transfer through the pair dressing}
\label{ss:ms-transfer}

\begin{lemma}[Mass-uniform kernel decay]
\label{lem:ms-tail}
Let $G_\sigma$ be as in Eq.~\eqref{eq:G} and fix $\sigma_0>0$.
\begin{enumerate}
\item[(a)] $|G_\sigma(\br)|\le C(1+|\br|)^{-2}$ for $\br\ne0$, uniformly
for $0\le\sigma\le\sigma_0$.
\item[(b)] $|C_F(\br)|\le C(1+|\br|)^{-2}$.
\item[(c)] For sufficiently small $\eps$, uniformly in $h\in\cI$,
$|w_2(\br)|\le C\eps^2(1+|\br|)^{-2}$; hence
$\sup_{|\eps\br|\ge D}|w_2(\br)|/\eps^4\le C/D^2$.
\item[(d)] $|A_h(\bx)|\le[16(\pi-2)|\bx|^2]^{-1}$ for $\bx\ne0$.
\end{enumerate}
\end{lemma}

\begin{proof}
(a) Use the periodic partition of the proof of
Lemma~\ref{lem:range-test-bounds}. On the annulus $|\bk|\asymp\kappa$ the
multiplier $\mathfrak d/[2(\lambda+\sigma)]$ has derivatives
$|\partial^\alpha|\le C_\alpha\kappa^{2-|\alpha|}/(\kappa^2+\sigma)\le
C_\alpha\kappa^{-|\alpha|}$, because $\mathfrak d=O(|\bk|^2)$. Fourier
integration by parts $N$ times bounds the annular kernel by
$C_N\kappa^2(1+\kappa|\br|)^{-N}$. Take $N=3$. The annuli with
$\kappa\le(1+|\br|)^{-1}$ contribute at most $C(1+|\br|)^{-2}$, and the
others at most $C|\br|^{-3}\sum_{\kappa>(1+|\br|)^{-1}}\kappa^{-1}\le
C(1+|\br|)^{-2}$. The smooth part away from the origin has uniformly
bounded derivatives and a kernel decaying faster than any power. No
continuum rotational invariance is used.

(b) Since $F$ is real and even, $C_F=F*F$ has symbol
$|\widehat F|^2=\mathfrak d^2/(4Z_0^2\lambda^2)$, by
Eq.~\eqref{eq:massless-defect} and $F=G_0/Z_0$. This symbol is bounded,
smooth away from the origin, and obeys
$|\partial^\alpha|\widehat F|^2|\le C_\alpha|\bk|^{-|\alpha|}$ near zero,
because $\mathfrak d^2=O(|\bk|^4)$ and $\lambda^2\ge c|\bk|^4$. The
argument of (a) with $\kappa^2/(\kappa^2+\sigma)$ replaced by one gives
the same annular bound and the claim.

(c) By Eq.~\eqref{eq:rankone}, $w_2=dG_\sigma/(a_\eps-\cos\eps\,
t_\sigma)$ with $d\le\eps^2/2$ and denominator at least $Z_0/2$ by
Eq.~\eqref{eq:range-defect-threshold}; $\sigma\le4h_+\eps^2$. Apply (a).

(d) $(z^2K_2(z))'=-z^2K_1(z)<0$ and $z^2K_2(z)\to2$ as $z\downarrow0$,
so $z^2K_2(z)\le2$. Insert $z=m_h|\bx|$ in
Eq.~\eqref{eq:profile-statement}.
\end{proof}

With $R=\lceil\eps^{-2}\rceil$ and the pair-dressed state $\widetilde
\omega$ of Sec.~\ref{ss:ms-sandwich}, put $S_R=\|F_R\|_2^2$,
$C_R(\br)=\sum_{\bk}F_R(\bk)F_R(\bk-\br)$,
\begin{equation}
 \eta_{\rm tr}=(1+2\sqrt3\,da_R)\sqrt{\widetilde p}+12d^2a_R^2,\qquad
 e_{\rm tr}=4d^2a_R^2,\qquad
 c_R=\sqrt{S_R+\widetilde pa_R^2}.
 \label{eq:ms-tr-parameters}
\end{equation}
We write $\|A\|_{\widetilde\omega}=\widetilde\omega(A^\dagger A)^{1/2}$
for the GNS seminorm.

\begin{lemma}[Transfer of $q$ and $v$]
\label{lem:ms-transfer}
For every translation-invariant state and every $\br\ne0$,
\begin{align}
 \sqrt{q_\br}\le d|F_R(\br)|+\eta_{\rm tr},\qquad
 |q_\br-d^2F_R(\br)^2|&\le2d|F_R(\br)|\eta_{\rm tr}+\eta_{\rm tr}^2,
 \label{eq:ms-tr-q}\\
 |v_\br-d^2C_R(\br)|\le\widetilde p+2dc_R\sqrt{\widetilde p}
 +12d^2a_R^2\sqrt{\widetilde p}
 &+2e_{\rm tr}(\sqrt{\widetilde p}+dc_R)+e_{\rm tr}^2,
 \label{eq:ms-tr-v}
\end{align}
and $\sup_{\br}|C_R(\br)-C_F(\br)|\le(\|F_R\|_2+\|F\|_2)\|F_R-F\|_2\le
CR^{-1}$. If moreover $\sqrt{\widetilde p}\le C_0\eps^2\Lambda^{-1}$ for
some $\Lambda\ge1$, then, uniformly in $\br\ne0$,
\begin{equation}
 \frac{q_\br}{\eps^4}\le C\bigl[(1+|\br|)^{-4}+\Lambda^{-2}
 +\eps^4\ell_\eps^4\bigr],\quad
 \left|\frac{v_\br}{\eps^4}-\frac{C_F(\br)}4\right|+
 \left|\frac p{\eps^4}-n_*\right|
 \le C\bigl[\Lambda^{-1}+\eps^2\ell_\eps^2\bigr].
 \label{eq:ms-tr-quant}
\end{equation}
\end{lemma}

\begin{proof}
For a local $A$ on $k$ sites, Eq.~\eqref{eq:pop-pair-comm-count} gives
$\|\operatorname{ad}_X^2A\|\le\|A\|(2a_R)^2k(k+1)$, a count of sites
independent of the geometric diameter. Hence the integral Taylor
remainder after the first commutator has norm at most $12d^2a_R^2\|A\|$
for $k=2$ and $4d^2a_R^2\|A\|$ for $k=1$. In particular, by
Eq.~\eqref{eq:pop-pair-local-transfer},
$B_i:=U_R^\dagger b_iU_R=b_i+dC_i+E_i$ with
$C_i=(1-2n_i)\sum_{\bk}F_R(\bk-i)b_{\bk}^\dagger$ and
$\|E_i\|\le e_{\rm tr}$. Also $\|b_i\|_{\widetilde\omega}=
\sqrt{\widetilde p}$ and, by Eq.~\eqref{eq:pop-C0-bound},
$\|C_i\|_{\widetilde\omega}\le c_R$; using the operator norm $a_R$ for
$C_i$ instead would lose a logarithm.

For distinct $i,j$ direct spin algebra gives
\begin{equation}
 [b_ib_j,X]=F_R(j-i)(1-n_i-n_j)
 +\sum_{\bk\ne i,j}\bigl[F_R(\bk-i)(1-2n_i)b_jb_{\bk}^\dagger
 +F_R(\bk-j)(1-2n_j)b_ib_{\bk}^\dagger\bigr].
 \label{eq:ms-tr-pair-commutator}
\end{equation}
After the scalar $F_R(j-i)$ is removed, each summand annihilates its
local vacuum, acts on at most three sites, and has norm at most its
coefficient. Its GNS seminorm is therefore at most the coefficient times
$\sqrt{3\widetilde p}$. The coefficients sum to at most $2a_R$, so
$\|[b_ib_j,X]-F_R(j-i)\|_{\widetilde\omega}\le2\sqrt3\,a_R
\sqrt{\widetilde p}$. With $\|b_ib_j\|_{\widetilde\omega}^2=
\widetilde\omega(n_in_j)\le\widetilde p$ and the two-site Taylor
remainder,
\begin{equation}
 \|U_R^\dagger b_ib_jU_R-dF_R(j-i)\|_{\widetilde\omega}\le
 \eta_{\rm tr}.
 \label{eq:ms-tr-two-site}
\end{equation}
Since $q_\br=\|U_R^\dagger b_0b_\br U_R\|_{\widetilde\omega}^2$, the
triangle inequality for the seminorm proves Eq.~\eqref{eq:ms-tr-q}.

For the normal coherence, $v_\br=\widetilde\omega(B_0^\dagger B_\br)$.
Expanding with $B_i=b_i+dC_i+E_i$ and Cauchy-Schwarz in the GNS
representation gives all terms of Eq.~\eqref{eq:ms-tr-v} except
$d^2|\widetilde\omega(C_0^\dagger C_\br)-C_R(\br)|$. On the vacuum,
$\langle0|C_0^\dagger C_\br|0\rangle=C_R(\br)$; the term with the
creator at the origin has coefficient $F_R(0)=0$. Each coefficient term
of $C_0^\dagger C_\br$ has norm at most one on at most four sites. After
its vacuum scalar is removed it has norm at most two and zero vacuum
diagonal, so Eq.~\eqref{eq:pop-vacuum-diagonal} costs
$3\cdot2\sqrt{4\widetilde p}=12\sqrt{\widetilde p}$ per unit of absolute
coefficient. The coefficients sum to at most $a_R^2$, which gives the
term $12d^2a_R^2\sqrt{\widetilde p}$. The bound on $C_R-C_F$ is
Cauchy-Schwarz, and $\|F_R-F\|_2=O(R^{-1})$ by Eq.~\eqref{eq:green-tail}.
At fixed $\eps$ these local identities hold on tori containing the
supports and pass to the thermodynamic state; their bounds are
independent of $\br$.

Under $\sqrt{\widetilde p}\le C_0\eps^2\Lambda^{-1}$ one has
$\eta_{\rm tr}\le C\eps^2[\Lambda^{-1}+\eps^2\ell_\eps^2]$, and
$|F_R(\br)|\le C(1+|\br|)^{-2}$ by Eq.~\eqref{eq:green-tail}; this gives
the bound on $q_\br$. The right side of Eq.~\eqref{eq:ms-tr-v} is at most
$C\eps^4[\Lambda^{-1}+\eps^2\ell_\eps^2]$, and
$|d^2/\eps^4-1/4|\le C\eps^2$, $|C_R|\le S_R\le C$ and
$|C_R-C_F|\le C\eps^2$ give the bound on $v_\br$. For the population,
Eq.~\eqref{eq:pop-final-local-bound} gives
$|\sqrt p-d\sqrt{S_R}|\le C\eps^2[\Lambda^{-1}+\eps^2\ell_\eps^2]$, and
$|d\sqrt{S_R}-\eps^2\sqrt{S_F}/2|\le C\eps^4$; recall $S_F/4=n_*$.
\end{proof}

\begin{lemma}[Joint population and normal coherence]
\label{lem:ms-qv}
Uniformly for $h\in\cI$ and the symmetric thermodynamic ground states,
Eq.~\eqref{eq:ms-main-qv} holds, and for every $r_\eps\to\infty$, in
particular $r_\eps=c/\eps$,
\begin{equation}
 \sup_{|\br|\ge r_\eps}\frac{q_\br+|v_\br|+|\zeta_\br|}{\eps^4}
 \longrightarrow0 .
 \label{eq:ms-qv}
\end{equation}
No reflection symmetry, spectral gap, or estimate of the anomalous
coherence is used.
\end{lemma}

\begin{proof}
Equation~\eqref{eq:dressed-density-sharp} gives $\widetilde
p=o(\eps^4)$, while $a_R=O(\ell_\eps)$ and $c_R=O(1)$. Thus
$\eta_{\rm tr}=o(\eps^2)$ uniformly in $\br$, and the five terms in
Eq.~\eqref{eq:ms-tr-v} are $o(\eps^4)$, $o(\eps^4)$,
$o(\eps^6\ell_\eps^2)$, $O(\eps^6\ell_\eps^2)$ and
$O(\eps^8\ell_\eps^4)$. Since $d/\eps^2\to1/2$, $F_R\to F$ uniformly and
$\sup|C_R-C_F|\to0$, Lemma~\ref{lem:ms-transfer} gives
Eq.~\eqref{eq:ms-main-qv}. For $|\br|\ge r_\eps\to\infty$,
Eq.~\eqref{eq:ms-tr-q} and $\sup_{|\br|\ge r_\eps}|F_R(\br)|\to0$ give
$q_\br=o(\eps^4)$. The autocorrelation of an $\ell^2$ sequence vanishes
at infinity: approximate $F$ in $\ell^2$ by a finitely supported
sequence and use the Cauchy-Schwarz bound for $C_R-C_F$; hence
$v_\br=o(\eps^4)$ there. Positivity of the principal $2\times2$ minor of
Eq.~\eqref{eq:rho} in the rows $\uparrow\downarrow,\downarrow\downarrow$
gives $|\zeta_\br|^2\le(p-q_\br)q_\br$, and $p=O(\eps^4)$ by
Theorem~\ref{thm:population}.
\end{proof}

\subsection{The mass-adjusted point probe}
\label{ss:ms-probe}

Use the quadrupole $g_\br$ and the massive probe $u_\br=A_2^{-1}g_\br$ of
Eq.~\eqref{eq:range-test}, with $A_2$ of
Eq.~\eqref{eq:range-pair-operator}. The probe is real, even, and odd
under the quarter-turn. Write $a_u,S_u,v_u$ for its budgets
\eqref{eq:range-kernel-budgets}, and
$\|\nabla_{\rm ret}u\|_2=(u,\cL u)^{1/2}$ for its retained-edge
gradient norm. Lemma~\ref{lem:range-test-bounds} gives bounds growing
with $|\br|$; the following bounds are uniform in the displacement.

\begin{lemma}[Uniform probe bounds and near-source cancellation]
\label{lem:ms-probe}
For sufficiently small $\eps$, uniformly in $h\in\cI$ and $\br\ne0$,
\begin{equation}
 a_u\le(2\Delta_\eps)^{-1}\le C\eps^{-2},\qquad
 \|u_\br\|_2+v_u\le C\eps^{-1},\qquad
 \|u_\br\|_\infty\le C\ell_\eps,\qquad
 \|\nabla_{\rm ret}u_\br\|_2^2\le C\ell_\eps .
 \label{eq:ms-probe-norms}
\end{equation}
If $|\br|\ge c/\eps$ and $1\le D\le c/(8\eps)$, then
\begin{equation}
 \sup_{|\mathbf z|\le D}|u_\br(\mathbf z)|\le C_c\eps^2(1+D)^2 .
 \label{eq:ms-probe-near}
\end{equation}
\end{lemma}

\begin{proof}
The operator $a_\eps\cL$ is a graph Laplacian on the punctured lattice
and $(d/2)P_\nn$ is a nonnegative multiplication. Their sum therefore
generates a positivity-preserving $\ell^1$-contraction semigroup, and
$\|A_2^{-1}\|_{1\to1}\le\int_0^\infty e^{-2\Delta_\eps t}\dd t=
(2\Delta_\eps)^{-1}$. For $\br\ne0$ the four points $\pm\br,\pm
\mathsf J\br$ are distinct, so $\|g_\br\|_1=1$; and
$\Delta_\eps\ge(h_--1/2)\eps^2\ge\eps^2$. This proves the $\ell^1$
bound, which removes the displacement-dependent factor of
Lemma~\ref{lem:range-test-bounds}.

For the other bounds use the defect formula
\eqref{eq:range-test-defect}: $u_\br=u_0+\alpha_\br G_\sigma$, with
$u_0=a_\eps^{-1}(\cL_0+\sigma)^{-1}g_\br$, $\sigma=2\Delta_\eps/a_\eps
\asymp\eps^2$, $u_0(e_x)=a_\eps^{-1}G_\sigma(\br)$, and
$|\alpha_\br|\le C|G_\sigma(\br)|\le C$ by
Eq.~\eqref{eq:range-defect-threshold}. Since $|\widehat g_\br|\le1$,
$\|(\cL_0+\sigma)^{-1}g_\br\|_2^2\le\int(\lambda+\sigma)^{-2}
\dd^2k/(2\pi)^2\le\pi/(16\sigma)$, as in Eq.~\eqref{eq:current-free-ltwo};
and $\|G_\sigma\|_2\le1/4$. For first differences write
$(\cL_0+\sigma)^{-1}=\int_0^\infty e^{-\sigma t}e^{-t\cL_0}\dd t$; by
Eq.~\eqref{eq:current-heat-gradient} each coordinate difference of the
heat kernel has $\ell^1$ norm at most $\sqrt\pi/(2\sqrt t)$. Hence the
first-difference $\ell^1$ norms of $u_0$ and $G_\sigma$ are at most
$C\sigma^{-1/2}$, using $\|g_\br\|_1=\|f\|_1=1$; the retained-edge sum
$v_u$ is no larger. The diagonal free Green function is
$\int(\lambda+\sigma)^{-1}\dd^2k/(2\pi)^2\le C\ell_\eps$ by
Eq.~\eqref{eq:current-two-dimensional-log}, and a positive-definite
kernel is maximal on the diagonal, so $\|u_0\|_\infty\le C\ell_\eps$;
also $\|G_\sigma\|_\infty\le1/4$. Finally $A_2\ge a_\eps\cL$ and
$(u_\br,A_2u_\br)=(u_\br,g_\br)\le\|u_\br\|_\infty$ give the energy
bound.

For Eq.~\eqref{eq:ms-probe-near}, let $\mathcal R_\sigma(\bx)$ be the
kernel of $(\cL_0+\sigma)^{-1}$. In the dyadic decomposition of the proof
of Lemma~\ref{lem:range-test-bounds}, every second nearest-neighbor
difference of the annular kernel at scale $\kappa$ is bounded by
$C\kappa^4(\kappa^2+\sigma)^{-1}(1+\kappa|\bx|)^{-6}$. For
$|\bx|\ge c/(2\eps)$, the scales $\kappa\le\eps$ contribute at most
$C\eps^4/\sigma\le C\eps^2$, and the scales $\kappa>\eps$ at most
$C|\bx|^{-6}\sum_{\kappa>\eps}\kappa^{-4}\le C_c\eps^2$; the smooth part
decays faster than any power. Thus all second differences of
$\mathcal R_\sigma$ are $O_c(\eps^2)$ beyond distance $c/(2\eps)$. Now
\[
 u_0(\mathbf z)=(4a_\eps)^{-1}[\mathcal R_\sigma(\mathbf z-\br)
 +\mathcal R_\sigma(\mathbf z+\br)-\mathcal R_\sigma(\mathbf z-\mathsf J\br)
 -\mathcal R_\sigma(\mathbf z+\mathsf J\br)],
\]
and for $|\mathbf z|\le D+2$ all four arguments have length at least
$|\br|-D-2\ge c/(2\eps)$ for small $\eps$. Hence every second difference
of $u_0$ in this ball is $O_c(\eps^2)$. The function $u_0$ is even and
vanishes at the origin. For $\mathbf z=z_1e_x+z_2e_y$ write
$2u_0(\mathbf z)=u_0(\mathbf z)+u_0(-\mathbf z)-2u_0(0)$ as the sum of two central second
differences along the coordinate segments, each a sum of at most
$z_j^2$ second differences, and two mixed rectangles, each a sum of at
most $|z_1z_2|$ mixed second differences. This gives
$|u_0(\mathbf z)|\le C_c\eps^2|\mathbf z|^2$. The defect term obeys
$|\alpha_\br G_\sigma|\le C|G_\sigma(\br)|\le C_c\eps^2$ by
Lemma~\ref{lem:ms-tail}(a). Adding proves Eq.~\eqref{eq:ms-probe-near}.
The same bound holds for nearest-neighbor differences on $|\mathbf z|\le D$,
with $D+1$ in place of $D$.
\end{proof}

The near-source bound is the decisive cancellation: any contact source
of diameter $D\ll\eps^{-1}$ meets the probe only through values of
order $\eps^2D^2$, two powers of $\eps$ below the probe's typical size
$\ell_\eps$ at the origin of a generic source.

\subsection{The stationary comparison}
\label{ss:ms-comparison}

For $u=u_\br$ the left side of the exact identity
\eqref{eq:range-stationarity} is $(w(\br)-w_2(\br))/2$: joint symmetry
gives $(g_\br,w)_{\rm b}=w(\br)$, and $d(u_\br,f)_{\rm
b}=(g_\br,dA_2^{-1}f)_{\rm b}=-w_2(\br)$. Hence
\begin{equation}
 |w(\br)-w_2(\br)|\le\frac{d|\omega(R_D)|+2d|\omega(R_P)|
 +2a_\eps|\omega(R_Q)|+2|s||\omega(R_V)|}{N},
 \qquad R_\bullet=R_\bullet(u_\br),
 \label{eq:ms-master}
\end{equation}
with the residuals of Eqs.~\eqref{eq:range-RQ}--\eqref{eq:range-RPV}.
The even residuals are estimated in the physical state; the exchange and
DM residuals are transferred through both dressings.

\begin{lemma}[Even residuals by a near/far split]
\label{lem:ms-even}
Let $|\br|\ge c/\eps$, $1\le D\le c/(8\eps)$, and
\begin{equation}
 \Xi_D=\sup_{|\bx|\ge D-2}\frac{q_\bx+|v_\bx|}{\eps^4}.
 \label{eq:ms-Xi}
\end{equation}
Then, uniformly in the states and $h\in\cI$,
\begin{equation}
 \frac{d|\omega(R_D(u_\br))|+2d|\omega(R_P(u_\br))|}{N}
 \le C_c\eps^4\bigl(\Xi_D+\sqrt{\Xi_D}\bigr)+C_c\eps^8(1+D)^4 .
 \label{eq:ms-even}
\end{equation}
\end{lemma}

\begin{proof}
For distinct sites $i,j,k$ and the physical state,
\begin{equation}
 |\omega(n_jb_ib_k)|\le\sqrt{p\,q_{j-k}},\qquad
 |\omega(n_0b_jb_k^\dagger)|\le\sqrt{q_{\bk}q_j}\le\sqrt{p\,q_j}.
 \label{eq:ms-conditional-cs}
\end{equation}
For the first, $n_jb_ib_k=(b_i^\dagger n_j)^\dagger(b_kn_j)$, and
Cauchy-Schwarz gives the factors $\omega(n_jb_ib_i^\dagger n_j)\le p$
and $\omega(n_jb_k^\dagger b_kn_j)=q_{j-k}$. For the second,
$n_0b_jb_k^\dagger=n_0b_k^\dagger b_j=(n_0b_k^\dagger)(n_0b_j)
=(b_kn_0)^\dagger(n_0b_j)$, and the two squared norms are
$\omega(n_0n_{\bk})=q_{\bk}\le p$ and $\omega(n_0n_j)=q_j$. These bounds
retain the distant occupancy factor that the estimate
$|\omega(n_jb_ib_k)|\le p$ of Lemma~\ref{lem:range-pair-identity}
discards.

In $R_D(u)$ the summand $u(\bk-i)n_jb_ib_k$ has $|j-i|=1$. If
$|\bk-i|\ge D$ then $|\bk-j|\ge D-1$, so by
Eq.~\eqref{eq:ms-conditional-cs} and $p\le C\eps^4$
(Theorem~\ref{thm:population}) its expectation is at most
$C\eps^4\sqrt{\Xi_D}$; these summands have total coefficient at most
$Ca_u\le C\eps^{-2}$ per site. If $|\bk-i|<D$, the expectation is at most
$p$, and Eq.~\eqref{eq:ms-probe-near} bounds the coefficient by
$C_c\eps^2(1+D)^2$ on $O((1+D)^2)$ sites. Multiplication by $d\le\eps^2/2$
gives the $R_D$ part of Eq.~\eqref{eq:ms-even}.

In $R_P(u)$ the equal pairs leave $-u(e_\mu)f(e_\mu)(n_i+n_j)$ after the
scalar is removed; by Eq.~\eqref{eq:ms-probe-near} with $D=1$ their
contribution after multiplication by $d$ is $O(\eps^2\cdot\eps^2\cdot
p)=O(\eps^8)$. Pairs sharing one site $0$ give
$u(j)f(\bk)(1-2n_0)b_jb_{\bk}^\dagger$ with $\bk\in\{\pm e_x,\pm e_y\}$.
For $|j|\ge D$,
\begin{equation}
 |\omega((1-2n_0)b_jb_{\bk}^\dagger)|\le|v_{j-\bk}|
 +2\sqrt{p\,q_j}\le C\eps^4(\Xi_D+\sqrt{\Xi_D}),
\end{equation}
by Eq.~\eqref{eq:ms-conditional-cs} and $|j-\bk|\ge D-1$; the
coefficients sum to at most $a_u\|f\|_1\le C\eps^{-2}$. For $|j|<D$ the
operator and its adjoint kill the local vacuum, so the expectation is
at most $3p$, and the coefficient is at most $C_c\eps^2(1+D)^2$. This
gives the $R_P$ part. No odd-coherence estimate and no X-state form of
the reduced density matrix is used.
\end{proof}

The transfer of $R_Q$ and $R_V$ needs a creator bound for more than four
particles. Extend Eqs.~\eqref{eq:current-configuration-space}
and~\eqref{eq:current-creator-convention} verbatim to any $n\ge2$,
with $m_n=n-1$ labeled relative coordinates.

\begin{lemma}[Anchored creator bound]
\label{lem:ms-gram}
For every $n\ge2$, every translation-invariant state and every
$S_n\in\ell^1(\Omega_n)$,
\begin{equation}
 |\mathfrak c_n^+(S_n;\omega)|\le\frac1{n!}\sqrt p
 \Bigl[\sqrt{(n-1)!}\,\|S_n\|_2+\sqrt p\,\|S_n\|_1\Bigr],
 \label{eq:ms-gram}
\end{equation}
where $\mathfrak c_n^+$ is the creator-only part of
Eq.~\eqref{eq:current-density-definition}.
\end{lemma}

\begin{proof}
Isolate the creator at the anchor:
$n!\,\mathfrak c_n^+=\omega(b_0^\dagger Z)$ with
$Z=\sum_{\br\in\Omega_n}S_n(\br)\prod_jb_{r_j}^\dagger$. Group the
labeled tuples by the unordered set $A$ of the $n-1$ remaining sites.
At most $(n-1)!$ tuples give a set, so its coefficient $t_A$ satisfies
$\sum_A|t_A|^2\le(n-1)!\|S_n\|_2^2$ and $\sum_A|t_A|\le\|S_n\|_1$. Then
$|\omega(b_0^\dagger Z)|\le\|b_0\Omega_\omega\|\|Z\Omega_\omega\|$, with
$\|b_0\Omega_\omega\|=\sqrt p$ and Eq.~\eqref{eq:current-creator-vector-bounds}
for $\|Z\Omega_\omega\|$. The Gram estimate of
Lemma~\ref{lem:current-creator-gram} holds for sets of any common
cardinality. No $n$-particle routing or resolvent is used.
\end{proof}

We now evaluate $\omega(R_Q)$ and $\omega(R_V)$ in the twice-dressed
state $\widehat\omega$ of Sec.~\ref{ss:ms-sandwich}, for a general scale
$L$. The hypotheses are
\begin{equation}
 L\ge\ell_\eps,\quad \eps(\ell_\eps+L^2)\le c_0,\quad
 \eps\sqrt{\ell_\eps}\le1,\quad \mathsf S_L+2\le\frac c{8\eps},\quad
 \sqrt{\widehat p}\le C_0\frac{\eps^2}L,\quad
 \sqrt{\widehat{\mathfrak q}}\le C_0\frac{\eps^3}L .
 \label{eq:ms-hyp}
\end{equation}
By Eq.~\eqref{eq:ms-M8} they hold for $L=\ell_\eps^8$ and small $\eps$.

\begin{lemma}[Exchange and DM residuals at the mass scale]
\label{lem:ms-budget}
Assume Eq.~\eqref{eq:ms-hyp} and $|\br|\ge c/\eps$. Then, uniformly in
the states and $h\in\cI$,
\begin{equation}
 \frac{2a_\eps|\omega(R_Q(u_\br))|+2|s||\omega(R_V(u_\br))|}{N\eps^4}
 \le C_c\,\mathfrak P(\eps,L),
 \label{eq:ms-budget}
\end{equation}
where $\mathfrak P$ is the sum of the right column of
Table~\ref{tab:ms-budget}. For $L=\ell_\eps^8$,
$\mathfrak P\le C[\ell_\eps^{-7}+\eps\ell_\eps^{52}]$.
\end{lemma}

\begin{table}[t]
\caption{Contributions to the right side of Eq.~\eqref{eq:ms-master}
from the exchange and DM residuals, divided by $\eps^4$, under
Eq.~\eqref{eq:ms-hyp}; $\ell=\ell_\eps$, $A\in\{R_Q,R_V\}$, and constants
are omitted. Rows 1--10 are the bounds of
Eq.~\eqref{eq:range-general-residuals} evaluated at
$(\widehat p,\widehat{\mathfrak q})$; rows 11--16 are the first
commutators with the triplet generator; rows 17--19 are Taylor
remainders in $Y_L$.}
\label{tab:ms-budget}
\begin{ruledtabular}
\begin{tabular}{rll}
 & Term & Normalized bound\\
\hline
1 & $R_Q$: $8v_u\sqrt{\widehat{\mathfrak q}\widehat p}$ & $L^{-2}$\\
2 & $d[R_Q,X]$, quartet source: $24dv_uv_R\mathsf T_4$ & $L^{-1}$\\
3 & $d[R_Q,X]$, remainder: $96dv_ua_R\widehat p$ & $\eps\ell L^{-2}$\\
4 & $\frac{d^2}2[[R_Q,X],X]$: $144\sqrt5d^2v_ua_R^2\sqrt{\widehat p}$
 & $\eps\ell^2L^{-1}$\\
5 & pair Taylor remainder of $R_Q$: $160d^3v_ua_R^3$ & $\eps\ell^3$\\
6 & $sR_V$: $|s|\mathcal V_0(u)$ & $\ell L^{-1}$\\
7 & $sd[R_V,X]$, triple sources: $17280|s|da_ua_R\mathsf T_3$
 & $\ell L^{-1}$\\
8 & $sd[R_V,X]$, remainder: $11904|s|da_ua_R\widehat p$
 & $\eps\ell L^{-2}$\\
9 & $s\frac{d^2}2[[R_V,X],X]$: $1152\sqrt5|s|d^2a_ua_R^2\sqrt{\widehat p}$
 & $\eps\ell^2L^{-1}$\\
10 & pair Taylor remainder of $sR_V$: $1280|s|d^3a_ua_R^3$ & $\eps\ell^3$\\
11 & $\tau[R_Q,Y_L]$, Eq.~\eqref{eq:ms-RQY} & $\eps L\sqrt\ell+\eps^2$\\
12 & $s\tau([R_V,Y_L]-c_0)$, Eq.~\eqref{eq:ms-RVY} & $\eps L+\eps^2$\\
13 & $s\tau c_0$, Eq.~\eqref{eq:ms-c0} & $\eps^2L^6\log^2(2L)$\\
14 & $d\tau[[R_Q,X],Y_L]$ & $\eps^2\ell L$\\
15 & $sd\tau c_1$, Eq.~\eqref{eq:ms-c1} & $\eps^4L^6\log^2(2L)$\\
16 & $sd\tau([[R_V,X],Y_L]-c_1)$ & $\eps^2\ell L$\\
17 & $\tau^2b_Ay^2$ & $\eps L^4$\\
18 & $d\tau^2b_Aa_Ry^2$ & $\eps^3\ell L^4$\\
19 & $d^2|\tau|b_Aa_R^2y$ & $\eps^2\ell^2L^2$
\end{tabular}
\end{ruledtabular}
\end{table}

\begin{proof}
\emph{Expansion.} Write $\omega(A)=\widehat\omega(V_L^\dagger U_R^\dagger
AU_RV_L)$. Expand $U_R^\dagger AU_R$ through two pair commutators,
Eq.~\eqref{eq:range-observable-taylor}, with the remainders
$\mathcal E_A$ of Eq.~\eqref{eq:range-observable-remainders}. Then
expand $V_L^\dagger AV_L$ and $V_L^\dagger[A,X]V_L$ through their first
$Y_L$ commutators and leave $[[A,X],X]$ at zeroth order in $Y_L$. With
the budgets $b_{R_Q}=2v_u$ and $b_{R_V}=16a_u$ of
Eq.~\eqref{eq:range-bch-budgets}, the triple-commutator count
\eqref{eq:pop-triple-comm-count} on at most eight actual sites and the
unitary integral remainder give the omitted norms
\begin{equation}
 C\{\tau^2b_Ay^2+d\tau^2b_Aa_Ry^2+d^2|\tau|b_Aa_R^2y\},
 \label{eq:ms-Ytaylor}
\end{equation}
which are rows 17--19 after the prefactor $a_\eps$ or $s$; note
$|s|b_{R_V}\le C\eps^{-1}$ and $b_{R_Q}\le C\eps^{-1}$ by
Lemma~\ref{lem:ms-probe}. The pair remainder $\mathcal E_A$ keeps its
norm under conjugation by $V_L$. No geometric volume of radius $\mathsf
S_L$ enters these counts.

\emph{Zeroth order in $Y_L$.} Equation~\eqref{eq:range-general-residuals}
bounds $\omega(A+d[A,X]+\frac{d^2}2[[A,X],X])$ in every
translation-invariant state, with $a_F,v_F$ replaced by $a_R,v_R$; in
particular the one-spin source of $[R_V,X]$ vanishes as an operator.
Evaluate these bounds in $\widehat\omega$. By Eq.~\eqref{eq:ms-hyp},
$\mathsf T_3+\mathsf T_4\le C\eps^3L^{-1}$, since
$H_4(\eps^2)\le C\sqrt{\ell_\eps}$ and $H_6$ is bounded. In
$\mathcal V_0(u)$ use $M_T\le3v_u\le C\eps^{-1}$ and
$I_T\le18I_2(\eps^2)(u,\cL u)\le C\ell_\eps^2$; its five terms are at
most $C\eps^3\ell_\eps L^{-1}$, $C\eps^3L^{-2}$,
$C\eps^3\sqrt{\ell_\eps}L^{-1}$, $C\eps^3L^{-2}$ and $C\eps^3L^{-2}$, the
first being the leading triple current $C_3\sqrt{\widehat{\mathfrak q}
I_T}$. With Eq.~\eqref{eq:ms-probe-norms} this gives rows 1--10.

\emph{The commutator $[R_Q,Y_L]$.} Both $R_Q$ and $R_Q^\dagger$ kill
$\Omega$, because each monomial contains a density at an empty site.
With $Y_L\Omega=\Psi_\Theta:=\mathcal C_3^+(\Theta_L)\Omega$ one has
$[R_Q,Y_L]\Omega=R_Q\Psi_\Theta$. This is a translation-invariant
one-magnon vector, hence a multiple of $\sum_ib_i^\dagger\Omega$; the
joint quarter-turn fixes $\Omega$, $R_Q(u_\br)$ and $Y_L$ but multiplies
$\sum_ib_i^\dagger\Omega$ by a nontrivial phase, so the multiple
vanishes. The vacuum scalar is zero, and since $Y_L^\dagger=-Y_L$,
\begin{equation}
 [R_Q,Y_L]^\dagger\Omega=[R_Q^\dagger,Y_L]\Omega
 =R_Q^\dagger\Psi_\Theta
 \label{eq:ms-RQY-column}
\end{equation}
lies in the five-magnon sector. For an unordered five-site set $\mathcal
C$, with $\Delta_{ij}u(\bk)=u(\bk-i)-u(\bk-j)$ and each bond counted
once, its coefficient is
\begin{equation}
 S_5(\mathcal C)=\sum_{\langle ij\rangle\subset\mathcal C}\
 \sum_{\bk\in\mathcal C\setminus\{i,j\}}
 \overline{\Delta_{ij}u(\bk)}\,
 \bigl[\Theta_L(\mathcal C\setminus\{j,\bk\})
 -\Theta_L(\mathcal C\setminus\{i,\bk\})\bigr].
 \label{eq:ms-S5}
\end{equation}
Indeed $R_Q^\dagger$ contains $\overline{\Delta_{ij}u(\bk)}\,
b_{\bk}^\dagger(b_j^\dagger n_i-b_i^\dagger n_j)$: the density selects
an endpoint already present in the triple, and the other endpoint and
$\bk$ are created. Fix a triple shape, one of its sites $i$, one of the
four neighbors $j$ of $i$, and the sign of the endpoint. The remaining
site $\bk$ is free, its coefficient is a retained nearest-neighbor
difference of $u$, and $\bk\mapsto\mathcal C$ is injective. Minkowski's
inequality over the triple shapes and the at most $2\cdot3\cdot4$ role
choices, followed by conversion to a symmetric anchored kernel (at most
$5!$ labelings), gives
\begin{equation}
 \|S_5\|_2\le Cy\|\nabla_{\rm ret}u\|_2,\qquad
 \|S_5\|_1\le Cyv_u .
 \label{eq:ms-S5-norms}
\end{equation}
All multiplicities count site roles and labels; none grows with $\mathsf
S_L$. Triples that meet only the remote site $\bk$ produce no vacuum
image, because the untouched endpoint density in $(n_ib_j-n_jb_i)$ kills
both the vacuum row and column. If $\bk$ also lies in the triple, the
union has at most four sites and the only possible pure source is the
one-spin column, which vanishes as shown. The local budget of
$[R_Q,Y_L]$ is at most $12v_uy$ on at most five sites. Termwise
extraction, Lemma~\ref{lem:pop-local-extraction}, leaves a
vacuum-annihilating remainder of budget $Cyv_u$, and
Lemma~\ref{lem:ms-gram} with $n=5$ gives
\begin{equation}
 \frac{|\widehat\omega([R_Q,Y_L])|}N\le
 Cy\bigl[\sqrt{\widehat p}\,\|\nabla_{\rm ret}u\|_2
 +\widehat p\,v_u\bigr].
 \label{eq:ms-RQY}
\end{equation}
After the prefactor $\tau$ this is row 11. Using $v_u$ in place of
$\|\nabla_{\rm ret}u\|_2$ would lose a factor $\eps^{-1}$ and is not
done.

\emph{The commutator $[R_V,Y_L]$.} One has $R_V\Omega=0$, and
$R_V^\dagger=[\VDM,B(u)^\dagger]$ gives $R_V^\dagger\Omega=\VDM
B(u)^\dagger\Omega=\Psi_{\mathcal T_{\bar u}}$: the three-magnon part is
the vertex~\eqref{eq:current-triplet-vertex}, and the one-magnon part is
uniform and vanishes by the quarter-turn. Therefore
\begin{equation}
 c_0(u,\Theta_L):=\frac{\langle\Omega,[R_V,Y_L]\Omega\rangle}N
 =\frac{\langle R_V^\dagger\Omega,\Psi_\Theta\rangle}N
 =\langle\mathcal V_{32}\overline u,\Theta_L\rangle_3 .
 \label{eq:ms-c0-def}
\end{equation}
The charges of $R_V$ are $-1,-3$ and those of $Y_L$ are $\pm3$, so
$[R_V,Y_L]$ has charges $2,0,-4,-6$. The charge $-6$ part vanishes
identically: the charge $-3$ part of $R_V$ is $[B(u),\VDM^{(-1)}]$, and
$[b_kb_l,n_ib_j]$ is nonzero only for $i\in\{k,l\}$, $j\notin\{k,l\}$,
where it is a product of three annihilators; similarly for
$[b_kb_l,b_in_j]$. Products of annihilators commute, so their commutator
with $-\mathcal C_3^+(\Theta_L)^\dagger$ is zero. Hence, after the scalar
$c_0$ is removed, the only pure vacuum sources are a pair column and a
quartet row.

Every local term of $R_V(u)$ is either a same-edge term with coefficient
$u(e_\mu)$ or a remote-site term $u(\bk-i)\,b_{\bk}A_{ij}$, with
$A_{ij}=[b_i,C_{ij}]$ on a bond and $\bk\notin\{i,j\}$, or the same with
$i,j$ exchanged. Consider its commutator with a local triplet term $Y_T$
on a triple $T$. If $T$ meets only $\bk$, the commutator is
$u(\bk-i)[b_{\bk},Y_T]A_{ij}$. Its vacuum column vanishes since
$A_{ij}\Omega=b_iC_{ij}\Omega-C_{ij}b_i\Omega=0$. Its vacuum row
vanishes since $A_{ij}^\dagger\Omega=c_\mu b_i^\dagger b_j^\dagger
\Omega$ and $[b_{\bk},Y_T]^\dagger$ contains the annihilators of the two
sites of $T\setminus\{\bk\}$, which are disjoint from $\{i,j\}$ and
empty. Thus remote-only overlaps give no source. If $T$ meets the bond
and $\bk\notin T$, the factor $b_{\bk}$ survives in every monomial, so
the source contains $\bk$ as a pure creator or annihilator with
coefficient $u(\bk-i)$ times a factor fixed by $T$ and the local role;
for fixed shape and role, $\bk\mapsto$ configuration is injective, so the
$\ell^2$ norm is at most $C|\Theta_L(T)|\|u\|_2$. If $\bk\in T$, given
$T$ and the bond there are at most three choices of $\bk$, each with
coefficient at most $\|u\|_\infty\le\|u\|_2$; the same-edge terms are
bounded likewise. Summing the finitely many roles and then the triple
shapes, the pair and quartet sources have $\ell^2$ norm at most
$Cy\|u\|_2$ and $\ell^1$ norm at most $Cya_u$. A conservative
multiplicity is $2^{20}\cdot6!$; it covers sixteen monomial slots, nine
overlapping site pairs, $64$ extraction masks, $6!$ labelings and the
remaining commutator and normalization factors. It is a count of sites,
not of a geometric radius. The extraction remainder has budget
$Cya_u$, and Lemma~\ref{lem:ms-gram} with $n=2,4$ gives
\begin{equation}
 \left|\frac{\widehat\omega([R_V,Y_L])}N-c_0(u,\Theta_L)\right|
 \le Cy\bigl[\sqrt{\widehat p}\,\|u\|_2+\widehat p\,a_u\bigr],
 \label{eq:ms-RVY}
\end{equation}
which is row 12 after the prefactor $s\tau$. Every pair separation that
$\mathcal V_{32}\overline u$ evaluates on the support of $\Theta_L$ is
at most $\mathsf S_L$, so Eq.~\eqref{eq:ms-probe-near} with
$D=\mathsf S_L$ gives
\begin{equation}
 |c_0(u,\Theta_L)|\le C_cy\,\eps^2(1+\mathsf S_L)^2 .
 \label{eq:ms-c0}
\end{equation}
With $y\le CL^2$ and $\mathsf S_L\le CL^2\log(2L)$ this is row 13. The
deterministic scalar is retained, not dropped because it is a triplet
term.

\emph{Mixed terms.} The operator $[[R_Q,X],Y_L]$ has odd charge, hence
zero local vacuum diagonal, and local budget $Cv_ua_Ry$ on at most six
sites; Eq.~\eqref{eq:pop-vacuum-diagonal} bounds its density by
$Cv_ua_Ry\sqrt{\widehat p}$, which gives row 14. For $A=[R_V,X]$ the
vector $A\Omega=R_V\Psi_F$, $\Psi_F=X\Omega$, is a translation-invariant
one-magnon vector and vanishes by the quarter-turn. Next
$A^\dagger\Omega=R_V^\dagger\Psi_F-X\Psi_{\mathcal T_{\bar u}}$, and
$R_V^\dagger\Psi_F=\VDM B(u)^\dagger\Psi_F-B(u)^\dagger\VDM\Psi_F$. The
one-magnon part of $\VDM\Psi_F$ vanishes by the quarter-turn, so the
second term lies in the five-magnon sector, and
$X\Psi_{\mathcal T_{\bar u}}$ lies in the one- and five-magnon sectors.
With $P_3$ the projection on three magnons and $\mathcal V_{34}$ the
four-to-three block of $\VDM$,
\begin{equation}
 P_3A^\dagger\Omega=\mathcal V_{34}B(u)^\dagger\Psi_F,\qquad
 c_1:=\frac{\langle\Omega,[[R_V,X],Y_L]\Omega\rangle}N
 =\frac{\langle B(u)^\dagger\Psi_F,\mathcal V_{43}\Psi_\Theta\rangle}N .
 \label{eq:ms-c1-def}
\end{equation}
No convolution with the long pair kernel survives in $c_1$: the block
$\mathcal V_{43}$ adds one nearest-neighbor site to a triple, so every
four-site configuration in the support has pair separations at most
$\mathsf S_L+1$. On such a configuration $\mathcal C$ the coefficient of
$B(u)^\dagger\Psi_F$ is $\sum_{A\subset\mathcal C,|A|=2}
\overline{u(A)}F_R(\mathcal C\setminus A)$, and every argument of $u$ is
a pair separation of $\mathcal C$. The $\ell^1$ budget of $\mathcal
V_{43}\Psi_\Theta$ per site is at most $Cy$, so
Eq.~\eqref{eq:ms-probe-near} with $D=\mathsf S_L+1$ gives
\begin{equation}
 |c_1|\le C_cy\,\eps^2(2+\mathsf S_L)^2\|F_R\|_\infty ,
 \label{eq:ms-c1}
\end{equation}
row 15 after the prefactor $sd\tau=O(\eps^6)$. After its local scalars
are removed, $[[R_V,X],Y_L]$ has zero vacuum diagonal and budget
$Ca_ua_Ry$, which gives row 16.

\emph{Assembly.} Insert Eq.~\eqref{eq:ms-probe-norms},
Eq.~\eqref{eq:ms-hyp}, $a_R\le C\ell_\eps$, $\|F_R\|_\infty\le C$,
$|s|\le\eps$, $d\le\eps^2/2$ and $|\tau|=\eps^3/2$ into the bounds above
and divide by $\eps^4$; for example row 11 is
$\eps^{-4}\eps^3\cdot L^2\cdot\eps^2L^{-1}\sqrt{\ell_\eps}=\eps
L\sqrt{\ell_\eps}$, and row 17 is $\eps^{-4}\tau^2\eps^{-1}y^2\le
C\eps L^4$. For $L=\ell_\eps^8$ the zeroth-order rows are dominated by
the leading triple current, $O(\ell_\eps^{-7})$, and every other row is
at most $C\eps\ell_\eps^{32}$ or $C\eps^2\ell_\eps^{50}$; the common bound
$C\eps\ell_\eps^{52}$ is conservative.
\end{proof}

Rows 13 and 15 explain why the near-source cancellation of
Lemma~\ref{lem:ms-probe} is essential: without it the scalars $c_0,c_1$
would be bounded only through $\|u\|_\infty\le C\ell_\eps$, and the
twice-dressed comparison would not resolve order $\eps^4$.

\subsection{Proof of Theorem~\ref{thm:massscale}(i)--(iii) and of
Corollary~\ref{cor:ms-range}}
\label{ss:ms-proof}

\begin{lemma}[Partial-transpose perturbation]
\label{lem:ms-pt}
Let $\rho$ have the form~\eqref{eq:rho} with $p\le1/4$. Then
\begin{equation}
 \bigl|\Neg(\rho)-(|w|-p+q)_+\bigr|\le8\sqrt2\,|\zeta|+2v^2,\qquad
 |\zeta|\le\sqrt{(p-q)q}.
 \label{eq:ms-pt}
\end{equation}
\end{lemma}

\begin{proof}
The $\zeta$ entries of the partial transpose~\eqref{eq:PT} form a
Hermitian matrix with eight entries of modulus $|\zeta|$; its operator
norm is at most its Frobenius norm $2\sqrt2|\zeta|$. By Weyl's
inequality each ordered eigenvalue moves by at most this amount when the
entries are deleted, so $\Neg$, the sum of the four negative parts,
changes by at most $8\sqrt2|\zeta|$. Without the $\zeta$ entries the
eigenvalues are~\eqref{eq:Xeigen}. The pair $b\pm|w|$ contributes
$(|w|-p+q)_+$. The smaller eigenvalue of $\bigl(\begin{smallmatrix}
a&v\\v&q\end{smallmatrix}\bigr)$ is at least $q-v^2/(a-q)$, and
$a-q=1-2p\ge1/2$, so its negative part is at most $2v^2$. The bound on
$\zeta$ is positivity of the principal minor of $\rho$ in the rows
$\uparrow\downarrow,\downarrow\downarrow$.
\end{proof}

\begin{proof}[Proof of Theorem~\ref{thm:massscale}(i)--(iii)]
Statement (ii) is Lemma~\ref{lem:ms-qv}.

For the first limit in (i), take $L=\ell_\eps^8$ and $D=\ell_\eps$. For
small $\eps$, Eq.~\eqref{eq:ms-hyp} holds by Eq.~\eqref{eq:ms-M8}, and
$D\le c/(8\eps)$. Equations~\eqref{eq:ms-master}, \eqref{eq:ms-even}
and~\eqref{eq:ms-budget} give
\begin{equation}
 \sup_{|\br|\ge c/\eps}\frac{|w(\br)-w_2(\br)|}{\eps^4}
 \le C_c\bigl[\Xi_{\ell_\eps}+\sqrt{\Xi_{\ell_\eps}}
 +\eps^4\ell_\eps^4+\ell_\eps^{-7}+\eps\ell_\eps^{52}\bigr].
 \label{eq:ms-log-bound}
\end{equation}
By (ii) with $r_\eps=\ell_\eps-2\to\infty$, $\Xi_{\ell_\eps}\to0$, so
the right side tends to zero uniformly in $h\in\cI$ and the states.

The limit order is that of Sec.~\ref{sm:range}. At fixed $\eps$ and
$\br$, replace $u_\br$ by $u^{(K)}=\chi(\cdot/K)u_\br$. All identities
above are finite-support identities for $u^{(K)}$ and for the compact
kernel $\Theta_L$; apply them on large tori and take the thermodynamic
limit at fixed $\eps,K$. Then let $K\to\infty$: by
Eq.~\eqref{eq:range-test-cutoff-limits} the norms $\|u\|_1$, $\|u\|_2$,
$v_u$ and $\|\nabla_{\rm ret}u\|_2$ converge, every extracted source
converges in $\ell^1$ and enters through
Lemma~\ref{lem:current-density-continuity}, the scalars $c_0,c_1$ are
finite sums, and the left side of Eq.~\eqref{eq:range-stationarity}
converges since $A_2$ is bounded on $\ell^1$. The bounds of
Lemmas~\ref{lem:ms-probe}--\ref{lem:ms-budget} are then applied to
$u_\br$ itself. The coefficient-current truncations of
Sec.~\ref{sm:current} are removed at fixed positive regulator after the
thermodynamic limit. Only then is $\eps$ decreased; the suprema use
bounds uniform in $\br$, and no limits are exchanged.

For the second limit in (i), fix $D>c$. On $c\le|\eps\br|\le D$,
Eq.~\eqref{eq:Alimit} with the compact set $\{c\le|\bx|\le D\}$ gives
uniform convergence of $\eps^{-4}w_2$ to $A_h$. On $|\eps\br|>D$,
Lemma~\ref{lem:ms-tail}(c,d) bounds both by $C/D^2$. Letting first
$\eps\downarrow0$ and then $D\to\infty$,
\begin{equation}
 \sup_{h\in\cI,\,|\eps\br|\ge c}
 \bigl|\eps^{-4}w_2(\br)-A_h(\eps\br)\bigr|\longrightarrow0 .
 \label{eq:ms-w2-unbounded}
\end{equation}
Together with the first limit this proves the second.

For (iii), Lemma~\ref{lem:ms-pt} applies since $p=O(\eps^4)$. With
$|(x)_+-(x')_+|\le|x-x'|$ and $||w|-|A||\le|w-A|$,
\begin{equation}
 \left|\frac{\Neg(\rho_{0\br})}{\eps^4}-\Phi_h(\eps\br)\right|
 \le\left|\frac w{\eps^4}-A_h\right|+\left|\frac p{\eps^4}-n_*\right|
 +\frac{q_\br}{\eps^4}+\frac{8\sqrt2\sqrt{pq_\br}+2v_\br^2}{\eps^4}.
 \label{eq:ms-neg-error}
\end{equation}
The first term tends to zero uniformly on $|\eps\br|\ge c$ by (i), the
second by Theorem~\ref{thm:population}, and the others by (ii) with
$r_\eps=c/\eps$ and $p=O(\eps^4)$.
\end{proof}

The odd coherence enters Eq.~\eqref{eq:ms-neg-error} only through
$|\zeta|\le\sqrt{pq}$; no $\zeta=o(\eps^6)$ statement is used. In the far
tail the controlled entries give more than Eq.~\eqref{eq:ms-main-neg}.
Let $R_E=\max_{h\in\cI}\rho_E(h,0)$, the largest radius of the contour
\eqref{eq:boundary}. Since $|A_h(\rho,\theta)|\le|A_h(\rho,0)|$, $K_2$ is
strictly decreasing, and Lemma~\ref{lem:ms-tail}(d) bounds the tail, for
every $\eta>0$ there is $\delta_\eta>0$ with
$|A_h(\bx)|\le n_*-\delta_\eta$ for all $h\in\cI$ and $|\bx|\ge
R_E+\eta$. Theorem~\ref{thm:massscale}(i),(ii) and
Theorem~\ref{thm:population} then give
$p-q_\br-|w(\br)|\ge\delta_\eta\eps^4/2$ for all $|\br|\ge(R_E+\eta)/
\eps$ and small $\eps$: the principal block~\eqref{eq:principal-pt} is
positive definite throughout the tail, so it detects no remote NPT pair.
Exact PPT there would still require the other Schur block.

\begin{proof}[Proof of Corollary~\ref{cor:ms-range}]
(a) On the ray, $\Phi_h(\rho,\theta)$ is continuous in $\rho>0$,
strictly decreasing where positive, tends to infinity as
$\rho\downarrow0$ and to zero as $\rho\to\infty$, since
$K_2(x)\sim2/x^2$ at zero, $K_2'<0$ and $K_2\to0$. Hence
Eq.~\eqref{eq:ms-rhokappa} has a unique solution. Fix $0<\eta<
\rho_\kappa$ and take $c<\rho_\kappa-\eta$ in Theorem~\ref{thm:massscale}.
There is $k\in\mathbb N$ with $\eps|k\mathbf a|\in[\rho_\kappa-\eta-
\eps|\mathbf a|,\rho_\kappa-\eta]$; there
$\Phi_h(\eps k\mathbf a)\ge\Phi_h(\rho_\kappa-\eta,\theta)>\kappa$, so
Eq.~\eqref{eq:ms-main-neg} gives $\Neg\ge\kappa\eps^4$ for small $\eps$
and $\eps R_\kappa\ge\rho_\kappa-\eta-\eps|\mathbf a|$. Every lattice
point of the ray with $\eps|k\mathbf a|\ge\rho_\kappa+\eta$, including
the whole infinite tail, has $\Phi_h\le\Phi_h(\rho_\kappa+\eta,\theta)
<\kappa$ by a fixed margin; the uniform convergence on the unbounded
region gives $\Neg<\kappa\eps^4$ there, so $\eps R_\kappa\le
\rho_\kappa+\eta$. Let $\eta\downarrow0$. No finite-angle monotonicity of
the negativity and no absence of holes is used.

(b) Since $|\cos2\theta|\le1$, every $\bx$ with $|\bx|\ge\rho_\kappa(h,0)
+\eta$ has $\Phi_h(\bx)\le\Phi_h(\rho_\kappa(h,0)+\eta,0)<\kappa$, with a
margin independent of the direction; this gives the upper bound. The
axial sequence $k e_x$ gives the lower bound by (a). Uniformity in
$h\in\cI$ in (a) and (b) follows because the margins
$\Phi_h(\rho_\kappa\mp\eta,\theta)-\kappa$ are continuous in $h$ and
hence uniformly positive on the compact $\cI$.

(c) At $h=2$ one has $m_h=\sqrt3$ and
$A_{h=2}(\rho,\theta)=3\cos2\theta\,K_2(\sqrt3\rho)/[32(\pi-2)]$, which on
$\mathsf K=\{0.9\le\rho\le1,\ |\theta|\le0.15\}$ is positive and
decreasing in $\rho$ and in $|\theta|$. Thus
\begin{equation}
 \inf_{\mathsf K}A_{h=2}=\frac{3\cos(0.3)K_2(\sqrt3)}{32(\pi-2)}
 =0.0306177748\ldots,\qquad
 \Delta:=\inf_{\mathsf K}A_{h=2}-n_*=0.0198680154\ldots .
 \label{eq:ms-sector}
\end{equation}
A rigorous lower bound follows from the representation
\begin{equation}
 K_2(z)=\int_0^\infty e^{-z\cosh t}\cosh2t\,\dd t .
 \label{eq:ms-K2}
\end{equation}
On $[a,b]\subset[0,\infty)$ the integrand is at least
$e^{-z\cosh b}\cosh2a$, and it is positive, so the lower Riemann sum over
$[0,6]$ with step $10^{-3}$ bounds $K_2(\sqrt3)$ from below. Evaluated in
interval arithmetic with 50 significant digits, together with interval
enclosures of $\cos(0.3)$ and $\pi$, it gives
$\Delta/2>0.0099052873>0.0099$. Now Eq.~\eqref{eq:population-npt} with
$q_\br\ge0$, statement (i) and Theorem~\ref{thm:population} give,
uniformly over the lattice points with $\eps\br\in\mathsf K$ and over
the states,
$\Neg(\rho_{0\br})/\eps^4\ge|A_{h=2}(\eps\br)|-n_*-o(1)\ge\Delta-o(1)$,
which exceeds $0.0099$ for small $\eps$.
\end{proof}

The existence of a sufficiently small angle in (c) is rigorous, but its
value is not explicit: it inherits the routing constants of
Sec.~\ref{sm:current}.

\subsection{The explicit rate}
\label{ss:ms-rate}

The rate (iv) uses the same physical state, Hamiltonian, limit order and
pair dressing, with a polynomially large triplet cutoff and two further
zero-scalar extractions. Take
\begin{equation}
 L=(\eps\ell_\eps)^{-4/13},\qquad D=L^{1/2},
 \label{eq:ms-rate-choice}
\end{equation}
with $L$ rounded upward. For small $\eps$: $L\ge\ell_\eps$,
$\eps L^2=\eps^{5/13}\ell_\eps^{-8/13}\to0$,
$\eps L^3=\eps^{1/13}\ell_\eps^{-12/13}\to0$, $\log(2L)\le\ell_\eps$, so
$\eps\mathsf S_L\le C\eps^{5/13}\ell_\eps^{5/13}\to0$, and
$D\le c/(8\eps)$.

\emph{Densities.} At the choice~\eqref{eq:ms-rate-choice} the four
ratios of Lemma~\ref{lem:ms-energyL} are
$\eps^{5/13}\ell_\eps^{31/13}$, $\eps^{2/13}\ell_\eps^{15/13}$,
$\eps^{12/13}\ell_\eps^{-40/13}$ and $\eps^{74/13}\ell_\eps^{-56/13}$; all
tend to zero. Hence
\begin{equation}
 \sqrt{\widehat p}\le C\eps^2L^{-1},\qquad
 \sqrt{\widehat{\mathfrak q}}\le C\eps^3L^{-1},\qquad
 \sqrt{\widetilde p}\le C(\eps^2L^{-1}+\eps^3L^2)\le C\eps^2L^{-1},
 \label{eq:ms-rate-densities}
\end{equation}
the last because $\eps L^3\to0$. Thus Eq.~\eqref{eq:ms-hyp} holds, and
Lemma~\ref{lem:ms-transfer} applies with $\Lambda=L$. For $|\bx|\ge
D-2$ one has $(1+|\bx|)^{-4}\le CL^{-2}$ and, by
Lemma~\ref{lem:ms-tail}(b), $|C_F(\bx)|\le CL^{-1}$; hence
\begin{equation}
 \Xi_D\le C\bigl[L^{-1}+\eps^2\ell_\eps^2\bigr],\qquad
 \sqrt{\Xi_D}\le C\bigl[L^{-1/2}+\eps\ell_\eps\bigr],\qquad
 \eps^4(1+D)^4\le C\eps^4L^2 .
 \label{eq:ms-rate-Xi}
\end{equation}

\begin{lemma}[Two further zero-scalar extractions]
\label{lem:ms-extraction}
Under Eq.~\eqref{eq:ms-hyp}, the bound~\eqref{eq:ms-budget} holds with
row 17 of Table~\ref{tab:ms-budget}, $\eps L^4$, replaced by
$\eps^3L^3+\eps^4L^6$.
\end{lemma}

\begin{proof}
Row 17 is the second-order Taylor remainder of $V_L^\dagger AV_L$,
$A\in\{R_Q(u_\br),R_V(u_\br)\}$. Retain instead
$\frac{\tau^2}2[[A,Y_L],Y_L]$ in the expectation. Its vacuum diagonal
vanishes term by term. Every monomial of $R_Q$ has charge $-2$, and two
$Y_L$ commutators add $-6$, $0$ or $6$, leaving $-8$, $-2$ or $4$. The
operator $R_V=[B(u),\VDM]$ has charges $-1,-3$, and two odd
commutators preserve odd charge. Local monomials of $X$ and $Y_L$ have
definite charges, so this classification applies to the actual local
decomposition, not only after extensive sums cancel. Each summand acts
on at most seven sites, independently of its diameter, and
Eq.~\eqref{eq:pop-vacuum-diagonal} bounds the retained term by
$C\tau^2b_Ay^2\sqrt{\widehat p}$. The next Taylor remainder has norm
at most $C|\tau|^3b_Ay^3$. After the prefactors $a_\eps$ or $s$, with
$b_{R_Q}\le C\eps^{-1}$ and $|s|b_{R_V}\le C\eps^{-1}$, and division by
$\eps^4$, these are $C\eps^{-4}\tau^2\eps^{-1}y^2\eps^2L^{-1}\le
C\eps^3L^3$ and $C\eps^{-4}|\tau|^3\eps^{-1}y^3\le C\eps^4L^6$. No other
row is changed; the scalars $c_0,c_1$ remain with their bounds
\eqref{eq:ms-c0} and~\eqref{eq:ms-c1}.
\end{proof}

\begin{lemma}[Projected profile with a rate]
\label{lem:ms-projected-rate}
For every $c>0$ and sufficiently small $\eps$,
\begin{equation}
 \sup_{h\in\cI,\,|\eps\br|\ge c}
 \bigl|\eps^{-4}w_2(\br)-A_h(\eps\br)\bigr|\le C_c\eps .
 \label{eq:ms-projected-rate}
\end{equation}
\end{lemma}

\begin{proof}
By Eq.~\eqref{eq:rankone}, $w_2=dG_\sigma/(a_\eps-\cos\eps\,t_\sigma)$
with $\sigma=\mu_\eps/a_\eps$, $\sigma/\eps^2=m_h^2+O(\eps^2)$,
$d/\eps^2=1/2+O(\eps^2)$, and $a_\eps-\cos\eps\,t_\sigma=Z_0+O(\eps^2)$,
since $0\le t_0-t_\sigma\le\sigma\langle\mathfrak d^2/\lambda^2\rangle_k
\le\sigma/4$. Split Eq.~\eqref{eq:G} with a smooth periodic cutoff
$\chi_{\rm B}$, equal to one near zero and supported in a disk inside
the Brillouin square. The complementary multiplier has uniformly bounded
derivatives; six integrations by parts along a coordinate with
$|r_j|\ge|\br|/\sqrt2$ bound its contribution to $\eps^{-2}G_\sigma(\br)$
by $C\eps^{-2}|\br|^{-6}\le C_c\eps^4$.

In the low part put $\bk=\eps\bp$, $\bx=\eps\br$; it equals the integral
\eqref{eq:scaledintegral} with symbol $q_{\eps,h}$, which tends to
$q_{0,h}$ of Eq.~\eqref{eq:qsymbol}. Insert a second radial cutoff
$\psi(\bp/\mathsf R)$, $\mathsf R=\eps^{-1/4}$, with $\psi$ as in the
proof of Eq.~\eqref{eq:Alimit}. On $|\bp|\le2\mathsf R$,
$\chi_{\rm B}(\eps\bp)=1$ for small $\eps$, and Taylor expansion of
$\mathfrak d(\eps\bp)$, $\lambda(\eps\bp)$ and $\sigma$, with
$m_h\ge m_->0$, gives $|q_{\eps,h}(\bp)-q_{0,h}(\bp)|\le
C\eps^2(1+|\bp|^2)$. The integral of this difference over
$|\bp|\le2\mathsf R$ is $O(\eps^2\mathsf R^4)=O(\eps)$, uniformly in
$\bx$. On each dyadic annulus $|\bp|\asymp\kappa\ge\mathsf R$ within the
support of $\chi_{\rm B}(\eps\,\cdot)$, Eq.~\eqref{eq:symbolestimate}
gives $|\partial^\nu q_{\eps,h}|+|\partial^\nu q_{0,h}|\le
C_\nu\kappa^{-|\nu|}$, and the derivatives of both cutoffs obey the same
bound. Six integrations by parts give the annular kernel bound
$C\kappa^2(1+\kappa|\bx|)^{-6}\le C_c\kappa^{-4}$ for $|\bx|\ge c$.
The geometric sum over $\kappa\ge\mathsf R$ is $C_c\mathsf R^{-4}=C_c
\eps$, separately for the lattice and the continuum symbol. The
continuum transform, defined by this convergent annular sum away from
the origin, is $m_h^2K_2(m_h|\bx|)\cos2\theta/(8\pi)$ by
Eq.~\eqref{eq:Hessian}, which is bounded on $|\bx|\ge c$. Multiplying by
$(d/\eps^2)/(a_\eps-\cos\eps\,t_\sigma)=(2Z_0)^{-1}+O(\eps^2)$ and using
$16\pi Z_0=32(\pi-2)$ proves Eq.~\eqref{eq:ms-projected-rate}. Only
$|\bx|\ge c$ is used, never an upper bound on $|\bx|$.
\end{proof}

\begin{proof}[Proof of Theorem~\ref{thm:massscale}(iv)]
At the choice~\eqref{eq:ms-rate-choice},
Eqs.~\eqref{eq:ms-master}, \eqref{eq:ms-even}, \eqref{eq:ms-rate-Xi},
Lemma~\ref{lem:ms-budget} and Lemma~\ref{lem:ms-extraction} give, for
$|\br|\ge c/\eps$,
\begin{equation}
\begin{split}
 \frac{|w(\br)-w_2(\br)|}{\eps^4}\le C_c\bigl\{&
 L^{-1/2}+\eps\ell_\eps+\eps^4L^2+\ell_\eps L^{-1}+\eps\ell_\eps^3
 +\eps L\sqrt{\ell_\eps}+\eps^2\\
 &+\eps^2L^6\log^2(2L)+\eps^4L^6\log^2(2L)+\eps^2\ell_\eps L\\
 &+\eps^3L^3+\eps^4L^6+\eps^3\ell_\eps L^4+\eps^2\ell_\eps^2L^2\bigr\}.
\end{split}
 \label{eq:ms-rate-budget}
\end{equation}
The first line contains the near/far split and the bounds of
Eq.~\eqref{eq:range-general-residuals} and of the five- and two-to-four
particle sources (rows 1--12 of Table~\ref{tab:ms-budget}, with
$L^{-2}+L^{-1}+\eps\ell_\eps L^{-2}+\eps\ell_\eps^2L^{-1}\le
C[\ell_\eps L^{-1}+\eps\ell_\eps^3]$); the second line retains the
deterministic scalars and the scalar-subtracted mixed terms (rows
13--16); the third line is the refined triplet Taylor budget (rows
17--19 with Lemma~\ref{lem:ms-extraction}). The two balanced terms are
\begin{equation}
 L^{-1/2}=(\eps\ell_\eps)^{2/13},\qquad
 \eps^2L^6\log^2(2L)\le\eps^{2/13}\ell_\eps^{-24/13}\ell_\eps^2
 =(\eps\ell_\eps)^{2/13}.
 \label{eq:ms-rate-balance}
\end{equation}
Every other term is $o((\eps\ell_\eps)^{2/13})$:
$\ell_\eps L^{-1}=\eps^{4/13}\ell_\eps^{17/13}$,
$\eps L\sqrt{\ell_\eps}=\eps^{9/13}\ell_\eps^{5/26}$,
$\eps^2\ell_\eps L=\eps^{22/13}\ell_\eps^{9/13}$,
$\eps^3L^3=\eps^{27/13}\ell_\eps^{-12/13}$,
$\eps^4L^6=\eps^{28/13}\ell_\eps^{-24/13}$,
$\eps^3\ell_\eps L^4=\eps^{23/13}\ell_\eps^{-3/13}$,
$\eps^2\ell_\eps^2L^2=\eps^{18/13}\ell_\eps^{18/13}$ and
$\eps^4L^2=\eps^{44/13}\ell_\eps^{-8/13}$, while $\eps\ell_\eps$,
$\eps\ell_\eps^3$, $\eps^2$ and $\eps^4L^6\log^2(2L)$ are smaller still.
This proves the first bound of (iv). Without
Lemma~\ref{lem:ms-extraction}, row 17 would contribute
$\eps L^4=\eps^{-3/13}\ell_\eps^{-16/13}$, which diverges; and without
Lemma~\ref{lem:ms-energyL} the density bounds~\eqref{eq:ms-rate-densities}
would fail at this $L$.

Lemma~\ref{lem:ms-projected-rate} adds $C_c\eps$, which proves the bound
for $|w/\eps^4-A_h|$. For the negativity insert
Eq.~\eqref{eq:ms-tr-quant} with $\Lambda=L$ into
Eq.~\eqref{eq:ms-neg-error}: on $|\br|\ge c/\eps$,
$|p/\eps^4-n_*|\le C[L^{-1}+\eps^2\ell_\eps^2]$,
$q_\br/\eps^4\le C_c[\eps^4+L^{-2}+\eps^4\ell_\eps^4]$,
$\sqrt{pq_\br}/\eps^4\le C\sqrt{q_\br/\eps^4}\le
C_c[\eps^2+L^{-1}+\eps^2\ell_\eps^2]$, and
$v_\br^2/\eps^4\le C\eps^4$. All are $O((\eps\ell_\eps)^{4/13})$.
\end{proof}

The exponent $2/13$ is the optimum of the displayed finite collection of
bounds, obtained by balancing the square-root loss of the near/far split
against the localized scalar $c_0$. It is not optimal even within the
present method: in Lemma~\ref{lem:ms-even} only the joint population
needs the square root, and the normal coherence enters linearly.

\section{Entanglement above a vanishing threshold}
\label{sm:threshold}

Theorem~\ref{thm:massscale} determines the normalized negativity
$\eps^{-4}\Neg(\rho_{0\br})$ up to an additive error $C_cE_\eps$,
$\varpi_\eps=(\eps\ell_\eps)^{2/13}$, on the whole unbounded region
$|\eps\br|\ge c$. Corollary~\ref{cor:ms-range} turns this into the
farthest distance at which the negativity exceeds $\kappa\eps^4$ for a
fixed $\kappa>0$. In this section the threshold $\kappa_\eps$ tends to
zero with the angle. The profile error alone would require
$\kappa_\eps\gg \varpi_\eps$. A direct bound on the negativity outside the
leading lobes, Proposition~\ref{prop:th-exterior}, weakens this
requirement to $\kappa_\eps\gg\eps^{4/3}\ell_\eps^2$.
Theorem~\ref{thm:threshold} then identifies the limits of the ray and
all-direction distances and of the full planar superlevel set, and
Proposition~\ref{prop:th-shift} gives the first-order displacement of
the boundary when $\kappa_\eps\gg \varpi_\eps$.

Fields lie in $\cI$, states are the translation- and
jointly-fourfold-invariant energy-density minimizers of
Sec.~\ref{sm:conventions}, and the thermodynamic limit is taken at fixed
angle before $\eps\downarrow0$. Uniformity in $h$ and $\omega$ refers to
$h\in\cI$ and to all such states. A constant $C_c$ depends only on
$\cI$, the displayed parameters and the fixed cutoffs, never on the
state, the displacement or the angle. Theorem~\ref{thm:massscale} is
used in the following form: for every $c>0$ there is $C_c$ such that,
for all sufficiently small $\eps$,
\begin{equation}
 \sup_{h,\omega,\ |\eps\br|\ge c}\left\{
 \left|\frac{w(\br)}{\eps^4}-A_h(\eps\br)\right|
 +\left|\frac{\Neg(\rho_{0\br})}{\eps^4}
        -\bigl(|A_h(\eps\br)|-n_*\bigr)_+\right|\right\}
 \le C_cE_\eps .
 \label{eq:th-ms}
\end{equation}

\subsection{The limit profile and its superlevel sets}
\label{ss:th-profile}

Write $\bx=\rho(\cos\theta,\sin\theta)$ for a scaled displacement and
$A_h$ for the profile of Eq.~\eqref{eq:Alimit}, with
$m_h=\sqrt{2h-1}$. A direction is nodal if $\cos2\theta=0$. For a nonnodal $\theta$ and $s>-n_*$, let
$\rho_s(h,\theta)$ be the positive solution of
\begin{equation}
 |A_h(\rho_s,\theta)|=n_*+s,\qquad\text{equivalently}\qquad
 K_2(m_h\rho_s)=\frac{32(\pi-2)(n_*+s)}{m_h^2|\cos2\theta|}.
 \label{eq:th-rhokappa}
\end{equation}
Thus $\rho_0=\rho_E$ is the boundary radius of Eq.~\eqref{eq:boundary},
and for fixed $\kappa>0$ the radius $\rho_\kappa$ is the limit in
Corollary~\ref{cor:ms-range}. We also use the radial slope of the
profile at the boundary and the limit set,
\begin{gather}
 D_{h,\theta}=-\partial_\rho|A_h|\bigl(\rho_E(h,\theta),\theta\bigr)
 =\frac{m_h^3|\cos2\theta|}{64(\pi-2)}
  \bigl[K_1+K_3\bigr]\bigl(m_h\rho_E(h,\theta)\bigr),
 \label{eq:th-D}\\
 \mathcal E_h=\{0\}\cup
 \overline{\{\bx\ne0:|A_h(\bx)|>n_*\}}.
 \label{eq:th-Eh}
\end{gather}

\begin{lemma}[Limit profile]\label{lem:th-profile}
The following hold uniformly for $h\in\cI$.
\begin{enumerate}
\item[(a)] On a nonnodal ray, $\rho\mapsto|A_h(\rho,\theta)|$ is smooth
 and strictly decreasing from $+\infty$ to $0$. Hence $\rho_s(h,\theta)$
 exists, is unique and strictly decreasing in $s$, and
 $(s,h)\mapsto\rho_s(h,\theta)$ is smooth with
 $\partial_s\rho_s(h,\theta)|_{s=0}=-1/D_{h,\theta}$ and
 $D_{h,\theta}>0$.
\item[(b)] $\rho_E(h,\theta)^2<16|\cos2\theta|/\pi$. The function
 $\bar\rho_E$ equal to $\rho_E$ off the nodes and to $0$ on them is
 continuous on $\cI\times[0,2\pi]$, and
 $\bar\rho_E(h,\theta)\le\rho_E(h,0)\le B:=\max_{h\in\cI}\rho_E(h,0)$.
\item[(c)] $\mathcal E_h=\{\bx:|\bx|\le\bar\rho_E(h,\arg\bx)\}$, with
 $0\in\mathcal E_h$. The set
 $\mathcal E=\{(h,\bx):h\in\cI,\ \bx\in\mathcal E_h\}$ is compact.
\item[(d)] For $\eta>0$ the number
 $\delta_\eta=\min_{h,\theta}\bigl[n_*-|A_h(\bar\rho_E(h,\theta)+\eta,
 \theta)|\bigr]$ is positive, and $n_*-|A_h(\bx)|\ge\delta_\eta$
 whenever $|\bx|\ge\bar\rho_E(h,\arg\bx)+\eta$. This includes every
 $\bx$ with $\operatorname{dist}(\bx,\mathcal E_h)\ge\eta$ and every
 $\bx$ with $|\bx|\ge\rho_E(h,0)+\eta$.
\end{enumerate}
\end{lemma}

\begin{proof}
(a) The functions $K_\nu$ are positive on $(0,\infty)$, and
$K_2'=-(K_1+K_3)/2<0$. Moreover $K_2(z)\sim2/z^2$ as $z\downarrow0$ and
$K_2(z)\to0$ as $z\to\infty$. Hence $|A_h(\cdot,\theta)|$ is a smooth
decreasing bijection of $(0,\infty)$ onto itself when
$\cos2\theta\ne0$, which gives existence, uniqueness and monotonicity
of $\rho_s$. Its radial derivative,
$-m_h^3|\cos2\theta|[K_1+K_3](m_h\rho)/[64(\pi-2)]$, never vanishes,
so the implicit function theorem gives smoothness in $(s,h)$.
Differentiating Eq.~\eqref{eq:th-rhokappa} in $s$ at $s=0$ gives
$\partial_s\rho_s=-1/D_{h,\theta}$ and the formula~\eqref{eq:th-D}.

(b) Since $(z^2K_2(z))'=-z^2K_1(z)<0$ and $z^2K_2(z)\to2$ as
$z\downarrow0$, one has $z^2K_2(z)<2$ for $z>0$. At $\rho=\rho_E$,
\begin{equation*}
 n_*=\frac{(m_h\rho_E)^2K_2(m_h\rho_E)\,|\cos2\theta|}
          {32(\pi-2)\rho_E^2}
 <\frac{|\cos2\theta|}{16(\pi-2)\rho_E^2},
\end{equation*}
and $n_*=\pi/[256(\pi-2)]$ gives the bound. Off the nodes,
$\rho_E(h,\theta)=m_h^{-1}K_2^{-1}\bigl(\pi/[8m_h^2|\cos2\theta|]\bigr)$,
where $K_2^{-1}$ is the continuous decreasing inverse of
$K_2:(0,\infty)\to(0,\infty)$. This is continuous in $(h,\theta)$ and
increasing in $|\cos2\theta|$, so it is maximal on the axes. Near a node
the bound forces $\rho_E\to0$ uniformly in $h$, which proves continuity
of $\bar\rho_E$.

(c) By (a), $\{\bx\ne0:|A_h(\bx)|>n_*\}=\{\bx:0<|\bx|<
\bar\rho_E(h,\arg\bx)\}$; it contains no nodal point. Let $G_h$ be the
set on the right of (c). It is closed: if $\bx_n\in G_h$ and
$\bx_n\to\bx\ne0$, then $\arg\bx_n\to\arg\bx$ and continuity of
$\bar\rho_E$ gives $|\bx|\le\bar\rho_E(h,\arg\bx)$. Every point of $G_h$
is a limit of points of the open lobes: for $\bx\ne0$ in $G_h$ one has
$\bar\rho_E(h,\arg\bx)\ge|\bx|>0$, so $(1-1/n)\bx$ lies in a lobe, and
$0$ is the limit of $(1/n,0)$. Therefore $\mathcal E_h=G_h$. The same
argument, with joint continuity of $\bar\rho_E$, shows that
$\mathcal E$ is closed; it is bounded by (b).

(d) The function $(h,\theta)\mapsto|A_h(\bar\rho_E(h,\theta)+\eta,
\theta)|$ is continuous on the compact set $\cI\times[0,2\pi]$, because
its radius is at least $\eta$. It is smaller than $n_*$ everywhere:
it vanishes on the nodes, and elsewhere the radius exceeds
$\rho_E(h,\theta)$, so (a) applies. Hence $\delta_\eta>0$. If
$|\bx|\ge\bar\rho_E(h,\theta)+\eta$ with $\theta=\arg\bx$, then either
$\theta$ is nodal and $A_h(\bx)=0$, or (a) gives
$|A_h(\bx)|\le|A_h(\bar\rho_E+\eta,\theta)|\le n_*-\delta_\eta$. If
$\operatorname{dist}(\bx,\mathcal E_h)\ge\eta$, then
$|\bx|-\bar\rho_E(h,\theta)\ge\eta$, because the point
$\bar\rho_E(h,\theta)(\cos\theta,\sin\theta)$ belongs to
$\mathcal E_h$. The last case follows from (b).
\end{proof}

Part~(b) is the uniform form of the nodal cusp~\eqref{eq:cusp}. By
part~(c), $\mathcal E_h$ is the closed radial region under the
four-lobed boundary curve, including its four nodal cusps at the origin.

\subsection{Exterior bounds}
\label{ss:th-exterior}

We use the reduced state~\eqref{eq:rho}, with $a=1-2p+q$, $b=p-q$ and
$\zeta=\langle b_0n_\br\rangle^*$, and its partial transpose in the
block form~\eqref{eq:PT}.

\begin{proposition}[Exterior bounds]\label{prop:th-exterior}
The following three statements hold.
\begin{enumerate}
\item[(i)] Let $\rho$ be a two-qubit state of the form~\eqref{eq:rho}
 with $a>0$ and $b>|w|$. Then
 \begin{equation}
  \Neg(\rho)\le\frac{2v^2}{a}+\frac{8|\zeta|^2}{b-|w|}.
  \label{eq:th-complete}
 \end{equation}
\item[(ii)] For every $c>0$ there is $C_c$ such that, for all
 sufficiently small $\eps$,
 \begin{equation}
  \sup_{h,\omega,\ |\eps\br|\ge c}|\zeta(\br)|
  \le C_c\,\eps^{14/3}\ell_\eps .
  \label{eq:th-zeta}
 \end{equation}
\item[(iii)] For every $c>0$ and $0<\delta<n_*$ there is $C_{c,\delta}$
 such that, for all sufficiently small $\eps$ and uniformly in $h$ and
 $\omega$,
 \begin{equation}
  \Neg(\rho_{0\br})\le C_{c,\delta}\,\eps^{16/3}\ell_\eps^2
  \qquad\text{whenever}\quad |\eps\br|\ge c,\quad
  n_*-|A_h(\eps\br)|\ge\delta .
  \label{eq:th-negext}
 \end{equation}
\end{enumerate}
\end{proposition}

The region in (iii) is unbounded. By Lemma~\ref{lem:th-profile}(d) it
contains every displacement with $|\eps\br|\ge\rho_E(h,0)+\eta$ once
$\delta\le\delta_\eta$. Statement (iii) bounds the negative spectral
weight. It does not decide whether the smallest partial-transpose
eigenvalue is negative.

\paragraph{Proof of (i).}
In the ordered basis of Eq.~\eqref{eq:PT} write
\begin{equation*}
 \rho^{T_2}=\begin{pmatrix}\mathsf A&\mathsf D\\
            \mathsf D^\dagger&M\end{pmatrix},\qquad
 \mathsf A=\begin{pmatrix}a&v\\v&q\end{pmatrix},\qquad
 \mathsf D=\mathbf 1u^\dagger,\qquad \mathbf 1=(1,1)^T .
\end{equation*}
Since $b>|w|$, $M\succ0$ and $\lambda_{\min}(M)=b-|w|$. Exact completion
of the square gives
\begin{equation*}
 \rho^{T_2}-\begin{pmatrix}\mathsf A-\mathsf DM^{-1}\mathsf D^\dagger
 &0\\0&0\end{pmatrix}
 =\begin{pmatrix}\mathsf DM^{-1/2}\\ M^{1/2}\end{pmatrix}
  \begin{pmatrix}M^{-1/2}\mathsf D^\dagger& M^{1/2}\end{pmatrix}
 \succeq0 .
\end{equation*}
Here $\mathsf DM^{-1}\mathsf D^\dagger=\chi\mathbf 1\mathbf 1^T$ with
the scalar $\chi=u^\dagger M^{-1}u$ of Eq.~\eqref{eq:Schur}, and
$0\le\chi\le\|u\|^2/\lambda_{\min}(M)=2|\zeta|^2/(b-|w|)$. Further
$\chi\mathbf 1\mathbf 1^T\preceq2\chi I_2$. For
$\xi=(\xi_1,\xi_2)\in\mathbb C^2$, $q\ge0$ gives
\begin{equation*}
 \xi^\dagger\mathsf A\xi
 =a\Bigl|\xi_1+\frac va\xi_2\Bigr|^2
  +\Bigl(q-\frac{v^2}a\Bigr)|\xi_2|^2
 \ge-\frac{v^2}{a}\|\xi\|^2 .
\end{equation*}
Therefore $\rho^{T_2}\succeq\operatorname{diag}(-tI_2,0)$ with
$t=v^2/a+2\chi$. By eigenvalue monotonicity, $\rho^{T_2}$ has at most
two negative eigenvalues, each at least $-t$, so
$\Neg(\rho)\le2t\le2v^2/a+8|\zeta|^2/(b-|w|)$.
No information on the number of negative eigenvalues is used.
\hfill$\square$

\begin{lemma}[Joint population at mass-scale separation]
\label{lem:th-joint}
For every $c>0$,
$\sup_{h,\omega,\ |\eps\br|\ge c}q_\br/\eps^4\to0$.
\end{lemma}

\begin{proof}
Let $\widetilde\omega(A)=\omega(U_RAU_R^\dagger)$ be the pair-dressed
state of Sec.~\ref{sm:population}, with population $\widetilde p$ and
$a_R=\|F_R\|_1=O(\ell_\eps)$. For distinct sites $i,j$, the identities
$[b_i,X]=(1-2n_i)\sum_kF_R(k-i)b_k^\dagger$,
$b_k^\dagger n_k=0$ and $b_ib_i^\dagger=1-n_i$ give
\begin{equation*}
 [b_ib_j,X]=F_R(j-i)(1-n_i-n_j)
 +\sum_{k\ne i,j}\bigl[F_R(k-i)(1-2n_i)b_jb_k^\dagger
 +F_R(k-j)(1-2n_j)b_ib_k^\dagger\bigr].
\end{equation*}
After the scalar $F_R(j-i)$ is removed, every summand annihilates the
local vacuum, acts on at most three sites, and has norm at most its
coefficient; the coefficients sum to at most $2a_R$. For such a
summand $Z$, $\widetilde\omega(Z^\dagger Z)\le\|Z\|^2\,3\widetilde p$.
The second commutator obeys
$\|[[b_ib_j,X],X]\|\le24a_R^2$ by Eq.~\eqref{eq:pop-pair-comm-count}.
In the GNS seminorm $\|Z\|_{\widetilde\omega}=
\widetilde\omega(Z^\dagger Z)^{1/2}$, the integral Taylor formula and
$\widetilde\omega(n_in_j)\le\widetilde p$ therefore give
\begin{equation*}
 \bigl\|U_R^\dagger b_ib_jU_R-dF_R(j-i)\bigr\|_{\widetilde\omega}
 \le\eta_R:=(1+2\sqrt3\,da_R)\sqrt{\widetilde p}+12d^2a_R^2 .
\end{equation*}
Since $q_\br=\omega(n_0n_\br)=\|U_R^\dagger b_0b_\br U_R\|_{\widetilde
\omega}^2$, this yields $\sqrt{q_\br}\le d|F_R(\br)|+\eta_R$. By
Eq.~\eqref{eq:dressed-density-sharp}, $\eta_R=o(\eps^2)$ uniformly. By
Eq.~\eqref{eq:green-tail}, $|F_R(\br)|\le|F(\br)|\le C\eps^2/c^2$ when
$|\eps\br|\ge c$, so $d|F_R(\br)|=O_c(\eps^4)$.
\end{proof}

\paragraph{Proof of (ii).}
The proof transfers $\zeta$ to a twice-dressed state in which the
magnon density is small, using the pair dressing $U_R=e^{dX}$ of
Sec.~\ref{sm:population} and the quantitative compact triplet dressing
used for Theorem~\ref{thm:massscale}, with a cutoff adapted to the odd
coherence.

\emph{Auxiliary dressing.} For an integer $L\ge2$ let
$T_L=\mathcal V_{32}(\chi(\cdot/L)F)$, $\beta=L^{-2}$,
$N_L=\lceil13L^2\log L\rceil$, and
\begin{equation}
 \Theta_L=-\frac1{12+\beta}\sum_{j=0}^{N_L-1}
 \Bigl(\frac{12}{12+\beta}\Bigr)^j\bigl(I-Q_3/12\bigr)^jT_L .
 \label{eq:th-theta}
\end{equation}
Put $Y_L=C^+_{\Theta_L}-C^{+\dagger}_{\Theta_L}$,
$V_L=e^{\tau Y_L}$ with $\tau=\eps^3/2$ as in
Eq.~\eqref{eq:pop-triplet-dressing}, and
$\widehat\omega_L(A)=\omega(U_RV_LAV_L^\dagger U_R^\dagger)$, with
population $\widehat p_L$ and exchange density $\widehat{\mathfrak q}_L$.
The kernel $I-Q_3/12$ is a lazy symmetric Markov kernel on $\Omega_3$
that commutes with the symmetries of $\mathcal T$. Hence $\Theta_L$ is
compact and has these symmetries, and $\widehat\omega_L$ is
translation and jointly fourfold invariant. Let $y_L$ be the norm sum of
the terms of $Y_L$ incident at one site. Each lazy step contracts
$\ell^1$ and moves each coordinate by at most one lattice step, and
$\|T_L\|_1\le C$; therefore
\begin{equation}
 y_L\le C\|\Theta_L\|_1\le C\beta^{-1}\|T_L\|_1\le CL^2,\qquad
 \operatorname{diam}\operatorname{supp}\Theta_L\le C(L+N_L)
 \le CL^2\log(2L).
 \label{eq:th-kernel}
\end{equation}
On a spectral value $\lambda\in[0,12]$ of $Q_3$ the multiplier in
Eq.~\eqref{eq:th-theta} has modulus at most $(\lambda+\beta)^{-1}$. The
Nash inequality on the physical three-particle exclusion graph, used for
Theorem~\ref{thm:massscale}, bounds the spectral measure of $T_L$ by
$\mu_{T_L}([0,\lambda])\le C\lambda^2$ uniformly in $L$. Integrating
below $\beta$ and over dyadic bands above it gives
\begin{equation}
 \|\Theta_L\|_2^2\le\int_0^{12}\frac{\dd\mu_{T_L}(\lambda)}
 {(\lambda+\beta)^2}\le C\log(2L).
 \label{eq:th-theta-l2}
\end{equation}

\emph{Energy and densities.} The proof of Theorem~\ref{thm:massscale}
establishes the following energy inequality for this dressing. For every
choice of cutoffs with $\ell_\eps\le L$ and $\eps L^2\to0$, uniformly in
$h$ and $\omega$ and for all sufficiently small $\eps$,
\begin{equation}
 \frac{g_*}4\widehat{\mathfrak q}_L+\frac{m_*}4\eps^2\widehat p_L
 \le C\bigl[\eps^6L^{-2}+\eps^7\ell_\eps^3+\eps^8\ell_\eps^3L^4
 +\eps^{10}L^8+\eps^{16}L^{12}\bigr].
 \label{eq:th-energy}
\end{equation}
It is the energy sandwich of Sec.~\ref{sm:population} with the kernel
$\Theta_L$, in which three further terms of the second conjugation,
each with zero vacuum scalar, are retained; the common scalar is
unchanged, and the first term on the right is the cost of the residual
triplet source. Choose
\begin{equation}
 L=L_*:=\bigl\lceil\eps^{-1/3}\ell_\eps^{-1/2}\bigr\rceil .
 \label{eq:th-Lstar}
\end{equation}
For small $\eps$, $\ell_\eps\le L_*\le2\eps^{-1/3}\ell_\eps^{-1/2}$ and
$\eps L_*^2\le4\eps^{1/3}\ell_\eps^{-1}\to0$. The four remainder terms
in Eq.~\eqref{eq:th-energy}, divided by $\eps^6L_*^{-2}$, are at most
$C$ times $\eps^{1/3}\ell_\eps^2$, $1$, $\eps^{2/3}\ell_\eps^{-5}$ and
$\eps^{16/3}\ell_\eps^{-7}$. A bounded ratio suffices, and
$L_*^{-2}\le\eps^{2/3}\ell_\eps$, so
\begin{equation}
 \widehat p_{L_*}\le C\eps^4L_*^{-2}\le C\eps^{14/3}\ell_\eps,\qquad
 \widehat{\mathfrak q}_{L_*}\le C\eps^6L_*^{-2}\le C\eps^{20/3}\ell_\eps .
 \label{eq:th-density}
\end{equation}
From now on $L=L_*$, $V=V_{L_*}$, $Y=Y_{L_*}$, $\Theta=\Theta_{L_*}$,
$y=y_{L_*}$, and the index is dropped from $\widehat\omega$. By
Eq.~\eqref{eq:th-kernel},
$\eps\operatorname{diam}\operatorname{supp}\Theta\le C\eps^{1/3}$,
$|\tau|y\le C\eps^{7/3}\ell_\eps^{-1}$ and $da_R\le C\eps^2\ell_\eps$
all tend to zero. In particular, for small $\eps$ every $\br$ with
$|\eps\br|\ge c$ satisfies $|\br|>\operatorname{diam}\operatorname{supp}
\Theta$, which we assume below.

\emph{Exact local identities.} Let $O_\br=b_0n_\br$, so that
$\zeta(\br)^*=\omega(O_\br)=\widehat\omega(V^\dagger U_R^\dagger O_\br
U_RV)$. Write $[O_\br,X]=[b_0,X]n_\br+b_0[n_\br,X]$, with
$[b_0,X]=(1-2n_0)\sum_kF_R(k)b_k^\dagger$ and
$[n_\br,X]=\sum_{k\ne\br}F_R(k-\br)(b_\br^\dagger b_k^\dagger
+b_\br b_k)$. The $k=\br$ term of the first part vanishes because
$b_\br^\dagger n_\br=0$, and the $k=0$ term of the second part equals
$F_R(\br)(1-n_0)b_\br^\dagger$. Hence
\begin{equation}
 \mathcal K_\br:=[O_\br,X]
 =F_R(\br)(1-n_0)b_\br^\dagger
 +\sum_{k\ne0,\br}F_R(k)(1-2n_0)n_\br b_k^\dagger
 +\sum_{k\ne0,\br}F_R(k-\br)\bigl(b_0b_\br^\dagger b_k^\dagger
 +b_0b_\br b_k\bigr).
 \label{eq:th-comm}
\end{equation}
Its vacuum column is $F_R(\br)b_\br^\dagger\Omega$, and its vacuum row
is that of
$\Xi_\br:=\sum_{k\ne0,\br}F_R(k-\br)\,b_0b_\br b_k$. Thus
$\mathcal K_\br=F_R(\br)b_\br^\dagger+\Xi_\br+\mathcal K_\br^{\rm ann}$,
where
$\mathcal K_\br^{\rm ann}=-F_R(\br)n_0b_\br^\dagger
+\sum_kF_R(k)(1-2n_0)n_\br b_k^\dagger
+\sum_kF_R(k-\br)b_0b_\br^\dagger b_k^\dagger$
is a sum of terms on at most three sites, each annihilating the local
vacuum together with its adjoint, of total norm at most $3a_R$.

Let $\mathcal M_\br=[O_\br,Y]$ and $\Psi_\Theta=C^+_\Theta\Omega$.
Since $O_\br\Omega=O_\br^\dagger\Omega=0$,
$\mathcal M_\br\Omega=O_\br\Psi_\Theta$ and
$\mathcal M_\br^\dagger\Omega=O_\br^\dagger\Psi_\Theta$. No triplet in
$\operatorname{supp}\Theta$ contains both $0$ and $\br$, so
$O_\br\Psi_\Theta=0$, while
$O_\br^\dagger\Psi_\Theta=\sum_{C\ni\br,\,0\notin C}\Theta(C)
|\{0\}\cup C\rangle$ consists of separated quartets. Let
$\Lambda_\br^+=\sum_{C\ni\br,\,0\notin C}\Theta(C)\,b_0^\dagger
\prod_{j\in C}b_j^\dagger$ be the four-creator operator with this vacuum
image. Extracting the vacuum scalar, pure creators and pure annihilators
of each local term of $\mathcal M_\br$ as in
Lemma~\ref{lem:pop-local-extraction}, whose proof applies verbatim to
non-Hermitian terms, and regrouping the extracted parts, which are
fixed by the two vacuum images, gives
$\mathcal M_\br=(\Lambda_\br^+)^\dagger+\mathcal M_\br^{\rm ann}$. Here
$\mathcal M_\br^{\rm ann}$ is a sum of terms on at most five sites, each
annihilating the local vacuum with its adjoint, of total norm at most
$Cy$. Finally, since $Y\Omega=\Psi_\Theta$,
$\mathcal K_\br^\dagger\Omega=\Xi_\br^\dagger\Omega$ and
$\mathcal K_\br\Omega=F_R(\br)b_\br^\dagger\Omega\perp Y^\dagger\Omega
=-\Psi_\Theta$,
\begin{equation}
 \langle\Omega,[\mathcal K_\br,Y]\Omega\rangle
 =\sum_{k\ne0,\br}F_R(k-\br)\,\Theta(\{0,\br,k\})=0,
 \label{eq:th-mixed-scalar}
\end{equation}
because every triplet $\{0,\br,k\}$ has diameter at least
$|\br|>\operatorname{diam}\operatorname{supp}\Theta$. This cancels a
scalar; the operators $\mathcal K_\br$, $[\mathcal K_\br,Y]$ and
$O_\br$ are not zero.

\emph{Expansion.} Let
$\mathcal X_\br=U_R^\dagger O_\br U_R-O_\br-d\mathcal K_\br
=\sum_{j\ge2}(d^j/j!)\operatorname{ad}^j(O_\br)$, where
$\operatorname{ad}(A)=[A,X]$; the series converges for $2da_R<1$ by
Eq.~\eqref{eq:pop-pair-comm-count}. Let
$\mathcal Y_\br=V^\dagger O_\br V-O_\br-\tau\mathcal M_\br
-(\tau^2/2)[\mathcal M_\br,Y]$. Then
\begin{gather}
 \zeta(\br)^*=\widehat\omega(O_\br)+d\,\widehat\omega(\mathcal K_\br)
 +\tau\widehat\omega(\mathcal M_\br)+\mathcal R_\br,
 \label{eq:th-odd-decomp}\\
 \mathcal R_\br=\frac{\tau^2}2\widehat\omega([\mathcal M_\br,Y])
 +\widehat\omega(\mathcal Y_\br)
 +d\,\widehat\omega(V^\dagger\mathcal K_\br V-\mathcal K_\br)
 +\widehat\omega(V^\dagger\mathcal X_\br V).\nonumber
\end{gather}
We show
\begin{equation}
 |\mathcal R_\br|\le C\bigl\{\tau^2y^2\sqrt{\widehat p}+|\tau|^3y^3
 +d|\tau|a_Ry\sqrt{\widehat p}+d\tau^2a_Ry^2
 +d^2a_R^2\sqrt{\widehat p}+d^2|\tau|a_R^2y\bigr\}.
 \label{eq:th-taylor}
\end{equation}
The operator $O_\br$ has magnon charge $-1$, $X$ has charges $\pm2$ and
$Y$ has charges $\pm3$. Thus every local monomial of
$[[O_\br,Y],Y]$ has charge $-7$, $-1$ or $5$, and every monomial of
$\operatorname{ad}^j(O_\br)$ has odd charge; all of them have zero vacuum
diagonal. By Eq.~\eqref{eq:pop-triple-comm-count} with two initial
sites, $[[O_\br,Y],Y]$ has norm budget at most $32y^2$, so
Eq.~\eqref{eq:pop-vacuum-diagonal} bounds its expectation by
$Cy^2\sqrt{\widehat p}$, and the third-order integral remainder obeys
$\|\mathcal Y_\br\|\le64|\tau|^3y^3$. Next,
$V^\dagger\mathcal K_\br V-\mathcal K_\br=\tau[\mathcal K_\br,Y]$ plus
a second-order remainder of norm at most $C\tau^2a_Ry^2$. Subtract from
every local term of $[\mathcal K_\br,Y]$ its vacuum scalar. These
scalars sum to zero by Eq.~\eqref{eq:th-mixed-scalar}, and each
subtracted term has zero vacuum diagonal and at most five sites, so
$|\widehat\omega([\mathcal K_\br,Y])|\le Ca_Ry\sqrt{\widehat p}$.
Finally, the $j$th term of $\mathcal X_\br$ has norm budget at most
$(2a_R)^j(j+1)!$ on at most $j+2$ sites by
Eq.~\eqref{eq:pop-pair-comm-count}. Hence, for $2da_R\le1/2$,
\begin{equation*}
 |\widehat\omega(\mathcal X_\br)|\le\sum_{j\ge2}\frac{d^j}{j!}
 (2a_R)^j(j+1)!\,3\sqrt{(j+2)\widehat p}\le Cd^2a_R^2\sqrt{\widehat p},
\end{equation*}
and $\|V^\dagger\mathcal X_\br V-\mathcal X_\br\|\le|\tau|\,
\|[\mathcal X_\br,Y]\|\le|\tau|\sum_{j\ge2}(d^j/j!)(2a_R)^j(j+1)!\,
2(j+2)y\le Cd^2|\tau|a_R^2y$. This proves Eq.~\eqref{eq:th-taylor}.

\emph{Retained terms.} Since $O_\br=(b_\br b_0^\dagger)^\dagger b_\br$,
Cauchy-Schwarz gives $|\widehat\omega(O_\br)|\le\widehat p$. The joint
quarter-turn multiplies a one-site transverse mean by $\pm i$, so
$\widehat\omega(b_\br^\dagger)=0$ and
$d\,\widehat\omega(\mathcal K_\br)=d\,\widehat\omega(\Xi_\br)
+d\,\widehat\omega(\mathcal K_\br^{\rm ann})$, with
$|\widehat\omega(\mathcal K_\br^{\rm ann})|\le9a_R\widehat p$ by
Eq.~\eqref{eq:vacuum-annihilation-bound}. Similarly
$|\widehat\omega(\mathcal M_\br^{\rm ann})|\le Cy\widehat p$.

The remaining two expectations are creation densities. In the
convention~\eqref{eq:current-creator-convention},
$\overline{\widehat\omega(\Xi_\br)}$ is the creator-only density of the
source
$R_\br(x_1,x_2)=3[\delta_{x_1,\br}F_R(x_2-\br)
+\delta_{x_2,\br}F_R(x_1-\br)]$ on $\Omega_3$, and
$\overline{\widehat\omega((\Lambda_\br^+)^\dagger)}$ is that of
$S_\br(x_1,x_2,x_3)=4\,\Theta(\{x_1,x_2,x_3\})\,
\mathbf 1[\br\in\{x_1,x_2,x_3\}]$ on $\Omega_4$. Thus
$M_3(R_\br)\le6a_R$ and $M_4(S_\br)\le Cy$. Both sources are sums of
anchored sources $S(x_1,\ldots,x_m)=\delta_{x_1,\br}
K(x_2-\br,\ldots,x_m-\br)$, up to a permutation of coordinates, which
leaves the free inverse form unchanged. For such a source
$|\widehat S(k_1,\ldots,k_m)|=|\widehat K(k_2,\ldots,k_m)|$. Integrating
first over $k_1$ and using Parseval in the other momenta,
Eq.~\eqref{eq:current-two-dimensional-log} gives
\begin{equation}
 I_\eta(S)\le\mathcal J_2(\eta)\|K\|_2^2
 \le\frac\pi{16}\log(1+8/\eta)\,\|K\|_2^2 .
 \label{eq:th-anchored}
\end{equation}
For $R_\br$ the kernel $K$ is a restriction of $F_R$, and for $S_\br$ it
is $\Theta$ in anchored relative coordinates. With $\eta=\eps^2$,
Eqs.~\eqref{eq:th-anchored} and \eqref{eq:th-theta-l2} give
$I_{\eps^2}(R_\br)\le C\ell_\eps S_R\le C\ell_\eps$ and
$I_{\eps^2}(S_\br)\le C\ell_\eps\log(2L)$. The absolute-sum bound
$I_\eta(S_\br)\le\mathcal G_6\|S_\br\|_1^2=O(y^2)$ of
Eq.~\eqref{eq:current-source-transience} would not suffice here.
Lemma~\ref{lem:creation-current}, with $H_4(\eps^2)\le C\ell_\eps^{1/2}$
and $H_6$ constant, now gives
\begin{align*}
 d|\widehat\omega(\Xi_\br)|&\le Cd\bigl\{\sqrt{\ell_\eps
 \widehat{\mathfrak q}}+\eps^{-1}a_R\sqrt{\widehat{\mathfrak q}
 \widehat p}+\eps^2\ell_\eps^{1/2}a_R\sqrt{\widehat p}
 +a_R\widehat p\bigr\},\\
 |\tau\widehat\omega(\mathcal M_\br)|&\le C|\tau|\bigl\{
 \sqrt{\ell_\eps\log(2L)\,\widehat{\mathfrak q}}
 +\eps^{-1}y\sqrt{\widehat{\mathfrak q}\widehat p}
 +\eps^2y\sqrt{\widehat p}+y\widehat p\bigr\}.
\end{align*}

\emph{Exponent count.} Insert Eq.~\eqref{eq:th-density},
$a_R\le C\ell_\eps$, $y\le CL^2$, $d\le\eps^2/2$, $|\tau|=\eps^3/2$ and
$\log(2L)\le C\ell_\eps$. Up to constants, the contributions are
\begin{equation}
 \renewcommand{\arraystretch}{1.15}
 \begin{array}{l@{\qquad}l@{\qquad}l}
 \hline
 \text{Contribution} & \text{Bound} &
 \text{Divided by }\eps^4L_*^{-2}\\
 \hline
 \widehat\omega(O_\br) & \eps^4L^{-2} & 1\\
 d\sqrt{\ell_\eps\widehat{\mathfrak q}} &
   \eps^5\ell_\eps^{1/2}L^{-1} & \eps^{2/3}\\
 d\eps^{-1}a_R\sqrt{\widehat{\mathfrak q}\widehat p},\
   da_R\widehat p & \eps^6\ell_\eps L^{-2} & \eps^2\ell_\eps\\
 d\eps^2\ell_\eps^{1/2}a_R\sqrt{\widehat p} &
   \eps^6\ell_\eps^{3/2}L^{-1} & \eps^{5/3}\ell_\eps\\
 |\tau|\sqrt{\ell_\eps\log(2L)\widehat{\mathfrak q}} &
   \eps^6\ell_\eps L^{-1} & \eps^{5/3}\ell_\eps^{1/2}\\
 |\tau|\eps^{-1}y\sqrt{\widehat{\mathfrak q}\widehat p},\
   |\tau|y\widehat p & \eps^7 & \eps^{7/3}\ell_\eps^{-1}\\
 |\tau|\eps^2y\sqrt{\widehat p} & \eps^7L & \eps^2\ell_\eps^{-3/2}\\
 \tau^2y^2\sqrt{\widehat p} & \eps^8L^3 & \eps^{7/3}\ell_\eps^{-5/2}\\
 |\tau|^3y^3 & \eps^9L^6 & \eps^{7/3}\ell_\eps^{-4}\\
 d|\tau|a_Ry\sqrt{\widehat p} & \eps^7\ell_\eps L &
   \eps^2\ell_\eps^{-1/2}\\
 d\tau^2a_Ry^2 & \eps^8\ell_\eps L^4 & \eps^2\ell_\eps^{-2}\\
 d^2a_R^2\sqrt{\widehat p} & \eps^6\ell_\eps^2L^{-1} &
   \eps^{5/3}\ell_\eps^{3/2}\\
 d^2|\tau|a_R^2y & \eps^7\ell_\eps^2L^2 & \eps^{5/3}\\
 \hline
 \end{array}
 \label{eq:th-odd-budget}
\end{equation}
The last column is evaluated at $L=L_*$. Every entry except the first
tends to zero, the largest being $\eps^{2/3}$. Therefore
$|\zeta(\br)|\le C\eps^4L_*^{-2}\le C_c\eps^{14/3}\ell_\eps$, uniformly
in $h$, $\omega$ and $|\eps\br|\ge c$.

All identities above concern finite local operators. They are proved
in finite volumes containing the supports and passed to the
thermodynamic limit at fixed angle; the current inequality is used with
the truncation order of Lemma~\ref{lem:creation-current}. The angle
limit is taken last.\hfill$\square$

\paragraph{Proof of (iii).}
Fix $c$ and $\delta$. By Theorem~\ref{thm:population},
Eq.~\eqref{eq:th-ms} and Lemma~\ref{lem:th-joint}, for all sufficiently
small $\eps$, uniformly in $h$, $\omega$ and the displacements in
Eq.~\eqref{eq:th-negext},
\begin{equation*}
 \Bigl|\frac p{\eps^4}-n_*\Bigr|\le\frac\delta8,\qquad
 q_\br\le\frac{\delta\eps^4}8,\qquad
 |w(\br)|\le\Bigl(|A_h(\eps\br)|+\frac\delta4\Bigr)\eps^4 .
\end{equation*}
Hence $a=1-2p+q\ge1/2$ and
$b-|w|=p-q-|w|\ge(n_*-|A_h(\eps\br)|-\delta/2)\eps^4\ge\delta\eps^4/2$.
Positivity of the one-flip block of Eq.~\eqref{eq:rho} gives
$|v|\le b\le p\le2n_*\eps^4$. Parts (i) and (ii) now give
\begin{equation*}
 \Neg(\rho_{0\br})\le4v^2+\frac{16|\zeta|^2}{\delta\eps^4}
 \le16n_*^2\eps^8+\frac{16C_c^2}{\delta}\eps^{16/3}\ell_\eps^2 ,
\end{equation*}
and $\eps^8\le\eps^{16/3}\ell_\eps^2$.\hfill$\square$

\paragraph{Example.}
At $h=2$ let $K$ be the off-axis sector
$27/20\le\rho\le3/2$, $\arctan(9/20)\le\theta\le\arctan(11/20)$. It is
separated from the axes and the nodes and lies outside the lobe; for
comparison, $\rho_E(2,\arctan\frac12)=1.2116$.
Since $A_2$ decreases in $\rho$ and in $\theta$ on $K$, its extreme
values are attained at two opposite corners: $A_2=0.0046571$ at
$(3/2,\arctan(11/20))$ and $A_2=0.0083325$ at $(27/20,\arctan(9/20))$.
Thus $23/5000<A_2<n_*-23/10000$ on $K$, and part (iii) with
$\delta=23/10000$ gives $\Neg(\rho_{0\br})\le C\eps^{16/3}\ell_\eps^2$
for $\eps\br\in K$. Whether the smallest partial-transpose eigenvalue is
negative on $K$ is decided at order $\eps^8$ by
Eq.~\eqref{eq:exterior}, which requires $\zeta=o(\eps^6)$;
Eq.~\eqref{eq:th-zeta} does not reach this precision.

\subsection{Distances and planar sets above a vanishing threshold}
\label{ss:th-theorem}

For a primitive vector $\mathbf a\in\mathbb Z^2\setminus\{0\}$ with
$\cos(2\arg\mathbf a)\ne0$ and for $\kappa>0$, let
\begin{align}
 R^{\mathbf a}_\kappa(\eps,h,\omega)&=\sup\bigl\{k|\mathbf a|:
 k\in\mathbb N,\ \Neg(\rho_{0,k\mathbf a})\ge\kappa\eps^4\bigr\},
 \label{eq:th-Ra}\\
 R^{\rm all}_\kappa(\eps,h,\omega)&=\sup\bigl\{|\br|:
 \br\in\mathbb Z^2\setminus\{0\},\ \Neg(\rho_{0\br})\ge\kappa\eps^4
 \bigr\}.\nonumber
\end{align}
For a threshold $\kappa_\eps$ define the scaled superlevel set
\begin{equation}
 \mathcal S_{\eps,h,\omega}=\bigl\{\eps\br:
 \br\in\mathbb Z^2\setminus\{0\},\
 \Neg(\rho_{0\br})\ge\kappa_\eps\eps^4\bigr\}.
 \label{eq:th-S}
\end{equation}
The origin is not a pair displacement; $\mathcal E_h$ contains it as the
limit of the shrinking microscopic neighborhood.

\begin{theorem}[Vanishing-threshold geometry]\label{thm:threshold}
Let $\kappa_\eps>0$ satisfy
\begin{equation}
 \kappa_\eps\longrightarrow0,\qquad
 \frac{\kappa_\eps}{\eps^{4/3}\ell_\eps^2}\longrightarrow\infty .
 \label{eq:th-window}
\end{equation}
Then, for all sufficiently small $\eps$ and uniformly in $h$ and
$\omega$, the sets defining $R^{\mathbf a}_{\kappa_\eps}$,
$R^{\rm all}_{\kappa_\eps}$ and $\mathcal S_{\eps,h,\omega}$ are finite
and nonempty, and
\begin{gather}
 \sup_{h,\omega}\bigl|\eps R^{\mathbf a}_{\kappa_\eps}
 -\rho_E(h,\arg\mathbf a)\bigr|\longrightarrow0
 \quad\text{for every primitive }\mathbf a
 \text{ with }\cos(2\arg\mathbf a)\ne0,
 \label{eq:th-ray}\\
 \sup_{h,\omega}\bigl|\eps R^{\rm all}_{\kappa_\eps}
 -\rho_E(h,0)\bigr|\longrightarrow0,
 \label{eq:th-all}\\
 \sup_{h,\omega}d_H\bigl(\mathcal S_{\eps,h,\omega},\mathcal E_h\bigr)
 \longrightarrow0 .
 \label{eq:th-hausdorff}
\end{gather}
\end{theorem}

Condition~\eqref{eq:th-window} holds for every inverse-logarithmic
threshold $\kappa_\eps=\ell_\eps^{-\gamma}$, $\gamma>0$, for every power
$\kappa_\eps=\eps^\gamma$ with $0<\gamma<4/3$, and for
$\kappa_\eps=\eps^{4/3}\ell_\eps^3$. Thresholds of the critical order
$\kappa_\eps\asymp\eps^{4/3}\ell_\eps^2$, and smaller ones, are not
covered: they need not dominate the exterior bound~\eqref{eq:th-negext},
whose constant is not explicit. The exponent $4/3$ is a property of the
present estimates, not a claimed physical exponent. Nothing is asserted
about holes in, or monotonicity of, the finite-angle superlevel sets.

\paragraph{Proof.}
Two consequences of the inputs are used.

(A) Let $c,m>0$. For all sufficiently small $\eps$, uniformly in $h$ and
$\omega$, every $\br$ with $|\eps\br|\ge c$ and
$|A_h(\eps\br)|\ge n_*+m$ satisfies
$\Neg(\rho_{0\br})\ge\kappa_\eps\eps^4$. Indeed Eq.~\eqref{eq:th-ms}
gives $\Neg/\eps^4\ge m-C_cE_\eps$, which exceeds $\kappa_\eps$ once
$C_cE_\eps+\kappa_\eps<m$. Only $\kappa_\eps\to0$ is used here.

(B) Let $\eta>0$. For all sufficiently small $\eps$, uniformly in $h$ and
$\omega$, every $\br$ with $|\eps\br|\ge\eta$ and
$|\eps\br|\ge\bar\rho_E(h,\arg\br)+\eta$ satisfies
$\Neg(\rho_{0\br})<\kappa_\eps\eps^4$. Indeed
Lemma~\ref{lem:th-profile}(d) gives $n_*-|A_h(\eps\br)|\ge\delta_\eta$,
and Proposition~\ref{prop:th-exterior}(iii) with
$(c,\delta)=(\eta,\delta_\eta)$ gives
\begin{equation*}
 \Neg(\rho_{0\br})\le C_{\eta,\delta_\eta}\eps^{16/3}\ell_\eps^2
 =C_{\eta,\delta_\eta}\,\frac{\eps^{4/3}\ell_\eps^2}{\kappa_\eps}\,
 \kappa_\eps\eps^4<\kappa_\eps\eps^4
\end{equation*}
once $\kappa_\eps/(\eps^{4/3}\ell_\eps^2)>C_{\eta,\delta_\eta}$. This
is the only place where Eq.~\eqref{eq:th-window} enters, and it covers
the whole unbounded tail.

\emph{Rays.} Fix $\mathbf a$, $\theta=\arg\mathbf a$,
$\rho_-=\min_{h\in\cI}\rho_E(h,\theta)>0$ and
$0<\eta<\rho_-/4$. If $\eps k|\mathbf a|\ge\rho_E(h,\theta)+\eta$, then
(B) applies, since $\bar\rho_E=\rho_E$ on this ray. Hence the qualifying
set is finite and $\eps R^{\mathbf a}_{\kappa_\eps}<\rho_E(h,\theta)
+\eta$. For the lower bound, the set
$\{(h,\rho):h\in\cI,\ \rho_E(h,\theta)-2\eta\le\rho\le
\rho_E(h,\theta)-\eta\}$ is compact, and by
Lemma~\ref{lem:th-profile}(a) the continuous function
$|A_h(\rho,\theta)|-n_*$ is positive on it; let $m_\eta$ be its minimum.
Put $k_h=\lfloor(\rho_E(h,\theta)-\eta)/(\eps|\mathbf a|)\rfloor$. For
$\eps|\mathbf a|\le\eta$ the scaled radius $\eps k_h|\mathbf a|$ lies in
$[\rho_E-2\eta,\rho_E-\eta]$ and is at least $\rho_-/2$. By (A) with
$(c,m)=(\rho_-/2,m_\eta)$, $k_h$ qualifies. Therefore
$\rho_E-2\eta\le\eps R^{\mathbf a}_{\kappa_\eps}<\rho_E+\eta$ for all
small $\eps$, uniformly, and $\eta$ is arbitrary.

\emph{All directions.} If $|\eps\br|\ge\rho_E(h,0)+\eta$, then
$|\eps\br|\ge\bar\rho_E(h,\arg\br)+\eta$ by
Lemma~\ref{lem:th-profile}(b), and (B) excludes $\br$. Hence
$\eps R^{\rm all}_{\kappa_\eps}<\rho_E(h,0)+\eta$. Since
$R^{\rm all}_{\kappa_\eps}\ge R^{(1,0)}_{\kappa_\eps}$, the axial ray
gives the lower bound.

\emph{Planar set, outer inclusion.} Let $\eta>0$ and
$\eps\br\in\mathcal S_{\eps,h,\omega}$. If
$\operatorname{dist}(\eps\br,\mathcal E_h)\ge\eta$, then
$|\eps\br|\ge\eta$ because $0\in\mathcal E_h$, and
$|\eps\br|\ge\bar\rho_E(h,\arg\br)+\eta$ by
Lemma~\ref{lem:th-profile}(d). This contradicts (B). Hence, for small
$\eps$ and uniformly, every point of $\mathcal S_{\eps,h,\omega}$ lies
within $\eta$ of $\mathcal E_h$.

\emph{Planar set, inner inclusion.} Let $B$ be as in
Lemma~\ref{lem:th-profile}(b), and fix $t\in(0,1/2)$ with $tB<\eta/4$.
First let $\bx\in\mathcal E_h$ with $|\bx|\ge\eta/3$, and put
$\mathbf y=(1-t)\bx$. The direction $\theta=\arg\bx$ is nonnodal,
because $\bar\rho_E(h,\theta)\ge|\bx|>0$, and
$|\mathbf y|<|\bx|\le\rho_E(h,\theta)$, so $|A_h(\mathbf y)|>n_*$. The
set $\{(h,\bx)\in\mathcal E:|\bx|\ge\eta/3\}$ is compact by
Lemma~\ref{lem:th-profile}(c), and $(h,\bx)\mapsto
|A_h((1-t)\bx)|-n_*$ is continuous and positive on it; let $2m_\eta$ be
its minimum. The map $(h,\mathbf z)\mapsto A_h(\mathbf z)$ is uniformly
continuous on $\cI\times\{\eta/12\le|\mathbf z|\le B+1\}$. Round both
coordinates of $\mathbf y$ simultaneously to a nearest point
$\eps\br\in\eps\mathbb Z^2$, so $|\eps\br-\mathbf y|\le\eps/\sqrt2$.
Since $|\mathbf y|\ge\eta/6$, for small $\eps$ one has
$|\eps\br|\ge\eta/12$ and $|A_h(\eps\br)-A_h(\mathbf y)|\le m_\eta$,
hence $|A_h(\eps\br)|\ge n_*+m_\eta$. By (A) with
$(c,m)=(\eta/12,m_\eta)$, $\eps\br\in\mathcal S_{\eps,h,\omega}$, and
$|\eps\br-\bx|\le\eps/\sqrt2+tB<\eta/2$.

Next let $\bx\in\mathcal E_h$ with $|\bx|<\eta/3$. Fix one axial point
$\mathbf y_0=(s_0,0)$ with $0<s_0<\min\{\eta/6,\rho_{\min}/2\}$, where
$\rho_{\min}=\min_{h\in\cI}\rho_E(h,0)>0$; the same point serves every
$h\in\cI$. Let $\br_0=(\lfloor s_0/\eps\rfloor,0)$. For small $\eps$,
$s_0/2\le|\eps\br_0|\le s_0$, and radial monotonicity gives
$|A_h(\eps\br_0)|\ge|A_h(\mathbf y_0)|\ge n_*+m_0$, with
$m_0=\min_h|A_h(\mathbf y_0)|-n_*>0$. By (A),
$\eps\br_0\in\mathcal S_{\eps,h,\omega}$, and
$|\bx-\eps\br_0|\le|\bx|+s_0<\eta/2$. This includes $\bx=0$.

Thus every point of $\mathcal E_h$ lies within $\eta$ of
$\mathcal S_{\eps,h,\omega}$. With the outer inclusion,
$d_H(\mathcal S_{\eps,h,\omega},\mathcal E_h)\le\eta$ for all
sufficiently small $\eps$, uniformly in $h$ and $\omega$.
\hfill$\square$

The inner inclusion samples the plane in two dimensions: it rounds both
coordinates of a contracted interior point and does not assemble the
set from ray limits. The neighborhood of the origin is reached through
a single axial point because Eq.~\eqref{eq:th-ms} holds only on
$|\eps\br|\ge c$.

\subsection{First-order displacement of the boundary}
\label{ss:th-shift}

\begin{proposition}[Displacement of the threshold boundary]
\label{prop:th-shift}
Let $\kappa_\eps\to0$ with $\kappa_\eps/\varpi_\eps\to\infty$. For every
primitive $\mathbf a$ with nonnodal $\theta=\arg\mathbf a$, uniformly in
$h$ and $\omega$,
\begin{equation}
 \eps R^{\mathbf a}_{\kappa_\eps}
 =\rho_E(h,\theta)-\frac{\kappa_\eps}{D_{h,\theta}}
 +O_{\cI,\mathbf a}\bigl(\kappa_\eps^2+\varpi_\eps+\eps\bigr)
 =\rho_E(h,\theta)-\frac{\kappa_\eps}{D_{h,\theta}}+o(\kappa_\eps).
 \label{eq:th-shift}
\end{equation}
The same holds for $\eps R^{\rm all}_{\kappa_\eps}$ with $\theta=0$.
\end{proposition}

Since $\varpi_\eps/(\eps^{4/3}\ell_\eps^2)\to\infty$, the hypothesis implies
Eq.~\eqref{eq:th-window}. It holds for every fixed inverse-logarithmic
threshold $\kappa_\eps=\ell_\eps^{-\gamma}$, $\gamma>0$, and for
$\kappa_\eps=\eps^\gamma$ exactly when $0<\gamma<2/13$.
Theorem~\ref{thm:threshold} covers the wider range $\gamma<4/3$ because
Eq.~\eqref{eq:th-negext} bounds the exterior negativity far below the
profile error. Near the boundary, where $|A_h|-n_*$ is small, only
Eq.~\eqref{eq:th-ms} is available, and the displacement must exceed the
profile error.

\paragraph{Proof.}
Fix $\mathbf a$ and $\theta$. Let
$c=\frac12\min_{h\in\cI}\rho_{n_*}(h,\theta)>0$ and $E=C_cE_\eps$ with
$C_c$ from Eq.~\eqref{eq:th-ms}. For small $\eps$, $E<\kappa_\eps$ and
$\kappa_\eps+E\le n_*$. We first show
\begin{equation}
 \rho_{\kappa_\eps+E}(h,\theta)-\eps|\mathbf a|
 <\eps R^{\mathbf a}_{\kappa_\eps}
 \le\rho_{\kappa_\eps-E}(h,\theta).
 \label{eq:th-bracket}
\end{equation}
Let $k$ qualify. If $\eps k|\mathbf a|<c$, then
$\eps k|\mathbf a|<\rho_{\kappa_\eps-E}$, because $\rho_s$ decreases in
$s$ and $\rho_{n_*}\ge2c$. Otherwise Eq.~\eqref{eq:th-ms} gives
$(|A_h(\eps k\mathbf a)|-n_*)_+\ge\kappa_\eps-E>0$, so
$|A_h(\eps k\mathbf a)|\ge n_*+\kappa_\eps-E$ and
$\eps k|\mathbf a|\le\rho_{\kappa_\eps-E}$ by
Lemma~\ref{lem:th-profile}(a). For the lower bound let
$k_0=\lfloor\rho_{\kappa_\eps+E}/(\eps|\mathbf a|)\rfloor$. Then
$\eps k_0|\mathbf a|\in(\rho_{\kappa_\eps+E}-\eps|\mathbf a|,
\rho_{\kappa_\eps+E}]$, which exceeds $c$ for small $\eps$. Monotonicity
gives $|A_h(\eps k_0\mathbf a)|\ge n_*+\kappa_\eps+E$, and
Eq.~\eqref{eq:th-ms} gives $\Neg/\eps^4\ge\kappa_\eps$, so $k_0$
qualifies. No monotonicity of the finite-angle negativity is used.

By Lemma~\ref{lem:th-profile}(a), $s\mapsto\rho_s(h,\theta)$ is smooth
on $[-n_*/2,n_*]$ with $\partial_s\rho_s|_{s=0}=-1/D_{h,\theta}$; its
second derivative is bounded on $\cI\times[-n_*/2,n_*]$ by continuity,
and $D_{h,\theta}$ is bounded below on $\cI$. Hence
$\rho_s=\rho_E-s/D_{h,\theta}+O(s^2)$ uniformly in $h$. Inserting
$s=\kappa_\eps\pm E$ into Eq.~\eqref{eq:th-bracket} gives
Eq.~\eqref{eq:th-shift} with error
$O(E+(\kappa_\eps+E)^2+\eps|\mathbf a|)=O(\kappa_\eps^2+\varpi_\eps+\eps)$.
Finally $\kappa_\eps^2=o(\kappa_\eps)$, $\varpi_\eps=o(\kappa_\eps)$ and
$\eps=o(\varpi_\eps)$. For $R^{\rm all}_{\kappa_\eps}$ take
$\mathbf a=(1,0)$, $\theta=0$. A qualifying $\br$ with $|\eps\br|<c$
lies below $\rho_{\kappa_\eps-E}(h,0)$ as before, and one with
$|\eps\br|\ge c$ satisfies
$|A_h(|\eps\br|,0)|\ge|A_h(\eps\br)|\ge n_*+\kappa_\eps-E$, so
$|\eps\br|\le\rho_{\kappa_\eps-E}(h,0)$; the lower bound is the axial
one.\hfill$\square$

Table~\ref{tab:th-radii} lists the limit radii at $h=2$ on the axis and
on the rays of slopes $1/2$ and $2/3$. The column $\kappa=0$ gives the
limits in Theorem~\ref{thm:threshold}; the axial value agrees with
Eq.~\eqref{eq:boundary}. The last column is the inward displacement of
the boundary per unit threshold in Eq.~\eqref{eq:th-shift}.

\begin{table}[htb]
\caption{Limit radii and boundary slopes at $h=2$ ($m_2=\sqrt3$,
$n_*=0.010750$) on three lattice rays $\mathbf a$. The radius
$\rho_\kappa(2,\theta)$ solves the limit equation
$|A_2(\rho,\theta)|=n_*+\kappa$, Eq.~\eqref{eq:th-rhokappa}. The column
$\kappa=0$ is the zero line $\rho_E$ of the leading profile, the limit
of $\eps R^{\mathbf a}_{\kappa_\eps}$ in Theorem~\ref{thm:threshold};
a fixed $\kappa>0$ gives the limit radius of
Corollary~\ref{cor:ms-range}. $D_{2,\theta}$ is defined in
Eq.~\eqref{eq:th-D}. All entries are solutions of the limit equations,
not finite-angle data.}
\label{tab:th-radii}
\begin{ruledtabular}
\begin{tabular}{lccccccc}
 & & \multicolumn{4}{c}{$\rho_\kappa(2,\theta)$} & & \\
\cline{3-6}
Ray & $|\cos2\theta|$ & $\kappa=0$ & $\kappa=0.001$ & $\kappa=0.005$
 & $\kappa=n_*$ & $D_{2,\theta}$ & $1/D_{2,\theta}$\\
\hline
Axis, $\mathbf a=(1,0)$ & $1$ & 1.412822 & 1.376850 & 1.261104
 & 1.143119 & 0.026444 & 37.815\\
Slope $1/2$, $\mathbf a=(2,1)$ & $3/5$ & 1.211595 & 1.177954
 & 1.070348 & 0.961780 & 0.028240 & 35.410\\
Slope $2/3$, $\mathbf a=(3,2)$ & $5/13$ & 1.047969 & 1.016662
 & 0.917134 & 0.817765 & 0.030304 & 32.999\\
\end{tabular}
\end{ruledtabular}
\end{table}

\subsection{What is not established}
\label{ss:th-scope}

The theorem does not give the farthest distance of nonzero
entanglement. Let $R^{\mathbf a}_{\rm NPT}$ and $R^{\rm all}_{\rm NPT}$
be defined as in Eq.~\eqref{eq:th-Ra} with the condition
$\Neg(\rho_{0\br})>0$. Since $\Neg\ge\kappa_\eps\eps^4>0$ implies NPT,
$R^{\mathbf a}_{\rm NPT}\ge R^{\mathbf a}_{\kappa_\eps}$, and
Theorem~\ref{thm:threshold} yields only
\begin{equation}
 \liminf_{\eps\downarrow0}\eps R^{\mathbf a}_{\rm NPT}
 \ge\rho_E(h,\arg\mathbf a),\qquad
 \liminf_{\eps\downarrow0}\eps R^{\rm all}_{\rm NPT}\ge\rho_E(h,0).
 \label{eq:th-liminf}
\end{equation}
No upper bound follows. Proposition~\ref{prop:th-exterior}(iii) permits
negativity up to $C\eps^{16/3}\ell_\eps^2$ throughout the exterior,
including its unbounded tail, so entanglement below every admissible
threshold at arbitrarily large distance is not excluded; even
finiteness of $R_{\rm NPT}$ at fixed small angle is not established.
Nor is separability proved for any exterior region. For two qubits PPT
is equivalent to separability~\cite{Horodecki}, but in the exterior the
sign of the smallest partial-transpose eigenvalue is decided at order
$\eps^8$ by Eq.~\eqref{eq:exterior}, which requires $\zeta=o(\eps^6)$
(Theorem~\ref{thm:rdmcriterion}); Eq.~\eqref{eq:th-zeta} gives
$\eps^{14/3}\ell_\eps$. Finally, nothing is asserted about the exact
sign of the partial-transpose spectrum on or near the nodal diagonals.
There the profile vanishes, Eq.~\eqref{eq:th-ray} excludes nodal
directions, and Eq.~\eqref{eq:th-hausdorff} states only that the
superlevel set contracts onto the nodal cusps of $\mathcal E_h$.

\section{Exact separability: partial results and open problems}
\label{sm:exterior}

Theorem~\ref{thm:massscale} gives the leading negativity profile
$\eps^{-4}\Neg(\rho_{0\br})\to(|A_h(\eps\br)|-n_*)_+$ uniformly on
$|\eps\br|\ge c$, and Theorem~\ref{thm:threshold} locates the
farthest pairs whose negativity exceeds a slowly vanishing threshold.
Neither result decides exact separability outside the four-lobed region
\begin{equation}
 \mathcal O_h=\{\bx\in\mathbb R^2\setminus\{0\}:|A_h(\bx)|>n_*\}
 =\{\rho<\rho_E(h,\theta)\},
 \label{eq:ex-lobes}
\end{equation}
with $\rho_E(h,\theta)$ defined by Eq.~\eqref{eq:boundary}. There the
sign of the smallest partial-transpose eigenvalue is decided at order
$\eps^8$ (Sec.~\ref{ss:massive-rdm}), while the available estimates of
the two-site data are of order $\eps^4$ times a vanishing factor. This
section collects what is proved about exact
positive partial transposition (PPT) outside $\mathcal O_h$, at the
nodes, and in the far tail, and it states precisely what remains open.
Section~\ref{ss:ex-schur} gives the exact criterion with odd coherences
and shows that leading-order data cannot decide it.
Sections~\ref{ss:ex-axial} and \ref{ss:ex-nodes} treat the axes and
the diagonals. Section~\ref{ss:ex-variance} reduces the general case to
a bound on a centered pair variance, and Sec.~\ref{ss:ex-open} lists
the open problems.

\subsection{Setting and inputs}
\label{ss:ex-inputs}

The states are the translation- and jointly-fourfold-invariant
thermodynamic ground states of Sec.~\ref{sm:conventions}, with
$h\in\cI\Subset(3/2,\infty)$ and the limit order stated there. The
two-site reduced density matrix (RDM) has the form~\eqref{eq:rho}; we
write $W=|w|$, $a=1-2p+q$, $b=p-q$, and use $\Neg$ for the sum of the
moduli of the negative eigenvalues of $\rho^{T_2}$. Put
\begin{equation}
 \varpi_\eps=(\eps\ell_\eps)^{2/13},\qquad
 R_E(h)=\rho_E(h,0)=\max_\theta\rho_E(h,\theta),\qquad
 m_+=\max_{h\in\cI}m_h .
 \label{eq:ex-notation}
\end{equation}
The maximum is attained on the axes because $K_2$ decreases and
$|\cos2\theta|\le1$ in Eq.~\eqref{eq:boundary}. For $c>0$ and
$0<\delta<n_*$ the \emph{fixed-margin exterior} is
\begin{equation}
 \mathcal X_{c,\delta}=\{\br\in\mathbb Z^2:\ |\eps\br|\ge c,\
 |A_h(\eps\br)|\le n_*-\delta\}.
 \label{eq:ex-margin-set}
\end{equation}
It depends on $\eps$ and $h$. It contains the macroscopic diagonals,
and for every $\eta>0$ it contains the unbounded region
$\eps|\br|\ge R_E(h)+\eta$ for suitable $c,\delta>0$
(Lemma~\ref{lem:ex-far}).

We use the following estimates from
Theorems~\ref{thm:population} and \ref{thm:massscale} and from
Proposition~\ref{prop:th-exterior}. They hold uniformly in $h\in\cI$,
in the states, and in $\br\in\mathcal X_{c,\delta}$ for every fixed
$c,\delta$, as $\eps\downarrow0$:
\begin{align}
 &\bigl|\eps^{-4}w(\br)-A_h(\eps\br)\bigr|\le C_cE_\eps,\qquad
 \frac p{\eps^4}\to n_*,\qquad q(\br)=o(\eps^4),\qquad
 v(\br)=o(\eps^4),\label{eq:ex-input-profile}\\
 &|\zeta(\br)|\le C_{c}\,\eps^{14/3}\ell_\eps .
 \label{eq:ex-input-odd}
\end{align}
We also use a rate for the normal coherence.

\begin{lemma}[Normal-coherence rate]\label{lem:ex-normal}
For every $c>0$, uniformly in $h$, the states and $|\eps\br|\ge c$,
\begin{equation}
 |v(\br)|\le C_c\,\eps^{13/3}\ell_\eps^{1/2}.
 \label{eq:ex-input-normal}
\end{equation}
\end{lemma}

\begin{proof}
The proof of Proposition~\ref{prop:th-exterior} uses an auxiliary
twice-dressed state with population $\widehat p\le C\eps^{14/3}\ell_\eps$,
Eq.~\eqref{eq:th-density}, and triplet budget
$|\tau|y\le C\eps^{7/3}\ell_\eps^{-1}$, which follows from
Eq.~\eqref{eq:th-kernel}. By
Eq.~\eqref{eq:pop-triplet-local-transfer}, the pair-dressed state
$\widetilde\omega(A)=\omega(U_RAU_R^\dagger)$ of Sec.~\ref{sm:population},
$R=\lceil\eps^{-2}\rceil$, therefore has population
$\sqrt{\widetilde p}\le\sqrt{\widehat p}+2|\tau|y\le
C\eps^{7/3}\ell_\eps^{1/2}$.
Write $\beta_j=U_R^\dagger b_jU_R=b_j+dC_j+\mathcal F_j$ as in
Eq.~\eqref{eq:pop-pair-local-transfer}, translated to site $j$, so that
$C_j=(1-2n_j)\sum_kF_R(k-j)b_k^\dagger$ and
$\|\mathcal F_j\|\le4d^2a_R^2$. In the GNS representation of
$\widetilde\omega$ with cyclic vector $\widetilde\Omega$,
$v(\br)=\langle\beta_0\widetilde\Omega,\beta_\br\widetilde\Omega\rangle$,
$\|b_j\widetilde\Omega\|^2=\widetilde p$, and
$\|C_j\widetilde\Omega\|^2=q_C\le S_R+\widetilde pa_R^2$ by
Eq.~\eqref{eq:pop-C0-bound}. Expanding the inner product,
\[
 |v(\br)|\le\widetilde p+2d\sqrt{\widetilde pq_C}
 +d^2|\widetilde\omega(C_0^\dagger C_\br)|
 +8d^2a_R^2\bigl(\sqrt{\widetilde p}+d\sqrt{q_C}\bigr)+16d^4a_R^4 .
\]
The vacuum expectation of $C_0^\dagger C_\br$ is
$C_R(\br)=\sum_kF_R(k)F_R(k-\br)$; the remainder has zero vacuum
diagonal, local terms on at most four sites, and coefficient sum
$a_R^2$, so Eq.~\eqref{eq:pop-vacuum-diagonal} bounds it by
$Ca_R^2\sqrt{\widetilde p}$. Since $|F(k)|\le C(1+|k|)^{-2}$ by
Eq.~\eqref{eq:green-tail}, splitting the sum at $|k|=|\br|/2$ gives
$|C_R(\br)|\le Ca_R|\br|^{-2}\le C_c\eps^2\ell_\eps$. With $d\le\eps^2/2$, $a_R=O(\ell_\eps)$ and
$S_R=O(1)$, the second term, $O(\eps^{13/3}\ell_\eps^{1/2})$, dominates.
\end{proof}

All constants below are independent of the state, of the displacement,
and of $h\in\cI$; they may depend on $\cI$ and on the fixed parameters
indicated by subscripts.

\subsection{Complete partial transpose with odd coherences}
\label{ss:ex-schur}

Section~\ref{ss:massive-rdm} reduced positivity of $\rho^{T_2}$ to the
$2\times2$ Schur complement \eqref{eq:Schur}. We use its determinant
\begin{equation}
 D_{\rm PT}=aq-v^2-\chi(a+q-2v),\qquad
 \chi=\frac{2[b|\zeta|^2+\Ree(w\zeta^2)]}{b^2-W^2},
 \label{eq:ex-DPT}
\end{equation}
and, when $b>v$, the corresponding cost already paid by positivity of
the state itself,
\begin{equation}
 \kappa=\frac{2|\zeta|^2}{b-v},\qquad
 E_{\rm ph}=aq-W^2-\frac{2|\zeta|^2(a+q)-4\Ree(w\zeta^2)}{b-v}.
 \label{eq:ex-kappa}
\end{equation}
Only the phase invariant $w\zeta^2$ enters. It is unchanged by the
common one-site rotation $\diag(1,e^{i\alpha})^{\otimes2}$, which maps
$w\mapsto e^{2i\alpha}w$, $\zeta\mapsto e^{-i\alpha}\zeta$, fixes
$a,b,q,v$, and preserves the spectra of $\rho$ and $\rho^{T_2}$.

\begin{lemma}[Complete partial transpose]\label{lem:ex-schur}
Let $\rho$ be a density matrix of the form~\eqref{eq:rho} with $b>W$.
\begin{enumerate}
\item[(a)] $\det\rho^{T_2}=(b^2-W^2)D_{\rm PT}$. If $a>\chi$, then
$\rho^{T_2}\succeq0$ if and only if $D_{\rm PT}\ge0$, and
$\rho^{T_2}\succ0$ if and only if $D_{\rm PT}>0$. If $a>\chi$ and
$D_{\rm PT}<0$, exactly one eigenvalue of $\rho^{T_2}$ is negative.
\item[(b)] If $b>v$, then $a\ge\kappa$, $q\ge\kappa$,
$E_{\rm ph}\ge0$, and
\begin{align}
 D_{\rm PT}&=E_{\rm ph}+(W^2-v^2)\Bigl(1+\frac{2\chi}{b-v}\Bigr)
 -(\chi-\kappa),\label{eq:ex-identity}\\
 (b^2-W^2)(b-v)(\chi-\kappa)
 &=2\bigl[(b-v)\Ree(w\zeta^2)+(W^2-bv)|\zeta|^2\bigr].
 \label{eq:ex-phase-cost}
\end{align}
\item[(c)] If $b>W>|v|$ and
\begin{equation}
 (b-v)\Ree(w\zeta^2)\le(bv-W^2)|\zeta|^2,
 \label{eq:ex-phase-condition}
\end{equation}
then $\rho^{T_2}\succ0$.
\item[(d)] If $b>v$ and $a>\chi$, then
\begin{equation}
 \Neg(\rho)\le(\chi-\kappa)_++\frac{(v-\chi)^2}{a-\chi}.
 \label{eq:ex-shared-cost}
\end{equation}
\end{enumerate}
\end{lemma}

The quantity $\chi-\kappa$ is the part of the Schur cost of the odd
coherence that positivity of $\rho$ does not already pay; by
Eq.~\eqref{eq:ex-phase-cost} it is fixed by the phase of $w\zeta^2$. We
call it the phase cost. Condition~\eqref{eq:ex-phase-condition} states
$\chi\le\kappa$. Equation~\eqref{eq:ex-identity} uses the trace
condition $a+2b+q=1$ and discards no phase.

\begin{proof}
(a) In the order $(00,11;01,10)$ of Eq.~\eqref{eq:PT},
$\rho^{T_2}$ has diagonal blocks $A_0=\bigl(\begin{smallmatrix}a&v\\
v&q\end{smallmatrix}\bigr)$ and $M\succ0$, and lower-left block
$u\mathbf 1^T$ with $\mathbf 1=(1,1)^T$. With $\mathsf L$ the block
lower-triangular matrix having identity diagonal blocks and lower-left
block $M^{-1}u\mathbf 1^T$,
\begin{equation}
 \rho^{T_2}=\mathsf L^\dagger(\mathsf S\oplus M)\mathsf L,\qquad
 \mathsf S=A_0-\chi\mathbf 1\mathbf 1^T
 =\begin{pmatrix}a-\chi&v-\chi\\v-\chi&q-\chi\end{pmatrix},
 \label{eq:ex-congruence}
\end{equation}
because $\mathbf 1u^\dagger M^{-1}u\mathbf 1^T=\chi\mathbf 1\mathbf 1^T$.
Hence $\det\rho^{T_2}=\det M\det\mathsf S$, and $\det\mathsf S=D_{\rm PT}$.
By Sylvester's law of inertia the inertia of $\rho^{T_2}$ is that of
$\mathsf S$ plus two positive eigenvalues. A $2\times2$ Hermitian
matrix with positive first diagonal entry is positive definite,
positive semidefinite with one zero eigenvalue, or indefinite according
to the sign of its determinant.

(b) The vector $(|01\rangle+|10\rangle)/\sqrt2$ is an eigenvector of
$\rho$ with eigenvalue $b+v$. In the orthonormal basis
$|00\rangle,|11\rangle,|t\rangle=(|01\rangle-|10\rangle)/\sqrt2$ of its
complement,
\begin{equation}
 \rho\simeq\begin{pmatrix}a&w^*&\sqrt2\zeta\\
 w&q&\sqrt2\zeta^*\\ \sqrt2\zeta^*&\sqrt2\zeta&b-v\end{pmatrix},
 \qquad
 \mathsf S_{\rm ph}=\begin{pmatrix}a-\kappa&w^*-2\zeta^2/(b-v)\\
 w-2\zeta^{*2}/(b-v)&q-\kappa\end{pmatrix}\succeq0 .
 \label{eq:ex-physical-block}
\end{equation}
The second matrix is the Schur complement of the positive entry
$b-v$. Its diagonal gives $a,q\ge\kappa$, and
$\det\mathsf S_{\rm ph}=E_{\rm ph}\ge0$. Equation~\eqref{eq:ex-phase-cost}
follows from $(b^2-W^2)\chi=2[b|\zeta|^2+\Ree(w\zeta^2)]$ and
$(b-v)\kappa=2|\zeta|^2$. For Eq.~\eqref{eq:ex-identity}, subtract,
substitute $a+q=1-2b$, and use
$2b\kappa+4\Ree(w\zeta^2)/(b-v)=2(b^2-W^2)\chi/(b-v)$:
\[
 D_{\rm PT}-E_{\rm ph}
 =W^2-v^2-(\chi-\kappa)
 +2\chi\Bigl[b+v-\frac{b^2-W^2}{b-v}\Bigr]
 =W^2-v^2-(\chi-\kappa)+\frac{2\chi(W^2-v^2)}{b-v}.
\]

(c) Here $b>v$, and Eq.~\eqref{eq:ex-phase-cost} shows that
\eqref{eq:ex-phase-condition} is equivalent to $\chi\le\kappa$. Since
$\chi\ge0$ and $E_{\rm ph}\ge0$, Eq.~\eqref{eq:ex-identity} gives
$D_{\rm PT}\ge W^2-v^2>0$. Part (b) gives $a\ge\kappa\ge\chi$, and
$a=\chi$ would force $D_{\rm PT}=-(v-\chi)^2\le0$. Thus $a>\chi$, and
(a) applies.

(d) By Eq.~\eqref{eq:ex-congruence},
$\rho^{T_2}-(\mathsf S\oplus0)=K^\dagger K$ with the $2\times4$ matrix
$K=(M^{-1/2}u\mathbf 1^T,\,M^{1/2})$, and hence it is positive
semidefinite. Moreover
$\mathsf S-\diag(0,\sigma)=(a-\chi)^{-1}\xi\xi^T\succeq0$ with
$\xi=(a-\chi,v-\chi)^T$ and $\sigma=q-\chi-(v-\chi)^2/(a-\chi)$. Thus
$\rho^{T_2}$ dominates $\sigma$ times a rank-one projector. The sum of
the negative eigenvalues of a Hermitian matrix $X$ equals
$\min_{0\preceq P\preceq I}\Tr(PX)$ and is therefore monotone under
$\succeq$. Hence $\Neg(\rho)\le(-\sigma)_+$, and $q\ge\kappa$ from (b)
gives Eq.~\eqref{eq:ex-shared-cost}.
\end{proof}

The axial phase $w\zeta^2=-W|\zeta|^2$ satisfies
\eqref{eq:ex-phase-condition} whenever $b>W>|v|$: the condition then
reads $(W-b)(W+v)\le0$. Section~\ref{ss:ex-axial} sharpens this to an
exact formula.

\begin{lemma}[Anomalous block in the far exterior]\label{lem:ex-far}
For every $\eta>0$ there are $c_\eta>0$ and $\eps_\eta>0$ such that,
for $0<\eps<\eps_\eta$, all $h\in\cI$, all states, and all $\br$ with
$\eps|\br|\ge R_E(h)+\eta$,
\begin{equation}
 b-W\ge c_\eta\eps^4,\qquad b-v\ge\tfrac12n_*\eps^4,\qquad
 \chi\le C_\eta\eps^{16/3}\ell_\eps^2,\qquad a-\chi\ge\tfrac12 .
 \label{eq:ex-far}
\end{equation}
Consequently, throughout this unbounded region, $\rho_{0\br}$ is PPT if
and only if $D_{\rm PT}\ge0$, strictly PPT if and only if
$D_{\rm PT}>0$, and $\rho_{0\br}^{T_2}$ has at most one negative
eigenvalue.
\end{lemma}

\begin{proof}
Since $|A_h(\rho,\theta)|\le A_h(\rho,0)$, $K_2$ is strictly
decreasing, and $\rho_E(h,0)$ depends continuously on $h$ by the
implicit-function theorem, compactness of $\cI$ gives $2c_\eta>0$ with
$|A_h(\eps\br)|\le n_*-2c_\eta$ in the region. It is therefore
contained in $\mathcal X_{R_E+\eta,2c_\eta}$. By
Eq.~\eqref{eq:ex-input-profile}, $W\le(n_*-2c_\eta+C \varpi_\eps)\eps^4$
and $b=(n_*+o(1))\eps^4$, while $|v|=o(\eps^4)$. Since
$\chi=u^\dagger M^{-1}u\le\|u\|^2/(b-W)=2|\zeta|^2/(b-W)$,
Eq.~\eqref{eq:ex-input-odd} bounds $\chi$, and $a\to1$. Lemma
\ref{lem:ex-schur}(a) gives the remaining statements.
\end{proof}

Exact separability in the far exterior is therefore equivalent to the
sign of a single scalar, $D_{\rm PT}$. The decisive scale is $\eps^8$:
positivity of the state gives $aq\ge|w|^2=O(\eps^8)$, while the
subtracted term $\chi(a+q-2v)$ is only known to be
$O(\eps^{16/3}\ell_\eps^2)$. The following algebraic examples show that
the leading-order data do not decide the sign. They are two-qubit density
matrices of the form~\eqref{eq:rho}, not ground states of the model.

\begin{proposition}[Leading-order data do not decide PPT]
\label{prop:ex-counter}
Let $0<t\le10^{-2}$.
\begin{enumerate}
\item[(i)] The matrix $\rho_{\rm I}(t)$ of the form~\eqref{eq:rho}
with $p=t$, $q=3t^2$, $w=t/2$, $v=0$, $\zeta=t^{3/2}$ is positive
definite, with $\lambda_{\min}(\rho_{\rm I})=\frac34t^2+O(t^3)$, and
\begin{equation}
 \det\rho_{\rm I}^{T_2}
 =\tfrac34t^4(2t-1)(1+16t-21t^2+54t^3)<0,\qquad
 \lambda_{\min}(\rho_{\rm I}^{T_2})=-t^2-16t^3+O(t^4).
 \label{eq:ex-counter-I}
\end{equation}
\item[(ii)] The matrix $\rho_{\rm II}(t)$ with $p=t$,
$w=q=\zeta=0$, $v=t^{3/2}$ is positive semidefinite, with eigenvalues
$1-2t$, $t\pm t^{3/2}$, $0$, and
\begin{equation}
 \lambda_{\min}(\rho_{\rm II}^{T_2})
 =\tfrac12\bigl[1-2t-\sqrt{(1-2t)^2+4t^3}\bigr]=-t^3+O(t^4).
 \label{eq:ex-counter-II}
\end{equation}
Replacing $q=0$ by $q=t^4$ gives a positive-definite matrix with
smallest eigenvalue $t^4$ and the same conclusion.
\end{enumerate}
\end{proposition}

With $t=n_*\eps^4$, family (i) has $p/\eps^4=n_*$,
$w/\eps^4=n_*/2$, $q,v=o(\eps^4)$ and
$\zeta=n_*^{3/2}\eps^6$. Its negativity
$n_*^2\eps^8+O(\eps^{12})$ is $o(\eps^4)$, so the leading negativity
$(|A|-n_*)_+=0$ is reproduced, yet the matrix is entangled for every
$\eps$. The hypothesis $\zeta=o(\eps^6)$ of
Theorem~\ref{thm:rdmcriterion} is therefore sharp: $\zeta=O(\eps^6)$
does not suffice. Family (ii) has the data of a node, $w=0$ with
$p/\eps^4=n_*$, $q=\zeta=0$ and $v=o(\eps^4)$, yet it is entangled
with $\Neg=n_*^3\eps^{12}+O(\eps^{16})$. Thus a vanishing leading
amplitude does not imply separability, and the strictly positive lower
bound on $|A_h|$ in Theorem~\ref{thm:rdmcriterion} is substantive.
Table~\ref{tab:ex-counter} lists the computed extreme eigenvalues.

\begin{proof}
(i) The trace is one, the populations are equal, and the one-site
transverse means vanish. The vector $(|01\rangle+|10\rangle)/\sqrt2$
has eigenvalue $b=t-3t^2>0$. In the order $|00\rangle,|t\rangle,
|11\rangle$ the leading principal minors of the complementary block
\eqref{eq:ex-physical-block} are
$1-2t+3t^2$, $t(1-5t+7t^2-9t^3)$ and
$\frac34t^3(1-11t+20t^2-36t^3)$, all positive for $0<t\le10^{-2}$, so
Sylvester's criterion gives $\rho_{\rm I}\succ0$. The smallest
eigenvalue of this block is at most $\langle11|\rho_{\rm I}|11\rangle=3t^2$,
so the product of the other two equals the sum of the principal
$2\times2$ minors up to $O(t^2)$, that is $t(1+O(t))$; dividing the
determinant by it gives $\lambda_{\min}(\rho_{\rm I})$. The determinant in
Eq.~\eqref{eq:ex-counter-I} follows from Lemma~\ref{lem:ex-schur}(a)
or by direct expansion. In the order of Eq.~\eqref{eq:PT}, the
diagonal blocks of $\rho_{\rm I}^{T_2}$ have eigenvalues $1+O(t)$,
$3t^2$, $\frac t2-3t^2$ and $\frac{3t}2-3t^2$, and the off-diagonal
entries are $O(t^{3/2})$. By Weyl's inequality three eigenvalues are
$1+O(t)$, $\frac t2+O(t^{3/2})$ and $\frac{3t}2+O(t^{3/2})$. Dividing
the determinant by their product gives
$\lambda_{\min}=-t^2(1+O(t^{1/2}))$. The eigenvalue is simple, so the
implicit-function theorem applied to the characteristic polynomial
gives the expansion in Eq.~\eqref{eq:ex-counter-I}.

(ii) The matrix is block diagonal with the stated eigenvalues. By
Eq.~\eqref{eq:Xeigen} the partial transpose has eigenvalues $t$, $t$,
and $\frac12[1-2t\pm\sqrt{(1-2t)^2+4t^3}]$. For $q=t^4$ the
eigenvalues of the state are $1-2t+t^4$, $t-t^4\pm t^{3/2}$ and $t^4$,
and the $(00,11)$ block of the partial transpose has determinant
$t^3[t(1-2t+t^4)-1]<0$.
\end{proof}

\begin{table}[htb]
\caption{Smallest eigenvalues of the matrices of
Proposition~\ref{prop:ex-counter}, computed in 60-digit arithmetic.
$\rho_{\rm II'}$ is $\rho_{\rm II}$ with $q=t^4$; its state has
$\lambda_{\min}=t^4$, and $\lambda_{\min}(\rho_{\rm II})=0$.}
\label{tab:ex-counter}
\begin{ruledtabular}
\begin{tabular}{ccccc}
$t$ & $\lambda_{\min}(\rho_{\rm I})$ &
$\lambda_{\min}(\rho_{\rm I}^{T_2})$ &
$\lambda_{\min}(\rho_{\rm II}^{T_2})$ &
$\lambda_{\min}(\rho_{\rm II'}^{T_2})$\\
\hline
$10^{-2}$ & $6.8905\times10^{-5}$ & $-1.1550\times10^{-4}$
          & $-1.0204\times10^{-6}$ & $-1.0104\times10^{-6}$\\
$10^{-3}$ & $7.4399\times10^{-7}$ & $-1.0160\times10^{-6}$
          & $-1.0020\times10^{-9}$ & $-1.0010\times10^{-9}$\\
$10^{-4}$ & $7.4940\times10^{-9}$ & $-1.0016\times10^{-8}$
          & $-1.0002\times10^{-12}$ & $-1.0001\times10^{-12}$
\end{tabular}
\end{ruledtabular}
\end{table}

\subsection{The parity-even part and one quadrature of the odd coherence}
\label{ss:ex-parity}

Let $Z=\sigma^z_0\sigma^z_\br$ and
\begin{equation}
 \rho^X=\tfrac12\bigl(\rho+Z\rho Z\bigr).
 \label{eq:ex-parity-part}
\end{equation}
Conjugation by $Z$ changes the sign of the entries between the even
states $|00\rangle,|11\rangle$ and the odd states $|01\rangle,|10\rangle$.
Hence $\rho^X$ has the form~\eqref{eq:rho} with the same $a,b,q,v,w$ and
with $\zeta=0$. It is the image of $\rho$ under the local operation
that rotates both spins by $\pi$ about the field axis with
probability $1/2$. For every two-site observable $O$ with $[O,Z]=0$,
$\Tr(O\rho)=\Tr(O\rho^X)$. This includes the products
$\sigma^\alpha_0\sigma^\beta_\br$ with $\alpha,\beta\in\{x,y\}$, the
product $\sigma^z_0\sigma^z_\br$, and the one-site observables
$\sigma^z_0$ and $\sigma^z_\br$, and therefore the three-setting
witness~\eqref{eq:three-setting-witness}. The correlations
$\langle\sigma^\alpha_0\sigma^z_\br\rangle$ with $\alpha\in\{x,y\}$,
which carry the odd coherence, are not of this kind.

\begin{theorem}[Parity-even part outside the clover]\label{thm:ex-parity}
Let $K\subset\mathbb R^2\setminus\{0\}$ be compact with
$0<\inf_{h\in\cI,\,\mathbf x\in K}|A_h(\mathbf x)|$ and
$\sup_{h\in\cI,\,\mathbf x\in K}|A_h(\mathbf x)|<n_*$. There are
$c_K,\eps_K>0$ such that for $0<\eps<\eps_K$, all $h\in\cI$, all
states, and all $\br$ with $\eps\br\in K$:
\begin{enumerate}
\item[(a)] $\lambda_{\min}\bigl[(\rho^X_{0\br})^{T_2}\bigr]\ge c_K\eps^8$;
in particular $\rho^X_{0\br}$ is separable;
\item[(b)] if a two-site observable $O$ with $[O,Z]=0$ satisfies
$\Tr(O\sigma)\ge0$ for every separable $\sigma$, then
$\Tr(O\rho_{0\br})\ge0$: no such observable detects entanglement of
$\rho_{0\br}$.
\end{enumerate}
\end{theorem}

The set $K$ may lie off the axes and have positive area; no reflection
symmetry is assumed. Any entanglement of $\rho_{0\br}$ on $K$ is
therefore carried by the odd coherence alone, and by
Proposition~\ref{prop:th-exterior} its negativity is at most
$C\eps^{16/3}\ell_\eps^2$.

\begin{proof}
On $K$ there are $A_->0$ and $\delta>0$ with
$A_-\le|A_h|\le n_*-\delta$. The partial transpose of $\rho^X$ is the
direct sum of $M=\bigl(\begin{smallmatrix}b&w^*\\w&b\end{smallmatrix}\bigr)$,
with eigenvalues $b\mp W$, and of
$A_0=\bigl(\begin{smallmatrix}a&v\\v&q\end{smallmatrix}\bigr)$
[Eq.~\eqref{eq:ex-congruence} with $u=0$]. By
Theorem~\ref{thm:massscale}, $b/\eps^4\to n_*$, $W/\eps^4-|A_h|\to0$,
and $q,v=o(\eps^4)$, uniformly; hence $b-W\ge(\delta/2)\eps^4$ and
$W\ge(A_-/2)\eps^4$ for small $\eps$. Positivity of the $\{00,11\}$
block of $\rho$ gives $aq\ge W^2$, so
$aq-v^2\ge W^2-v^2\ge A_-^2\eps^8/8$, and the smaller eigenvalue of
$A_0$ is at least $(aq-v^2)/(a+q)\ge A_-^2\eps^8/8$. For two qubits a
positive partial transpose is equivalent to
separability~\cite{Horodecki}. For (b), $\Tr(O\rho_{0\br})
=\Tr(O\rho^X_{0\br})\ge0$ because $\rho^X_{0\br}$ is separable.
\end{proof}

\begin{corollary}[Growing sector window]\label{cor:ex-parity-window}
Let $m_+=\max_{h\in\cI}m_h$, and fix $\gamma,\eta>0$ and $\beta>0$ with
$\beta m_+<2/13$. For small $\eps$, uniformly in $h\in\cI$ and in the
states, $\rho^X_{0\br}$ is strictly PPT, with
$\lambda_{\min}[(\rho^X_{0\br})^{T_2}]\ge c\,\eps^{8+2\beta m_+}/\log(1/\eps)$,
at every $\br$ with $|\cos2\theta_\br|\ge\gamma$ and
$\rho_E(h,\theta_\br)+\eta\le\eps|\br|\le\beta\log(1/\eps)$; statement
(b) of Theorem~\ref{thm:ex-parity} holds there as well.
\end{corollary}

\begin{proof}
In the sector, $n_*-|A_h|\ge\delta(\eta,\gamma,\cI)>0$ for
$\eps|\br|\ge\rho_E+\eta$, since $K_2$ decreases, and
Lemma~\ref{lem:ex-far} gives $b-W\ge c\eps^4$. By
Eq.~\eqref{eq:ex-bessel-lower}, $|A_h(\eps\br)|=|\cos2\theta_\br|\,
A_h(\eps|\br|,0)\ge\gamma c\,t^{-1/2}\eps^{\beta m_+}$ with
$t=\log(1/\eps)$. Since $\beta m_+<2/13$, this dominates the rate
$\varpi_\eps$ of Theorem~\ref{thm:massscale}(iv), so
$W\ge c\,\eps^{4+\beta m_+}t^{-1/2}$, and since $2/13<1/3$ it dominates
$|v|/\eps^4\le C\eps^{1/3}\ell_\eps^{1/2}$ (Lemma~\ref{lem:ex-normal}).
Hence $|v|\le W/2$ and $aq-v^2\ge\frac34W^2$, and the argument of
Theorem~\ref{thm:ex-parity} applies.
\end{proof}

The full state differs from $\rho^X$ by the odd coherence. Its effect
is carried by one real correlation. Let $b>W>0$, $\phi_w=\arg w$, and
choose $\alpha=(\pi-\phi_w)/2$, so that $e^{2i\alpha}w=-W$. Write
\begin{equation}
 \zeta=e^{i\alpha}(x+iy),\qquad
 y=-\omega\bigl(S^{(\phi_w/2)}_0n_\br\bigr),\quad
 x=-\omega\bigl(S^{(\phi_w/2+\pi/2)}_0n_\br\bigr),
 \label{eq:ex-quadratures}
\end{equation}
with $S^{(\beta)}=\cos\beta\,S^x+\sin\beta\,S^y$. The two expressions
for $x,y$ follow from $\omega(b_0n_\br)=\zeta^*$ and
$e^{-i\beta}(S^x+iS^y)=S^{(\beta)}+iS^{(\beta+\pi/2)}$. The angle
$\phi_w/2$ is the principal axis of the pair correlation
$\omega(S^+_0S^+_\br)=w$; $y$ is the correlation of the transverse spin
along this axis with the flip at $\br$, and $x$ the one across it.

\begin{proposition}[One quadrature decides]\label{prop:ex-quadrature}
Let $\rho$ have the form~\eqref{eq:rho} with $b>W>0$ and $b>v$.
\begin{enumerate}
\item[(a)] $\chi=\dfrac{2x^2}{b+W}+\dfrac{2y^2}{b-W}$ and
$\chi-\kappa=\dfrac{2}{b-v}\Bigl[\dfrac{y^2(W-v)}{b-W}
-\dfrac{x^2(W+v)}{b+W}\Bigr]$.
\item[(b)] If $W>|v|$, $a>\chi$, and
\begin{equation}
 y^2<\tfrac12(b-v)(b-W)(W+v),
 \label{eq:ex-quadrature-condition}
\end{equation}
then $\rho^{T_2}\succ0$, with
$D_{\rm PT}\ge(W-v)\bigl[W+v-2y^2/((b-v)(b-W))\bigr]$, whatever $x$ is.
\item[(c)] Let $K$ be as in Theorem~\ref{thm:ex-parity} and
$0<\vartheta<1$. For small $\eps$, if $\eps\br\in K$ and
\begin{equation}
 y(\br)^2\le(1-\vartheta)\,\eps^{12}\,
 \frac{n_*\bigl(n_*-|A_h(\eps\br)|\bigr)|A_h(\eps\br)|}{2},
 \label{eq:ex-quadrature-target}
\end{equation}
then $\lambda_{\min}(\rho_{0\br}^{T_2})\ge c_K\vartheta\eps^8$.
\end{enumerate}
\end{proposition}

\begin{proof}
(a) Since $w\zeta^2=-W(x+iy)^2$,
$b|\zeta|^2+\Ree(w\zeta^2)=x^2(b-W)+y^2(b+W)$, which gives $\chi$.
Inserting $\Ree(w\zeta^2)=-W(x^2-y^2)$ and $|\zeta|^2=x^2+y^2$ in
Eq.~\eqref{eq:ex-phase-cost} gives
$(b^2-W^2)(b-v)(\chi-\kappa)=2[y^2(b+W)(W-v)-x^2(b-W)(W+v)]$.

(b) By (a), $-(\chi-\kappa)\ge-2y^2(W-v)/[(b-v)(b-W)]$. Insert this in
Eq.~\eqref{eq:ex-identity} and drop $E_{\rm ph}\ge0$ and
$2\chi(W^2-v^2)/(b-v)\ge0$. The resulting lower bound on $D_{\rm PT}$
is positive under Eq.~\eqref{eq:ex-quadrature-condition}, and
Lemma~\ref{lem:ex-schur}(a) gives $\rho^{T_2}\succ0$.

(c) On $K$, $b/\eps^4\to n_*$, $W/\eps^4-|A_h|\to0$ and $v=o(\eps^4)$,
so the right side of Eq.~\eqref{eq:ex-quadrature-condition} is
$\eps^{12}n_*(n_*-|A_h|)|A_h|/2\,(1+o(1))$, and
Eq.~\eqref{eq:ex-quadrature-target} gives
$D_{\rm PT}\ge\frac{\vartheta}{2}W^2$ for small $\eps$. Moreover
$\chi\le2|\zeta|^2/(b-W)\to0<a$ by Proposition~\ref{prop:th-exterior}.
In Eq.~\eqref{eq:ex-congruence}, $\mathsf S$ has determinant
$D_{\rm PT}$ and trace at most $1$, so
$\lambda_{\min}(\mathsf S)\ge\frac{\vartheta}{2}W^2$;
$M\succeq(b-W)\ge c\eps^4$; and
$\|\mathsf L-I\|\le\|M^{-1}u\|\sqrt2\le C|\zeta|/(b-W)\to0$. Hence
$\lambda_{\min}(\rho^{T_2})\ge\frac12\min\{\lambda_{\min}(\mathsf S),
\lambda_{\min}(M)\}\ge c_K\vartheta\eps^8$.
\end{proof}

For the axial phase of Sec.~\ref{ss:ex-axial}, $y=0$ and part (b) holds
for every $x$. Condition~\eqref{eq:ex-quadrature-target} replaces the
hypothesis $\zeta=o(\eps^6)$ of Theorem~\ref{thm:rdmcriterion} by a
bound on a single real three-point function of the physical state,
with an explicit constant; the other quadrature $x$ may be as large as
positivity of $\rho$ allows. The two-site entries alone do not
suffice:

\begin{proposition}[Two-site data do not decide]\label{prop:ex-twosite}
Let $b>W>|v|$ and $a>\chi$.
\begin{enumerate}
\item[(i)] If $aq-W^2<(W+v)(b-W)(a+q-2W)$, then $\rho^{T_2}\succ0$ for
every $\zeta$ for which Eq.~\eqref{eq:rho} is a density matrix.
\item[(ii)] If $b+v\ge0$, $aq>W^2$, and
$(aq-W^2)(1-2W-2v)>(W+v)(b-W)(a+q-2W)$, then there is a $\zeta$ for
which Eq.~\eqref{eq:rho} is a positive-definite density matrix with a
negative partial-transpose eigenvalue.
\end{enumerate}
\end{proposition}

\begin{proof}
(i) Positivity of $\mathsf S_{\rm ph}$ in Eq.~\eqref{eq:ex-physical-block}
and $|2\zeta^{*2}/(b-v)|=\kappa$ give
$(a-\kappa)(q-\kappa)\ge(W-\kappa)^2$, that is,
$\kappa(a+q-2W)\le aq-W^2$. Since $2y^2/(b-v)\le\kappa$,
Proposition~\ref{prop:ex-quadrature}(a) gives
$\chi-\kappa\le\kappa(W-v)/(b-W)<W^2-v^2$, and
Eq.~\eqref{eq:ex-identity} gives $D_{\rm PT}>0$.
(ii) Take $x=0$ and $y^2=\kappa_{\max}(b-v)/2$ with
$\kappa_{\max}=(aq-W^2)/(a+q-2W)\le\min(a,q)$. Then
$\zeta^{*2}=y^2w/W$, the matrix $\mathsf S_{\rm ph}$ has diagonal
$a-\kappa_{\max},q-\kappa_{\max}\ge0$ and determinant zero, so $\rho\succeq0$,
and $\chi=\kappa_{\max}(b-v)/(b-W)$ gives
$D_{\rm PT}=(W-v)[(W+v)-\kappa_{\max}(1-2W-2v)/(b-W)]<0$. By
Lemma~\ref{lem:ex-schur}(a) the partial transpose has a negative
eigenvalue. Replacing $\zeta$ by $(1-s)\zeta$ with small $s>0$ keeps
$D_{\rm PT}<0$ and makes $\rho$ positive definite.
\end{proof}

With $a\to1$ and $v=o(W)$, part (i) reads $q<bW(1+o(1))$ and part (ii)
$q>bW(1+o(1))$. For a quasi-free state $q\simeq p^2+W^2>pW$, so a proof
of exterior separability cannot use only $p,q,v,w$ and positivity of
the two-site state; information on the odd coherence beyond positivity
is needed. All identities of this subsection were also checked on
$2\times10^4$ random density matrices of the form~\eqref{eq:rho}.

No symmetry supplies that information off the axes and diagonals. A
transformation built from complex conjugation or not, on-site unitaries,
and a lattice isometry commutes with $H$ only if, up to a translation
and a global phase, it is a joint rotation by $k\pi/2$ or one of the
four antiunitary mirrors $b_j\mapsto e^{i(\pi-2\beta)}b_{m_\beta j}$ across
the lines at angle $\beta\in\{0,\pi/4,\pi/2,3\pi/4\}$; the group is the
magnetic point group $4m'm'$. The reason is the chiral form factor of
the DM term: in the hard-core form its single-flip vertex is
$\sum_{i,\bm\delta}c(\bm\delta)n_ib^\dagger_{i+\bm\delta}$ with
$c(\bm\delta)=\frac i2(\delta_x+i\delta_y)$, which a unitary mirror or an
antiunitary rotation cannot preserve, while the real pair source and
the real hopping fix the on-site phases to be uniform. At a generic
separation only the joint half turn maps the pair to itself, so the
phase of $w\zeta^2$, which enters $y$, is not constrained; on the axes
and diagonals one antiunitary mirror makes $w\zeta^2$ real but does not
fix its sign. Exact diagonalization of $3\times4$, $4\times4$ and
$4\times5$ tori at $h=2$ confirms these constraints to $10^{-14}$ and
shows unconstrained phases at generic separations.

\subsection{Axes under an additional antiunitary reflection}
\label{ss:ex-axial}

The antiunitary $\mathsf A_x=\mathsf CP_x$ of
Eq.~\eqref{eq:antiunitary-reflection} preserves $H_{\eps,h}$. With the
joint quarter-turn $U$ it satisfies
$\mathsf A_xU\mathsf A_x^{-1}=U^{-1}$, so the two generate a dihedral
group of order eight acting on states; the antiunitary image of a
state, $A\mapsto\overline{\omega(\mathsf A_xA\mathsf A_x^{-1})}$, is
again a state. The image of a minimizer in our class is a minimizer in
the class, because the energy density is preserved and the dihedral
relation keeps the image fourfold invariant. Averaging a minimizer
with its image therefore produces a minimizer that is also
$\mathsf A_x$ invariant, and periodic Gibbs states and their symmetric
limits are $\mathsf A_x$ invariant. It is not known that every
fourfold-invariant minimizer is $\mathsf A_x$ invariant, and the
partial transpose of an average is not controlled by those of its
components. The extra invariance is therefore an additional assumption
on the state. It is used only in Lemma~\ref{lem:ex-phases}(a,b),
Lemma~\ref{lem:ex-diagonal}, and Theorem~\ref{thm:ex-axial}.

\begin{lemma}[Phases fixed by symmetry]\label{lem:ex-phases}
Let $\omega$ be translation invariant and invariant under the joint
quarter-turn, and let $J$ rotate lattice vectors by a quarter-turn.
Then $w(J\br)=-w(\br)$, $\zeta(J\br)=-i\zeta(\br)$, $w(-\br)=w(\br)$,
$\zeta(-\br)=-\zeta(\br)$, and $p,q,v$ are invariant. The spectrum of
$\rho_{0\br}^{T_2}$ is invariant under $\br\mapsto J\br$ and
$\br\mapsto-\br$. If $\omega$ is also $\mathsf A_x$ invariant, then
$w(P_x\br)=\overline{w(\br)}$ and $\zeta(P_x\br)=\overline{\zeta(\br)}$;
hence
\begin{enumerate}
\item[(a)] on the $x$ axis $w\in\mathbb R$ and $\zeta\in i\mathbb R$;
on the $y$ axis $w,\zeta\in\mathbb R$; and $w(0,j)=-w(j,0)$;
\item[(b)] at $\br=(j,j)$ one has $w=is$ and
$\zeta=e^{i\pi/4}z$ with $s,z\in\mathbb R$.
\end{enumerate}
Joint quarter-turn invariance alone imposes no pointwise condition on
a diagonal, since it maps a diagonal to the other diagonal.
\end{lemma}

\begin{proof}
The joint quarter-turn acts as $b_i\mapsto-ib_{Ji}$ and
$n_i\mapsto n_{Ji}$. Invariance gives $w(\br)=-w(J\br)$ and
$\omega(b_0n_\br)=-i\omega(b_0n_{J\br})$; since
$\zeta=\overline{\omega(b_0n_\br)}$, this is $\zeta(J\br)=-i\zeta(\br)$.
Applying the rotation twice gives the parity relations. On the two-site
RDM the rotation acts by conjugation with $V\otimes V$ for a diagonal
one-site unitary $V$, and
$[(V\otimes V)\rho(V\otimes V)^\dagger]^{T_2}
=(V\otimes\overline V)\rho^{T_2}(V\otimes\overline V)^\dagger$.
By translation invariance $\rho_{0,-\br}$ is $\rho_{0\br}$ with the
two sites exchanged, whose partial transpose is unitarily equivalent to
$(\rho_{0\br}^{T_2})^T$. Complex conjugation in the $S^z$ basis fixes
$b_i$ and $n_i$, so $\mathsf A_xb_i\mathsf A_x^{-1}=b_{P_xi}$, and
invariance of the state under an antiunitary means
$\omega(\mathsf A_xA\mathsf A_x^{-1})=\overline{\omega(A)}$. On the $x$
axis $P_x\br=-\br$, on the $y$ axis $P_x\br=\br$, and at $(j,j)$ one
has $P_x\br=J\br$. Combining these relations gives (a) and (b); for
instance at $(j,j)$, $\overline\zeta=\zeta(J\br)=-i\zeta$.
\end{proof}

\begin{lemma}[Exact axial negativity]\label{lem:ex-axial-exact}
Let $\rho$ be a density matrix of the form~\eqref{eq:rho} with
\begin{equation}
 b>v,\qquad W>|v|,\qquad w\zeta^2=-W|\zeta|^2 .
 \label{eq:ex-axial-hyp}
\end{equation}
Then the spectrum of $\rho^{T_2}$ consists of $b-W$ and the three
eigenvalues of a positive-definite matrix $B_{\rm ax}$. Consequently
\begin{equation}
 \Neg(\rho)=(W-b)_+=(|w|-p+q)_+ ,
 \label{eq:ex-axial-formula}
\end{equation}
and $\rho$ is strictly PPT if and only if $W<b$. Moreover the Schur
complement $\mathsf S_{\rm ax}$ of the last entry of $B_{\rm ax}$ obeys
$\det\mathsf S_{\rm ax}\ge W^2-v^2$.
\end{lemma}

No smallness of $\zeta$ is assumed, and $W\ge b$ is allowed, where $M$
is not positive definite and Lemma~\ref{lem:ex-schur} does not apply.
Equation~\eqref{eq:ex-axial-formula} says that the principal-block
bound~\eqref{eq:population-npt} is an equality.

\begin{proof}
A common one-site rotation makes $w=W>0$; the phase condition then
gives $\zeta=iz$ with $z\in\mathbb R$. In the basis
$|00\rangle,|11\rangle,|s\rangle,|t\rangle$ with
$|s\rangle,|t\rangle=(|01\rangle\pm|10\rangle)/\sqrt2$,
Eq.~\eqref{eq:PT} shows that $|t\rangle$ is an eigenvector of
$\rho^{T_2}$ with eigenvalue $b-W$, and that on the span of the other
three vectors
\begin{equation}
 B_{\rm ax}=\begin{pmatrix}a&v&-i\sqrt2z\\v&q&-i\sqrt2z\\
 i\sqrt2z&i\sqrt2z&b+W\end{pmatrix},\qquad
 \mathsf S_{\rm ax}=\begin{pmatrix}a-\chi&v-\chi\\v-\chi&q-\chi
 \end{pmatrix},\quad \chi=\frac{2z^2}{b+W}.
 \label{eq:ex-axial-block}
\end{equation}
Positivity of the one-flip block of $\rho$ gives $b\ge|v|$, so
$b+W>0$. In Eq.~\eqref{eq:ex-physical-block} the axial phase gives
$\mathsf S_{\rm ph}=\bigl(\begin{smallmatrix}a-\kappa&W+\kappa\\
W+\kappa&q-\kappa\end{smallmatrix}\bigr)\succeq0$ with
$\kappa=2z^2/(b-v)$, that is, $a\ge\kappa$ and
$aq-W^2\ge\kappa(a+q+2W)$. Since $W+v>0$, $b-v<b+W$ and
$\kappa\ge\chi$. Therefore
\[
 \det\mathsf S_{\rm ax}=aq-v^2-\chi(a+q-2v)
 \ge W^2-v^2+(\kappa-\chi)(a+q+2W)+2\chi(W+v)\ge W^2-v^2>0 .
\]
Together with $a\ge\kappa\ge\chi$ this forces $a>\chi$, so
$\mathsf S_{\rm ax}\succ0$ and $B_{\rm ax}\succ0$. The only possibly
nonpositive eigenvalue of $\rho^{T_2}$ is $b-W$.
\end{proof}

If $b=v$, positivity forces $\zeta=0$ and $\rho$ is an X state; the
conclusion persists by Eq.~\eqref{eq:Xeigen}.

\begin{theorem}[Axial partition and a growing PPT window]
\label{thm:ex-axial}
Consider the minimizers of Sec.~\ref{sm:conventions} that are also
$\mathsf A_x$ invariant, and axial lattice points
$\br\in(\mathbb Z\times\{0\})\cup(\{0\}\times\mathbb Z)$.
\begin{enumerate}
\item[(i)] For fixed $0<c<C'$ and all sufficiently small $\eps$,
uniformly for $c\le\eps|\br|\le C'$,
\begin{equation}
 \Neg(\rho_{0\br})=\bigl(|w(\br)|-p+q(\br)\bigr)_+ .
 \label{eq:ex-axial-exact}
\end{equation}
Hence $\eps^{-4}\Neg\to(A_h(\eps|\br|,0)-n_*)_+$ uniformly; for fixed
$\eta>0$ the pair is NPT where $\eps|\br|\le R_E(h)-\eta$ and strictly
PPT where $\eps|\br|\ge R_E(h)+\eta$.
\item[(ii)] Let $\eta>0$ and $\beta>0$ with $\beta m_+<2/13$. For all
sufficiently small $\eps$, every axial $\br$ with
\begin{equation}
 R_E(h)+\eta\le\eps|\br|\le\beta\log(1/\eps)
 \label{eq:ex-window}
\end{equation}
gives a strictly PPT pair, and
\begin{equation}
 \lambda_{\min}(\rho_{0\br}^{T_2})\ge
 c_{\eta,\beta}\,\frac{\eps^{8+2\beta m_+}}{\log(1/\eps)} .
 \label{eq:ex-window-gap}
\end{equation}
\end{enumerate}
\end{theorem}

\begin{proof}
By Lemma~\ref{lem:ex-phases} it suffices to take $\br=(j,0)$ with
$j\ge1$, where $w$ is real, $\zeta$ is purely imaginary, and
$A_h(\eps\br)=A_h(\eps j,0)>0$. If $w>0$, then
$w\zeta^2=-W|\zeta|^2$.

(i) On $[c,C']$, $A_h(\rho,0)\ge A_{\min}>0$ uniformly in $h$. By
Eq.~\eqref{eq:ex-input-profile}, $w/\eps^4\ge A_{\min}/2$ for small
$\eps$, so $w=W>0$ and the phase condition holds. Also
$|v|=o(\eps^4)<W$ and $b=(n_*+o(1))\eps^4>v$.
Lemma~\ref{lem:ex-axial-exact} gives Eq.~\eqref{eq:ex-axial-exact}, and
Eq.~\eqref{eq:ex-input-profile} gives its limit; the signs follow from
the strict monotonicity of $A_h(\rho,0)$ in $\rho$ and compactness of
$\cI$.

(ii) Put $t=\log(1/\eps)$ and $\rho=\eps|\br|$. The representation
$K_\nu(x)=\int_0^\infty e^{-x\cosh u}\cosh(\nu u)\,\dd u$ shows that
$K_\nu$ increases with $\nu\ge0$, so
$K_2(x)\ge K_{1/2}(x)=\sqrt{\pi/(2x)}\,e^{-x}$. On the window,
\begin{equation}
 A_h(\rho,0)\ge c_\cI\,\rho^{-1/2}e^{-m_+\rho}
 \ge c_{\cI,\beta}\,t^{-1/2}\eps^{\beta m_+} .
 \label{eq:ex-bessel-lower}
\end{equation}
Since $\beta m_+<2/13$, $A_h(\rho,0)/\varpi_\eps\to\infty$, and
Eq.~\eqref{eq:ex-input-profile} gives $w\ge\frac12A_h\eps^4$; thus
$w=W>0$ and $W\ge c\,\eps^{4+\beta m_+}t^{-1/2}$. By
Lemma~\ref{lem:ex-normal},
$|v|/W\le C\eps^{1/3-\beta m_+}t^{1/2}\ell_\eps^{1/2}\to0$, because
$\beta m_+<1/3$. The window lies in the region of
Lemma~\ref{lem:ex-far}, so $b-W\ge c_\eta\eps^4$ and $b>v$.
Lemma~\ref{lem:ex-axial-exact} gives strict PPT, the spectrum being
$b-W$ together with that of $B_{\rm ax}$.

For the spectral bound, $\det\mathsf S_{\rm ax}\ge W^2-v^2\ge W^2/2$
and $\Tr\mathsf S_{\rm ax}\le a+q\le1$, so
$\lambda_{\min}(\mathsf S_{\rm ax})\ge W^2/2$. The congruence that
restores the last entry of $B_{\rm ax}$, as in
Eq.~\eqref{eq:ex-congruence}, has off-diagonal block of norm
$2|z|/(b+W)\le2|\zeta|/b=O(\eps^{2/3}\ell_\eps)$ by
Eq.~\eqref{eq:ex-input-odd}, so its inverse has norm $1+o(1)$. Since
$b+W\ge b\gg W^2$ and $b-W\ge c_\eta\eps^4\gg W^2$, one obtains
$\lambda_{\min}(\rho^{T_2})\ge W^2/4$ for small $\eps$, which is
Eq.~\eqref{eq:ex-window-gap}.
\end{proof}

The same argument reaches the upper end
$\eps|\br|\le T_\eps(h)=[\frac2{13}t-\gamma\log t]/m_h$ for any fixed
$\gamma>17/26$, with
$\lambda_{\min}(\rho_{0\br}^{T_2})\ge c_{\eta,\gamma}\eps^{108/13}
t^{2\gamma-1}$: at this end
$A_h(\rho,0)\ge c\,\eps^{2/13}t^{\gamma-1/2}$, which dominates
$\varpi_\eps\asymp\eps^{2/13}t^{2/13}$ exactly when
$\gamma>\frac12+\frac2{13}$. Every such window still has a finite upper
end at fixed $\eps$. Where $\eps|\br|\ge\beta'\log(1/\eps)$ with
$\beta'm_h>2/13$, the profile $A_h$ is smaller than the error $\varpi_\eps$
of Eq.~\eqref{eq:ex-input-profile}, so even the sign of the axial
amplitude is not determined by the present estimates.

\subsection{Diagonal rays and their neighbourhoods}
\label{ss:ex-nodes}

The leading profile vanishes on the diagonals: $F(j,\pm j)=0$ by
coordinate exchange (Sec.~\ref{sm:green}), and $A_h\propto\cos2\theta$.
Theorem~\ref{thm:rdmcriterion} therefore does not apply there. For
$\mathsf A_x$-invariant states the diagonal phases are nevertheless
fixed, and the sign test reduces to one scalar.

\begin{lemma}[Diagonal sign test]\label{lem:ex-diagonal}
Let $\omega$ be $\mathsf A_x$ invariant and $\br=(j,j)$, so that
$w=is$ and $\zeta=e^{i\pi/4}z$ with $s,z$ real
(Lemma~\ref{lem:ex-phases}). If $b>|s|$, the spectrum of $\rho^{T_2}$
consists of $b-s>0$ and the spectrum of the matrix $B_{\rm ax}$ of
Eq.~\eqref{eq:ex-axial-block} with $W$ replaced by $s$. If moreover
$a>\chi=2z^2/(b+s)$, then the pair is strictly PPT, PPT with exactly
one zero eigenvalue, or NPT with exactly one negative eigenvalue,
according as
\begin{equation}
 D_{\rm diag}=aq-v^2-\frac{2z^2}{b+s}(a+q-2v)
 \label{eq:ex-Ddiag}
\end{equation}
is positive, zero, or negative. If $s>|v|$ and $b>v$, then
$\Neg=(s-b)_+$. By Lemma~\ref{lem:ex-phases} the same holds on the
other diagonal half-rays.
\end{lemma}

\begin{proof}
The one-site rotation with $e^{2i\alpha}=-i$ maps $(w,\zeta)$ to
$(s,iz)$. The computation in the proof of
Lemma~\ref{lem:ex-axial-exact} used no sign of $W$ up to
Eq.~\eqref{eq:ex-axial-block}; it gives the eigenvalue $b-s$ and the
block $B_{\rm ax}$ with $b+s>0$. Here $w\zeta^2=-sz^2$, so
$\chi=2z^2/(b+s)$ coincides with Eq.~\eqref{eq:ex-DPT} and
$D_{\rm diag}=D_{\rm PT}$. The inertia of $B_{\rm ax}$ is that of the
entry $b+s$ plus that of its Schur complement, whose first diagonal
entry is $a-\chi>0$ and whose determinant is $D_{\rm diag}$. For
$s>|v|$, $b>v$, Lemma~\ref{lem:ex-axial-exact} applies.
\end{proof}

Neither the sign of $s$ nor $s>|v|$ follows from symmetry, and the
leading profile gives $s=o(\eps^4)$. The following bounds hold without
the additional reflection.

\begin{theorem}[Nodal bounds]\label{thm:ex-nodes}
Fix $c>0$ and $0<\delta<n_*$. For all sufficiently small $\eps$,
uniformly in $h\in\cI$ and the states,
\begin{align}
 \Neg(\rho_{0\br})&\le C_{c,\delta}\,\eps^{16/3}\ell_\eps^2
 \bigl[|\cos2\theta_\br|+\varpi_\eps\bigr],
 &&\br\in\mathcal X_{c,\delta},\label{eq:ex-node-nbhd}\\
 \Neg(\rho_{0\br})&\le C_c\,\eps^{214/39}\ell_\eps^{28/13},
 &&\br=(j,\pm j),\ \eps|\br|\ge c .\label{eq:ex-node-diag}
\end{align}
No upper bound on $|\br|$ is imposed.
\end{theorem}

\begin{proof}
On $\mathcal X_{c,\delta}$, Eqs.~\eqref{eq:ex-input-profile} and
\eqref{eq:ex-input-odd} give, for small $\eps$,
$b=(n_*+o(1))\eps^4$, $b-W\ge\frac\delta2\eps^4\ge\delta_0b$ with
$\delta_0=\delta/(4n_*)$, $|v|\le b/2$, and
$\chi\le2|\zeta|^2/(b-W)=O(\eps^{16/3}\ell_\eps^2)$, hence
$a-\chi\ge\frac12$. Lemma~\ref{lem:ex-schur}(d) applies. In
Eq.~\eqref{eq:ex-phase-cost}, $|\Ree(w\zeta^2)|\le W|\zeta|^2$,
$W<b$ and $b-v\le\frac32b$ bound the numerator by
$10\,b|\zeta|^2(W+|v|)$, while the denominator is at least
$\frac12\delta_0b^3$. Also
$(v-\chi)^2/(a-\chi)\le4v^2+4\chi^2$ and $\chi\le2|\zeta|^2/(\delta_0b)$.
Therefore
\begin{equation}
 \Neg(\rho)\le C_\delta\Bigl[\frac{|\zeta|^2(W+|v|)}{b^2}+v^2
 +\frac{|\zeta|^4}{b^2}\Bigr].
 \label{eq:ex-node-master}
\end{equation}
Insert $|\zeta|^2/b^2\le C\eps^{4/3}\ell_\eps^2$,
$W\le\eps^4(|A_h(\eps\br)|+CE_\eps)$, and Lemma~\ref{lem:ex-normal}:
\[
 \Neg\le C\eps^{16/3}\ell_\eps^2\bigl[|A_h|+\varpi_\eps
 +\eps^{1/3}\ell_\eps^{1/2}\bigr]
 +C\eps^{26/3}\ell_\eps+C\eps^{32/3}\ell_\eps^4 .
\]
Since $\eps^{1/3}\ell_\eps^{1/2}\le CE_\eps$, the last two terms are
smaller than $\eps^{16/3}\ell_\eps^2\varpi_\eps$, and $m_h^2>2$ gives
$|A_h(\rho,\theta)|\le C_c|\cos2\theta|$ for $\rho\ge c$. This proves
Eq.~\eqref{eq:ex-node-nbhd}. The diagonals lie in
$\mathcal X_{c,\delta}$ for every $\delta<n_*$, $\cos2\theta=0$ there,
and $\eps^{16/3}\ell_\eps^2\varpi_\eps=\eps^{214/39}\ell_\eps^{28/13}$.
\end{proof}

Equation~\eqref{eq:ex-node-nbhd} improves the fixed-margin bound of
Proposition~\ref{prop:th-exterior} by the factor
$|\cos2\theta|+\varpi_\eps$, and it covers a fixed angular neighbourhood of
the diagonals together with its unbounded tail: for
$|\cos2\theta|\le n_*/(2C_c)$ the margin holds with $\delta=n_*/2$. By
Eq.~\eqref{eq:ex-node-diag}, a negativity threshold $\kappa_\eps\eps^4$
detects no pair on the macroscopic diagonals once
$\kappa_\eps/(\eps^{58/39}\ell_\eps^{28/13})\to\infty$. These are upper
bounds. They do not decide the sign, and Sec.~\ref{ss:ex-open} records
that all three signs are compatible with the known nodal data.

\subsection{Centered pair variance and a conditional criterion}
\label{ss:ex-variance}

Off the symmetry rays, deciding the sign of $D_{\rm PT}$ in general
needs $q$ to precision $o(\eps^8)$ and $\zeta$ to precision
$o(\eps^6)$. Both are controlled by one fluctuation of the residual
state. Let
\[
 \widehat\omega(A)=\omega(U_RVAV^\dagger U_R^\dagger)
\]
be the twice-dressed state used for the limit statements of
Theorem~\ref{thm:massscale}: $U_R$ is the pair dressing of
Eq.~\eqref{eq:pop-cutoff} with $R=\lceil\eps^{-2}\rceil$, and
$V=e^{\tau Y}$, $\tau=\eps^3/2$, is the triplet dressing with the
polynomial kernel at scale $L=\ell_\eps^8$, whose incident norm budget
is $y\le C\ell_\eps^{16}$. Its population obeys
$\sqrt{\widehat p}\le C\eps^2\ell_\eps^{-8}$. Let $\widehat\Omega$ be
its GNS vector, $B_\br=b_0b_\br$, and $\widehat w(\br)=
\widehat\omega(B_\br)$. The \emph{centered pair variance} is
\begin{equation}
 \mathcal V_\br=\widehat\omega(n_0n_\br)-|\widehat w(\br)|^2
 =\|X_\br\|^2,\qquad X_\br=(B_\br-\widehat w(\br))\widehat\Omega .
 \label{eq:ex-variance-def}
\end{equation}
It vanishes when the residual pair amplitude is deterministic, and it
satisfies $\mathcal V_\br\le\widehat p$.

\begin{proposition}[Variance bound]\label{prop:ex-variance}
There is $\vartheta_0>0$ such that for every $c>0$ and
$0<\vartheta<\min\{\vartheta_0,1/2\}$,
\begin{equation}
 \sup_{h\in\cI,\ \omega,\ |\eps\br|\ge c}\mathcal V_\br
 \le C_{c,\vartheta}\,\eps^{(14+2\vartheta)/3}\ell_\eps^{1+\vartheta}
 \le C'_{c,\vartheta}\,\eps^{14/3+\vartheta/3}.
 \label{eq:ex-variance-bound}
\end{equation}
Here $\vartheta_0$ is the H\"older exponent of the lazy random walk on
the three-magnon exclusion graph, which is positive by the parabolic
Harnack inequality of Ref.~\cite{Delmotte1999}; no numerical value is
claimed.
\end{proposition}

\begin{proof}[Structure of the proof]
The argument has five steps. Step~1 is not reproduced here; Steps 2--5
are complete given Step~1 and Proposition~\ref{prop:ex-gram}(i).

\emph{Step 1: an auxiliary triplet and its energy inequality.} For a
parameter $L\ge\ell_\eps$ let
\[
 \Theta_L=-\frac1{12+\beta}\sum_{j=0}^{N-1}
 \Bigl(\frac{12}{12+\beta}\Bigr)^j\Bigl(I-\frac{Q_3}{12}\Bigr)^j
 \mathcal T_L,\qquad \beta=L^{-2/(1+\vartheta)},\quad
 N=\lceil13\beta^{-1}\log L\rceil,
\]
be the polynomial approximation of the three-magnon coefficient
$-Q_3^{-1}\mathcal T$ of Theorem~\ref{thm:energy}, with the source
$\mathcal T_L$ cut off at radius $L$. The H\"older estimate gives the
low-energy spectral bound $\mu_{\mathcal T_L}([0,\lambda])\le
C\lambda^{2+\vartheta}$, uniformly in $L$, and hence an inverse-form
error $O(L^{-2})$, an incident norm budget $y_L\le CL^k$ with
$k=(2-\vartheta)/(1+\vartheta)$, a residual source of bounded absolute
sum, and support diameter $O(L^{2/(1+\vartheta)}\log L)$. The
sixth-order energy inequality used in the proof of
Theorem~\ref{thm:massscale}, in which three odd terms with vanishing
vacuum diagonal are retained so that the common energy scalar is
unchanged, then holds with $y_L$ in place of the budget $CL^2$ of that
proof. For the auxiliary state dressed by $U_R$ and
$V_L=e^{\tau Y_L}$ it reads
\begin{equation}
 \kappa_6\widehat{\mathfrak q}_L+\mu_6\eps^2\widehat p_L
 \le C\eps^6L^{-2}+C\bigl[\eps^7\ell_\eps^3+\eps^8\ell_\eps^3L^{2k}
 +\eps^{10}L^{4k}+\eps^{16}L^{6k}\bigr].
 \label{eq:ex-energy-L}
\end{equation}

\emph{Step 2: choice of $L$.} Take
$L=\eps^{-(1+\vartheta)/3}\ell_\eps^{-(1+\vartheta)/2}$, rounded
upward. Divided by $\eps^6L^{-2}$, the four bracketed terms become
$\eps^{(1-2\vartheta)/3}\ell_\eps^{2-\vartheta}$, $1$,
$\eps^{(2+2\vartheta)/3}\ell_\eps^{\vartheta-5}$ and
$\eps^{(16+4\vartheta)/3}\ell_\eps^{2\vartheta-7}$, which are bounded
for $\vartheta<1/2$. Also $\eps(\ell_\eps+y_L)\to0$ and
$\eps^4\ell_\eps^2(1+y_L)^2\to0$, the conditions under which
Eq.~\eqref{eq:ex-energy-L} holds. Hence $\widehat p_L\le C\eps^4L^{-2}$.

\emph{Step 3: removal of the auxiliary triplet.} By
Eq.~\eqref{eq:pop-triplet-local-transfer} with $a_\Theta$ replaced by
$y_L$, $\sqrt{\widetilde p}\le\sqrt{\widehat p_L}+\eps^3y_L\le
C\eps^2L^{-1}$, since $\eps L^{k+1}=\ell_\eps^{-3/2}$.

\emph{Step 4: physical joint population.} Expanding
$U_R^\dagger b_0b_\br U_R$ with Eq.~\eqref{eq:pop-pair-local-transfer}
and separating the term $F_R(\br)(1-n_0)(1-2n_\br)$,
\[
 \sqrt{q(\br)}\le d|F_R(\br)|+(1+2da_R)\sqrt{\widetilde p}+Cd^2a_R^2 .
\]
On $|\eps\br|\ge c$, $d|F_R(\br)|\le C_c\eps^4$ by
Eq.~\eqref{eq:green-tail}, and $d^2a_R^2\le C\eps^4\ell_\eps^2$; both
are smaller than $\eps^2L^{-1}$. Thus
$q(\br)\le C_c\eps^4L^{-2}=C_c\eps^{(14+2\vartheta)/3}
\ell_\eps^{1+\vartheta}$ for the physical state. The auxiliary state
does not enter this conclusion.

\emph{Step 5: return to the residual state.} By
Proposition~\ref{prop:ex-gram}(i), the vector
$(V^\dagger U_R^\dagger B_\br U_RV-w(\br))\widehat\Omega$, whose norm
is exactly $\sqrt{q(\br)-|w(\br)|^2}$, equals
$X_\br+\mathcal Z_\br+E_\br$ with $\|\mathcal Z_\br\|=O(\eps^4)$ and
$\|E_\br\|=o(\eps^4)$. Hence
$\sqrt{\mathcal V_\br}\le\sqrt{q(\br)}+C_c\eps^4$, and squaring gives
Eq.~\eqref{eq:ex-variance-bound}; the last inequality there uses
$\eps^{\vartheta/3}\ell_\eps^{1+\vartheta}\to0$.
\end{proof}

The bound covers all directions and all macroscopic distances, but it
is far above the scale $\eps^8$ needed below.

\begin{proposition}[Gram representation]\label{prop:ex-gram}
Let $\Phi_j^\dagger=\sum_kF_R(k-j)b_k^\dagger$, with $F_R(0)=0$, and
let $\Phi_{j,\br}^\dagger$ be the same sum restricted to
$k\notin\{0,\br\}$. Put
\begin{equation}
 \mathcal Z_\br=d^2\bigl(\Phi_0^\dagger\Phi_\br^\dagger
 -\widehat\omega(\Phi_0^\dagger\Phi_\br^\dagger)\bigr)\widehat\Omega,
 \qquad
 Y_\br=\tfrac12\bigl[b_\br-b_0+d(\Phi_{\br,\br}^\dagger
 -\Phi_{0,\br}^\dagger)\bigr]\widehat\Omega .
 \label{eq:ex-gram-vectors}
\end{equation}
Uniformly in $h\in\cI$, the states, and $|\eps\br|\ge c$:
\begin{enumerate}
\item[(i)] $(V^\dagger U_R^\dagger B_\br U_RV-w(\br))\widehat\Omega
=X_\br+\mathcal Z_\br+E_\br$ with $\|E_\br\|=o(\eps^4)$; in particular
$q(\br)-|w(\br)|^2=\|X_\br+\mathcal Z_\br+E_\br\|^2$;
\item[(ii)] $\overline{\zeta(\br)}=\langle Y_\br,X_\br\rangle
+o(\eps^6)$;
\item[(iii)] $\eps^{-8}\|\mathcal Z_\br\|^2\to n_*^2$,
$\eps^{-4}\|Y_\br\|^2\to n_*/2$, and
$\eps^{-6}\langle Y_\br,\mathcal Z_\br\rangle\to0$.
\end{enumerate}
\end{proposition}

\begin{proof}[Structure of the proof]
The remainder bounds in (i) and (ii) are not reproduced here; the
exact identities they rest on, and part (iii), are proved below.

(i) Expand $U_R^\dagger B_\br U_R$ to second order in $d$ and conjugate
by $V$ to first order in $\tau$. The first commutator is
\[
 [B_\br,X]=F_R(\br)(1-n_0-n_\br)
 +\sum_{k\notin\{0,\br\}}\bigl[F_R(k)(1-2n_0)b_\br
 +F_R(k-\br)(1-2n_\br)b_0\bigr]b_k^\dagger ,
\]
with vacuum column $F_R(\br)\Omega_0$, where $\Omega_0$ is the
fully polarized vector. The second has vacuum column
\[
 \tfrac12[[B_\br,X],X]\Omega_0
 =\bigl[\Phi_0^\dagger\Phi_\br^\dagger
 -2F_R(\br)(b_0^\dagger\Phi_0^\dagger+b_\br^\dagger\Phi_\br^\dagger)
 +2F_R(\br)^2b_0^\dagger b_\br^\dagger\bigr]\Omega_0 .
\]
After the vacuum columns are subtracted, the local terms have zero
vacuum column, and Lemma~\ref{lem:pop-local-extraction} bounds their
GNS norms by their coefficient sums times $\sqrt{\widehat p}$. The
commutator $[B_\br,Y]$ has zero vacuum column when $|\br|$ exceeds the
support diameter of $Y$. The remainder $E_\br$ is bounded by a fixed
multiple of $da_R\sqrt{\widehat p}+d^2a_R^2\sqrt{\widehat p}+d^3a_R^3
+\tau y\sqrt{\widehat p}+\tau^2y^2+d\tau a_Ry+d^2\tau a_R^2y
+d^2|F_R(\br)|a_R+d^2F_R(\br)^2$; with the bounds on
$\widehat p$, $y$ and $a_R$ above, the largest term is
$O(\eps^4\ell_\eps^{-7})$, so $\|E_\br\|=o(\eps^4)$ on $|\eps\br|\ge c$.
Taking the
mean gives $w(\br)=dF_R(\br)+\widehat w(\br)+o(\eps^4)$; subtracting it
gives (i).

(ii) The same expansion of $b_0n_\br$ gives
$\overline{\zeta(\br)}=\omega(b_0n_\br)=\widehat\omega(b_0n_\br)
+d\,\widehat\omega(T_\br)+o(\eps^6)$ with
$T_\br=\sum_{k\notin\{0,\br\}}F_R(k-\br)b_kb_0b_\br$. Since one-site
transverse means vanish in $\widehat\omega$ and distinct sites commute,
$\widehat\omega(b_0n_\br)=\langle b_\br\widehat\Omega,X_\br\rangle$
and $\widehat\omega(T_\br)=\langle\Phi_{\br,\br}^\dagger
\widehat\Omega,X_\br\rangle$ exactly. The joint half-turn followed by
translation by $\br$ exchanges the two sites and reverses the sign of
every $b_j$, so $\omega(n_0b_\br)=-\omega(b_0n_\br)$; the analogous
representation of $\omega(n_0b_\br)$ uses $b_0$ and
$\Phi_{0,\br}^\dagger$. Half the difference of the two representations
is (ii).

(iii) Counting the two matchings of each unordered pair, and the
excluded coincident sites, gives the vacuum norms
\[
 \|\Phi_0^\dagger\Phi_\br^\dagger\Omega_0\|^2
 =S_R^2+C_R(\br)^2-2H_R(\br),\qquad
 \|(\Phi_{\br,\br}^\dagger-\Phi_{0,\br}^\dagger)\Omega_0\|^2
 =2[S_R-F_R(\br)^2-C_R(\br)],
\]
with $C_R(\br)=\sum_kF_R(k)F_R(k-\br)$ and
$H_R(\br)=\sum_kF_R(k)^2F_R(k-\br)^2$. The vacuum overlap of the two
vectors vanishes because they have different magnon numbers, and
$\widehat\omega(\Phi_{\br,\br}-\Phi_{0,\br})=0$. Each residual-state
quantity differs from its vacuum value by a zero-vacuum-diagonal local
sum with coefficient sum at most $Ca_R^4$, $Ca_R^2$ or $Ca_R^3$; by
Lemma~\ref{lem:pop-local-extraction} the differences are
$O(a_R^4\sqrt{\widehat p})=o(1)$. Also
$|\widehat\omega(\Phi_0^\dagger\Phi_\br^\dagger)|\le Ca_R^2
\sqrt{\widehat p}$ and $\|(b_\br-b_0)\widehat\Omega\|\le2\sqrt{\widehat
p}=o(\eps^2)$. By Eq.~\eqref{eq:green-tail}, $F_R(\br)$, $C_R(\br)$ and
$H_R(\br)$ tend to zero uniformly on $|\eps\br|\ge c$; finally
$S_R\to S_F=4n_*$ and $d/\eps^2\to\frac12$.
\end{proof}

\begin{lemma}[Gram minimization]\label{lem:ex-gram}
Let $0\le A<n$ and $\Lambda\ge0$, and let $\mathsf X,\mathsf Y,\mathsf Z$
be vectors in a complex Hilbert space with $\|\mathsf Z\|=n$,
$\|\mathsf Y\|^2=n/2$, $\langle\mathsf Y,\mathsf Z\rangle=0$ and
$\|\mathsf X\|^2\le\Lambda$. Then
\begin{equation}
 A^2+\|\mathsf X+\mathsf Z\|^2-\frac2{n-A}|\langle\mathsf Y,\mathsf X
 \rangle|^2\ge g_n(A,\Lambda)=
 \begin{cases}
 A^2+(n-\sqrt\Lambda)^2, & \Lambda\le(n-A)^2,\\[1mm]
 \dfrac{A}{n-A}\,(n^2-A^2-\Lambda), & \Lambda\ge(n-A)^2,
 \end{cases}
 \label{eq:ex-gn}
\end{equation}
and equality is attained. If $0<\delta<n$ and
$\Lambda<\delta(2n-\delta)$, then
$\inf_{0\le A\le n-\delta}g_n(A,\Lambda)>0$.
\end{lemma}

\begin{proof}
Let $\hat{\mathsf y}=\mathsf Y/\|\mathsf Y\|$,
$s=|\langle\hat{\mathsf y},\mathsf X\rangle|$, and let $t$ be the norm
of the component $\mathsf X_\perp$ of $\mathsf X$ orthogonal to
$\hat{\mathsf y}$, so $s^2+t^2\le\Lambda$. Since
$\mathsf Z\perp\hat{\mathsf y}$,
$\|\mathsf X+\mathsf Z\|^2=s^2+\|\mathsf X_\perp+\mathsf Z\|^2\ge
s^2+(n-t)^2$, and $2|\langle\mathsf Y,\mathsf X\rangle|^2/(n-A)
=ns^2/(n-A)$. The left side of Eq.~\eqref{eq:ex-gn} is therefore at
least $A^2+(n-t)^2-As^2/(n-A)\ge\varphi(t)=A^2+(n-t)^2
-A(\Lambda-t^2)/(n-A)$ for $0\le t\le\sqrt\Lambda$. The quadratic
$\varphi$ has $\varphi'(t)=2n(t-n+A)/(n-A)$, so its minimum on
$[0,\sqrt\Lambda]$ is at $t=\min\{\sqrt\Lambda,n-A\}$, which gives
$g_n$. Equality holds for $\mathsf Z=ne_1$,
$\mathsf Y=\sqrt{n/2}\,e_2$ and $\mathsf X=-te_1+\sqrt{\Lambda-t^2}\,e_2$
at that $t$. If $\Lambda<\delta(2n-\delta)<n^2$, the first branch is at
least $(n-\sqrt\Lambda)^2$; the second occurs only for
$A\ge n-\sqrt\Lambda$, where $A/(n-A)\ge(n-\sqrt\Lambda)/\sqrt\Lambda$
and $n^2-A^2-\Lambda\ge\delta(2n-\delta)-\Lambda$.
\end{proof}

\begin{proposition}[Conditional complete-PT positivity]
\label{prop:ex-conditional}
Fix $c>0$ and $0<\delta<n_*$, let $\mathcal D_\eps\subset
\mathcal X_{c,\delta}$ (with $h$ and the state allowed to vary), and
suppose
\begin{equation}
 \limsup_{\eps\downarrow0}\ \sup_{\mathcal D_\eps}\
 \eps^{-8}\mathcal V_\br\le\Lambda<\infty .
 \label{eq:ex-variance-premise}
\end{equation}
Then
\begin{equation}
 \liminf_{\eps\downarrow0}\ \inf_{\mathcal D_\eps}
 \Bigl[\eps^{-8}D_{\rm PT}-g_{n_*}\bigl(|A_h(\eps\br)|,\Lambda\bigr)
 \Bigr]\ge0 .
 \label{eq:ex-conditional}
\end{equation}
If $g_{n_*}(|A_h(\eps\br)|,\Lambda)\ge m>0$ on $\mathcal D_\eps$, then
$\liminf_{\eps\downarrow0}\inf_{\mathcal D_\eps}\eps^{-8}
\lambda_{\min}(\rho_{0\br}^{T_2})\ge m$. In particular,
$\Lambda<\delta(2n_*-\delta)$ implies strict PPT throughout
$\mathcal D_\eps$ for all sufficiently small $\eps$, including
macroscopic nodes and unbounded tails contained in $\mathcal D_\eps$.
\end{proposition}

\begin{proof}
Put $x_\br=\eps^{-4}X_\br$, $y_\br=\eps^{-2}Y_\br$ and
$z_\br=\eps^{-4}\mathcal Z_\br$; by the premise $\|x_\br\|^2\le
\Lambda+o(1)$. Because $\|x_\br\|$ is bounded, the identity in
Proposition~\ref{prop:ex-gram}(i) may be squared:
$q/\eps^8=|w/\eps^4|^2+\|x_\br+z_\br\|^2+o(1)$. Part (ii) gives
$|\zeta|/\eps^6=|\langle y_\br,x_\br\rangle|+o(1)$, so $\zeta=O(\eps^6)$
and $q=O(\eps^8)$. On $\mathcal X_{c,\delta}$,
$b-W=(n_*-A+o(1))\eps^4\ge\frac\delta2\eps^4$ with
$A=|A_h(\eps\br)|$, so $\chi\le2|\zeta|^2/(b-W)=O(\eps^8)$,
$a-\chi\to1$, $a+q-2v\to1$, $aq=q+O(\eps^{12})$ and $v^2=o(\eps^8)$.
Hence
\[
 \eps^{-8}D_{\rm PT}\ge A^2+\|x_\br+z_\br\|^2
 -\frac2{n_*-A}|\langle y_\br,x_\br\rangle|^2+o(1).
\]
By (iii), normalizing $y_\br$, removing from $z_\br$ its component
along $y_\br$ and normalizing it, and scaling $x_\br$ down to norm at
most $\sqrt\Lambda$ change the right side by $o(1)$, uniformly, since
all vectors are bounded and $n_*-A\ge\delta$. Lemma~\ref{lem:ex-gram}
and the continuity of $g_{n_*}$ give Eq.~\eqref{eq:ex-conditional}.

For the eigenvalue, $\Tr\mathsf S=1+o(1)$ and $\det\mathsf S=D_{\rm PT}$,
so $\lambda_{\min}(\mathsf S)\ge(m-o(1))\eps^8$. In the congruence
\eqref{eq:ex-congruence} the off-diagonal block has norm
$\le2|\zeta|/(b-W)=O(\eps^2)$, so $\|\mathsf L^{-1}\|=1+O(\eps^2)$,
while $\lambda_{\min}(M)=b-W\ge\frac\delta2\eps^4$. Hence
$\lambda_{\min}(\rho^{T_2})\ge(m-o(1))\eps^8$. The last statement is
Lemma~\ref{lem:ex-gram}.
\end{proof}

\paragraph{An explicit off-axis sector.} Let $h=2$, so $m_2=\sqrt3$,
and
\begin{equation}
 \mathcal K=\Bigl\{(\rho,\theta):\ \tfrac{27}{20}\le\rho\le\tfrac32,\
 \arctan\tfrac9{20}\le\theta\le\arctan\tfrac{11}{20}\Bigr\}.
 \label{eq:ex-sector}
\end{equation}
On $\mathcal K$, $A_2$ decreases in $\rho$ and in $\theta$, so its
extremes are at the corners:
$4.6571\times10^{-3}\le A_2\le8.3325\times10^{-3}$. Thus
$23/5000<A_2<n_*-23/10000$ on $\mathcal K$, which keeps a fixed
distance from the axes and the nodes and lies outside $\mathcal O_2$,
since $\rho_E(2,\theta)\le1.2500$ on its angular range. With
$\delta=23/10000$, Proposition~\ref{prop:ex-conditional} applies on
$\mathcal K$ with the sufficient constant
\begin{equation}
 \Lambda_{\mathcal K}=\delta(2n_*-\delta)
 =\frac{23}{10000}\Bigl(\frac{\pi}{128(\pi-2)}-\frac{23}{10000}\Bigr)
 =4.4158893015\times10^{-5}.
 \label{eq:ex-sector-constant}
\end{equation}
For example, $\Lambda=4\times10^{-5}$ would give
$\lambda_{\min}(\rho^{T_2})\ge(1.5279\times10^{-5}-o(1))\eps^8$ on
$\mathcal K$. A Cauchy-Schwarz bound on $\zeta$ that ignores the joint
Gram structure of Proposition~\ref{prop:ex-gram} requires instead
$\Lambda<\delta A_-^2/(2n_*)=2.2637\times10^{-6}$ with $A_-=23/5000$, a
constant smaller by a factor $19.51$. Proposition~\ref{prop:ex-variance}
gives only $\eps^{-8}\mathcal V_\br\le C\eps^{-(10-2\vartheta)/3}
\ell_\eps^{1+\vartheta}$, so the premise~\eqref{eq:ex-variance-premise}
is not established on $\mathcal K$ or elsewhere.

\subsection{Open problems}
\label{ss:ex-open}

The following statements are not proved. They refer to the minimizers
of Sec.~\ref{sm:conventions}, without the additional reflection unless
stated, uniformly in $h\in\cI$, and to the limit order stated there.

\begin{enumerate}
\item[(P1)] \emph{Off-axis exterior.} Strict PPT, for all sufficiently
small $\eps$, on a positive-area set of scaled positions outside
$\mathcal O_h$ and away from the axes, for instance on $\mathcal K$ at
$h=2$. Known: the anomalous block is uniformly positive,
$\Neg\le C\eps^{16/3}\ell_\eps^2[|\cos2\theta|+\varpi_\eps]$
(Theorem~\ref{thm:ex-nodes}), strict PPT is equivalent to
$D_{\rm PT}>0$ (Lemma~\ref{lem:ex-schur}), the parity-even part
$\rho^X$ is strictly separable (Theorem~\ref{thm:ex-parity}), and strict
PPT of $\rho$ follows from the bound~\eqref{eq:ex-quadrature-target} on
the single correlation $y$ (Proposition~\ref{prop:ex-quadrature}). Off the symmetry rays
neither the joint quarter-turn nor $\mathsf A_x$ fixes the phase of
$w\zeta^2$, which decides the phase cost~\eqref{eq:ex-phase-cost}, and
the bound $|\zeta|\le C\eps^{14/3}\ell_\eps$ is far from the precision
$o(\eps^6)$ required by Theorem~\ref{thm:rdmcriterion};
Proposition~\ref{prop:ex-counter}(i) shows that $\zeta=O(\eps^6)$ is
not enough in general.

\item[(P2)] \emph{Far tail.} PPT for all pairs with
$\eps|\br|\ge R_E(h)+\eta$, in every direction and without an upper
limit on $|\br|$. By Lemma~\ref{lem:ex-far} this is the statement
$D_{\rm PT}\ge0$ on that region. It is proved only on the axes, for
$\mathsf A_x$-invariant minimizers, and only for
$\eps|\br|\le\beta\log(1/\eps)$ (Theorem~\ref{thm:ex-axial}). The
sufficient tail condition~\eqref{eq:tailcondition} needs a lower bound
$q\ge c\,p^2$ on the joint population, whereas only $q=o(p)$ is known.

\item[(P3)] \emph{Nodes.} The sign of the smallest partial-transpose
eigenvalue on the diagonal rays. Known: on the macroscopic diagonals
$\Neg\le C\eps^{214/39}\ell_\eps^{28/13}$ (Theorem~\ref{thm:ex-nodes}),
and for $\mathsf A_x$-invariant minimizers the sign is that of
$D_{\rm diag}$ (Lemma~\ref{lem:ex-diagonal}). The known data do not
decide this sign: with $p=t$ and $w=\zeta=0$, which satisfy the phase
constraints of Lemma~\ref{lem:ex-phases}(b), the choices
$(q,v)=(t^2,0)$, $(0,0)$ and $(t^4,t^{3/2})$ give a product state
(strictly PPT), a classical state (PPT with a zero eigenvalue) and a
positive-definite NPT state (Proposition~\ref{prop:ex-counter}(ii));
with $t=n_*\eps^4$ all three reproduce the known nodal asymptotics. At
microscopic and intermediate diagonal separations the massless kernel
$F$ vanishes as well (Sec.~\ref{sm:green}), and the sign is decided by
higher-order coefficients that are not available.

\item[(P4)] \emph{Zero-threshold farthest distance.} For a primitive
lattice vector $\mathbf a$ with $\theta=\arg\mathbf a$, let
$R^{\mathbf a}_{\rm NPT}(\eps)=\sup\{|k\mathbf a|:k\in\mathbb N,\
\rho_{0,k\mathbf a}\ \text{NPT}\}$, and let $R^{\rm all}_{\rm NPT}(\eps)$ be the
supremum of $|\br|$ over all NPT pairs; both may be infinite.
Theorem~\ref{thm:massscale} gives
$\liminf_{\eps\downarrow0}\eps R^{\mathbf a}_{\rm NPT}\ge
\rho_E(h,\theta)$ and $\liminf_{\eps\downarrow0}\eps R^{\rm all}_{\rm NPT}\ge
R_E(h)$, and Theorem~\ref{thm:threshold} gives the corresponding limits
for thresholds that vanish slowly. The upper bounds
$\limsup\eps R^{\mathbf a}_{\rm NPT}\le\rho_E(h,\theta)$ and
$\limsup\eps R^{\rm all}_{\rm NPT}\le R_E(h)$ are open, and finiteness of
$R^{\rm all}_{\rm NPT}(\eps)$ at fixed $\eps$ is not known. A clustering bound
$\|\rho_{0\br}-\rho_1\otimes\rho_1\|_F\le C_\eps e^{-|\br|/\xi_\eps}$,
with $\rho_1=\diag(1-p,p)$, would give PPT for
$|\br|>\xi_\eps\log(C_\eps/p^2)$, because partial transposition
preserves the Frobenius norm and
$\lambda_{\min}[(\rho_1\otimes\rho_1)^{T_2}]=p^2$. No such bound is
proved for this model, and a generic correlation length
$\xi_\eps\sim\eps^{-2}$ would give a cutoff of order
$\eps^{-2}\log(1/\eps)$ rather than the scale $\eps^{-1}$.
\end{enumerate}

\paragraph{The missing estimate.} Each of the following would close
these problems on the sets where it holds.

\emph{(E1) Centered pair variance.} For every $c>0$,
\begin{equation}
 \sup_{h\in\cI,\ \omega,\ |\eps\br|\ge c}\mathcal V_\br=o(\eps^8).
 \label{eq:ex-target}
\end{equation}
Proposition~\ref{prop:ex-gram} then gives $\zeta=o(\eps^6)$ and
$q=(n_*^2+|A_h|^2+o(1))\eps^8$, and Proposition~\ref{prop:ex-conditional}
with $\Lambda=0$ gives $\lambda_{\min}(\rho^{T_2})\ge
(n_*^2+|A_h(\eps\br)|^2-o(1))\eps^8$ uniformly on every
$\mathcal X_{c,\delta}$. This settles (P1)--(P3) at macroscopic
separations, and together with the interior statement of
Theorem~\ref{thm:massscale} it gives
$\eps R^{\mathbf a}_{\rm NPT}\to\rho_E(h,\arg\mathbf a)$ and
$\eps R^{\rm all}_{\rm NPT}\to R_E(h)$. On a given fixed-margin set the bound
$\limsup\sup\eps^{-8}\mathcal V_\br<\delta(2n_*-\delta)$ suffices; on
$\mathcal K$ this is Eq.~\eqref{eq:ex-sector-constant}.

\emph{(E2) A signed source inequality.} The route to (E1) through the
ground-state property starts from the following inequality.

\begin{lemma}[Local passivity]\label{lem:ex-passivity}
Every minimizer $\omega$ of Sec.~\ref{sm:conventions} satisfies
$\omega(B^\dagger[H,B])\ge0$ for every local operator $B$. Hence
$\mathfrak K(A,B)=\omega(A^\dagger[H,B])$ is a positive-semidefinite
Hermitian form on local operators.
\end{lemma}

\begin{proof}
Let $B$ be supported in $S$, let $0\le t<\|B\|^{-2}$, and let
$\gamma_t(C)=K_tCK_t+tB^\dagger CB$ with $K_t=(I-tB^\dagger B)^{1/2}$.
This map is unital and completely positive. Apply it on the disjoint
translates of $S$ used in the proof of
Lemma~\ref{lem:stationarity-density} and average over cell translations
and the imposed symmetry group. The result is an admissible
competitor, whose energy density exceeds that of $\omega$ by
$M^{-2}\sum_{Z\cap S\ne\varnothing}\omega(\gamma_t(h_Z)-h_Z)\ge0$.
This vanishes at $t=0$, so its right derivative there,
$\omega(B^\dagger HB-\frac12\{B^\dagger B,H\})=\omega(B^\dagger[H,B])
+\frac12\omega([B^\dagger B,H])$, is nonnegative; the last term vanishes
by Lemma~\ref{lem:stationarity-density}. Stationarity applied to
$A^\dagger B$ shows that $\mathfrak K$ is Hermitian.
\end{proof}

Applied to the centered dressed pair operator, this inequality and the
exact spin algebra of the pair hops give a scalar inequality of the
form
\begin{equation}
 \bigl(m_\eps+\tfrac{a_\eps}2\cL_0\bigr)\mathcal V_\br\le\Sigma_\br ,
 \qquad m_\eps\ge c_\cI\,\eps^2,
 \label{eq:ex-hierarchy}
\end{equation}
on noncontact pairs, where $\cL_0$ is the lattice Laplacian in the
relative coordinate. The mass $m_\eps$ is what remains of $2\Delta_\eps$
after the coercive conditional-singlet part of the DM torque is
absorbed; its positivity uses $h>3/2$, and absorbing the whole
conditional-singlet term leaves $m_\eps\ge(2h-3)\eps^2$. The source
$\Sigma_\br$ collects signed terms: conditional three-point occupations
$\widehat\omega(n_0n_kn_\br)$ with $k$ adjacent to an endpoint, with
coefficient $(3\cos\eps-1)/4$; squared lattice gradients of
$\widehat w$; a coherent compensation term; a conditional DM flow; and
the dressing remainder. The derivation is not reproduced here. The
missing estimate is a bound
\begin{equation}
 \Sigma_\br\le\alpha_\eps\mathcal V_\br+f_\eps(\br),\qquad
 m_\eps-\alpha_\eps\ge c_1\eps^2,\qquad
 \sup_{|\eps\br|\ge c}(\mathcal G_\eps*f_{\eps,+})(\br)+B_\eps=o(\eps^8),
 \label{eq:ex-source}
\end{equation}
where $\mathcal G_\eps$ is the Green function of
$m_\eps-\alpha_\eps+\frac{a_\eps}2\cL_0$ and $B_\eps$ is the
contribution of the contact pairs. By the comparison principle it
implies Eq.~\eqref{eq:ex-target}; a uniform bound
$f_\eps=o(\eps^{10})$ suffices. The three-point terms cannot be
absorbed into the mass by a state-independent operator bound, since
their coefficient tends to $1/2$, and the gradient terms must be kept
together with the conditional singlet terms, with which they partly
cancel.

\emph{(E3) An off-axis phase-cost bound.} On a set where
$A_-\le|A_h|\le n_*-\delta$ with $A_->0$, a bound
\begin{equation}
 \chi-\kappa\le(1-\eta_1)|w|^2+o(\eps^8),\qquad 0<\eta_1\le1,
 \label{eq:ex-phase-target}
\end{equation}
suffices. By Eq.~\eqref{eq:ex-identity}, $E_{\rm ph}\ge0$,
$\chi\ge0$ and $v^2=o(\eps^8)$, it gives
$D_{\rm PT}\ge(\eta_1A_-^2-o(1))\eps^8$; since $a-\chi\to1$ by
Eq.~\eqref{eq:ex-input-odd}, Lemma~\ref{lem:ex-schur}(a) gives strict
PPT there. It closes (P1) but not (P2) or (P3), where
$|w|/\eps^4\to0$. By Proposition~\ref{prop:ex-quadrature}(a) the phase
cost involves only $y^2$ with a positive coefficient, so (E3) follows
from the bound~\eqref{eq:ex-quadrature-target} on the single real
correlation $y$; the quadrature $x$ never needs to be small.

\emph{(E4) A lower bound on the joint population.} On the diagonals and
in the far tail $W$ is not bounded below, positivity of $\rho$ gives no
margin, and strict PPT requires $q>\chi+(v-\chi)^2/(a-\chi)$. This
needs a lower bound $q\ge c\,\eps^8$, which is open even for the
parity-even part $\rho^X$, together with the odd-coherence bound
$\chi\le(1-\vartheta)c\,\eps^8$; both follow from (E1).

\paragraph{Formal evidence and a conditional improvement.} A leading
linked-cluster evaluation of the odd coherence, which is formal and not
a proof, gives $\zeta^*\propto s\,d\,(\partial_x-i\partial_y)w$ at the
scale $1/\eps$. Since $s\sim\eps$, $d\sim\eps^2$, and each derivative
of $w$ costs a factor $\eps$ there, this is of order $\eps^8$, two
powers below the threshold~\eqref{eq:ex-quadrature-target}, and the same
evaluation reproduces the odd coherence of exact diagonalization on
small tori, where it is of order $\eps^5$ at microscopic separations.
Separately, if the kernel budgets used in Proposition~\ref{prop:ex-variance}
are inserted in the proof of Proposition~\ref{prop:th-exterior}, the
exterior bounds improve to $|\zeta|\le C\eps^{(14+2\eta)/3}
\ell_\eps^{1+\eta}$ and $\Neg\le C\eps^{(16+4\eta)/3}\ell_\eps^{2+2\eta}$,
and the window of Theorem~\ref{thm:threshold} extends to
$\kappa_\eps/(\eps^{(4+4\eta)/3}\ell_\eps^{2+2\eta})\to\infty$, for a
small fixed $\eta>0$. Because the proof of that proposition is only
outlined here, the theorems above do not use this improvement.

\paragraph{Numerical evidence on cylinders.} The criterion of
Proposition~\ref{prop:ex-quadrature} was evaluated on the measured
two-site density matrices of the converged cylinder runs, after a local
rotation that removes the one-site transverse means caused by the open
ends; the rotation leaves the spectrum of the partial transpose
unchanged. At the $32$ off-axis pairs with $0.20\le\eps\le0.30$ that lie
outside the entangled region of the pair problem on the same cylinder,
the hypotheses $b>W>|v|$ and $a>\chi$ hold. At $29$ of them the ratio
of $y^2$ to the right side of Eq.~\eqref{eq:ex-quadrature-condition} is
below one (median $0.08$), and all $29$ have a positive partial
transpose with smallest eigenvalue between $0.64$ and $1.41$ times
$n_*^2\eps^8$. At $462$ axial pairs beyond $\rho_E$ the ratio is at most
$0.023$ and all are PPT. The exceptions are short diagonal pairs. At
$\br=(1,1)$ on the cylinder of circumference $8$ at $\eps=0.25$ the
ratio is $70$ and the measured state has a partial-transpose eigenvalue
$-17\,n_*^2\eps^8$, although its parity-even part is PPT: the
entanglement is carried by the odd coherence alone. Cylinders break the
diagonal mirror of the plane, so this does not decide (P3), but it
shows that the mechanism of Proposition~\ref{prop:ex-twosite}(ii) occurs
in ground states at microscopic separations. The changes between bond
dimensions, initial states and positions are smaller than these
margins, by a factor $2.3$ in the least favourable case. The data cover
three angles and carry no certified error bound; the analysis is in the
ancillary folder \texttt{mass\_scale/cylinder\_dmrg\_offaxis/exterior\_criterion/}.

\section{Symmetry selection and other lattices}
\label{sm:symmetry}

\subsection{Rotational character of a pair}
Let a joint symmetry with $\phi=2\pi/n$ act as
$U_\phi S_i^+U_\phi^\dagger=e^{-i\phi}S_{R_\phi i}^+$.
In a translation- and $U_\phi$-invariant state,
\begin{equation}
 w(R_\phi\br)=e^{2i\phi}w(\br),\qquad w(-\br)=w(\br).
 \label{eq:rotation}
\end{equation}
The first identity follows by applying $U_\phi$ inside the expectation;
the second follows by translation and commutation at distinct sites.
If the pair kernel commutes with the joint rotation, its inverse in a
fixed source character obeys the same rule. Every angular term
$e^{il\theta}$ in the scalar exchange-even far field must therefore satisfy
\begin{equation}
 \boxed{l\equiv2\pmod n,\qquad l\text{ even}.}
 \label{eq:selection}
\end{equation}
The monopole is forbidden for $n\ge3$. For a lattice with a basis,
Eq.~\eqref{eq:selection} applies to the exchange-even acoustic
projection; individual entries instead satisfy
$w_{\alpha\beta}(-\br)=w_{\beta\alpha}(\br)$ and can be permuted by
the point rotation.

For $C_4$, both $l=2$ and $l=-2$ are allowed. A real leading kernel
requires conjugate coefficients for these two harmonics. The square
source is additionally reflection even and antisymmetric under
$x\leftrightarrow y$, fixing $\cos2\theta$.
$C_4$ alone would also permit a single winding harmonic and does not
force nodes. For $C_3$ or $C_6$, evenness leaves
$l=\ldots,-10,-4,2,8,\ldots$; the unique lowest absolute harmonic
is $+2$. A nonzero leading coefficient then has phase $e^{2i\theta}$
and isotropic magnitude. Symmetry does not itself guarantee that
coefficient is nonzero; the following kernels establish it explicitly.

\subsection{Triangular lattice: source, contact solution, and coefficient}
Use the physical vectors
$a_1=(1,0)$, $a_2=(1/2,\sqrt3/2)$, $a_3=a_2-a_1$,
with angles $\theta_j=(j-1)\pi/3$. Count each of the three positive
bond directions once in
\begin{equation}
 H_\triangle/J=\sum_i\sum_{j=1}^3
 \left[\frac14-\mathbf S_i^T
 e^{\eps[\widehat a_j]_\times}\mathbf S_{i+a_j}\right]
 -h\eps^2\sum_iS_i^z.
 \label{eq:Htriangle}
\end{equation}
The pair-creation coefficient on a bond is
$-Jd e^{2i\theta_j}/4$. Set $\omega=e^{2\pi i/3}$ and define
the even shell function $q_\triangle(\pm a_1)=1$,
$q_\triangle(\pm a_2)=\omega$,
$q_\triangle(\pm a_3)=\omega^2$; the source is
$f_\triangle=-q_\triangle/4$.
For reciprocal coordinates $x=\bk\cdot a_1$, $y=\bk\cdot a_2$,
\begin{equation}
 \lambda_\triangle=6-2\cos x-2\cos y-2\cos(y-x),\qquad
 \widehat q_\triangle=2[\cos x+\omega\cos y+\omega^2\cos(y-x)].
 \label{eq:trisymbol}
\end{equation}
The full-plane inverse $G=L_{0,\triangle}^{-1}$ is used through
zero-sum sources or potential differences
$\mathcal P(\br)=G(0)-G(\br)$. Rotational covariance restricts
$Gq_\triangle$ on the six-site shell to $k_\triangle q_\triangle$.
Writing $\mathcal P_1,\mathcal P_{\sqrt3},\mathcal P_2$ for its
three required potential differences,
\begin{align}
 k_\triangle&=\mathcal P_1+\mathcal P_{\sqrt3}-\mathcal P_2,
 &\mathcal P_1&=1/6,
 &\mathcal P_2+2\mathcal P_{\sqrt3}&=2/3,\label{eq:tripotential}\\
 \mathcal P_2&=\frac4\pi\int_0^1
 \frac{u\sqrt{1-u^2}}{\sqrt{3+u^2}}\,\dd u
 =\frac43-\frac{2\sqrt3}{\pi},
 &k_\triangle&=\frac{3\sqrt3}{\pi}-\frac32.\nonumber
\end{align}
The first two relations are the origin and nearest-neighbor Laplace
equations; integrating one momentum gives the displayed elementary
integral. These potential differences are half the electrical
two-point resistance convention~\cite{Cserti}.

At spin $1/2$, the punctured operator is
$L_{p,\triangle}=L_{0,\triangle}-P_\nn$ in this character,
with degree five at the six neighbors. If
$v_\triangle=L_{0,\triangle}^{-1}f_\triangle$, the exact response is
\begin{equation}
 F_\triangle=-\frac{v_\triangle}{1-k_\triangle},\qquad
 F_\triangle(a_1)=\frac{k_\triangle}{4(1-k_\triangle)},\qquad
 w_2=dF_\triangle\quad\hbox{at zero pair mass}.
 \label{eq:triF}
\end{equation}
The unit-cell area is $\mathcal A_c=\sqrt3/2$; thus the physical
Fourier measure is $\mathcal A_c\dd^2k/(2\pi)^2$.
At small physical momentum,
\begin{equation}
 \lambda_\triangle=\frac32|\bk|^2+O(k^4),\qquad
 \widehat q_\triangle=-\frac34(k_x+ik_y)^2+O(k^4).
 \label{eq:trisoft}
\end{equation}
Using
$\mathcal F^{-1}[(k_x+ik_y)^2/k^2]=-e^{2i\theta}/(\pi r^2)$
away from the origin gives
\begin{equation}
 F_\triangle(\br)=
 \frac{\sqrt3}{8(5\pi-6\sqrt3)}\frac{e^{2i\theta}}{r^2}
 +O(r^{-4}),\qquad
 C_\triangle=\frac{\sqrt3}{16(5\pi-6\sqrt3)},
 \label{eq:triC}
\end{equation}
where $w_2\sim\eps^2C_\triangle e^{2i\theta}/r^2$.

The finite-angle projected block and its contact solution are
\begin{equation}
 A_{2,\triangle}=a_\eps L_{p,\triangle}
 +(2h\eps^2-3d)I+\frac d2P_\nn,\qquad
 w_{2,\triangle}=\frac{d g_\sigma}{4(a_\eps-ck_\sigma)},
 \label{eq:trimassive}
\end{equation}
where $g_\sigma=(L_{0,\triangle}+\sigma)^{-1}q_\triangle$,
$k_\sigma=g_\sigma(a_1)$, and
$\sigma=(2h\eps^2-3d)/a_\eps$.
For compact field sets above $3/4$, the proof of
Theorem~\ref{thm:massive} applies with the triangular symbol, yielding
\begin{equation}
 \eps^{-4}w_{2,\triangle}(\rho/\eps,\theta)
 \longrightarrow\frac{C_\triangle m_\triangle^2}{2}
 K_2(m_\triangle\rho)e^{2i\theta},\qquad
 m_\triangle^2=\frac43(h-3/4).
 \label{eq:triA}
\end{equation}
The notation on the left means lattice sequences tending to the
specified scaled displacement.

The pair population is also explicit:
\begin{equation}
 n_\triangle^{\rm pair}
 =\frac{1}{64(1-k_\triangle)^2}
 \int\frac{|\widehat q_\triangle|^2}{\lambda_\triangle^2}
            \frac{\dd x\,\dd y}{(2\pi)^2}
 =\frac{\pi(2\pi-3\sqrt3)}{32(5\pi-6\sqrt3)^2}.
 \label{eq:trin}
\end{equation}
For an explicit one-dimensional evaluation of the integral, put
$t=\cos x$ after integrating over $y-x/2$:
\begin{equation}
 I_\triangle=\frac1\pi\int_{-1}^1
 \frac{-49+112t-29t^2+2t^3+
 24(1-t)\sqrt{(1-t)(7-t)}}
 {2(7-t)^2(1+t)\sqrt{1-t^2}}\,\dd t
 =1-\frac{3\sqrt3}{2\pi}.
 \label{eq:triIntegral}
\end{equation}
The endpoint singularities of the measure are integrable; the
quotient before $1/\sqrt{1-t^2}$ has finite limiting values.
Under the RDM comparison with population coefficient
$n_\triangle^{\rm pair}$, the boundary is the circle
\begin{equation}
 K_2(m_\triangle\rho_E)
 =\frac{2n_\triangle^{\rm pair}}{C_\triangle m_\triangle^2}.
 \label{eq:tricircle}
\end{equation}

\subsection{Honeycomb lattice: four channels and physical phases}
Take an A site at zero and its B neighbors at
$\delta_0=(1,0)$,
$\delta_1=(-1/2,\sqrt3/2)$,
$\delta_2=(-1/2,-\sqrt3/2)$.
Orient each bond A to B and rotate about its physical bond axis.
Use $t_1=\delta_0-\delta_1$, $t_2=\delta_0-\delta_2$ as Bravais
vectors, with cell area $3\sqrt3/2$.
The source phases are $1,\omega^2,\omega$.
With $x=\bk\cdot t_1$, $y=\bk\cdot t_2$, define
\begin{equation}
 g_+=1+e^{ix}+e^{iy},\quad g_-=g_+^*,\quad
 D_{\rm hc}(\bk)=\begin{pmatrix}3&-g_+\\-g_-&3\end{pmatrix},
 \quad
 L_{0,\rm hc}(\bk)=\frac{D_{\rm hc}(\bk)\otimes I+
                       I\otimes D_{\rm hc}(-\bk)}2.
 \label{eq:hcL}
\end{equation}
All four channels AA, AB, BA, BB must be retained.
Let $f_{AB}(\bk)=\mathfrak f_\bk
=-(1+\omega^2e^{ix}+\omega e^{iy})/4$,
$f_{BA}(\bk)=\mathfrak f_{-\bk}$, and $f_{AA}=f_{BB}=0$.
For $v=L_{0,\rm hc}^{-1}f$, eliminating AA and BB gives
\begin{equation}
 v_{AB}(\bk)=\frac{\mathfrak f_\bk}{6}
 +\frac{9\mathfrak f_\bk+g_+^2\mathfrak f_{-\bk}}
              {6(9-|g_+|^2)},\qquad
 v_{BA}(\bk)=v_{AB}(-\bk),\qquad
 v_{AA}=v_{BB}=\frac{g_-v_{AB}+g_+v_{BA}}6.
 \label{eq:hcEliminate}
\end{equation}
This elimination includes the optical channels; discarding them before
inverting would change the source's acoustic quadrupole coefficient.
The denominator $9-|g_+|^2$ is the triangular symbol of
Eq.~\eqref{eq:trisymbol}. Restricting the inverse to the three AB
contacts and using Eq.~\eqref{eq:tripotential} gives
\begin{equation}
 k_{\rm hc}=\frac16+
 \frac{11\mathcal P_1-2\mathcal P_2-\mathcal P_{\sqrt3}}6
 =\frac13+\frac{k_\triangle}{6}
 =\frac{\pi+6\sqrt3}{12\pi},\qquad
 F_{\rm hc}=-\frac{v}{1-k_{\rm hc}}.
 \label{eq:hcK}
\end{equation}
The coincident AA and BB amplitudes vanish by the nontrivial local
rotational character, as required by the spin-$1/2$ constraint.

The cell-coordinate AB displacement is $R$ but its physical
displacement is $\br=R+\delta_0$. Thus its physical Fourier multiplier
is $e^{-i\bk\cdot\delta_0}v_{AB}(\bk)$.
In Eq.~\eqref{eq:hcEliminate}, the linear terms of the singular numerator
cancel, and its quadratic terms are
\begin{equation}
 9\mathfrak f_\bk+g_+^2\mathfrak f_{-\bk}
 =\frac{27}{16}(k_x+ik_y)^2+O(k^3),\qquad
 9-|g_+|^2=\frac92|\bk|^2+O(k^4).
 \label{eq:hcsoft}
\end{equation}
Consequently the nonanalytic leading part of each physical-channel
multiplier $v$ is $(k_x+ik_y)^2/(16|\bk|^2)$.
Using the cell area and Eq.~\eqref{eq:hcK} yields
\begin{equation}
 F_{\alpha\beta}(\br)=
 \frac{9\sqrt3}{8(11\pi-6\sqrt3)}\frac{e^{2i\theta}}{r^2}
 +o(r^{-2}),\qquad
 C_{\rm hc}=\frac{9\sqrt3}{16(11\pi-6\sqrt3)},
 \quad\alpha,\beta\in\{A,B\}.
 \label{eq:hcC}
\end{equation}
Its nearest AB value is $k_{\rm hc}/[4(1-k_{\rm hc})]$.
The acoustic relative-Laplacian coefficient is $3/4$, giving
$m_{\rm hc}^2=(8h-3)/3$ at spin $1/2$.
The isotropic leading magnitude would yield a circular threshold with
a scalar depletion coefficient, but no honeycomb population coefficient
or numerical radius is assigned here.

\subsection{Exact spin-\texorpdfstring{$S$}{S} interaction enhancement}
For
$H_{\rm FM}=J\sum_{\langle ij\rangle}(S^2-\mathbf S_i\cdot\mathbf S_j)$,
use normalized separated states
$|ij\rangle=S_i^-S_j^-|\Omega\rangle/(2S)$ and, when $S\ge1$,
coincident states
$|ii\rangle=(S_i^-)^2|\Omega\rangle/[2\sqrt{S(2S-1)}]$.
In this subsection $n_i=S-S_i^z$.
The exact occupation-basis diagonal and one-magnon-transfer matrix
elements on a bond are
\begin{equation}
 J[S(n_i+n_j)-n_i n_j],\qquad
 -\frac J2\sqrt{n_i(2S-n_i+1)(n_j+1)(2S-n_j)}.
 \label{eq:spinSmatrix}
\end{equation}
In particular, separated hopping is $-JS$ and the coupling from
$(1,1)$ to $(0,2)$ is $-J\sqrt{S(2S-1)}$.
These are the interacting two-spin-wave matrix elements~\cite{Wortis}.

On a regular Bravais lattice of coordination $z$, let $A(\br)$ and
$B$ be the separated and coincident translation-reduced amplitudes.
Their equations are
\begin{align}
 (\mathsf H_{2,S}A)(\br)
 &=(2JSz-J\mathbf1_\nn(\br))A(\br)
 -2JS\!\sum_{\substack{\delta\in\Nop_0\\\br+\delta\ne0}}
 A(\br+\delta)
 -2J\sqrt{S(2S-1)}\,\mathbf1_\nn(\br)B,\label{eq:spinSequations}\\
 (\mathsf H_{2,S}B)&=2JSzB
 -J\sqrt{S(2S-1)}\sum_{\delta\in\Nop_0}A(\delta).\nonumber
\end{align}
The reduced norm is proportional to
$\frac12\sum_{\br\ne0}|A(\br)|^2+|B|^2$, explaining the unequal
coupling factors. In the nontrivial quadrupole character, rotational
covariance forces $B=0$ and $\sum_\delta A(\delta)=0$.
On honeycomb the same argument applies separately to the coincident
AA and BB channels. The source sector therefore reduces to
\begin{equation}
 \mathsf H_{2,S}=2JS L_0-JP_\nn.
 \label{eq:spinSoperator}
\end{equation}
For a bond source $\Gamma=Jdf$, the source in normalized pair
coordinates is $2SJdf$. Set $v=L_0^{-1}f$ and suppose its restriction
to the single contact-shell character is $P_\nn v=k_\Lambda f$.
The resolvent identity gives
\begin{equation}
 A_S=-\frac{d v}{1-k_\Lambda/(2S)},\qquad
 w_S=2SA_S,\qquad
 \boxed{w_S=\mathcal R_\Lambda(S)w_{\rm LSW},\quad
 \mathcal R_\Lambda(S)=\left[1-\frac{k_\Lambda}{2S}\right]^{-1}.}
 \label{eq:enhance}
\end{equation}
Here $w_{\rm LSW}=-2Sd v$ is the linear-spin-wave response.
Equation~\eqref{eq:enhance} is a pointwise massless response identity,
including common zeros, for this single-shell source. A finite scalar
mass replaces $k_\Lambda$ by its massive value; additional shells or
multiple contact channels require a matrix inverse.

\begin{table}[htb]
\caption{Exact spin-$1/2$ response data. The coefficient convention is
$w_2\sim\eps^2 C_\Lambda\mathcal Y_\Lambda(\theta)/r^2$.}
\label{tab:lattices}
\begin{ruledtabular}
\begin{tabular}{lcccc}
Lattice & $k_\Lambda$ & $C_\Lambda$ & $\mathcal R_\Lambda(1/2)$ &
$\mathcal Y_\Lambda$ \\
\hline
Square & $4/\pi-1$ & $1/[16(\pi-2)]$ & $1.375969197$ & $\cos2\theta$ \\
Triangular & $3\sqrt3/\pi-3/2$ & $\sqrt3/[16(5\pi-6\sqrt3)]$ &
$1.182014495$ & $e^{2i\theta}$ \\
Honeycomb & $(\pi+6\sqrt3)/(12\pi)$ & $9\sqrt3/[16(11\pi-6\sqrt3)]$ &
$1.560057002$ & $e^{2i\theta}$
\end{tabular}
\end{ruledtabular}
\end{table}

For general $S$, the vacuum-to-$|11\rangle$ coherence in the local
$|S,S\rangle,|S,S-1\rangle$ basis is $w/(2S)$ at leading pair order.
If its two single-flip probabilities are equal, the associated
negative principal-block contribution is
$(|w|/(2S)-P_{10})_+$; with unequal probabilities it is
\begin{equation}
 \frac12\left[
 \sqrt{(P_{10}-P_{01})^2+4|w/(2S)|^2}-P_{10}-P_{01}\right]_+.
 \label{eq:spinSNeg}
\end{equation}
Equality with the full negativity also requires control of other
local-occupation sectors. At fixed separation in the quadrupolar
dilute pair state, $\Neg=|A_S|+o(\eps^2)$.
The enhancement comes from the Heisenberg contact attraction and the
exact spin matrix elements: deleting the coincident site alone leaves
the free quadrupole, already zero there, unchanged.

The triangular and honeycomb results above are exact pair-kernel
statements. Their transfer to thermodynamic spin-state negativity
requires $w-w_2=o(\eps^2/r^2)$ and negligible population in the
massless window, or the full RDM and exterior estimates at the mass
scale. In particular, neither lattice symmetry nor the spin-$S$
resolvent identity supplies those many-body bounds.

\section{Numerical methods and convergence}
\label{sm:numerics}

\subsection{Massive relative-coordinate solver}
The square-lattice calculation evaluates Eq.~\eqref{eq:rankone}
by discrete Fourier inversion on periodic $L\times L$ lattices.
For $\eps>0$ the denominator has positive mass; the numerator vanishes
at $\bk=0$, whose contribution is set to zero exactly. At zero mass
the constant Fourier mode is removed before inversion. The primary
grid uses $L=801$, $h=2,3$, and
$\eps=0.3,0.2,0.1,0.05,0.025,0.0125$.
For each solution we compute
\begin{equation}
 p_2=\sum_{\br\ne0}|w_2(\br)|^2,\qquad
 \Nop_{\rm pair}(\br)=(|w_2(\br)|-p_2)_+,
 \label{eq:pairnumerics}
\end{equation}
and the relative residual $\|A_2w_2+df\|_2/\|df\|_2$.
The quantity $\Nop_{\rm pair}$ is a population-subtracted pair
prediction, not the negativity of a full many-body RDM.

An independent sparse retained-edge matrix has off-diagonal entries
$-a_\eps$ on retained edges and diagonal
$a_\eps\deg(\br)+\mu_\eps+(d/2)\mathbf1_\nn(\br)$.
At $L=801$, $\eps=0.05$, $h=2$, its $641600$ active sites are
solved by conjugate gradients with relative tolerance $2\times10^{-12}$
and zero absolute tolerance. A full-plane massive Fourier inverse
serves only as preconditioner. The measured equation residual is
$7.23\times10^{-16}$ and the relative difference from the Fourier
solution is $1.51\times10^{-15}$. One preconditioned step suffices
in this source sector because the defect is rank one.
Across the twelve Fourier runs the largest relative residual is
$9.16\times10^{-16}$.

Finite-size error is checked independently of solver residual.
At $\eps=0.0125$, $h=2$, $r=80$ on the axis, the amplitudes for
$L=401,801,1601$ are respectively
$(7.83487625,7.82562489,7.82562402)\times10^{-10}$.
The last doubling changes the result by $1.11\times10^{-7}$ relatively.
The population and axial threshold converge as shown in
Table~\ref{tab:pairsolver}. Thresholds use interpolation between
neighboring lattice separations; polar profiles use interpolation
along each ray. No amplitude or mass is fitted.

\begin{table}[htb]
\caption{Square projected-pair convergence at $h=2$, $L=801$.
The last row is the analytic scaling limit.}\label{tab:pairsolver}
\begin{ruledtabular}
\begin{tabular}{ccc}
$\eps$ & $p_2/\eps^4$ & Axial zero $\eps r_E^{(2)}$ \\
\hline
$0.1$ & $0.01026859156$ & $1.436072299$ \\
$0.05$ & $0.01060402042$ & $1.419552399$ \\
$0.025$ & $0.01070712933$ & $1.414737930$ \\
$0.0125$ & $0.01073756639$ & $1.413319445$ \\
$0$ & $0.01074975935$ & $1.412821696$
\end{tabular}
\end{ruledtabular}
\end{table}

\subsection{Interacting pair-triplet channel}
Translation orbits of spin configurations are represented by normalized
states $|\mathcal O\rangle=|\mathcal O|^{-1/2}
\sum_{C\in\mathcal O}|C\rangle$. Matrix elements assembled from one
representative retain the ratio of orbit sizes, including configurations
with nontrivial stabilizers. The exchange operators are built directly
on distinct-site configurations, retaining all contact interactions.
The anchored triplet inner product is
$\langle u,v\rangle_3=\frac16\sum_{x,y\ne0,\,x\ne y}\bar u(x,y)v(x,y)$.
For an oriented bond $(a,b)$ and a third site $t$, the vertex from an
even pair wavefunction is
\begin{equation}
 (\mathcal V_{32}\phi)(\{a,b,t\})\supset
 c_{ab}[\phi(a-t)-\phi(b-t)],\qquad c_x=i/2,\quad c_y=-1/2.
 \label{eq:tripletvertex}
\end{equation}
The difference, rather than either term alone, is retained in the
matrix construction. The triplet block is
$A_3=a_\eps Q_3+3(h\eps^2-d)I+(d/2)D_3$;
the pair correction uses its interacting resolvent.

For $h=2$ the sequence $L=9,13,21,29,41$ with $L\eps=3$ tests
the axial displacement closest to $1/\eps$. At $L=21,29,41$,
$|\delta w_{\rm DM}|/|w_2|$ is
$0.00499366,0.00256905,0.00130670$, approximately $0.24\eps^2$.
The corresponding $|\langle b_0n_r\rangle|/\eps^8$ is
$0.00149683,0.00138428,0.00147295$.
The largest basis contains $470120$ triplet orbits.
Separate $h=3$ checks use $L=21,29$, and the fixed-distance
$L=41$, $\eps=0.005$ solve has relative residuals about
$2\times10^{-12}$. Holding $L\eps$ fixed tests powers and vertex
cancellations in a finite continuum box; it does not extract an
infinite-plane crossover coefficient.

\subsection{Projected many-body calculations and full-spin checks}
The projected spin Hamiltonian retains all flip sectors from zero
through $K_{\rm cut}=2,3,4$, including the odd sectors coupled by DM.
The primary grid is
$L=3,4,6,8,10,12$, $\eps=0.05,0.1,0.2,0.3$, and $h=2,3$,
giving $144$ cases. Twelve additional small-angle cases give $156$
projected calculations in total. The $L=12$, $K_{\rm cut}=4$
translation-zero basis has $122816$ orbit states, compared with
$17676661$ unreduced configurations. Lanczos residuals are at most
$3.10\times10^{-13}$. Random initial vectors check the lowest
translation-zero eigenvalue at the largest sizes.

Two-site RDMs are obtained by a full environment partial trace and
their negativities by diagonalizing all four eigenvalues of
$\rho^{T_2}$. All complex coherences are retained. An independently
implemented Pauli-basis Hamiltonian action and partial trace agree
with the orbit implementation within $1.8\times10^{-15}$.
Sixteen full-Hilbert-space calculations on $L=3,4$ cover the same
angle/field grid without a momentum restriction.
Table~\ref{tab:cutoff} shows the maximum nearest-neighbor/diagonal
negativity errors relative to full-spin ED at $L=4,h=2$.

\begin{table}[htb]
\caption{Finite-torus truncation errors in negativity relative to
full-spin ED on the $4\times4$ square lattice at $h=2$.}
\label{tab:cutoff}
\begin{ruledtabular}
\begin{tabular}{cccc}
$\eps$ & $K_{\rm cut}=2$ & $K_{\rm cut}=3$ & $K_{\rm cut}=4$ \\
\hline
$0.05$ & $4.425\times10^{-8}$ & $2.382\times10^{-10}$ & $1.549\times10^{-13}$ \\
$0.1$ & $6.924\times10^{-7}$ & $1.413\times10^{-8}$ & $3.695\times10^{-11}$ \\
$0.2$ & $1.014\times10^{-5}$ & $7.616\times10^{-7}$ & $7.219\times10^{-9}$ \\
$0.3$ & $4.427\times10^{-5}$ & $6.696\times10^{-6}$ & $1.206\times10^{-7}$
\end{tabular}
\end{ruledtabular}
\end{table}

There are two distinct finite-volume restrictions. First, a cutoff in
the \emph{total} number of flips is not uniform in area, since a
thermodynamic state has an extensive number $Np$ of rare flips.
Second, at fixed $N$ the isotropic ferromagnet has Dicke zero modes.
Their leading field/anisotropy coefficients are
\begin{equation}
 a_k=k\left[h-\frac{N-k}{2(N-1)}\right],\qquad
 \langle D_4|\Psi\rangle=\eps^2y_4+o(\eps^2),\qquad
 y_4=\frac{NF_L(\mathbf e_x)}{8a_4\sqrt{\binom N4}}.
 \label{eq:Dicke}
\end{equation}
Two pair insertions reach $D_4$ with an order-$\eps^4$ numerator
but an order-$\eps^2$ denominator. Its population contribution is
$p_{D_4}/\eps^4=4y_4^2/N\sim
3F(\mathbf e_x)^2/(32g^2N^3)$.
Thus blanket fixed-volume amplitude counting by the number of flips
would be incorrect. Comparisons are made to the pair kernel on the
\emph{same} torus before comparing to the plane.

Four sectors are also needed for the exterior RDM sign. At
$L=12,h=2,\eps=0.1,\br=(1,1)$, the two-flip truncation gives
an artificial negativity $7.31\times10^{-14}$ when its vanishing
joint population fails to compensate the normal coherence.
The four-flip RDM instead has smallest PT eigenvalue
$+4.75\times10^{-11}$. A last entangled axial separation of six on
an $L=12$ torus is the largest distinct tested displacement, not a
plane-range estimate. Signs at the floating-point noise floor are
not positivity certificates.

For the triangular lattice, full-spin ED uses the $4\times4$ torus
with $16$ spins and dimension $65536$, without a sector restriction.
The exact finite-torus weak-angle coefficients for $|w|/\eps^2$ at
physical lengths $1,\sqrt3,2$ are respectively
$5/216,1/216,1/108$.
At $h=3$, all $15$ nonzero separations are NPT at the tested angles
$0.2,0.1,0.05,0.025,0.0125$, with maximum eigenvector residual
$4.68\times10^{-14}$. The nearest-neighbor coefficient tends to
$0.023148148148$, rather than the plane value
$0.0227518119049481$; the difference is about $1.74\%$.

\subsection{Cylinder DMRG: Hamiltonian and measurements}
The full spin calculation uses finite complex matrix-product states
with no conserved charge, implemented with TeNPy
1.1.1~\cite{TeNPy2018,TeNPy2024}. Cylinders have open axial length
$L_x$ and periodic circumference $L_y$. The implemented bonds are
\begin{align}
 H_x/J&=-S_i^xS_j^x-c(S_i^yS_j^y+S_i^zS_j^z)
       +s(S_i^yS_j^z-S_i^zS_j^y),\nonumber\\
 H_y/J&=-S_i^yS_j^y-c(S_i^xS_j^x+S_i^zS_j^z)
       +s(S_i^zS_j^x-S_i^xS_j^z).
 \label{eq:DMRGbonds}
\end{align}
Each positive bond, including the periodic wrap, occurs once, and
the on-site field is $-Jh\eps^2S_i^z$. The constant $J/4$ per bond
is omitted during minimization and restored to the reported energy;
there are $(2L_x-1)L_y$ bonds. In the open direction the linear DM
term leaves a boundary torque, which is retained.
The primary widths are $4,6$, with additional width-$8$ calculations;
lengths are $24,48$. The principal fields are $h=2,3$, and the
angle grid is $0.1,0.15,0.2,0.3,0.4,0.6$, with coverage and
convergence assessed for each individual parameter set.

An independent $2\times4$ Kronecker-product Hamiltonian of dimension
$256$ agrees with the matrix-product operator within
$6.66\times10^{-16}$ entrywise. Its ground-state energy agrees
with DMRG within $3.11\times10^{-15}$. The check is repeated under
the proper joint rotation $(x,y)=(-b,a)$ of cylinder coordinates:
the axial axis is $+y$ and the transverse axis is $-x$.
This preserves the sign of the directed rotation, unlike simply
interchanging the bond labels.

For axial separation $r$ the sites are centered at
$a=\lfloor(L_x-1-r)/2\rfloor$ and $a+r$, with periodic coordinate
$b=0$. Transverse and both diagonal slopes are sampled through
$L_y/2$, with extra center-shift and rung-translation checks.
Every pair measurement retains its full complex $4\times4$ RDM,
both local populations and transverse means, $q$, $w$, the normal
coherence, and all RDM and PT eigenvalues. No plane-symmetry
projection or X-state formula is imposed on a cylinder.

\subsection{DMRG convergence filters}
Two-site sweeps use Lanczos diagonalization without a mixer,
minimum singular value $10^{-13}$, truncation cutoff $10^{-14}$,
and increasing maximum bond dimension $\chi_{\rm MPS}$.
The recorded comparisons span bond dimensions from $32$ through
$128$. Early energy tolerances are $2\times10^{-12}$ in total
energy; most later runs use $10^{-10}$ and entropy-change tolerance
$2\times10^{-7}$. Continuations normally perform at least three
sweeps. The actual final changes and discarded weights, rather than
nominal tolerances, determine convergence.

Most initial states are all-up product states. The weak-gap
$L_y=8,\eps=0.1$ case also uses an optimized classical canted
product as initialization, followed by unrestricted complex DMRG.
At $L_y=8,\eps=0.6$, two-site optimization is followed by single-site
polishing at fixed bond dimension $96$ or $128$, with total-energy
tolerance $10^{-8}$. For this point, increased-bond-dimension and two
additional fixed-dimension sweeps must both pass the observable
filters; the larger change sets the resolution. A zero single-site
discarded weight is not a bound on finite-bond-dimension error.

Amplitude acceptance is independent of agreement with the pair
prediction. Both compared runs must satisfy their convergence
criteria. For each measured pair, require
\begin{equation}
 \delta\Neg\le\max(3\times10^{-8},0.003\Neg),\qquad
 \max_{ij}|\delta\rho_{ij}|<3\times10^{-6}.
 \label{eq:DMRGamplitude}
\end{equation}
The sign test is separate: a positive negativity is resolved only if
\begin{equation}
 \Neg>\max\{5\times10^{-10},\,5\max(\delta\Neg,\delta\lambda_{\rm PT})\},
 \label{eq:DMRGsign}
\end{equation}
where $\delta\lambda_{\rm PT}$ is the largest change among the four
ordered PT eigenvalues. A sign may be resolved even when a small
amplitude fails the relative-accuracy test. Unaccepted amplitudes
are omitted individually. These are empirical resolution filters,
not rigorous RDM error bounds.
Table~\ref{tab:DMRG} gives representative accepted comparisons.
The ranges in Fig.~3 of the Letter use $L_y=6$ and the longest
cylinder available at each angle: $L_x=48$ for $\eps\le0.2$ and, at
$h=2$, for $\eps=0.3$, and $L_x=24$ otherwise. The pair problem on a
cylinder is solved with periodic circumference $L_y$ and a length of
at least $512$ sites, and its range is the largest $r$ with
$|w_2|>\sum_{\br}|w_2|^2$.

\begin{table}[htb]
\caption{Representative cylinder convergence at $h=2$.
The last column is the final maximum discarded weight in two-site sweeps;
each listed axial profile passes the amplitude filter.}
\label{tab:DMRG}
\begin{ruledtabular}
\begin{tabular}{cccccc}
$L_x\times L_y$ & $\eps$ & $\chi_{\rm MPS}$ &
$\max\delta\Neg$ & $\max|\delta\rho_{ij}|$ & Discarded weight \\
\hline
$48\times4$ & $0.1$ & $32\to48$ & $6.16\times10^{-9}$ &
$6.14\times10^{-9}$ & $9.02\times10^{-17}$ \\
$48\times6$ & $0.15$ & $48\to64$ & $9.13\times10^{-9}$ &
$1.64\times10^{-8}$ & $2.65\times10^{-15}$ \\
$24\times6$ & $0.3$ & $64\to80$ & $1.27\times10^{-9}$ &
$2.52\times10^{-8}$ & $2.06\times10^{-13}$
\end{tabular}
\end{ruledtabular}
\end{table}

\subsection{Cylinder geometry and the limits of the comparison}
The normalized profile discrepancy is
$\|\Neg/\eps^4-P\|_2/\|P\|_2$ on the stated set of separations,
where $P$ is the specified continuum or lattice prediction.
At $24\times6,h=2,\eps=0.3$, using $r=2,3,4$, it is
$13.57\%$ for the plane continuum profile, $5.09\%$ for the massive
plane-lattice pair profile, and $0.54\%$ for the pair profile on
the matched periodic circumference. The last reference removes
axial ends; the finite-cylinder DMRG retains them.
At $24\times8$ with the same field, angle and separations,
the corresponding discrepancies are $17.47\%,1.85\%,0.39\%$.
No mass, population, amplitude, or length is fitted.
Table~\ref{tab:profiles} lists these discrepancies for all accepted
profiles at $\eps=0.2$ and $0.3$.

Physical size effects exceed these numerical convergence errors.
At $L_y=4,h=2,\eps=0.1$, increasing $L_x$ from $24$ to $48$
changes the nearest-neighbor coefficient $\Neg/\eps^2$ from
$0.04276541$ to $0.02742221$, a $35.878\%$ decrease caused by
end-induced canting. At $L_x=24,h=2,\eps=0.3$, increasing width
from $6$ to $8$ changes the negativities at $r=1,2,3,4$ by
$1.27\%,3.13\%,5.69\%,9.79\%$.
These comparisons assess finite geometry separately from bond dimension.
The cylinders break the plane's quarter-turn symmetry and can lift
its leading diagonal node.

Holding width fixed also changes the weak-angle limit of the pair
operator. Its zero transverse momentum channel gives an axial tail
proportional to $\eps^3e^{-m_h|\eps|r}$ and a pair population of order
$\eps^4$, leading to
\begin{equation}
 r_E^{\rm cyl,pair}
 =\frac{\log(1/|\eps|)+O_{L_y,h}(1)}{m_h|\eps|},
 \qquad L_y\text{ fixed}.
 \label{eq:cylrange}
\end{equation}
The same channel allows an even, nonanalytic relative $|\eps|$
correction at fixed displacement. For example,
\begin{equation}
 \int_{-\Lambda}^{\Lambda}\frac{k^2}{k^2+a^2}\,\dd k
 =2\Lambda-2a\arctan(\Lambda/a)
 =2\Lambda-\pi a+O(a^2),\qquad a\propto|\eps|.
 \label{eq:cylcusp}
\end{equation}
Consequently the converged data establish finite-cylinder benchmarks;
they do not establish the plane weak-angle coefficient, exact plane
diagonal separability, or a plane supremal range.

The ancillary data also contain runs at $h=1$, outside the field range
of the square-lattice ground-state theorems; they are exploratory and
are not used in the Letter. The ancillary file
\texttt{gap\_linearization.json} tabulates the minimum of the quadratic
band $\sqrt{A_\bk^2-B_\bk^2}$ of the polarized state, where
$A_\bk=\Delta_\eps+a_\eps(2-\cos k_x-\cos k_y)$ and
$B_\bk=d(\cos k_y-\cos k_x)/2$. These coefficients follow by retaining
the quadratic terms in Eq.~\eqref{eq:hardcore-decomposition}; this
band minimum is not an excited-state DMRG gap.

\subsection{Other-lattice and thermal checks}
Triangular response kernels are inverted on Fourier grids through
$4096^2$ and checked against an independently assembled $511^2$
retained-edge sparse matrix. The sparse relative residual is
$2.85\times10^{-12}$, and the maximum same-torus field difference
from Fourier inversion is $1.27\times10^{-13}$.
The honeycomb calculation inverts the four-channel symbol
\eqref{eq:hcL}, retains the physical displacement phases, and uses
$1024^2$ cells. Its contact eigenvalue agrees with
Eq.~\eqref{eq:hcK} within $2\times10^{-16}$.
Independent normalized occupation-basis two-magnon matrices include
double occupation for $S\ge1$. On square and triangular $5\times5$
tori, $S=1/2,1,3/2,2,5$ gives agreement with the massive-shell
resummation to $5.7\times10^{-17}$; honeycomb $4\times4$-cell
checks, including $S=1,2$, agree to $1.9\times10^{-16}$.

The thermal calculation compares Eq.~\eqref{eq:Bose} with independent
radial quadrature and lattice momentum sums through $1601^2$ points,
and evaluates the convergent thermal integral in
Eq.~\eqref{eq:thermalA}. The maximum density-quadrature discrepancy
is $7.63\times10^{-17}$. At $h=2,\rho=1$ on the axis, the
unexpanded effective-density equation gives
$t_E/\eps^2=0.2805010253,0.193615999999,0.1101273818$ at
$\eps=0.1,0.025,0.001$. The anomalous thermal correction relative
to the zero-temperature amplitude decreases from
$3.51\times10^{-5}$ to $7.26\times10^{-10}$ between the first
and last points. These calculations check the effective gas;
no interacting finite-temperature spin simulation is used.

\subsection{Entanglement range on cylinders}
\label{sm:cylrangedata}
Table~\ref{tab:range} lists the data of Fig.~3 of the Letter. For
each field and angle it gives the largest axial separation $r_E$ at
which the sign test~\eqref{eq:DMRGsign} resolves a negative partial
transpose, on the longest cylinder of circumference $L_y=6$ available
at that angle, with the pair placed symmetrically about the center.
At $r_E+1$ the computed partial transpose has no negative eigenvalue.
The negativity at $r_E$ exceeds the resolution threshold
$5\times10^{-10}$ by at least two orders of magnitude in every row.

The open ends cant the spins at first order in $\eps$, and the canting
decays into the cylinder over the length $1/(m_h\eps)$. Where it
reaches the pair, the one-site population exceeds its bulk value and
the pair is no longer a bulk pair. The last column of the table is the
population at the two sites of the pair divided by the population at
the center of the cylinder. A point is marked as affected by the ends
when this ratio differs from one by more than $5\%$; these are the
open symbols of Fig.~3(a). At $\eps=0.1$ the length $1/(m_h\eps)$,
which is $5.8$ at $h=2$ and $4.5$ at $h=3$, is also comparable
to the circumference, so these points probe the quasi-one-dimensional
regime of Eq.~\eqref{eq:cylrange}.

The column $r_E^{\rm pair}$ is the range of the pair problem of
Eq.~\eqref{eq:pairnumerics} solved on a cylinder of the same
circumference without ends (periodic length at least $512$), that is,
the largest integer axial separation with $|w_2|>p_2$. It contains no
adjustable parameter. At the nine points that are not affected by the
ends the DMRG value coincides with it. At the affected point $h=2$,
$\eps=0.15$ the DMRG range is $9$ and the pair range is $10$. For all
ten points with $\eps\ge0.15$, $r_E$ also equals the plane value
$\rho_E/\eps$ rounded up or down to an integer, so these data
are consistent with $\rho_E$ but cannot separate it from the range of
the cylinder, which grows as in Eq.~\eqref{eq:cylrange} when
$\eps\to0$ at fixed circumference.

\begin{table}[htb]
\caption{Entanglement range on cylinders of circumference $L_y=6$
[Fig.~3 of the Letter]. $\chi_{\rm MPS}$ gives the two bond
dimensions that are compared in the sign test. $r_E$ is the DMRG
range, $r_E^{\rm pair}$ the range of the pair problem on the same
cylinder without ends, and $\rho_E/\eps$ the leading-order prediction
for the plane, with $\rho_E=1.4128$ at $h=2$ and $1.2597$ at $h=3$.
Rows marked with a dagger are affected by the ends of the cylinder.}
\label{tab:range}
\begin{ruledtabular}
\begin{tabular}{ccccccccc}
$h$ & $\eps$ & $L_x$ & $\chi_{\rm MPS}$ & $r_E$ & $r_E^{\rm pair}$ &
$\rho_E/\eps$ & $\Neg$ at $r_E$ & Population ratio \\
\hline
$2$ & $0.6$ & $24$ & $64\to80$ & $3$ & $3$ & $2.35$ & $2.41\times10^{-5}$ & $1.000$ \\
$2$ & $0.4$ & $24$ & $64\to80$ & $4$ & $4$ & $3.53$ & $4.23\times10^{-6}$ & $1.000$ \\
$2$ & $0.3$ & $48$ & $48\to64$ & $5$ & $5$ & $4.71$ & $2.17\times10^{-6}$ & $1.000$ \\
$2$ & $0.2$ & $48$ & $48\to64$ & $7$ & $7$ & $7.06$ & $2.96\times10^{-6}$ & $1.002$ \\
$2$ & $0.15$ & $48$ & $48\to64$ & $9$ & $10$ & $9.42$ & $7.42\times10^{-7}$ & $1.479^\dagger$ \\
$2$ & $0.1$ & $48$ & $32\to48$ & $19$ & $17$ & $14.13$ & $4.96\times10^{-7}$ & $250^\dagger$ \\[3pt]
$3$ & $0.6$ & $24$ & $32\to48$ & $2$ & $2$ & $2.10$ & $1.44\times10^{-3}$ & $1.000$ \\
$3$ & $0.4$ & $24$ & $32\to48$ & $3$ & $3$ & $3.15$ & $2.37\times10^{-4}$ & $1.000$ \\
$3$ & $0.3$ & $24$ & $48\to64$ & $4$ & $4$ & $4.20$ & $5.29\times10^{-5}$ & $1.000$ \\
$3$ & $0.2$ & $48$ & $32\to48$ & $6$ & $6$ & $6.30$ & $6.61\times10^{-6}$ & $1.000$ \\
$3$ & $0.15$ & $48$ & $48\to64$ & $9$ & $9$ & $8.40$ & $1.88\times10^{-7}$ & $1.017$ \\
$3$ & $0.1$ & $48$ & $32\to48$ & $8$ & $15$ & $12.60$ & $8.06\times10^{-8}$ & $6.39^\dagger$
\end{tabular}
\end{ruledtabular}
\end{table}

\emph{Profiles and nearest neighbors.} Table~\ref{tab:profiles} gives
the normalized discrepancy $\|\Neg/\eps^4-P\|_2/\|P\|_2$ of the DMRG
axial profile for every accepted profile at $\eps=0.2$ and $0.3$,
evaluated on the separations with $r\ge2$ and
$0.6\le\eps r\le1.2$. The three references $P$ contain no adjustable
parameter: the plane continuum profile $(|A_h|-n_*)_+$, the pair
problem on the plane lattice, and the pair problem on a cylinder of
the same circumference without ends. The discrepancy from the last
lies between $0.39\%$ and $1.89\%$. The discrepancies from the plane
lattice and from the continuum profile reach $19.7\%$ and $21.2\%$;
they measure the circumference, lattice, and finite-angle corrections
at these short separations.

For nearest neighbors at $h=2$ and $\eps=0.4$ on the $24\times6$
cylinder, the axial negativity is $0.0385279\,\eps^2$ and the
circumferential one $0.0395154\,\eps^2$. The $\eps\to0$ pair kernel
of a cylinder of circumference $6$ gives $0.0423196\,\eps^2$ on the
axis, and the plane value is $F_1\eps^2/2=0.0469961\,\eps^2$.

\begin{table}[htb]
\caption{Normalized root-mean-square discrepancy, in percent, between
the DMRG axial negativity profile and three parameter-free references,
for all accepted profiles at $\eps=0.2$ and $0.3$. The references are
the plane continuum profile, the pair problem on the plane lattice,
and the pair problem on a cylinder of the same circumference without
ends. The column $r$ gives the separations used.}
\label{tab:profiles}
\begin{ruledtabular}
\begin{tabular}{ccccccc}
$h$ & $\eps$ & $L_x\times L_y$ & $r$ & Continuum & Plane lattice &
Same cylinder \\
\hline
$2$ & $0.2$ & $48\times4$ & $3$--$6$ & $7.84$ & $19.69$ & $0.57$ \\
$2$ & $0.2$ & $48\times6$ & $3$--$6$ & $3.06$ & $13.25$ & $0.50$ \\
$2$ & $0.2$ & $48\times8$ & $3$--$5$ & $9.27$ & $6.91$ & $0.56$ \\
$2$ & $0.3$ & $24\times4$ & $2$--$4$ & $4.14$ & $12.89$ & $1.26$ \\
$2$ & $0.3$ & $48\times4$ & $2$--$4$ & $7.12$ & $10.41$ & $1.83$ \\
$2$ & $0.3$ & $24\times6$ & $2$--$4$ & $13.57$ & $5.09$ & $0.54$ \\
$2$ & $0.3$ & $48\times6$ & $2$--$4$ & $15.96$ & $3.10$ & $1.89$ \\
$2$ & $0.3$ & $24\times8$ & $2$--$4$ & $17.47$ & $1.85$ & $0.39$ \\[3pt]
$3$ & $0.2$ & $48\times4$ & $3$--$6$ & $6.11$ & $14.19$ & $0.86$ \\
$3$ & $0.2$ & $48\times6$ & $3$--$6$ & $8.41$ & $8.60$ & $0.93$ \\
$3$ & $0.3$ & $24\times4$ & $2$--$4$ & $12.10$ & $7.21$ & $1.25$ \\
$3$ & $0.3$ & $24\times6$ & $2$--$4$ & $18.84$ & $1.55$ & $1.33$ \\
$3$ & $0.3$ & $24\times8$ & $2$--$4$ & $21.22$ & $0.74$ & $1.33$
\end{tabular}
\end{ruledtabular}
\end{table}

\subsection{Off-axis separations and the circumference}
\label{sm:offaxis}
On a cylinder the pair problem has its own kernel, and away from the
axes its boundary depends strongly on the circumference. The point
$\br=(3,2)$ at $\eps=0.25$ and $h=2$ lies inside the clover of the
plane: $\eps\br=(0.75,0.5)$, $\rho=0.901$, $\theta=33.7^\circ$, and
$N_h(\eps\br)=0.00577$, while the lattice pair problem of the plane
at the same angle gives $(|w_2|-\sum_{\br}|w_2|^2)/\eps^4=0.00438$.
Table~\ref{tab:offaxis} gives the DMRG result on cylinders of length
$L_x=48$ and circumference $L_y=6$ to $12$, with the pair placed at
the center and $2|\Delta y|<L_y$. Each row uses the largest bond
dimension measured at that circumference.

At $L_y=6$ the partial transpose is positive. Its smallest eigenvalue,
$2.03\times10^{-9}$, changes by $2\times10^{-13}$ between bond
dimensions $96$ and $128$, by $4\times10^{-12}$ when the length is
increased from $48$ to $64$, and agrees between two different initial
product states. The pair problem on a cylinder of circumference $6$
also predicts no entanglement at this point, with margin
$-0.00203\,\eps^4$. At $L_y=8$, $10$, and $12$ the partial transpose
has a negative eigenvalue, and $\Neg/\eps^4$ follows the pair problem
on the same cylinder within $0.00007$ while approaching the value of
the plane. The runs at $L_y=10$ and $12$ did not reach the stopping
criterion of the DMRG sweeps; they are included to show the trend
with circumference, not as converged values. The quantity
$m_h\eps L_y$, which compares the circumference with the correlation
length of the pair, is $2.60$, $3.46$, $4.33$, and $5.20$.

\begin{table}[htb]
\caption{Two-spin state at $\br=(3,2)$, $\eps=0.25$, $h=2$ on
cylinders of length $48$ and circumference $L_y$. $\chi_{\rm MPS}$ is
the bond dimension, $\lambda_{\min}$ the smallest eigenvalue of the
partial transpose of the measured density matrix, and the last column
the margin $(|w_2|-\sum_{\br}|w_2|^2)/\eps^4$ of the pair problem on a
cylinder of the same circumference without ends. For comparison, the
plane gives $0.00438$ (lattice pair problem at $\eps=0.25$) and
$N_h=0.00577$ ($\eps\to0$). Rows marked with a dagger did not reach
the stopping criterion.}
\label{tab:offaxis}
\begin{ruledtabular}
\begin{tabular}{ccccc}
$L_y$ & $\chi_{\rm MPS}$ & $\Neg/\eps^4$ & $\lambda_{\min}$ &
Pair problem, same cylinder \\
\hline
$6$ & $128$ & $0$ & $+2.03\times10^{-9}$ & $-0.00203$ \\
$8$ & $96$ & $0.00038$ & $-1.50\times10^{-6}$ & $0.00042$ \\
$10$ & $64^\dagger$ & $0.00262$ & $-1.02\times10^{-5}$ & $0.00260$ \\
$12$ & $32^\dagger$ & $0.00372$ & $-1.45\times10^{-5}$ & $0.00365$
\end{tabular}
\end{ruledtabular}
\end{table}

The density matrices are used as measured, with complex entries, both
populations, the one-site transverse moments, and the odd coherences;
no symmetry is imposed. A positive smallest eigenvalue of order
$\eps^8$, as at $L_y=6$, is a numerical statement about the computed
state. The DMRG data carry no certified error bound with respect to
the exact ground state, and they say nothing about exact separability
in the plane, which is decided by the open problem of
Sec.~\ref{sm:exterior}.

\subsection{Independent re-derivations and reimplementations}
\label{sm:independent}
The results were checked in two ways that do not reuse the
derivations or the code described above, with the tools named in the
Letter; the scripts are in the ancillary folder
\texttt{independent\_checks/}.

\emph{Leading-order results.} Starting from Eq.~\eqref{eq:model}
alone, and before the original derivation was consulted, the
following were derived again and found to agree: the one-magnon gap
$\Delta_\eps=h\eps^2-(1-\cos\eps)$ of the number-conserving part and
the absence of a Dzyaloshinskii-Moriya shift of the uniform magnon;
the relative-motion operator~\eqref{eq:massive-block-statement} and
its rank-one solution, with $F(e_x)=(4-\pi)/[8(\pi-2)]$ and the far
field $\cos2\theta/[8(\pi-2)r^2]$; the profile
$A_h$ of Eq.~\eqref{eq:profile-statement}; the population coefficient
$n_*=\pi/[256(\pi-2)]$; the boundary equation
$K_2(m_h\rho_E)=\pi/(8m_h^2|\cos2\theta|)$ with
$\rho_E=1.4128216957$ at $h=2$ and $1.2596646347$ at $h=3$ on the
axes; and the thermal density and the Lambert-$W$ form of the thermal
boundary. The accompanying numerics are independent as well. Exact
diagonalization of the $4\times4$ torus in the full Hilbert space
confirms the gap $(h-\tfrac12)\eps^2+O(\eps^4)$ and that the lowest
excitation is the uniform magnon. A sparse solution of the pair
problem in a box of radius at least $15/(m_h\eps)$ reproduces
$\eps^4A_h$ on the axis within $4\times10^{-4}$ for
$0.5\le\rho\le3$ at $\eps=0.01$. The axial zero of the
population-subtracted pair amplitude lies at
$\eps r=1.426585$, $1.416854$, $1.413974$, and $1.413147$ for
$\eps=0.08$, $0.04$, $0.02$, and $0.01$ at $h=2$, and at $1.277089$,
$1.264831$, $1.261144$, and $1.260084$ at $h=3$; both sequences
converge to $\rho_E$, the deviation decreasing by a factor of about
$3.5$ each time $\eps$ is halved.

\emph{Proofs.} The arguments of
Secs.~\ref{sm:normalform}--\ref{sm:range} were examined line by line,
each decisive inequality was derived again, and the constants were
tested with code written from Eq.~\eqref{eq:model}. On the
$3\times3$ torus the decomposition of the Hamiltonian into exchange,
Dzyaloshinskii-Moriya, pair, and field parts used in
Sec.~\ref{sm:normalform} holds with residual $2\times10^{-15}$ in
the full Hilbert space. The closed forms $4/\pi-1$, $F_1$, and $S_F$
agree with two-dimensional quadrature within $2\times10^{-16}$.
Exact diagonalization of the zero-momentum sector of the $3\times3$
and $4\times4$ tori, for $h=1.75$, $2$, and $3$ and eight angles in
$0.04\le\eps\le0.14$, reproduces the finite-size counterparts of the
coefficients of Theorem~\ref{thm:energy}: the fourth-order coefficient
within $2\times10^{-11}$, the sixth-order coefficient within
$2\times10^{-8}$, and the population $p/\eps^4$ within
$7\times10^{-9}$. An independent three-magnon solver gives the
finite-size inverse forms on tori of linear size $12$ to $32$; they
converge as the inverse fourth power of the size to
$\Gamma_3\simeq0.0058244$, so that $g_6(2)\simeq0.0147938$. These
finite-size checks take $\eps\to0$ before the volume; the order of
limits of the theorems, volume first, is covered by the proofs only.

\emph{Mass-scale results.} The proofs of
Secs.~\ref{sm:massscale}--\ref{sm:exterior} were checked in the same
way. One check re-derived every step of the proof of
Theorem~\ref{thm:massscale} and of its rate; it found no incorrect step
and supplied three arguments that are now written out in
Sec.~\ref{sm:massscale}. It also computed the heat kernel on a piece
of the three-magnon exclusion graph with $704\,760$ configurations,
where $t^2p_t(z,z)$ is at most $0.045$ at contact configurations and
falls to $0.009$ at $t=30$, close to the free value
$1/(12\pi^2)=0.0084$, and the point-probe norms, which are independent
of $\eps$ and $r$ within the range tested. In
Secs.~\ref{sm:threshold} and \ref{sm:exterior} every step was checked
while the proofs were written, the limit radii and the boundary slopes
were recomputed, both two-qubit counterexamples were verified with
exact determinants, and the complex Schur identity was tested on
$3000$ random physical density matrices. The compact checks of the
ancillary folder \texttt{mass\_scale/} reproduce the finite algebra,
the interval certificates, the lattice kernels, and the checks of the
measured density matrices.
The thermal argument of Sec.~\ref{sm:thermal} was examined in the
same way. Each inequality was derived again, and the field-pressure
inequality of Lemma~\ref{lem:field-pressure} was confirmed by exact
evaluation of both partition functions on three graphs of four to
six sites with random couplings, temperatures, and fields.

\emph{Model without the DM term.} The transfer of
Secs.~\ref{sm:normalform}--\ref{sm:threshold} to $H_{\rm sym}$
(Sec.~\ref{sm:nodm}) was checked twice, independently. Each check
traced every occurrence of the DM coupling and of the field bound
$h>3/2$ along the chain. Both found that no step needs a new argument
once the triplet dressing is removed, that the field range extends to
$h>1/2$, and that the sixth-order energy loses the term $-\Gamma_3/4$;
the second check corrected the bookkeeping of the triplet matching,
which must be removed rather than evaluated at zero coupling, and is
followed in Lemma~\ref{lem:nodm-sandwich}. Exact diagonalization of $3\times3$ and
$4\times4$ tori in all magnon sectors confirms the decomposition
identity, the coercivity~\eqref{eq:nodm-coercivity}, the parity
structure and the scaling of the energy residual
(\texttt{nodm/}).

\subsection{Ancillary files}
The code and data are provided as ancillary files in the folder \texttt{anc/}.
Table~\ref{tab:anc} lists its parts; each folder is distributed as a zip archive of
the same name, and unpacking all archives inside \texttt{anc/} restores the layout of
the table. \texttt{anc/README.md} gives the software versions and the commands that
regenerate the figures and the tables of this section.

\begin{table}[htb]
\caption{Ancillary files. Paths are relative to \texttt{anc/}.}
\label{tab:anc}
\begin{ruledtabular}
\begin{tabular}{p{0.24\textwidth}p{0.70\textwidth}}
Path & Content \\
\hline
\texttt{make\_figures.py}, \texttt{figure\_data/} &
Script and data that generate Figs.~1--4 of the Letter and Figs.~S1 and S2, and the
constants quoted in the text. \\[3pt]
\texttt{pair\_crossover/} &
Massive pair problem on the punctured square lattice (Fourier and sparse solvers),
thermal boundary, exact diagonalization with at most four spin flips and in the
full Hilbert space of small tori, and the channel-resolved check of the
Dzyaloshinskii-Moriya vertex. \\[3pt]
\texttt{lattices\_spin/} &
Triangular and honeycomb kernels and constants, the spin-$S$ two-magnon kernel,
the rotational selection rule, and exact diagonalization of triangular and square
clusters. \\[3pt]
\texttt{dmrg/} &
Cylinder DMRG code, run specifications, convergence filters, comparison with the
pair problem on the plane and on the cylinder, and, in two archives, the measured
two-site density matrices of every run. \\[3pt]
\texttt{proof\_checks/} &
Numerical and symbolic checks of the identities and constants used in
Secs.~\ref{sm:normalform}--\ref{sm:range}. \\[3pt]
\texttt{independent\_checks/} &
Code of the independent re-derivations of Sec.~\ref{sm:independent}. \\[3pt]
\texttt{mass\_scale/} &
Checks of Secs.~\ref{sm:massscale}--\ref{sm:exterior} (finite algebra,
interval certificates, lattice kernels, Schur and Gram identities),
and the off-axis cylinder DMRG of Sec.~\ref{sm:offaxis}: code, run
specifications, complex two-site density matrices, and comparison
tables; \texttt{run\_checks.py} runs all compact checks. \\[3pt]
\texttt{experiment/} &
Code and data of Sec.~\ref{sm:experiment}: exact diagonalization in sectors with a fixed number of
flips, the pair theory on finite arrays and its time evolution, the validation against exact
diagonalization and DMRG, the XXZ factor, the predictions for a $10\times10$ array, the margins at
strong pair drive (\texttt{strong\_drive.py}), the collective witness and the constants of the
entanglement budget (\texttt{collective\_witness.py}), and, in \texttt{independent/}, independently
written checks. \\[3pt]
\texttt{nodm/} &
Exact checks on small tori for the model without the DM term (Sec.~\ref{sm:nodm}).
\end{tabular}
\end{ruledtabular}
\end{table}

\section{Variants of the model and a proposed measurement}
\label{sm:experiment}

This section supports the statements of the Letter about variants of Eq.~(1) and about an
experimental test. They rest on the two-magnon theory and on numerical calculations, not on the
theorems of Secs.~\ref{sm:normalform}--\ref{sm:exterior}. Code and data are in the ancillary folder
\texttt{anc/experiment}.

\subsection{Number of flips}
\label{sm:exp-parity}
For a bond along $\be_x$,
\begin{equation}
 \mathbf S_i^{T}e^{\eps[\be_x]_\times}\mathbf S_j
 =\cos\eps\,\mathbf S_i\cdot\mathbf S_j+(1-\cos\eps)\,S_i^xS_j^x
 +\sin\eps\,(S_i^zS_j^y-S_i^yS_j^z),
 \label{eq:exp-bond}
\end{equation}
and along $\be_y$ the same holds with $S^x\to S^y$ in the second term and the last term replaced by
$\sin\eps\,(S_i^xS_j^z-S_i^zS_j^x)$. Written with $S^\pm=S^x\pm iS^y$, the first two terms contain
$S^zS^z$, $S^\pm S^\mp$ and $S^\pm S^\pm$ and change the number of flipped spins
$\Nop=\sum_in_i$ by $0$ or $\pm2$; the Dzyaloshinskii-Moriya (DM) term is linear in $S^\pm$ and
changes $\Nop$ by $\pm1$. Hence the blocks of $H$ with zero, one and two flips, and its matrix
elements between zero and two flips, are the same with and without the DM term. In particular the
one-magnon dispersion $\omega(\bk)=\Delta_\eps+a_\eps(2-\cos k_x-\cos k_y)$, the two-flip block
$A_2$ of Eq.~(A5) of the Letter and the pair source $df$ do not involve it. An independent exact
diagonalization of the $4\times4$ torus confirms that these blocks coincide to machine precision,
whereas the ground-state energies per site differ, by about $-1.6\times10^{-3}\eps^6$ on this torus,
through the one- and three-flip sectors (\texttt{independent/v1\_flip\_sectors.py}); in the
thermodynamic limit this is the three-magnon term $-\Gamma_3\eps^6/4$ of Theorem~\ref{thm:energy},
which the DM term produces. We write $H_{\rm sym}$ for Eq.~(1) without the DM term. It conserves the
parity of $\Nop$, so the odd coherence $\zeta$ vanishes in parity-invariant states, and its
two-magnon theory coincides with that of Eq.~(1). Without the anisotropy as well, the polarized state
is an exact product eigenstate of the bulk, since the single-flip terms of the DM interaction cancel;
quantum treatments of chiral magnets with Heisenberg exchange, DM interaction and field accordingly
find the field-polarized phase nearly unentangled~\cite{Haller2022,Haller2024}. Section~\ref{sm:nodm}
proves that the theorems of this Supplemental Material hold for $H_{\rm sym}$ at all fields
$h>1/2$ and that, off the diagonals, the clover is then exactly the entangled region
(Theorem~\ref{thm:nodm}). The predictions for finite arrays below use the two-magnon theory and exact
diagonalization.

On an open array the single-flip terms of the DM interaction no longer cancel at the edges. There
the DM term acts on the spins as a transverse field of order $J\eps$ and tilts them; the tilt
decays into the bulk over the correlation length. This is the quantum counterpart of the edge
twist of confined chiral magnets~\cite{Rohart2013}. On an open $4\times4$ array at $\eps=0.4$ and
$h=2$, exact diagonalization in the full Hilbert space gives the negativity $2.15\times10^{-3}$ for
the central nearest-neighbor pair with the DM term and $6.41\times10^{-3}$ without it. A test on a
finite array is therefore best done with $H_{\rm sym}$.

\subsection{XXZ exchange and antiferromagnets near saturation}
\label{sm:exp-xxz}
Replace $\cos\eps\,\mathbf S_i\cdot\mathbf S_j$ by
$\cos\eps\,(S_i^xS_j^x+S_i^yS_j^y+\Delta S_i^zS_j^z)$ and raise the field by $2\cos\eps(1-\Delta)$,
so that the one-magnon gap is unchanged. The two-flip operator in the relative coordinate is then
$\cL+(1-\Delta)P_\nn$ at leading order. A $d$-wave amplitude vanishes at the origin, where also
$(\cL_0\psi)(0)=0$, so that $\cL=\cL_0-P_\nn$ on such amplitudes, with $\cL_0$ the Laplacian of the
full lattice, and the operator becomes $\cL_0-\Delta P_\nn$. On the four neighbors of the origin the
$d_{x^2-y^2}$ symmetry gives $P_\nn F=-4F(\be_x)f$, so the perturbation has rank one in this channel.
With the spin-wave kernel $F_0=-\cL_0^{-1}f$, the Sherman-Morrison formula gives
\begin{equation}
 F_\Delta=\frac{F_0}{1-\Delta k_\square},\qquad k_\square=4F_0(\be_x)=\frac4\pi-1,
 \label{eq:exp-xxz}
\end{equation}
at every separation. For $\Delta=1$ this is the enhancement $\pi/[2(\pi-2)]$ of
Sec.~\ref{sm:green}; for $\Delta=0$ the XY magnet reproduces linear spin-wave theory, since the hard
core alone does not act on a $d$-wave amplitude. The factor diverges at $\Delta=\pi/(4-\pi)=3.66$,
where a $d$-wave two-magnon bound state reaches threshold. Direct solutions of the lattice problem
on a $241\times241$ torus give $F_\Delta/F_0=0.78540$, $0.87980$, $1.15824$, $1.32611$ and
$1.37597$ for $\Delta=-1$, $-0.5$, $0.5$, $0.9$ and $1$, the same at the separations $(1,0)$,
$(3,0)$ and $(2,1)$, in agreement with Eq.~\eqref{eq:exp-xxz} (\texttt{xxz\_check.py}). The
amplitude scales with $1/(1-\Delta k_\square)$ and the depletion with its square, so the boundary of
the leading negativity solves
\begin{equation}
 K_2\bigl(m\rho_E(\theta)\bigr)=\frac{\pi-2}{4m^2|\cos2\theta|\,(1-\Delta k_\square)} .
 \label{eq:exp-boundary}
\end{equation}
On the axes at $h=2$ this gives $\rho_E=1.4128$, $1.5449$ and $1.6476$ for $\Delta=1$, $0$ and $-1$.

A bipartite spin-$1/2$ Heisenberg antiferromagnet $J\sum\mathbf S_i\cdot\mathbf S_j-B\sum S_i^z$
becomes, after a rotation of the spins of one sublattice by $\pi$ about the field axis,
$-J\sum(S_i^xS_j^x+S_i^yS_j^y-S_i^zS_j^z)-B\sum S_i^z$: ferromagnetic transverse exchange with
$\Delta=-1$. Every nearest-neighbor bond joins the two sublattices, so a bond-directional anisotropy
changes sign on every bond and keeps its $d$-wave pattern. Above the saturation field $B_s=4J$ of
the square lattice the polarized state is the ground state, and with the gap measured from the
bottom of the magnon band the two-magnon theory applies with the factor $1/(1+k_\square)=\pi/4$. In
the original frame the pair amplitude alternates in sign between the two sublattices,
$w(\br)\propto(-1)^{r_x+r_y}$, so the witness of Sec.~\ref{sm:rdm} must use the staggered sign
$\eta=\sgn w(\br)$.

\subsection{Pair theory on finite geometries}
\label{sm:exp-pair}
For a finite array we use the two-magnon theory in size-consistent form. Let $H_{20}$ be the
coupling of the polarized state $|\Omega\rangle$ to the sector with two flips, $H_{22}$ the block of
$H$ in that sector and $E_\Omega=\langle\Omega|H|\Omega\rangle$. The pair field $\phi$ solves
$(H_{22}-E_\Omega)\phi=-H_{20}$, and the two-spin quantities are those of independent pairs,
\begin{equation}
 w_{ab}=\phi_{ab},\quad p_a=\sum_c|\phi_{ac}|^2,\quad q_{ab}=|\phi_{ab}|^2,\quad
 v_{ab}=\sum_c\phi_{ac}^*\phi_{bc},
 \label{eq:exp-pair}
\end{equation}
with $\zeta=0$; the two-spin state is the $X$ state built from them. On the plane this is the pair
problem of Sec.~\ref{sm:numerics}. A variational ground state restricted to a few flips is not size
consistent: its normalization and its correlation energy, which grows with the number of sites,
reduce the pair amplitudes. On the $24\times6$ cylinders of the Letter a truncation to at most three
flips underestimates the negativity at the DMRG range by $10$ to $17\%$ for $h=3$ and misses the
outermost entangled separation for $h=2$ (\texttt{validate\_cylinder.py}). Equation
\eqref{eq:exp-pair} instead reproduces the DMRG range $r_E$ at all six points with $h=2,3$ and
$\eps=0.3,0.4,0.6$, and the negativity at $r_E$ to within $2.5\%$ for $h=3$; for $h=2$ the
negativities at $r_E$ are close to threshold ($\lesssim4\times10^{-5}$) and agree to within a factor
$1.8$. Table~\ref{tab:exp-ed} compares Eq.~\eqref{eq:exp-pair} with exact diagonalization of open
$4\times4$ arrays in the full Hilbert space for $H_{\rm sym}$ (\texttt{validate\_pairtheory.py}).

\begin{table}[t]
\caption{Negativity of selected pairs of an open $4\times4$ array of $H_{\rm sym}$ at $h=2$:
exact diagonalization in the full Hilbert space (ED) and the pair theory of
Eq.~\eqref{eq:exp-pair}. Sites are labeled $(x,y)$ with $x,y\in\{0,\dots,3\}$.}
\label{tab:exp-ed}
\begin{ruledtabular}
\begin{tabular}{lcccccc}
 & \multicolumn{2}{c}{$\eps=0.3$} & \multicolumn{2}{c}{$\eps=0.4$} & \multicolumn{2}{c}{$\eps=0.6$}\\
 Pair & ED & pair theory & ED & pair theory & ED & pair theory\\
\hline
 $(1,1)$--$(2,1)$ & $4.007\times10^{-3}$ & $4.007\times10^{-3}$ & $6.407\times10^{-3}$ & $6.409\times10^{-3}$ & $1.145\times10^{-2}$ & $1.146\times10^{-2}$\\
 $(0,1)$--$(2,1)$ & $1.439\times10^{-3}$ & $1.439\times10^{-3}$ & $1.963\times10^{-3}$ & $1.965\times10^{-3}$ & $2.375\times10^{-3}$ & $2.383\times10^{-3}$\\
 $(0,1)$--$(3,1)$ & $7.160\times10^{-4}$ & $7.166\times10^{-4}$ & $7.191\times10^{-4}$ & $7.209\times10^{-4}$ & $2.112\times10^{-4}$ & $2.143\times10^{-4}$\\
 $(0,0)$--$(2,1)$ & $7.747\times10^{-4}$ & $7.761\times10^{-4}$ & $8.262\times10^{-4}$ & $8.313\times10^{-4}$ & $3.660\times10^{-4}$ & $3.873\times10^{-4}$\\
 $(1,1)$--$(2,2)$ & $0$ & $9\times10^{-10}$ & $0$ & $6\times10^{-9}$ & $0$ & $7\times10^{-8}$\\
\end{tabular}
\end{ruledtabular}
\end{table}

\subsection{Predictions for a \texorpdfstring{$10\times10$}{10 x 10} array}
\label{sm:exp-array}
Table~\ref{tab:exp-array} lists, for an open $10\times10$ array of $H_{\rm sym}$ at $\eps=0.4$ and
$h=2$, the bulk witness and negativity at each separation, the witness averaged over the pairs used in
Fig.~4(a) of the Letter, and the number of snapshots for a $3\sigma$ detection of $\overline W>0$
(Sec.~\ref{sm:exp-errors}). The negativities of the most central pairs agree with those of a
$16\times16$ array and of a $24\times24$ torus to three digits, so the array is large enough for the
bulk. Pairs at the edges are not: they have fewer bonds and less depletion, and their witness is
larger. Averaged over all pairs of the array, $\overline W$ exceeds the bulk value by $13\%$, $26\%$,
$50\%$ and a factor $4.9$ at $r=1$ to $4$ on the axes. Excluding the outermost ring of sites, as in
Fig.~4(a), brings the averages within $1\%$, $4\%$ and $11\%$ of the bulk at $r=1$ to $3$; we combine
each separation with its quarter-turn partner, whose witness has the opposite $\eta$
(\texttt{fig4\_map.py}, \texttt{edge\_cutoff.py}). The pair at $r=4$ is entangled in the bulk, beyond
the leading-order boundary $\rho_E/\eps=3.53$, in agreement with the DMRG range $r_E=4$ on cylinders
at this angle [Fig.~3 of the Letter]; its averaged witness is still dominated by the pairs next to
the edge. For other parameters the central axial negativities at $r=1,2,3$ are $3.75\times10^{-3}$,
$1.07\times10^{-3}$, $3.36\times10^{-4}$ ($\eps=0.3$), $8.73\times10^{-3}$, $1.87\times10^{-3}$,
$2.67\times10^{-4}$ ($\eps=0.5$), and $1.13\times10^{-2}$, $2.02\times10^{-3}$, $4.6\times10^{-5}$
($\eps=0.6$), all at $h=2$. With XY exchange ($\Delta=0$) at $\eps=0.4$ they are
$4.78\times10^{-3}$, $1.22\times10^{-3}$, $3.12\times10^{-4}$, and with $\Delta=-1$ they are
$3.91\times10^{-3}$, $1.01\times10^{-3}$, $2.70\times10^{-4}$; the weaker interaction lowers the
amplitude and moves the boundary outward, raising the negativity at $r=4$ from $1.9\times10^{-5}$ to
$4.4\times10^{-5}$ and $5.1\times10^{-5}$ (\texttt{array\_pairs.py}).

\begin{table}[t]
\caption{Open $10\times10$ array of $H_{\rm sym}$, $\eps=0.4$, $h=2$, pair theory. For each
separation: bulk witness and negativity ($24\times24$ torus); number of pairs of the array with this
separation or its quarter turn whose sites avoid the outermost ring, the witness $\overline W$
averaged over them, and the snapshots for a $3\sigma$ detection of $\overline W>0$ (a dash means
$\overline W<0$), with the optimal fraction $f_z\simeq\sqrt{2p}$ of $z$ snapshots; a tenth of the
snapshots in $z$, as quoted in the Letter, raises these numbers by $8\%$.}
\label{tab:exp-array}
\begin{ruledtabular}
\begin{tabular}{cccccc}
 Separation & Bulk $W$ & Bulk $\Neg$ & Pairs & $\overline W$ & Snapshots\\
\hline
 $(1,0)$ & $2.44\times10^{-2}$ & $6.15\times10^{-3}$ & 112 & $2.46\times10^{-2}$ & $5.5\times10^{2}$\\
 $(2,0)$ & $6.14\times10^{-3}$ & $1.54\times10^{-3}$ & 96 & $6.37\times10^{-3}$ & $9.6\times10^{3}$\\
 $(3,0)$ & $1.46\times10^{-3}$ & $3.65\times10^{-4}$ & 80 & $1.62\times10^{-3}$ & $1.8\times10^{5}$\\
 $(4,0)$ & $7.3\times10^{-5}$ & $1.8\times10^{-5}$ & 64 & $1.7\times10^{-4}$ & $2\times10^{7}$\\
 $(1,1)$ & $-6.9\times10^{-4}$ & $0$ & 98 & $-7.1\times10^{-4}$ & --\\
 $(2,1)$ & $1.06\times10^{-3}$ & $2.64\times10^{-4}$ & 84 & $1.12\times10^{-3}$ & $3.6\times10^{5}$\\
 $(3,1)$ & $3.5\times10^{-4}$ & $8.7\times10^{-5}$ & 70 & $4.1\times10^{-4}$ & $3.2\times10^{6}$\\
 $(2,2)$ & $-6.9\times10^{-4}$ & $0$ & 72 & $-7.1\times10^{-4}$ & --\\
 $(3,2)$ & $-4.1\times10^{-4}$ & $0$ & 60 & $-4.1\times10^{-4}$ & --\\
\end{tabular}
\end{ruledtabular}
\end{table}

\subsection{Adiabatic preparation}
\label{sm:exp-ramp}
\begin{figure}[t]
 \centering
 \includegraphics[width=0.5\textwidth]{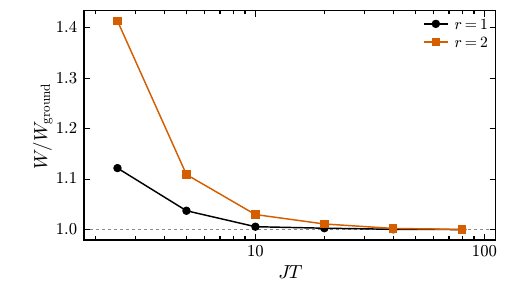}
 \caption{Witness of the central axial pairs of the open $10\times10$ array ($\eps=0.4$, $h=2$, pair
 theory) after the anisotropy is switched on in a time $T$ from the polarized product state, relative
 to its ground-state value.}
 \label{fig:sm-ramp}
\end{figure}

Without the anisotropy, $H_{\rm sym}$ reduces to the Heisenberg ferromagnet in a field, whose ground
state is the polarized product state. Switching the anisotropy on at fixed field,
$H(s)=H_{s=0}+s\,(H_{s=1}-H_{s=0})$ with $s(u)=3u^2-2u^3$ and $u=t/T$, lowers the gap monotonically
from $Jh\eps^2$ to $J(h\eps^2-d)$, so it stays open. In the linked form of the two-magnon theory the
pair field obeys $i\partial_t\phi=(H_{22}-E_\Omega)\phi+H_{20}$ with $\phi=0$ at $t=0$; this is the
time-dependent version of Eq.~\eqref{eq:exp-pair}. On the $10\times10$ array at $\eps=0.4$ and $h=2$
the witness of the central axial pairs at the end of the switch-on exceeds its ground-state value by
$12\%$, $3.7\%$, $0.6\%$, $0.3\%$ and $0.06\%$ at $r=1$, and by $41\%$, $11\%$, $3\%$, $1.1\%$ and
$0.2\%$ at $r=2$, for $JT=2.5$, $5$, $10$, $20$ and $40$ (\texttt{ramp2.py}). A faster switch-on
leaves pairs that have not yet spread, which raises the coherence at short distance; the overshoot is
physical, not an artifact of the linked theory. Exact time evolution in the full Hilbert space of an
open $4\times4$ array reproduces the linked dynamics to $0.1\%$ for $JT=5$ to $40$ and both
separations (\texttt{ramp\_exact\_check.py}). Lowering the field with the anisotropy already on is not
equivalent: the initial product state is then not an eigenstate, and in a closed system the resulting
oscillations of the pair field do not decay. Figure~\ref{fig:sm-ramp} shows the witness after the
switch-on.

\subsection{Qubit form and strong pair drives}
\label{sm:exp-strong}
With $\sigma_i^\pm=(\sigma_i^x\pm i\sigma_i^y)/2$, a bond of $H_{\rm sym}$ along $\be_\mu$ reads, up to a
constant,
\begin{equation}
 -J_\perp(\sigma_i^+\sigma_j^-+\mathrm{h.c.})-J_z\sigma_i^z\sigma_j^z
 -\eta_\mu J_p(\sigma_i^+\sigma_j^++\mathrm{h.c.}),\qquad \eta_x=1,\ \eta_y=-1,
 \label{eq:exp-qubit}
\end{equation}
with hopping $J_\perp=J(1+\cos\eps)/4$, Ising coupling $J_z=J\cos\eps/4$ and pair creation
$J_p=J(1-\cos\eps)/4$, and the field term is $-(B/2)\sum_i\sigma_i^z$ with $B=hJ\eps^2$. For XXZ
exchange $J_z\to\Delta J\cos\eps/4$ (Sec.~\ref{sm:exp-xxz}). In a simulator $J_\perp$, $J_z$, $J_p$
and $B$ are set independently; Eq.~(1) fixes $J_p/J_\perp=\tan^2(\eps/2)$, which is $0.041$ at
$\eps=0.4$. At $J_p=J_\perp$ the Ising coupling vanishes and the bonds become $-(J/4)\sigma_i^x
\sigma_j^x$ on $x$ bonds and $-(J/4)\sigma_i^y\sigma_j^y$ on $y$ bonds: the quantum compass model in a
transverse field.

Table~\ref{tab:exp-strong} follows the margins from weak to strong pair drive at the detuning
$B=6J_p$, for which the classical gap $B-4J_p$ is $2J_p$ (\texttt{strong\_drive.py}). The pair theory
of an open $10\times10$ array and exact diagonalization of an open $4\times4$ array agree within $8\%$
for the nearest-neighbor pair at all drives; for the collective witness of Sec.~\ref{sm:exp-collective}
the $4\times4$ array, dominated by its edges, gives up to $23\%$ less. The sign change of the coherence
between $x$ and $y$ bonds and its absence on the diagonals hold at every drive by the joint $C_4$
symmetry, and the diagonal pairs are separable in all cases computed. A single plaquette suffices to
see them: on an open $2\times2$ array at $J_p=J_\perp$ and $B=1.5J$, exact diagonalization gives
$4\,\mathrm{Re}\,w=\pm0.32$ on the $x$ and $y$ bonds, negativity $0.075$ and $W/4=0.068$ there, and a
separable diagonal pair; at $J_p/J_\perp=0.30$ these are $\pm0.24$, $0.056$ and $0.052$. The range
shrinks as the drive grows: the margin at $r=2$ is largest, $0.45\%$, near $J_p/J_\perp=0.18$ and
vanishes near $0.45$. Strong drives therefore test the pair source and its sign pattern, weak drives
the range and the clover. Leading-order formulas do not apply at strong drive; the numbers rest on
exact diagonalization and the pair theory.

\begin{table}[t]
\caption{Tolerated rate of incoherent spin flips per spin, in percent, versus pair drive at
detuning $B=6J_p$: $W/4$ of the central pair at $r=1$ and $r=2$ and the margin of the collective
witness. Pair theory on an open $10\times10$ array (PT) and exact diagonalization of an open
$4\times4$ array (ED); $\Neg_1$ is the central nearest-neighbor negativity. A negative entry means
that the witness is negative.}
\label{tab:exp-strong}
\begin{ruledtabular}
\begin{tabular}{cccccccc}
 $\eps$ & $J_p/J_\perp$ & $\Neg_1$ (PT) & $r=1$ (PT) & $r=2$ (PT) & Collective (PT) & $r=1$ (ED) & Collective (ED)\\
\hline
 $0.4$ & $0.041$ & $0.0072$ & $0.71$ & $0.21$ & $1.35$ & $0.77$ & $1.37$\\
 $0.6$ & $0.096$ & $0.0151$ & $1.49$ & $0.38$ & $2.73$ & $1.57$ & $2.67$\\
 $0.8$ & $0.179$ & $0.0249$ & $2.41$ & $0.45$ & $4.32$ & $2.47$ & $4.01$\\
 $1.0$ & $0.298$ & $0.0354$ & $3.37$ & $0.35$ & $5.97$ & $3.37$ & $5.25$\\
 $1.2$ & $0.468$ & $0.0458$ & $4.27$ & $-0.02$ & $7.55$ & $4.18$ & $6.32$\\
 $\pi/2$ & $1$ & $0.0627$ & $5.53$ & $-1.49$ & $10.1$ & $5.32$ & $7.79$\\
\end{tabular}
\end{ruledtabular}
\end{table}

\subsection{A collective witness}
\label{sm:exp-collective}
Let $M$ be a real symmetric $N\times N$ matrix with zero diagonal and operator norm $\|M\|$. Every
separable state of $N$ qubits obeys
\begin{equation}
 Q_M\equiv\sum_{i\ne j}M_{ij}\langle\sigma_i^x\sigma_j^x-\sigma_i^y\sigma_j^y\rangle
 \le\|M\|\sum_i\bigl(1-\langle\sigma_i^z\rangle^2\bigr).
 \label{eq:exp-collective}
\end{equation}
For a product state with Bloch vectors $(x_i,y_i,z_i)$, $Q_M=x^TMx-y^TMy\le\|M\|(|x|^2+|y|^2)\le
\|M\|\sum_i(1-z_i^2)$. The left side is linear in the state and the right side is concave, since
$z_i^2$ is convex, so the bound holds for mixtures. For two sites it is a slightly sharper form of
Eq.~(9) of the Letter, and it belongs to the family of collective two-body
witnesses~\cite{Toth2007,Guhne2009}. Incoherent flips at rate $e$ leave $Q_M$ unchanged at first order
and raise the right side by $4e\|M\|$ per spin, as for the pair witness.

In the ground state $\langle\sigma_i^x\sigma_j^x-\sigma_i^y\sigma_j^y\rangle=4\,\mathrm{Re}\,w_{ij}$,
and the best kernel is the sign of $\mathrm{Re}\,w$. We take the lattice kernel with Fourier
transform $M_{\bk}=\sgn(\cos k_x-\cos k_y)$, so that $\|M\|=1$, $M(\pm\be_x)=-M(\pm\be_y)=4/\pi^2$, and
$M$ vanishes on the diagonals. Per spin, $Q_M/(4N)$ is then the collective coherence
$\bar w=\int\!\frac{d^2k}{(2\pi)^2}|\hat w(\bk)|$, and the witness detects entanglement when
$\bar w>p(1-p)$. At leading order $\hat w(\bk)=\eps^2\hat F(\bk)/2$ with
$\hat F(\bk)=(\cos k_x-\cos k_y)/[2\lambda(\bk)(1-k_\square)]$ for $|\bk|\gg m\eps$. The substitution
$s=(k_x+k_y)/2$, $t=(k_x-k_y)/2$ turns $|\cos k_x-\cos k_y|/[2\lambda(\bk)]$ into
$|\sin s\sin t|/[4(1-\cos s\cos t)]$, whose Brillouin-zone average is $1/8$, so
\begin{equation}
 \frac{\bar w}{\eps^2}\longrightarrow\frac{\pi}{32(\pi-2)}=8n_*=0.0860,\qquad
 \frac{\bar w}{|w(\be_x)|}\longrightarrow\frac{\pi}{2(4-\pi)}=1.83 .
 \label{eq:exp-collective-constants}
\end{equation}
Since $p=O(\eps^4)$, the collective witness tolerates flip rates $1.83$ times larger than the
nearest-neighbor witness. On the open $10\times10$ array at $\eps=0.4$ and $h=2$, with the kernel
restricted to the array, it tolerates $1.07\%$, against $0.61\%$ for the central pair at $r=1$; the
variance of its estimate is $8\sum_{ij}M_{ij}^2/[N_s(1-f_z)]+16\|M\|^2\sum_ip_i/(N_sf_z)$ for $N_s$
snapshots, so a $3\sigma$ detection needs about $330$ snapshots with a tenth of them in $z$ ($310$ at
the optimal fraction). Table~\ref{tab:exp-strong} gives it at strong drive.

In momentum space the same numbers describe squeezing. On the $k_x$ axis and for $k\gg m\eps$,
$\hat F=-\pi/[8(\pi-2)]=-0.344$, so the transverse structure factors are
$4S^{xx}(\bk)=1-0.344\eps^2$ and $4S^{yy}(\bk)=1+0.344\eps^2$ up to $O(\eps^4)$; on the $k_y$ axis the
two are exchanged, and on the diagonals they are equal. Near $\bk=0$ the squeezing varies as
$\cos2\phi_{\bk}$ with the direction of $\bk$, the image of the inverse-square law, and it is cut off
at $k\sim m\eps$. With $\langle S^z\rangle=1/2-p$ this is spin squeezing at wave vector $\bk$, and the
quantum Fisher information density of the antisqueezed quadrature exceeds the separable bound by
$0.344\eps^2$, the quantity that neutron spectra give access to~\cite{Scheie2021,Simeth2026}. For
$\eps=0.1$ to $0.2$ it is a $0.3\%$ to $1.4\%$ effect
(\texttt{collective\_witness.py}).

\subsection{Measurement errors and statistics}
\label{sm:exp-errors}
For the witness $W$ of Eq.~(9) of the Letter, a bit flip ($X$ error) with probability $e$ on each
qubit before the measurement leaves $\langle\sigma^x\sigma^x\rangle$ unchanged and multiplies
$\langle\sigma^y\sigma^y\rangle$ by $(1-2e)^2$ and $\langle\sigma^z\rangle$ by $1-2e$, so that
\begin{equation}
 W_e=W-4e\,(1-p_a-p_b)+4\eta\,e(1-e)\,\langle\sigma_a^y\sigma_b^y\rangle
 =W-4e+O\bigl(e\,(p+|w|+|v|)\bigr).
 \label{eq:exp-bitflip}
\end{equation}
A readout error with probability $e$ in the $z$ basis alone gives $W_e=W-4e(1-p_a-p_b)$ exactly.
Incoherent flips therefore add to the population one for one, and $W>0$ requires their rate to stay
below $\eta\,\mathrm{Re}\,w-p$, which in the ground state is the negativity itself at leading order.
All these errors are local operations, so they cannot make a separable state look entangled.
Dephasing ($Z$ errors) and readout errors in the $x$ and $y$ bases multiply the coherence term by
$(1-2e)^2$ but leave the population term, so $W$ stays positive while $(1-2e)^2|w|>p$. This
tolerates errors of tens of percent at $r=1$ and $2$ ($e<0.40$ and $0.34$ at $\eps=0.4$) and becomes
restrictive only near the boundary, where $|w|-p\ll|w|$. Errors that add flips are the critical
ones. At a flip rate $e$ the margin $\overline W$ falls by $4e$, and the number of snapshots grows by
$[\overline W/(\overline W-4e)]^2$: for example from $5.5\times10^2$ to about $2.1\times10^3$ at
$r=1$ for $e=0.3\%$, and from $9.6\times10^3$ to about $7\times10^4$ at $r=2$ for $e=0.1\%$. Present
arrays have preparation and readout errors of a few percent~\cite{Karamlou2024,Andersen2025};
readout errors are routinely removed by inverting a calibrated error matrix, which makes the
certificate depend on that calibration, whereas flips from the preparation must be removed by
active reset.

The witness measures $\mathrm{Re}\,w$ in the frame in which the pair source is real. With pair drives
of phase $\varphi$ the readout axes must be rotated by $\varphi/2$ about $z$, and the relative phases
of the drives must be stable from shot to shot; otherwise $\mathrm{Re}\,w$ averages out. If the phase
is not known, two further settings rotated by $\pi/4$ about $z$ measure $\mathrm{Im}\,w$. The
coherence term $\langle\sigma^x\sigma^x-\sigma^y\sigma^y\rangle=4\,\mathrm{Re}\,w$ alone changes sign
between bonds along $x$ and $y$ and vanishes on the diagonals. Incoherent flips leave
$\langle\sigma^x\sigma^x\rangle$ unchanged and reduce $\langle\sigma^y\sigma^y\rangle$ only by
$(1-2e)^2$, so this sign pattern is a robust signature of the $d$-wave source even where the witness
is not positive. In a superconducting $10\times10$ array the source requires phase-controlled drives
on all $180$ couplers; so far pairing and hopping drives have been run together on the couplers of a
six-qubit chain~\cite{Liang2024}, and leakage out of the qubit subspace under strong parametric
drives must be kept below the flip budget above.

For the statistics, let $N$ snapshots be taken, each measuring all spins in one setting, with fixed
fractions $f_x=f_y=(1-f_z)/2$ of $x$, $y$ and $z$ snapshots, and let the estimates of
$\langle\sigma^x\sigma^x\rangle$, $\langle\sigma^y\sigma^y\rangle$ and
$\langle\sigma^z_a\rangle+\langle\sigma^z_b\rangle$ be averaged over the $P$ pairs of a snapshot
with the same separation or its quarter turn (with the opposite $\eta$). Treating these pairs as
independent, the variance of the estimate of
$W$ is $[2/f_x+8p/f_z]/(NP)$, which is smallest for $f_z\simeq\sqrt{2p}$, where it is close to
$4.2/(NP)$; a $3\sigma$ detection then needs $N\simeq38/(PW^2)$. Pairs that share a spin are
correlated through $\langle\sigma^x_a\sigma^x_c\rangle=O(\eps^2)$, and disjoint pairs only through
connected four-point functions of order $p$. Exact diagonalization of the $4\times4$ torus at
$\eps=0.4$ gives an effective number of independent nearest-neighbor pairs of $32.0$ out of $32$ for
the $x$ and $y$ settings, so the independence assumption holds for the coherence terms
(\texttt{independent/v45\_measurement.py}). If the setting is instead drawn at random for each
snapshot and the single-shot estimator of Sec.~\ref{sm:rdm} is used, the setting is common to all
pairs and the variance of that estimator cannot fall below about $4/N$; with per-setting sample
means there is no such floor. The analysis assumes the snapshots per setting are fixed or the
per-setting means are used.

\section{The model without the Dzyaloshinskii-Moriya term}
\label{sm:nodm}

Let $H_{\rm sym}$ be Eq.~\eqref{eq:model} with the antisymmetric part of each bond matrix removed,
\begin{equation}
 \frac{H_{\rm sym}}J=\sum_{i,\mu=x,y}\Bigl[\frac14-\cos\eps\,
 \mathbf S_i\cdot\mathbf S_{i+\be_\mu}
 -(1-\cos\eps)S^\mu_iS^\mu_{i+\be_\mu}\Bigr]-h\eps^2\sum_iS^z_i ,
 \label{eq:nodm-model}
\end{equation}
that is, the multiplier of $\sin\eps\,\VDM$ is set to zero while $d=1-\cos\eps$ is kept. In this
section $\cI=[h_-,h_+]\Subset(1/2,\infty)$ and $m_\cI=h_--\tfrac12>0$. States are the symmetric
thermodynamic ground states of Sec.~\ref{sm:conventions}, defined with $H_{\rm sym}$; limits are
taken in the same order. All other notation is that of
Secs.~\ref{sm:conventions}--\ref{sm:massscale}. The chain of proofs of
Secs.~\ref{sm:normalform}--\ref{sm:massscale} carries over once the parts that exist only to treat
the DM term are removed; this section records what changes. An independent check of the chain gave
the same conclusions (Sec.~\ref{sm:numerics}).

\subsection{Decomposition, parity and coercivity}
Equation~\eqref{eq:hardcore-decomposition} holds without its DM term,
\begin{equation}
 H_{\rm sym}/J-Ne_{\rm pol}=Q+d[B+W(f)]+h\eps^2\Nop
 =a_\eps Q+\Delta_\eps\Nop+\frac d2\Ncontact+dW(f).
 \label{eq:nodm-decomposition}
\end{equation}
Every term changes $\Nop$ by $0$ or $\pm2$. Hence $H_{\rm sym}$ commutes, term by term, with the
on-site automorphism $\alpha_\pi$ that rotates every spin by $\pi$ about the field, $b_i\mapsto-b_i$
(on a torus, conjugation by $\Pi=\prod_i\sigma^z_i$). The number-conserving part obeys
\begin{equation}
 H_0=a_\eps Q+\Delta_\eps\Nop+\frac d2\Ncontact
 \succeq a_\eps Q+m_\cI\eps^2\Nop,\qquad
 H_0\preceq Q+h_+\eps^2\Nop ,
 \label{eq:nodm-coercivity}
\end{equation}
because $0\preceq\Ncontact\preceq2\Nop$ and $\Delta_\eps=h\eps^2-d\ge(h-\tfrac12)\eps^2$. This
replaces Eq.~\eqref{eq:coercivity}: in its notation $g_*=1$, so that the exchange coefficient is
$a_\eps=1-d/2\ge3/4$, and $m_*=m_\cI$; no bound on $\VDM$ is needed. Apart from this, the bound
$h>3/2$ enters Secs.~\ref{sm:normalform}--\ref{sm:threshold} only through the inequality
$6\Delta_\eps/a_\eps\ge6\eps^2$ in the triplet channel of Sec.~\ref{sm:current}, which is not used
here, and $\Delta_\eps\ge\eps^2$ in the proof of Lemma~\ref{lem:ms-probe}, which becomes
$\Delta_\eps\ge m_\cI\eps^2$. Each bond of Eq.~\eqref{eq:nodm-model} is positive semidefinite, with
spectrum $0,0,d/2,a_\eps$.

\begin{lemma}[Parity-invariant ground states]\label{lem:nodm-parity}
If $\omega$ is a symmetric thermodynamic ground state of $H_{\rm sym}$, then so is
$\omega_\pi=\tfrac12(\omega+\omega\circ\alpha_\pi)$. For an $\alpha_\pi$-invariant state the entries
$\zeta$ of Eq.~\eqref{eq:rho} vanish at every separation, and $\rho_{0\br}$ commutes with
$\sigma^z_0\sigma^z_\br$.
\end{lemma}
\begin{proof}
$\alpha_\pi$ fixes every local term of $H_{\rm sym}$ and commutes with translations and with the joint
quarter turn. The energy density is affine in the state, so $\omega_\pi$ has the energy density of
$\omega$ and the same invariances. For an invariant state,
$\omega(b_0n_\br)=\omega(\alpha_\pi(b_0n_\br))=-\omega(b_0n_\br)$, and likewise for the other three odd
entries.
\end{proof}
An $\alpha_\pi$-invariant minimizer is a minimizer in the class of Sec.~\ref{sm:conventions}. Every
statement below about all symmetric thermodynamic ground states therefore applies to it.

\subsection{Energy, population and dressed densities}
Let $R=\lceil\eps^{-2}\rceil$, $X=X_{F_R}$, $U_R=e^{dX}$ and $\widetilde\omega(A)=\omega(U_RAU_R^\dagger)$,
as in Eq.~\eqref{eq:pop-cutoff}.

\begin{lemma}[Energy sandwich without a triplet correction]
\label{lem:nodm-sandwich}
Uniformly for $h\in\cI$ and all symmetric thermodynamic ground states of $H_{\rm sym}$, for
sufficiently small $\eps$, with $\widetilde{\mathfrak q}$ the exchange density and $\widetilde p$ the
flip density of $\widetilde\omega$,
\begin{gather}
 \tfrac14\widetilde{\mathfrak q}+\tfrac14m_\cI\eps^2\widetilde p
 \le C\eps^8\ell_\eps^3,\qquad
 \bigl|e_0-e_{\rm pol}-V_R\bigr|\le C\eps^8\ell_\eps^3,
 \label{eq:nodm-sandwich}\\
 e_0-e_{\rm pol}=-c_*\eps^4+\eps^6g_6^{\rm sym}(h)+O_\cI(\eps^8\ell_\eps^3),
 \qquad g_6^{\rm sym}(h)=(h-\tfrac12)n_*-\frac{F_1}{96}+\frac{F_1^2}8,
 \label{eq:nodm-energy}\\
 \left|\frac p{\eps^4}-n_*\right|\le C\eps\ell_\eps^{3/2}.
 \label{eq:nodm-population}
\end{gather}
Here $V_R$ is the scalar~\eqref{eq:pop-VR}.
\end{lemma}
\begin{proof}
The first conjugation~\eqref{eq:pop-first-full-expansion} is linear in the Hamiltonian. For
$H_{\rm sym}$ it is the same expansion without $sdD_R$ and $\frac{sd^2}2\operatorname{ad}_X^2\VDM$,
with $H_0$ of Eq.~\eqref{eq:nodm-coercivity}, and the remainder~\eqref{eq:pop-first-remainder} loses
its DM term: $\|\mathcal E_R\|/N\le C\eps^8\ell_\eps^3$. All retained operators have even charge, and
the odd rows of Table~\ref{tab:pop-first-orders} vanish; the vacuum scalar is $V_R$. The triplet
dressing~\eqref{eq:pop-triplet-dressing} and the matching of its coefficient are omitted: they exist
only to remove the cubic DM source, and with that source absent there is nothing to match. Hence the
contribution $-\Gamma_3/4$ to $g_6$ disappears. The nonscalar terms are bounded as in
Sec.~\ref{sm:population}, for an arbitrary translation-invariant state with densities
$p,\mathfrak q$. The quadratic terms $dA_R$ and $dW(q_R)$ are absorbed by
Eq.~\eqref{eq:pop-quadratic-absorption} with $\beta=1/8$ each, at a cost $\mathfrak q/4$, a number
coefficient $O(\eps^4)$ and a scalar $O(\eps^8)$. Lemma~\ref{lem:current-quartic-absorption}, applied
separately to the pair source~\eqref{eq:pop-combined-fourth-pair} and to the quartet
source~\eqref{eq:pop-quartic-four} with $\kappa=1/8$ and $\nu=m_\cI/8$, costs another
$\mathfrak q/4$, a number coefficient $m_\cI\eps^2/4$ and a scalar $C\eps^8\ell_\eps^3$. The
extraction remainder of $Z_R$ has number coefficient $128d^2z_R=O(\eps^4\ell_\eps)$. The four
sixth-order terms, with their scalars removed, are bounded by $\mathfrak h_R\sqrt p$ with
$\mathfrak h_R=O(\eps^6\ell_\eps^2)$ [Eq.~\eqref{eq:pop-first-higher-bound} without its DM term],
which Young's inequality turns into $m_\cI\eps^2p/8+O(\eps^{10}\ell_\eps^4)$. The four allocations of
$\mathfrak q$ total $1/2$, and $a_\eps\ge3/4$; this proves, for small $\eps$,
\[
 \frac{\omega(U_R^\dagger(H_{\rm sym}/J-Ne_{\rm pol})U_R)}N
 \ge V_R-C\eps^8\ell_\eps^3+\tfrac14\mathfrak q+\tfrac14m_\cI\eps^2p .
\]
The vacuum trial $U_R\Omega$ gives $e_0-e_{\rm pol}\le V_R+\|\mathcal E_R\|/N$. Applying the lower
bound to $\widetilde\omega$ proves Eq.~\eqref{eq:nodm-sandwich}; the passage to thermodynamic
densities is that of Sec.~\ref{sm:population}. Equation~\eqref{eq:pop-pair-sixth-scalar} gives
Eq.~\eqref{eq:nodm-energy}; for its remainder, $S_R-S_F=O(R^{-2})$ by the squared tail of $F$ and
the cutoff error of Eq.~\eqref{eq:green-cutoff-scalar} is $O(R^{-4})$, both $O(\eps^8)$ at
$R=\lceil\eps^{-2}\rceil$. Finally Eq.~\eqref{eq:pop-final-local-bound} with
$\sqrt{\widetilde p}\le C\eps^3\ell_\eps^{3/2}$ and $a_R\le C\ell_\eps$ gives
$|\sqrt p-d\sqrt{S_R}|\le C\eps^3\ell_\eps^{3/2}$, which is Eq.~\eqref{eq:nodm-population}.
\end{proof}
The sixth-order coefficient differs from Eq.~\eqref{eq:pop-g6} by the term $\Gamma_3/4$, which the DM
term produces. The dressed densities obey stronger upper bounds than in
Eq.~\eqref{eq:ms-densities-log}, because no odd source remains.

\subsection{Coherences at all separations}
\begin{lemma}[Pair coherence]\label{lem:nodm-w}
Uniformly for $h\in\cI$ and the symmetric thermodynamic ground states of $H_{\rm sym}$,
\begin{align}
 |w(\br)-w_2(\br)|&\le C\bigl[\eps^5(1+|\br|)\ell_\eps^4
 +\eps^6(1+|\br|)^2\ell_\eps^6\bigr],&&1\le|\br|\le\eps^{-1},
 \label{eq:nodm-short}\\
 |w(\br)-w_2(\br)|&\le C_c\,\eps^4(\eps\ell_\eps^{3/2})^{1/2},
 &&|\br|\ge c/\eps .
 \label{eq:nodm-far}
\end{align}
\end{lemma}
\begin{proof}
The identity~\eqref{eq:range-stationarity} is linear in the Hamiltonian; for $H_{\rm sym}$ it holds
without the term $-s\omega(R_V(u))/N$, its contact tubes and its commutators. Hence
Eqs.~\eqref{eq:range-master-bound} and~\eqref{eq:ms-master} hold without their DM terms. The bound
$\mathcal E_Q$ of Eq.~\eqref{eq:range-general-residuals} and the even residuals
\eqref{eq:range-even-residuals} do not involve the DM term, and Lemmas~\ref{lem:range-test-bounds}
and~\ref{lem:ms-probe} concern $A_2$ alone and use only $h_->1/2$. Equation~\eqref{eq:nodm-short} is
Eq.~\eqref{eq:range-pointwise} without the contribution of the DM term, which is the first term of
Eq.~(A3) of the Letter.

For $|\br|\ge c/\eps$ evaluate $\mathcal E_Q(u_\br)$ in $\widetilde\omega$, with the dressing radius
$R$. By Eqs.~\eqref{eq:nodm-sandwich} and~\eqref{eq:ms-probe-norms}, $v_u\le C\eps^{-1}$,
$a_R\le C\ell_\eps$, $v_R\le C$ and $\mathsf T_4\le C\eps^4\ell_\eps^{3/2}$, and the five exchange terms
of $\mathcal E_Q$ are at most $C\eps^6\ell_\eps^3$, $C\eps^5\ell_\eps^{3/2}$, $C\eps^7\ell_\eps^4$,
$C\eps^6\ell_\eps^{7/2}$ and $C\eps^5\ell_\eps^3$. The near-source probe estimate is still needed for the
even residuals. Lemma~\ref{lem:ms-transfer} with $\Lambda^{-1}=C\eps\ell_\eps^{3/2}$ and
Lemma~\ref{lem:ms-tail}(b) give $\Xi_D\le C[D^{-2}+\eps\ell_\eps^{3/2}]$ in Eq.~\eqref{eq:ms-Xi}.
Lemma~\ref{lem:ms-even} with $D=\lceil\eps^{-1/2}\rceil$ then gives
\[
 \frac{|w(\br)-w_2(\br)|}{\eps^4}\le C_c\bigl[\Xi_D+\sqrt{\Xi_D}
 +\eps^4(1+D)^4+\eps\ell_\eps^3\bigr]
 \le C_c(\eps\ell_\eps^{3/2})^{1/2}.
\]
The removal of the probe cutoff and the order of limits are those of Sec.~\ref{ss:ms-proof}.
\end{proof}

\subsection{Results}
\begin{theorem}[Model without the DM term]\label{thm:nodm}
Let $\cI\Subset(1/2,\infty)$. Uniformly for $h\in\cI$ and all symmetric thermodynamic ground states of
$H_{\rm sym}$:
\begin{enumerate}
\item[(i)] Theorems~\ref{thm:range} and \ref{thm:massive} hold, and Theorem~\ref{thm:energy} holds
with $g_6$ replaced by $g_6^{\rm sym}$ of Eq.~\eqref{eq:nodm-energy}; Theorem~\ref{thm:population}
holds with the rate~\eqref{eq:nodm-population}.
\item[(ii)] Theorem~\ref{thm:massscale}(i)--(iii) and Corollary~\ref{cor:ms-range} hold, with
$(\eps\ell_\eps)^{2/13}$ replaced by $(\eps\ell_\eps^{3/2})^{1/2}$ in all three estimates of
Theorem~\ref{thm:massscale}(iv). Moreover $q_\br\le C_c\eps^6\ell_\eps^3$,
$|v_\br|\le C_c\eps^5\ell_\eps^{3/2}$ and $|\zeta_\br|\le\sqrt{pq_\br}\le C_c\eps^5\ell_\eps^{3/2}$
for $|\br|\ge c/\eps$, and on every set where $p-q_\br-|w(\br)|\ge\delta\eps^4/2$,
$\Neg(\rho_{0\br})\le C_{c,\delta}\eps^6\ell_\eps^3$ by Proposition~\ref{prop:th-exterior}(i).
\item[(iii)] Let the state be $\alpha_\pi$-invariant (Lemma~\ref{lem:nodm-parity}), and let
$K\subset\mathbb R^2\setminus\{0\}$ be compact with $A_-:=\inf_{h\in\cI,\bx\in K}|A_h(\bx)|>0$ and
$\delta:=\inf_{h\in\cI,\bx\in K}\bigl||A_h(\bx)|-n_*\bigr|>0$. There are $c_K,\eps_K>0$ such that for
$0<\eps<\eps_K$ and $\eps\br\in K$: if $|A_h(\eps\br)|>n_*$, then $\rho_{0\br}^{T_2}$ has exactly one
negative eigenvalue, $p-q_\br-|w(\br)|\le-\delta\eps^4/2$, and $\Neg(\rho_{0\br})=|w(\br)|-p+q_\br$;
if $|A_h(\eps\br)|<n_*$, then $\lambda_{\min}(\rho_{0\br}^{T_2})\ge c_K\eps^8$ and $\rho_{0\br}$ is
separable. Thus $\rho_{0\br}$ is entangled if and only if $|A_h(\eps\br)|>n_*$.
\end{enumerate}
\end{theorem}
\begin{proof}
(i) Sections~\ref{sm:normalform} and~\ref{sm:range} with Eq.~\eqref{eq:nodm-coercivity} in place of
Eq.~\eqref{eq:coercivity} and without the DM terms give Eq.~\eqref{eq:apriori} and then
Theorem~\ref{thm:range}, by Eq.~\eqref{eq:nodm-short} and Lemma~\ref{lem:rdm-seven}.
Theorem~\ref{thm:massive} concerns $A_2$ alone, which does not involve the DM term. The energy and
the population are Lemma~\ref{lem:nodm-sandwich}.

(ii) The first limit of Theorem~\ref{thm:massscale}(i) is Eq.~\eqref{eq:nodm-far}; the second follows
with Lemma~\ref{lem:ms-projected-rate}, which concerns $w_2$ alone. Statement (ii) of that theorem is
Lemma~\ref{lem:ms-transfer}, valid for every translation-invariant state, with
$\sqrt{\widetilde p}\le C\eps^3\ell_\eps^{3/2}$; with Lemma~\ref{lem:ms-tail}(b) it gives the bounds on
$q$ and $v$. The bound on $\zeta$ is the Cauchy-Schwarz inequality for the state, and the bound on the
negativity is Proposition~\ref{prop:th-exterior}(i) with these inputs. Statement (iii) of
Theorem~\ref{thm:massscale} and the corollary follow as in Sec.~\ref{ss:ms-proof}, from
Lemma~\ref{lem:ms-pt}.

(iii) By Lemma~\ref{lem:nodm-parity}, $\zeta=0$, and in the basis of Eq.~\eqref{eq:PT}
\[
 \rho_{0\br}^{T_2}=\begin{pmatrix}a&v\\v&q\end{pmatrix}\oplus
 \begin{pmatrix}b&w^*\\w&b\end{pmatrix},
\]
with eigenvalues~\eqref{eq:Xeigen}. Positivity of the block of $\rho_{0\br}$ on
$\uparrow\uparrow,\downarrow\downarrow$ gives $aq\ge|w|^2$. By (i) and (ii), uniformly on $K$,
$b=\eps^4[n_*+o(1)]$, $|w|=\eps^4[|A_h(\eps\br)|+o(1)]$ and $v=o(\eps^4)$. Hence
$aq-v^2\ge|w|^2-v^2\ge A_-^2\eps^8/2$ for small $\eps$, and since $a+q\le1$ the first block has
smallest eigenvalue at least $A_-^2\eps^8/2$, inside and outside the lobes. The second block has
eigenvalues $b\pm|w|$, and $b-|w|=\eps^4[n_*-|A_h(\eps\br)|+o(1)]$ has the sign of
$n_*-|A_h(\eps\br)|$ and modulus at least $\delta\eps^4/2$. For two qubits a positive partial
transpose is equivalent to separability~\cite{Horodecki}.
\end{proof}

\paragraph{Scope.}
Statement (iii) does not cover the diagonals, where $A_h=0$, separations with $\eps|\br|\to\infty$, or
the contour $|A_h|=n_*$ itself. On the diagonals and in the far tail, positivity still gives
$q_\br\ge|w(\br)|^2/a$, but this can be too weak to imply $aq_\br\ge v_\br^2$, which is what
separability requires there. For states that are also invariant under the reflections about the
coordinate axes, which together with the joint quarter turn are unitary symmetries of $H_{\rm sym}$,
$w$ vanishes identically on the diagonals, and $\rho_{0\br}$ is separable there if and only if
$aq_\br\ge v_\br^2$. For the same reason only $\liminf\eps R_{\rm NPT}\ge\rho_E$ is known along a ray.
In Theorem~\ref{thm:threshold}, step (B) holds for $\alpha_\pi$-invariant states with
$\Neg\le2v^2/a\le C_c\eps^{10}\ell_\eps^3$ [Proposition~\ref{prop:th-exterior}(i)], so its conclusions
hold for every $\kappa_\eps\to0$ with $\kappa_\eps/(\eps^6\ell_\eps^3)\to\infty$.

\clearpage
\bibliographystyle{apsrev4-2}
\makeatletter
\immediate\write\@auxout{\string\citation{SMBibControl}}
\makeatother
\bibliography{sm_refs}